\documentclass[preprint,superscriptaddress]{revtex4-1}
\pdfoutput=1

\usepackage{graphicx}
\usepackage{bmpsize}
\usepackage{bm}        % for math
\usepackage{hyperref}
\usepackage{amsmath}
\usepackage{amssymb}   % for math
\usepackage{amsopn}
\usepackage{url}
\usepackage{longtable}
\usepackage{color}
\usepackage{listings}
\usepackage[caption=false]{subfig}
\usepackage{filecontents}
\usepackage{lineno}
\IfFileExists{scalerel.sty}
             {
               \usepackage{scalerel}
               \newlength{\heightnu}
               \AtBeginDocument{\settoheight{\heightnu}{$\nu$}}
               \newcommand{\nupbar}{\,\llap{\raisebox{\heightnu+0.2pt}{\scaleobj{0.4}{\hstretch{1.3}(}}}  \overline{\nu} \rlap{\raisebox{\heightnu+0.2pt}{\scaleobj{0.4}{\hstretch{1.3})}}}}
             }
             {
               \newcommand{\nupbar}{\vphantom{\overline{\nu}}^{(}\overline{\nu}^{)}}
             }

\makeatletter
\def\l@paragraph{\@dottedtocline{4}{5.3em}{2.1em}}
\makeatother

\definecolor{bp_red}{RGB}{153, 0, 0}
\definecolor{bp_green}{RGB}{0, 153, 0}
\definecolor{bp_blue}{RGB}{0, 0, 153}
\newcommand{\tlog}{\textrm{log}}
\newcommand{\tMIN}{MIN}
\newcommand{\dcp}{\delta_{CP}}
\newcommand{\nova}{NO$\nu$A}

\begin{document}

%\newcommand{\qOneThree}{sin$^2(2\theta_{13})$ }
%\newcommand{\dmSq}{$\mid\Delta m^2_{ee}\mid$ }

% The following information is for internal review, please remove them
% for submission
%\widetext
%\leftline{ Version 0 as of \today}
%\leftline{Primary authors: Joe E. Physics}
%\leftline{ To be submitted to PRD }
%\leftline{ Comment to {\tt shseo@phya.snu.ac.kr} by Jan. 31st, 2016 }
%\centerline{\em RENO INTERNAL DOCUMENT -- NOT FOR PUBLIC DISTRIBUTION }

% the following line is for submission, including submission to the arXiv!!
%\hspace{5.2in} \mbox{Fermilab-Pub-04/xxx-E}

\title{Physics Potentials with the Second Hyper-Kamiokande Detector in Korea\\
{\normalsize (Hyper-Kamiokande Proto-Collaboration)}}




\newcommand{\BOSTON}{\affiliation{Boston University, Department of Physics, Boston, Massachusetts, U.S.A.}}
\newcommand{\UBC}{\affiliation{University of British Columbia, Department of Physics and Astronomy, Vancouver, British Columbia, Canada }}
\newcommand{\UCDAVIS}{\affiliation{University of California, Davis, Department of Physics, Davis, California, U.S.A.}}
\newcommand{\UCI}{\affiliation{University of California, Irvine, Department of Physics and Astronomy, Irvine, California, U.S.A.}}
\newcommand{\CHUNGNAM}{\affiliation{Chungnam National University, Department of Physics, Daejeon 34134, Korea}}
\newcommand{\CSU}{\affiliation{California State University, Department of Physics, Carson, California, U.S.A.}}
\newcommand{\SACLAY}{\affiliation{IRFU, CEA Saclay, Gif-sur-Yvette, France}}
\newcommand{\CHONNAM}{\affiliation{Chonnam National University, Department of Physics, Gwangju, Korea}}
\newcommand{\DONGSHIN}{\affiliation{Dongshin University, Department of Physics, Naju, Korea}}
\newcommand{\LLR}{\affiliation{Ecole Polytechnique, IN2P3-CNRS, Laboratoire Leprince-Ringuet, Palaiseau, France}}
\newcommand{\LPNHE}{\affiliation{Laboratoire de Physique Nucleaire et de Hautes Energies, UPMC and Universite Paris-Diderot and CNRS/IN2P3, Paris, France}}
\newcommand{\EDINBURGH}{\affiliation{University of Edinburgh, School of Physics and Astronomy, Edinburgh, United Kingdom}}
\newcommand{\GENEVA}{\affiliation{University of Geneva, Section de Physique, DPNC, Geneva, Switzerland}}
\newcommand{\GIST}{\affiliation{GIST College, Gwangju Institute of Science and Technology, Gwangju 500-712, Korea}}
\newcommand{\HANYANG}{\affiliation{Hanyang University, Department of Physics, Seoul 04763, Korea}}
\newcommand{\HAWAII}{\affiliation{University of Hawaii, Department of Physics and Astronomy, Honolulu, Hawaii, U.S.A.}}
\newcommand{\IMPERIAL}{\affiliation{Imperial College London, Department of Physics, London, United Kingdom}}
\newcommand{\BARI}{\affiliation{INFN Sezione di Bari and Universit\`a e Politecnico di Bari, Dipartimento Interuniversitario di Fisica, Bari, Italy}}
\newcommand{\NAPOLI}{\affiliation{INFN Sezione di Napoli and Universit\`a di Napoli, Dipartimento di Fisica, Napoli, Italy}}
\newcommand{\PADOVA}{\affiliation{INFN Sezione di Padova and Universit\`a di Padova, Dipartimento di Fisica, Padova, Italy}}
\newcommand{\ROME}{\affiliation{INFN Sezione di Roma, Roma, Italy}}
\newcommand{\INR}{\affiliation{Institute for Nuclear Research of the Russian Academy of Sciences, Moscow, Russia}}
\newcommand{\NRNU}{\affiliation{National Research Nuclear University (MEPhI), Moscow, Russia}}
\newcommand{\MIPT}{\affiliation{Moscow Institute of Physics and Technology, Moscow region,  Russia}}
\newcommand{\ISU}{\affiliation{Iowa State University, Department of Physics and Astronomy, Ames, Iowa, U.S.A.}}
\newcommand{\KAIST}{\affiliation{Korea Advanced Institute of Science and Technology (KAIST), Department of Physics, Daejeon 34141, Korea}}
\newcommand{\KASI}{\affiliation{Korea Astronomy and Space Science Institute, Daejeon 34055, Korea}}
\newcommand{\KEK}{\affiliation{High Energy Accelerator Research Organization (KEK), Tsukuba, Ibaraki, Japan}}
\newcommand{\KISTI}{\affiliation{Korea Institute of Science and Technology Information, Daejeon 34141, Korea}}
\newcommand{\KNU}{\affiliation{Kyungpook National University, Department of Physics, Daegu 41566, Korea}}
\newcommand{\KNUASTRO}{\affiliation{Kyungpook National University, Department of Astronomy and Atmospheric Sciences, Daegu 41566, Korea}}
\newcommand{\KOBE}{\affiliation{Kobe University, Department of Physics, Kobe, Japan}}
\newcommand{\KOREAU}{\affiliation{Korea University, Department of Physics, Seoul 02841, Korea}}
\newcommand{\KUST}{\affiliation{Korea University of Science and Technology, Daejeon 34113, Korea}}
\newcommand{\KYIV}{\affiliation{Kyiv National University, Department of Nuclear  Physics, Kyiv, Ukraine}}
\newcommand{\KYOTO}{\affiliation{Kyoto University, Department of Physics, Kyoto, Japan}}
\newcommand{\YITP}{\affiliation{Kyoto University, Yukawa Institute for Theoretical Physics, Kyoto, Japan}}
\newcommand{\LNF}{\affiliation{Laboratori Nazionali di Frascati, Frascati, Italy}}
\newcommand{\LANCASTER}{\affiliation{Lancaster University, Physics Department, Lancaster, United Kingdom}}
\newcommand{\LIVERPOOL}{\affiliation{University of Liverpool, Department of Physics, Liverpool, United Kingdom}}
\newcommand{\LANL}{\affiliation{Los Alamos National Laboratory, New Mexico, U.S.A.}}
\newcommand{\LSU}{\affiliation{Louisiana State University, Department of Physics and Astronomy, Baton Rouge, Louisiana, U.S.A. }}
\newcommand{\MADRID}{\affiliation{University Autonoma Madrid, Department of Theoretical Physics, Madrid, Spain}}
\newcommand{\MADRIDIFT}{\affiliation{Instituto de F\'{\i}sica Te\'orica, UAM/CSIC, Madrid, Spain}} 
\newcommand{\MSU}{\affiliation{Michigan State University, Department of Physics and Astronomy,  East Lansing, Michigan, U.S.A.}}
\newcommand{\MIYAGI}{\affiliation{Miyagi University of Education, Department of Physics, Sendai, Japan}}
\newcommand{\NAGOYA}{\affiliation{Nagoya University, Graduate School of Science, Nagoya, Japan}}
\newcommand{\KIAS}{\affiliation{Korean Institute for Advanced Studies, Seoul, Korea}}
\newcommand{\KMI}{\affiliation{Nagoya University, Kobayashi-Maskawa Institute for the Origin of Particles and the Universe, Nagoya, Japan}}
\newcommand{\STELAB}{\affiliation{Nagoya University, Institute for Space-Earth Environmental Research, Nagoya, Japan}}
\newcommand{\NCBJ}{\affiliation{National Centre for Nuclear Research, Warsaw, Poland}}
\newcommand{\OKAYAMA}{\affiliation{Okayama University, Department of Physics, Okayama, Japan}}
\newcommand{\OCU}{\affiliation{Osaka City University, Department of Physics, Osaka, Japan}}
\newcommand{\OXFORD}{\affiliation{Oxford University, Department of Physics, Oxford, United Kingdom}}
\newcommand{\PENN}{\affiliation{Pennsylvania State University, Department of Physics, University Park, PA 16802, U.S.A.}}
\newcommand{\PITTSBURGH}{\affiliation{University of Pittsburgh, Department of Physics and Astronomy, Pittsburgh, Pennsylvania, U.S.A.}}
\newcommand{\PNU}{\affiliation{Pusan National University, Department of Physics, Busan 46241, Korea}}
\newcommand{\RAL}{\affiliation{STFC, Rutherford Appleton Laboratory, Harwell Oxford, and Daresbury Laboratory, Warrington, United Kingdom}}
\newcommand{\REGINA}{\affiliation{University of Regina, Department of Physics, Regina, Saskatchewan, Canada}}
\newcommand{\RHUL}{\affiliation{Royal Holloway University of London, Department of Physics, Egham, Surrey, United Kingdom}}
\newcommand{\RIO}{\affiliation{Pontif{\'\i}cia Universidade Cat{\'o}lica do Rio de Janeiro, Departamento de F\'{\i}sica, Rio de Janeiro, Brazil}}
\newcommand{\ROCHESTER}{\affiliation{University of Rochester, Department of Physics and Astronomy, Rochester, New York, U.S.A.}}
\newcommand{\QMUL}{\affiliation{Queen Mary University of London, School of Physics and Astronomy, London, United Kingdom}}
\newcommand{\SHEFFIELD}{\affiliation{University of Sheffield, Department of Physics and Astronomy, Sheffield, United Kingdom}}
\newcommand{\SNU}{\affiliation{Seoul National University, Department of Physics and Astronomy, Seoul 08826, Korea}}
\newcommand{\SNUST}{\affiliation{Seoul National University of Science and Technology, School of Liberal Arts, Seoul 01811, Korea}}
\newcommand{\SEOYEONG}{\affiliation{Seoyeong University, Department of Fire Safety, Gwangju, Korea }}
\newcommand{\SOONGSIL}{\affiliation{Soongsil University, Origin of Matter and Evolution of Galaxy (OMEG) Institute and Department of Physics, Seoul 156-743, Korea}}
\newcommand{\STOCKHOLM}{\affiliation{Stockholm University, Oskar Klein Centre and Dept. of Physics,  Stockholm, Sweden}}
\newcommand{\STONYBROOK}{\affiliation{State University of New York at Stony Brook, Department of Physics and Astronomy, Stony Brook, New York, U.S.A.}}
\newcommand{\SKKU}{\affiliation{Sungkyunkwan University, Department of Physics, Suwon 16419, Korea}}
\newcommand{\TOHOKU}{\affiliation{Research Center for Neutrino Science, Tohoku University, Sendai, Japan}}
\newcommand{\ERI}{\affiliation{University of Tokyo, Earthquake Research Institute, Tokyo, Japan}}
\newcommand{\KAMIOKA}{\affiliation{University of Tokyo, Institute for Cosmic Ray Research, Kamioka Observatory, Kamioka, Japan}}
\newcommand{\RCCN}{\affiliation{University of Tokyo, Institute for Cosmic Ray Research, Research Center for Cosmic Neutrinos, Kashiwa, Japan}}
\newcommand{\IPMU}{\affiliation{University of Tokyo, Kavli Institute for the Physics and Mathematics of the Universe (WPI), Todai Institutes for Advanced Study, Kashiwa, Chiba, Japan}}
\newcommand{\TOKYO}{\affiliation{University of Tokyo, Department of Physics, Tokyo, Japan}}
\newcommand{\TITECH}{\affiliation{Tokyo Institute of Technology, Department of Physics, Tokyo, Japan}}
\newcommand{\TMU}{\affiliation{Tokyo Metropolitan University, Department of Physics, Tokyo, Japan}}
\newcommand{\TRIUMF }{\affiliation{TRIUMF, Vancouver, British Columbia, Canada}}
\newcommand{\TORONTO}{\affiliation{University of Toronto, Department of Physics, Toronto, Ontario, Canada}}
\newcommand{\WARSAW}{\affiliation{University of Warsaw, Faculty of Physics, Warsaw, Poland}}
\newcommand{\WUT}{\affiliation{Warsaw University of Technology, Institute of Radioelectronics and Multimedia Technology, Warsaw, Poland}}
\newcommand{\WARWICK}{\affiliation{University of Warwick, Department of Physics, Coventry, United Kingdom}}
\newcommand{\WASHINGTON}{\affiliation{University of Washington, Department of Physics, Seattle, Washington, U.S.A.}}
\newcommand{\WINNIPEG}{\affiliation{University of Winnipeg, Department of Physics, Winnipeg, Manitoba, Canada}}
\newcommand{\VT}{\affiliation{Virginia Tech, Center for Neutrino Physics, Blacksburg, Virginia, U.S.A.}}
\newcommand{\WROCLAW}{\affiliation{Wroclaw University, Faculty of Physics and Astronomy, Wroclaw, Poland}}
\newcommand{\YEREVAN}{\affiliation{Yerevan Institute for Theoretical Physics and Modeling, Halabian Str. 34; Yerevan 0036, Armenia}}
\newcommand{\YONSEI}{\affiliation{Yonsei University, Department of Physics and IPAP, Seoul 03722, Korea}}
\newcommand{\YORK}{\affiliation{York University, Department of Physics and Astronomy, Toronto, Ontario, Canada}}
\newcommand{\YOKOHAMA}{\affiliation{Yokohama National University, Faculty of Engineering, Yokohama, Japan}}
\newcommand{\TUS}{\affiliation{Tokyo University of Science, Department of Physics, Chiba, Japan}}
\newcommand{\UNIST}{\affiliation{UNIST, Department of Physics, Ulsan, Korea}}
\newcommand{\KNO}{\affiliation{Author as part of the Korean Neutrino Observatory (KNO) interest group.}}

\BOSTON
\UBC
\UCDAVIS
\UCI
\CHUNGNAM
\CSU
\SACLAY
\CHONNAM
\DONGSHIN
\LLR
\EDINBURGH
\GENEVA
\GIST
\HANYANG
\HAWAII
\IMPERIAL
\BARI
\NAPOLI
\PADOVA
\ROME
\INR
\NRNU
\MIPT
\ISU
\KAIST
\KASI
\KEK
\KISTI
\KNU
\KNUASTRO
\KOBE
\KOREAU
\KUST
\KYIV
\KYOTO
\YITP
\LPNHE
\LNF
\LANCASTER
\LIVERPOOL
\LANL
\LSU
\MADRID
\MADRIDIFT
\MSU
\MIYAGI
\NAGOYA
\KIAS
\KMI
\STELAB
\NCBJ
\OKAYAMA
\OCU
\OXFORD
\PENN
\PITTSBURGH
\PNU
\RAL
\REGINA
\RHUL
\RIO
\ROCHESTER
\QMUL
\SHEFFIELD
\SNU
\SNUST
\SEOYEONG
\SOONGSIL
\STOCKHOLM
\STONYBROOK
\SKKU
\TOHOKU
\ERI
\KAMIOKA
\RCCN
\IPMU
\TOKYO
\TITECH
\TMU
\TUS
\TRIUMF
\TORONTO
\UNIST
\VT
\WARSAW
\WUT
\WARWICK
\WASHINGTON
\WINNIPEG
\WROCLAW
\YEREVAN
\YONSEI
\YORK
\YOKOHAMA
\KNO

\author{K.~Abe}\KAMIOKA\IPMU
\author{Ke.~Abe}\KOBE
\author{S.H.~Ahn}\KASI\KNO
\author{H.~Aihara}\IPMU\TOKYO
\author{A.~Aimi}\PADOVA
\author{R.~Akutsu}\RCCN
\author{C.~Andreopoulos}\LIVERPOOL\RAL
\author{I.~Anghel}\ISU
\author{L.H.V.~Anthony}\LIVERPOOL
\author{M.~Antonova}\INR
% ...

%option in \documentclass). \noaffiliationiation is required (may also be

\collaboration{Hyper-Kamiokande proto-collaboration}

\date{\today}

\vspace{2.cm}

\begin{abstract}
Hyper-Kamiokande consists of two identical
water-Cherenkov detectors of total 520~kt with the first one in Japan
at 295~km from the J-PARC neutrino beam with 2.5$^{\textrm{o}}$ Off-Axis Angles (OAAs),
and the second one possibly in Korea in a later stage.
Having the second detector in Korea would benefit almost all areas of
neutrino oscillation physics mainly due to longer baselines.
There are several candidate sites in Korea with baselines of
1,000$\sim$1,300~km and OAAs of 1$^{\textrm{o}}$$\sim$3$^{\textrm{o}}$.

We conducted sensitivity studies on neutrino oscillation physics
for a second detector, either in Japan (JD $\times$ 2) or
Korea (JD + KD) and compared the results with a single detector in Japan.
Leptonic CP violation sensitivity is improved especially when the CP
is non-maximally violated. The larger matter effect at Korean candidate sites
significantly enhances sensitivities to non-standard interactions of neutrinos
and mass ordering determination.
Current studies indicate the best sensitivity is obtained at
Mt. Bisul (1,088~km baseline, $1.3^\circ$ OAA). 
Thanks to a larger (1,000~m) overburden than the first detector site,
clear improvements to sensitivities for solar and supernova relic neutrino
searches are expected.
\end{abstract}

%\begin{keyword}
%Neutrino oscillation, 2$^{\rm nd}$ oscillation maximum, CP violation,
%Neutrino mass ordering, Non-standard neutrino interaction,
%Matter effect, J-PARC, Off-axis beam, Hyper-Kamiokande
%\end{keyword}

%\pacs{14.60.Pq, 29.40.Mc, 28.50.Hw, 13.15.+g}

\maketitle

\tableofcontents

\clearpage

%======================================================================
\graphicspath{{motivation/plots}}
%=======================================================================
\section{Introduction}
\label{sec:moti}
The proposed Hyper-Kamiokande (Hyper-K or HK) experiment~\cite{HK} builds upon
the highly successful Super-Kamiokande (Super-K or SK)
detector~\cite{SK} by constructing two large water-Cherenkov detectors
with 16.8 times the fiducial volume of SK to pursue a rich program of
neutrino (astro)physics and proton decay. The Hyper-K design
proposes the construction of two identical detectors in stage
with 187\,kt (fiducial mass) per detector. The first one will be constructed
near the current Super-K site, 295\,km away and 2.5$^\circ$ off-axis from
the J-PARC neutrino beam used by the T2K experiment. 
The second one is currently considered to be built in Korea where the J-PARC neutrino beam is still reachable.
The long-baseline neutrino program observing the J-PARC neutrino beam at Hyper-K aims
for a definitive observation of CP violation (CPV) in neutrino
oscillations, that may result from an irreducible phase $\delta_{CP}$
in the neutrino mixing matrix.  Hyper-K will make precise measurements
of $\delta_{CP}$ and other oscillation parameters, such as
$\theta_{23}$ and $\Delta m^2_{32}$, and will have sufficient
statistics to make `shape' tests of the three-flavor mixing paradigm.
These measurements are valuable towards elucidating the new physics
responsible for neutrino mass and mixing and its potential connections
to leptogenesis in the early universe.
%the mystery of the matter/antimatter asymmetry of the universe.
%%% TODO {Check Nomenclature: T2HK vs HK etc}

Especially placing the second detector in Korea rather than in Japan will 
enhance physics sensitivities to almost all searches and measurements. 
In the case of proton decay it would not matter where to locate 
the second detector since the improvement primarily depends on the 
detector mass. In other cases---particularly neutrino oscillation measurements 
using the J-PARC beam---the location of the detector is a significant factor in
determining the expected benefits. In this document, we explore
the possibility of placing the second detector in Korea at a
baseline of 1000$\sim$1300\,km; we will refer to this as T2HKK in
contrast to one detector in Kamioka with 295\,km baseline (T2HK).  
A configuration of the second detector in Korea
provides the opportunity for Hyper-K to
probe oscillation physics at both the first and second oscillation
maxima.

\begin{figure}
\begin{center}
\captionsetup{justification=centering}
\includegraphics[width=1.0\textwidth]{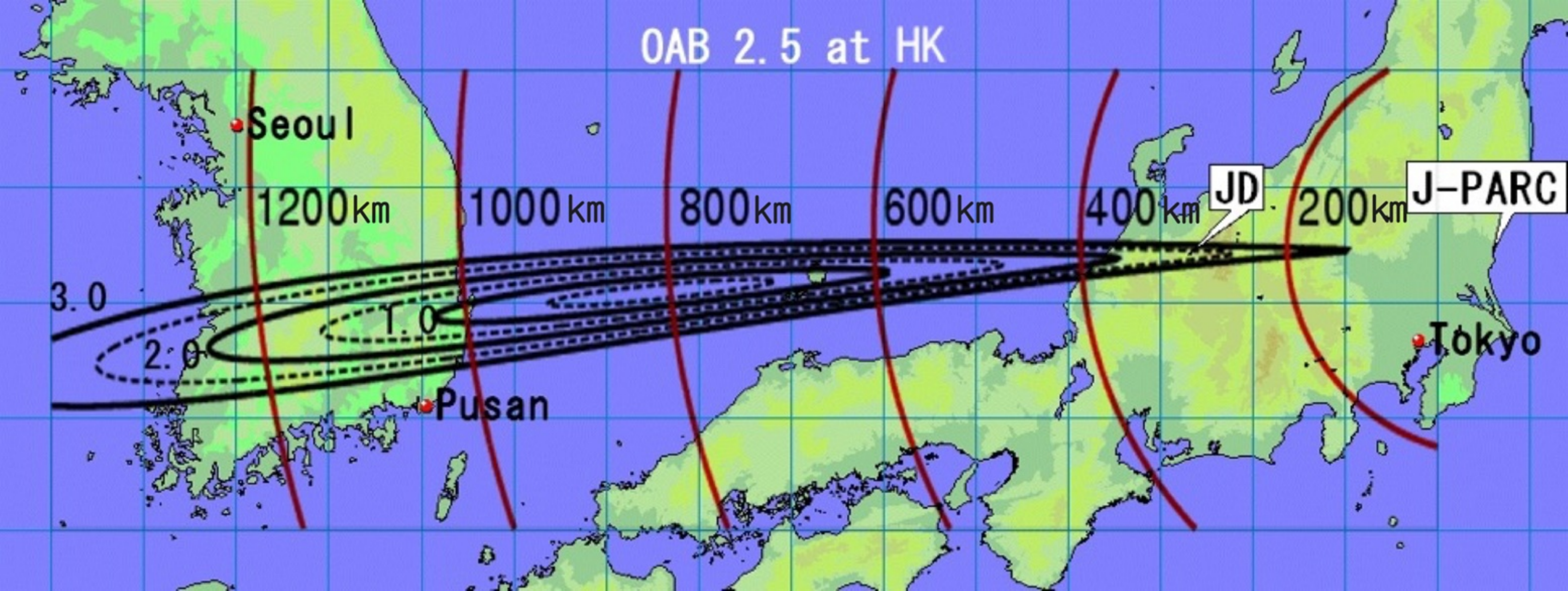}
\end{center}
\caption{Map showing the baseline and off-axis angle of the J-PARC beam in Japan and Korea~\cite{Hagiwara2006,Hagiwara2007}. }
\label{f:OAB}
\end{figure}

South Korea covers a range of angles from the axis of the J-PARC
neutrino beam, from 1 to $3^\circ$ (see Fig.~\ref{f:OAB}), which allows
for tuning of both the baseline and neutrino energy spectrum to
maximize the physics reach of the combined two-baseline experiment.
Such a configuration can improve neutrino oscillation physics sensitivities in Hyper-K
in a number of ways: it can break degeneracies related to the unknown
mass ordering, the mixing parameter $\theta_{23}$, and the
CP-violating phase $\delta_{CP}$; it has better precision (especially
on $\delta_{CP}$) in important regions of parameter space; and it can
serve to mutually reduce the impact of systematic uncertainties (both
known and unknown) across all measurements.  It also provides an
opportunity to test the preferred oscillation model in a regime not
probed with existing experiments.  Constraints on (or evidence of)
exotic neutrino models, such as non-standard interactions with matter,
are also expected to be significantly enhanced by the use of a
longer baseline configuration for a second detector.

%The possibility of operating two similar detectors at two meaningfully
%different baselines is a unique capability of putting a Korean detector
%in the J-PARC beam. No other neutrino beamline, either in
%operation or proposed, allows for this configuration.

Although the use of longer-baseline in conjunction with the J-PARC
beam is the primary feature distinguishing the use of a detector in
Korea from a second detector at Kamioka, there are several
mountains over 1\,km in height that could provide suitable sites.
This allows for greater overburden than the site selected for the
first Hyper-K detector and would enhance the program of low energy
physics that are impacted by cosmic-ray backgrounds.  This includes
solar neutrinos, supernova relic neutrinos, and dark matter neutrino detection studies,
and neutrino geophysics.  In the case of supernova neutrinos there is
some benefit from the separation of detector locations.

Further enhancements are possible but not considered in this document.
Recent developments in gadolinium doping of water and water-based
liquid scintillators could allow for a program based on reactor
neutrinos if these technologies are deployed in the detector. %at a later stage.

There were earlier studies of a large water-Cherenkov detector in
Korea using a J-PARC-based neutrino beam~\cite{SBK,Hagiwara2004}.
Originally an idea for a two-baseline experiment with a second
detector in Korea has been discussed by several authors, pointing out
possible improvements for measurements of CP violation and mass
hierarchy~\cite{Ishitsuka2005, Kajita2007, Dufour2010, Hagiwara2006,
Hagiwara2007}.  Three international workshops were held in Korea and
Japan in 2005, 2006 and 2007~\cite{T2KKWS}.  At the time, the mixing
angle of $\theta_{13}$ was not yet known, and therefore the required
detector size and mass could not be determined.  Now more realistic
studies and a detector design are possible due to the precisely
measured value of $\theta_{13}$~\cite{T2K_q13, DC, DB, RENO,
T2K_q13_2, DC2, DB2, RENO2}.

%---------------------------------------------------------------

%======================================================================
\graphicspath{{detector/plots}}
\section{\label{sec:detector} The Hyper-Kamiokande Experiment \& Extension to Korea }

In this section we present a summary of the proposed Hyper-Kamiokande
experiment, with particular reference to the long-baseline neutrino
oscillation physics program using the J-PARC neutrino beam.  We then
consider the simplest addition of an identical second detector in
Korea to this configuration, focussing on the practical considerations
such as the site selection.  The physics case for a Korean detector,
and studies of the experiment's capability with different configurations
is then considered in Section~\ref{sec:moti} onward.

%The J-PARC neutrino beam and the Hyper-K detector with the near and
%intermediate detectors will be described in the next subsections.

\graphicspath{{detector/plots}}
%======================================================================
\subsection{J-PARC neutrino beam }

The neutrino beam for Hyper-K is produced at J-PARC (Japan Proton
Accelerator Research Complex), located in Tokai Village, Ibaraki
prefecture on the east coast of Japan, which is 295~km from the Kamioka
detector sites.
The 30~GeV (kinetic energy) proton beam is extracted from the J-PARC
Main Ring (MR) by single-turn fast extraction and transported to the
production target after being deflected about 90$^\circ$ by 28
superconducting combined-function magnets to direct the beam towards
Kamioka.  The beam pulse consists of 8 bunches spaced 581~ns apart to
give a pulse of duration 4.2~$\mu$s.  As of 2017, the repetition
period of the pulses is 2.48~s.
The production target is a 26~mm diameter and 90~cm long graphite rod
(corresponding to 2 interaction lengths). About 80\% of incoming
protons interact in the target.
The secondary pions (and kaons) from the target are focused by three
consecutive electromagnetic horns operated by a 250~kA pulsed current.
It is expected that by the time of Hyper-K, the horn currents will be
increased to 320~kA.
The focused pions and kaons enter a 96~m length decay volume (DV)
filled with helium gas and decay in flight into neutrinos and charged
leptons.
The beam dump, which consists of graphite blocks of about 3.15~m
thickness followed by iron plates of 2.5~m total thickness, is placed
at the end of the DV to absorb remnant hadrons.
Muon monitors (MUMONs) are placed just behind the beam dump to monitor
on a spill-by-spill basis the intensity and the profile of muons with
initial energy over 5~GeV which pass through the beam dump.

T2K adopted the first ever off-axis scheme to produce a narrow energy
neutrino spectrum centered on the oscillation maximum to maximize the
physics sensitivity.  The J-PARC beam is aligned to provide a
$2.5^\circ$ degree off-axis beam to the Super-Kamiokande detector.
%, and
%directed equidistant between the Super-Kamioknade and proposed
%Hyper-Kamiokande site, so that both see the same flux.

%[ RPL: TODO TODO This must be updated but I don't have the official time table to hand ]
As of summer 2017, stable operation of the MR at 470~kW beam power has
been achieved.  The design power of the J-PARC main ring will be realized through
the upgrade of the magnet power supplies, RF core and other components. These upgrades
will increase the repetition rate of the beam from 0.40~Hz to 0.77~Hz.
 Preparation
for the upgrade has begun and the upgrade may be completed as early as 2019.
Further beam power increases will require upgrades to secondary
beamline components such as the beam window, target, and horns.
Upgrades primarily to the RF power supply will gradually increase the
number of protons per pulse (ppp) and repetition rate further to
$3.3\times10^{14}$p and 1/1.16~s, respectively, to reach $>$ 1.3~MW by
around 2025 before Hyper-K becomes operational.

%In 2018, the design power of 750~kW will be realized
%by increasing the repetition rate from 1/2.48~s to 1/1.3~s by
%upgrading magnet power supplies, RF core and other components.
%Further beam power increases will require upgrades to secondary
%beamline components such as the beam window, target, and horns.
%Upgrades primarily to the RF power supply will gradually increase the
%number of protons per pulse (ppp) and repetition rate further to
%$3.3\times10^{14}$p and 1/1.16~s, respectively, to reach $>$ 1.3~MW by
%around 2025 before Hyper-K becomes operational.

\graphicspath{{detector/plots}}
%%%%%%%%%%%%%%%%%%%%%%%%%%%%%%%%%%%%%
\subsection{Hyper-Kamiokande tank configuration}
%%%%%%%%%%%%%%%%%%%%%%%%%%%%%%%%%%%%%

%[RPL: TODO TODO baseline does not have two tanks ]
The Hyper-K experiment employs a ring-imaging water-Cherenkov detector
technique to detect rare interactions of neutrinos and the possible
spontaneous decay of protons and bound neutrons. The baseline configuration
of the experiment consists of a single cylindrical tank built in the Tochibora
mine at a baseline of 295~km from the J-PARC neutrino source.  
% {\bf \color{red}The
%baseline detector configuration consists of two cylindrical tanks with
%the second tank commencing operation later than the first tank.  The
%first priority is to perform a CP violation measurement at the
%earliest opportunity with the first tank.}

\begin{figure}[htbp]
  \begin{center}
  \includegraphics[width=0.9\textwidth]{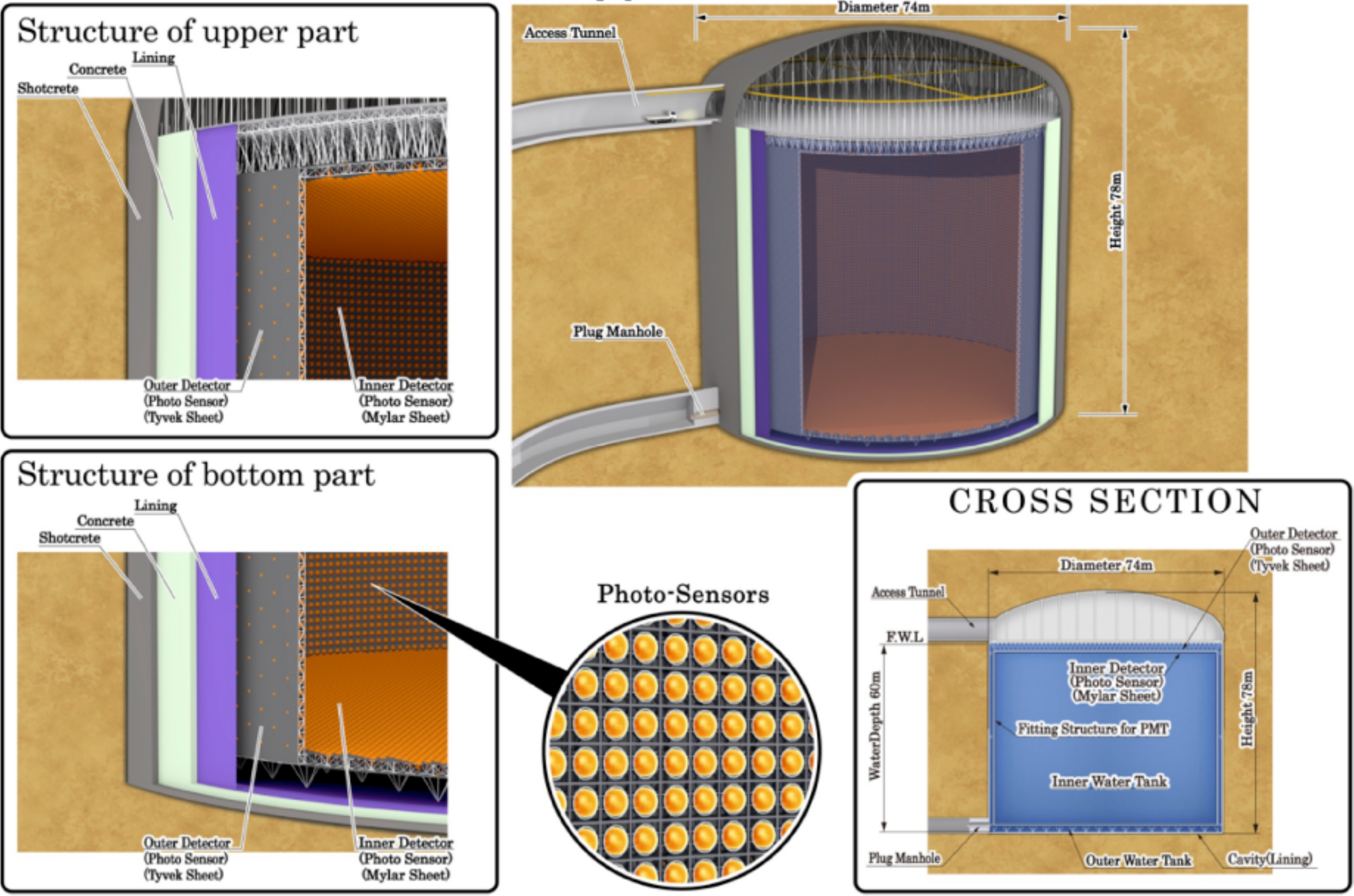}
  \caption{Schematic view for the first tank.}
  \label{fig:hk-schematic}
  \end{center}
\end{figure}

A full overview of the cavern and detector design R\&D, upgraded beam
and near detector suite, and expected physics sensitivities can be
found in the Hyper-Kamiokande Design Report~\cite{design-report}.  The
schematic view of the tank is shown in Fig.~\ref{fig:hk-schematic}.
The design is a upright cylindrical tank with a diameter of 74\,m and
height of 60\,m.  The total (fiducial) mass of the detector is 258
(187)\,kilotonnes, giving a fiducial mass that is 8 times larger than Super-K.
%  {\bf \color{red}Two tanks together will provide a
%fiducial volume about 20 times larger than that of Super-K.}  
The Hyper-K detector candidate site is in the Tochibora mine, which is
used by the Kamioka Mining and Smelting Company, near Kamioka town in
Gifu Prefecture, Japan.  This is 8\,km south of Super-K and 295\,km
away from J-PARC.  The J-PARC neutrino beamline is directed so that
the existing Super-Kamiokande detector in the Mozumi mine and the
Hyper-K candidate site in the Tochibora mine have the same off-axis
angle.  The detector will lie under the peak of Nijuugo-yama, with an
overburden of 650\,m of rock or 1,750\,meters-water-equivalent
(m.w.e.), at geographic coordinates
$\mathrm{N}36^\circ\,21^\prime\,20\vphantom{.105}^{\prime\prime}$,
$\mathrm{E}137^\circ\,18^\prime\,49\vphantom{.137}^{\prime\prime}$
(world geographical coordinate system), and an altitude of 514\,m
above sea level (a.s.l.).

The Hyper-K detector is designed to employ newly developed
high-efficiency and high-resolution (timing) PMTs (Hamamatsu R12860) which will
collect more photons compared to the current Super-K PMTs, and improve
sensitivity to low energy events such as neutron captures and nuclear de-excitations.
%amplify faint light signals such as those produced by neutrons
%associated with neutrino interactions, nuclear de-excitation gammas
%and $\pi^+$ in proton decays into kaons, and so on.  
This increased
sensitivity contributes significantly to the major goals of the
Hyper-K experiment such as clean proton decay searches via
$p\rightarrow e^+ + \pi^0$ and $p\rightarrow \bar{\nu} + K^+$ decay
modes and the observation of supernova electron anti-neutrinos.  The
inner detector region of the single tank is viewed by 40,000 PMTs,
corresponding to the PMT density of 40\% photo-cathode coverage (the
same as that of Super-K).  The detector will be instrumented with
front-end electronics and a readout network/computer system that is
capable of high-efficiency data acquisition for two successive events
in which Michel electron events follow muon events with a mean
interval of 2\,$\mu$s.  It is also able to recording data from the
vast number of neutrinos that would come from a nearby supernova in a
nominal time period of 10\,sec.  Similar to Super-K, an outer detector
(OD) with a layer width of 1--2\,m is envisaged. In addition to
enabling additional physics, this would help to constrain the external
background.  The photo-coverage of the OD can be sparser, and use
smaller PMTs than those used for the ID.

\subsection{Near and intermediate detector complex}
The neutrino flux and cross-section models can be probed by data
collected at near detectors, situated close enough to the neutrino
production point so that oscillation effects are negligible.  Near
detector data is extremely important to constrain uncertainties in
these models.

The existing T2K ND280 detector suite comprises two
detectors~\cite{T2K_ND280_nim}: INGRID, which consists of 14
iron-scintillator modules in a cross pattern centered on the neutrino
beam axis, and ND280, a multi-component detector at an angle of
2.5\,degree from the beam direction. The primary purpose of the INGRID
detector is to monitor the neutrino beam direction, whilst the
off-axis detector is used to characterize the spectrum and
interactions of neutrinos in the beam before oscillation.  T2K has
successfully applied a method of fitting to ND280 data with
parametrized models of the neutrino flux and interaction
cross-sections. Using the ND280 measurements, the systematic
uncertainties on the parts of the models constrained by ND280 have
been reduced to a typical 3\% uncertainty on the Super-K (SK)
predicted event rates.  An upgrade of the current ND280 detector is
planned before the start of Hyper-K.

Moreover, it is proposed to build a water-Cherenkov detector at 
$\sim$1$\sim$2\,km, before Hyper-K becomes
operational~\cite{nuPRISM}.  A water-Cherenkov intermediate detector can be
used to measure the cross section on H$_2$O directly, with the same
acceptance of lepton scattering angle as the far detector with no need
to account for the different target nuclei in a heterogeneous detector.
Additionally, water-Cherenkov detectors have shown excellent particle
identification capabilities, allowing for the detection of pure
$\nu_{\mu}$-CC, $\nu_{e}$-CC and NC$\pi^{0}$ samples.  The CC$\pi^{0}$
rate and kaon production in neutrino interactions, which are
backgrounds to nucleon decay searches, can also be measured.
The proposed water-Cherenkov detector will have the capability to make
measurements at OAAs of 1.0-4.0$^{\circ}$, covering the potential OAAs 
of detectors in Korea. 

These additional water-Cherenkov measurements are essential to achieve
the low systematic errors required by Hyper-K, but are complemented by
the magnetized ND280 tracking detector, which is capable of tracking
particles below the Cherenkov light threshold in water and of
separating neutrino and antineutrino beam components via the lepton
charge measurement.  The design anticipates that a combination of a
magnetized tracking detector such as ND280 and an `intermediate'
water-Cherenkov detector will provide the greatest reduction in
systematic uncertainties affecting the oscillation measurement.

\graphicspath{{detector/plots}}
%======================================================================
\subsection{T2HKK experimental configuration}
\label{s:expt-setup}
The axis of the J-PARC neutrino beam emerges upwards out of the sea
between Japan and Korea.  The southern part of the Korean peninsula is
exposed to the beam at a 1--3 degree range of off-axis angles. From
east to west, baselines of 1000--1300\,km are possible, as shown in
Fig.~\ref{f:OAB}.  The topography of South Korea is quite mountainous,
especially in the east of the country, and provides plenty of suitable
candidate sites.

\begin{figure}
\centering
\captionsetup{justification=centering}
\includegraphics[width=0.85\textwidth]{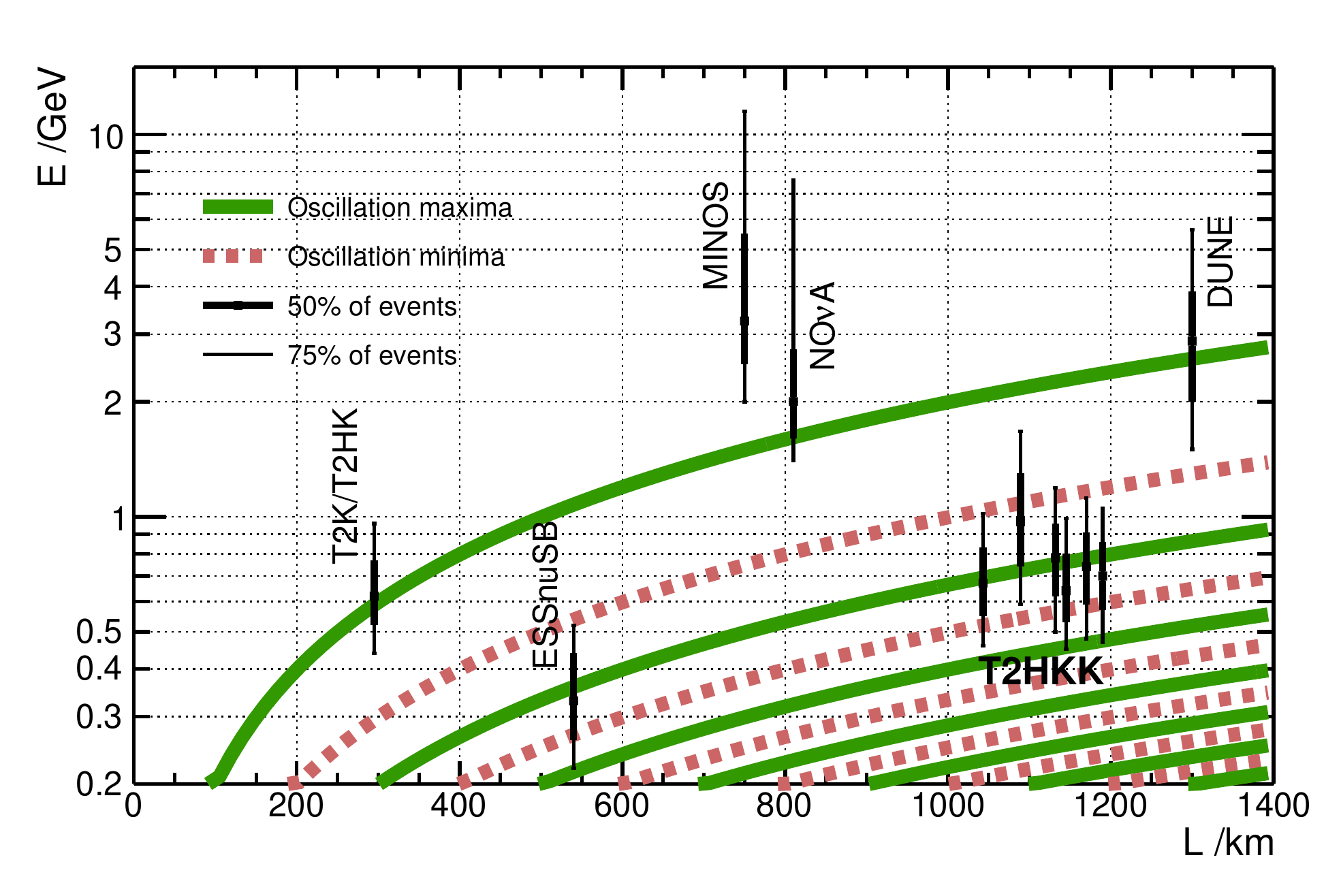}
\caption{Comparison of baseline and energy regime of various recent and
proposed long-baseline experiments.  Event rates for Kamioka and
Korean sites are based on calculated fluxes, using the quasi-elastic
charged-current cross-section from NEUT.  The ESSnuSB event rates are
calculated from publicly available flux histograms~\cite{ESS_WP},
and the NEUT cross-section. Event rates from MINOS~\cite{minos_spectrum},
NO$\nu$A~\cite{nova_spectrum}, and the DUNE~\cite{dune_spectrum}
optimized design use publicly available spectra, which typically
assume inclusive charged-current cross-sections.
%The six Korean sites under consideration are listed in
%section \ref{s:expt-setup}.
}
\label{f:baseline_comparison}
\end{figure}

The Korean rocks are in general made of granite, hard enough to build
a large cavern.  A search for mountains higher than 1000\,m has been
made to find several candidates for the Korean detector. Mountains in the national or
provincial parks were not considered in the search.  Six suitable
sites are listed in Table~\ref{t:six_sites}, along with their location
with respect to the J-PARC neutrino beam. The baselines and energy options 
of the six sites are shown in Fig.~\ref{f:baseline_comparison}.
All sites would provide a significant flux at the second oscillation
maximum, and depending on the site it is possible to sample neutrinos
from as far apart as the first to third maxima.

\begin{table}[hbt]
\captionsetup{justification=raggedright,singlelinecheck=false}
\small
 \caption{Detector candidate sites with off-axis angles between 1 and 2.5 degrees. The baseline is the distance from the production point of the J-PARC neutrino beam.}
 \centering
 \begin{tabular*}{0.85\textwidth}{@{\extracolsep{\fill}} l c c c l}

  \hline \hline
Site           & Height & Baseline & Off-axis
               &    Composition of rock \\[-1.4ex]
               & (m)    & (km)     & angle & \\ 
\hline
 Mt. Bisul     & 1084   &  1088    & 1.3$^{\circ}$
               &    Granite porphyry,\\[-1.4ex] %1.30
               &&&& andesitic breccia \\[0.4ex]
 Mt. Hwangmae  & 1113   &  1141    & 1.9$^{\circ}$
               &    Flake granite, \\[-1.4ex] %1.94
               &&&& porphyritic gneiss  \\[0.4ex]
 Mt. Sambong   & 1186   &  1169    & 2.1$^{\circ}$
               &    Porphyritic granite, \\[-1.4ex] %2.06
               &&&& biotite gneiss \\[0.4ex]
 Mt. Bohyun    & 1124   &  1043    & 2.3$^{\circ}$
               &    Granite, volcanic rocks, \\[-1.4ex] %2.29
               &&&& volcanic breccia \\[0.4ex] 
 Mt. Minjuji   & 1242   &  1145    & 2.4$^{\circ}$
               &    Granite, biotite gneiss \\[0.4ex] %2.38
 Mt. Unjang    & 1125   &  1190    & 2.2$^{\circ}$
               &    Rhyolite, granite porphyry, \\[-1.4ex] %2.21
               &&&& quartz porphyry  \\
  \hline \hline
  \end{tabular*}
 \label{t:six_sites}
 \end{table}

At an off-axis angle similar to that of the Kamioka site (2.5$^\circ$)
the neutrino interaction rate peaks at an energy of around
0.7\,GeV. At this energy the second oscillation maximum occurs at a
baseline of roughly 1100\,km, with longer baselines corresponding to
maxima for higher energies and vice versa.

The novel aspect of a detector built in Korea becomes clear if we
compare it to similar long-baseline neutrino experiments.
Fig.~\ref{f:baseline_comparison} shows the regime of baseline
($L$) and neutrino energy ($E$) covered by recent and proposed
experiments.  T2HKK provides a baseline almost as long as the proposed
DUNE experiment but in a similar energy band to the existing T2K
experiment, which allows it to probe oscillations at the second
oscillation maximum, a capability not available to any existing
experiment, and only shared by the proposed ESS neutrino
beam~\cite{ESS_WP, ESS_2016}.  Furthermore, the fact that oscillations
become more rapid at higher-order maxima means that the T2HKK
configuration can probe more of the oscillation shape than existing
experiments. It is also worth noting that a double-baseline
configuration using the similar fluxes and different (non-trivial)
baselines within one experiment is only possible because the axis of
the J-PARC beam passes below the Kamioka site.  Equivalent
configurations using the NuMI or proposed LBNF and ESS beamlines do
not exist.  

\subsubsection{Investigation of candidate sites}
We can roughly partition the candidate sites into two groups. For five of
the six sites the (unoscillated) interaction rate is expected to peak
near or slightly below the energy of the second oscillation maximum.
The exception is Mt. Bisul; at ($1.3^\circ$) it is much closer to the
beam axis, so the typical neutrino is more
energetic and the spectrum overall is broader.  A detector at this
site could still sample the second maximum but also sample a
significant part of the first oscillation maximum and the region between the
first and second oscillation maxima.  Physics studies
therefore treat Bisul as a distinct case.

The variation in $L/E$ between the other sites is less substantial,
and discriminating between the physics reach of each requires detailed
simulations.  The Bohyun site, being closest to J-PARC, is expected to
provide the highest event rate after Bisul.  Based on these
considerations, Mt. Bisul and Mt. Bohyun are the first sites
considered for more detailed investigation of their suitability.

\begin{figure}[hbt]
\captionsetup{justification=raggedright,singlelinecheck=false}
\begin{center}
\includegraphics[width=1.0\textwidth]{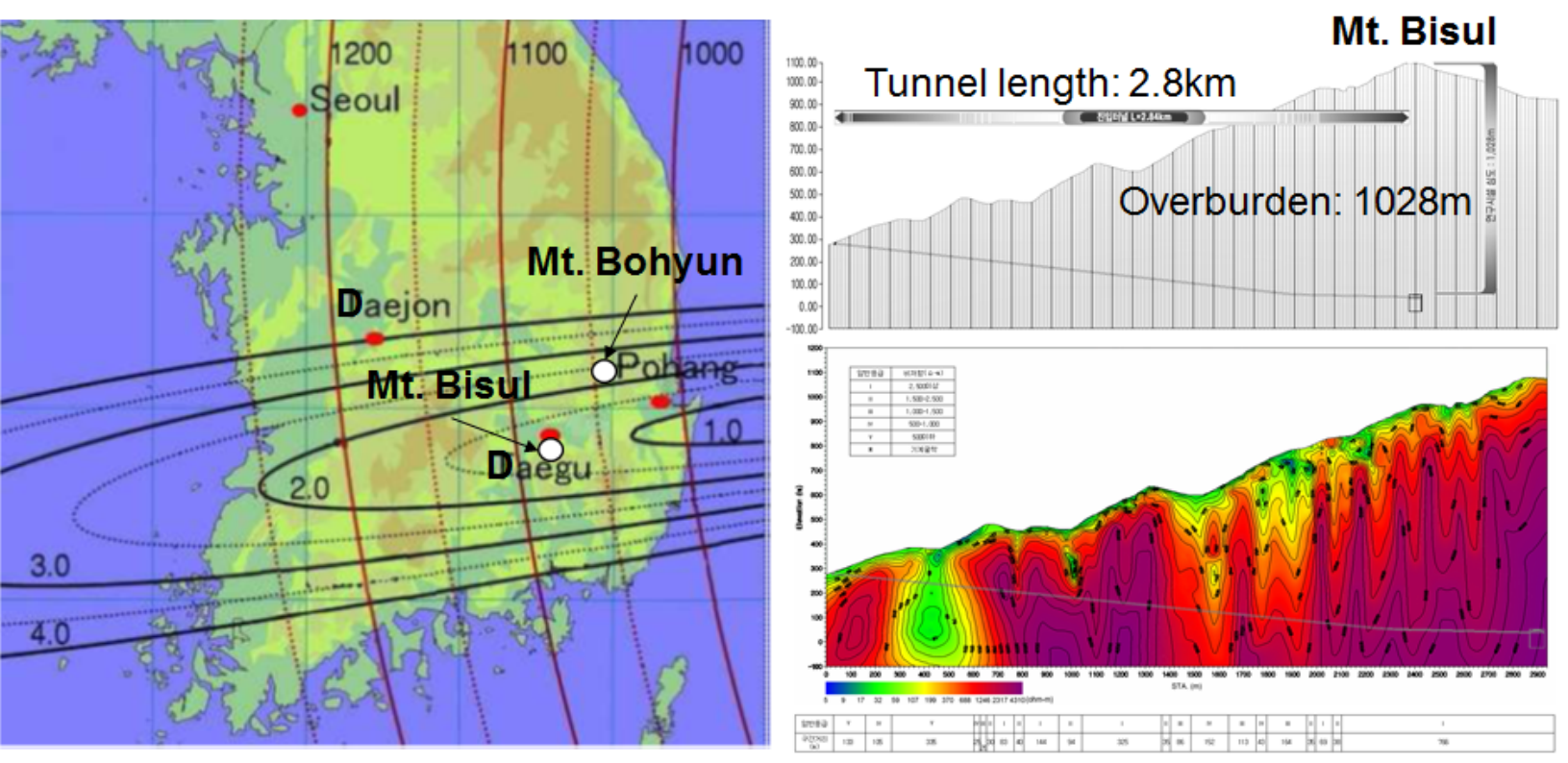}
\caption{
Two candidate sites for the second Hyper-K detector in Korea. 
Mt. Bisul is located near the city of Daegu at 1.3$^{\circ}$ off-axis, 
and Mt. Bohyun at Youngcheon at 2.3$^{\circ}$ off-axis. Mt. Biusl is 
1084~m high and provides excellent accessibility with an existing 
highway nearby. Mt. Bohyun is 1124~m high and accommodates an optical 
telescope on the top. A detector at Bisul is expected to have $\sim$1000~m 
overburden with a slightly inclined access tunnel of 2.8~km long. 
An electromagnetic geological-survey shows an excellent bedrock of the 
candidate site, suitable for a large cavern.
}
\label{candidate-sites}
\end{center}
\end{figure}

Mt. Bisul is located at Dalseong in the city of Daegu, the fourth largest city in population
in South Korea as shown in Fig.~\ref{candidate-sites}. Its accessibility is 
excellent. It takes one hour and forty minutes to get to Daegu from Seoul by the 
KTX (Korean bullet train). The mountain is 1084~m high and made of hard rocks: 
granite porphyry and andesitic breccia. 
A detector at Bisul is expected to have $\sim$1000~m overburden with a slightly 
inclined access tunnel and to be exposed to a 1.3 degree off-axis neutrino beam. 
The site coordinates are 
$\mathrm{N}35^\circ\,43^\prime\,00\vphantom{.105}^{\prime\prime}$ in latitude and
$\mathrm{E}128^\circ\,31^\prime\,28\vphantom{.105}^{\prime\prime}$ in longitude.
The baseline from J-PARC is 1088~km. 
A recent geological survey using a magnetotelluric method shows an excellent bedrock 
of the candidate site, say, belonging to the hardest rock classes 1 or 2.
Based on nearby lakes and rivers, sufficient underground water could be available in the site.
The survey result obtains an estimate of $2.7 \sim 3.2\, \textrm{m}^3/\textrm{km}/\textrm{min}$ 
underground-water flow into the access tunnel through rock fractures. The expected water would 
be sufficient enough to be supplied for the detector.
We find excellent access roads up to the candidate location of tunnel entrance, and easy 
access to electricity supply lines.

\begin{figure}[hbt]
\captionsetup{justification=raggedright,singlelinecheck=false}
\begin{center}
\includegraphics[width=0.8\textwidth]{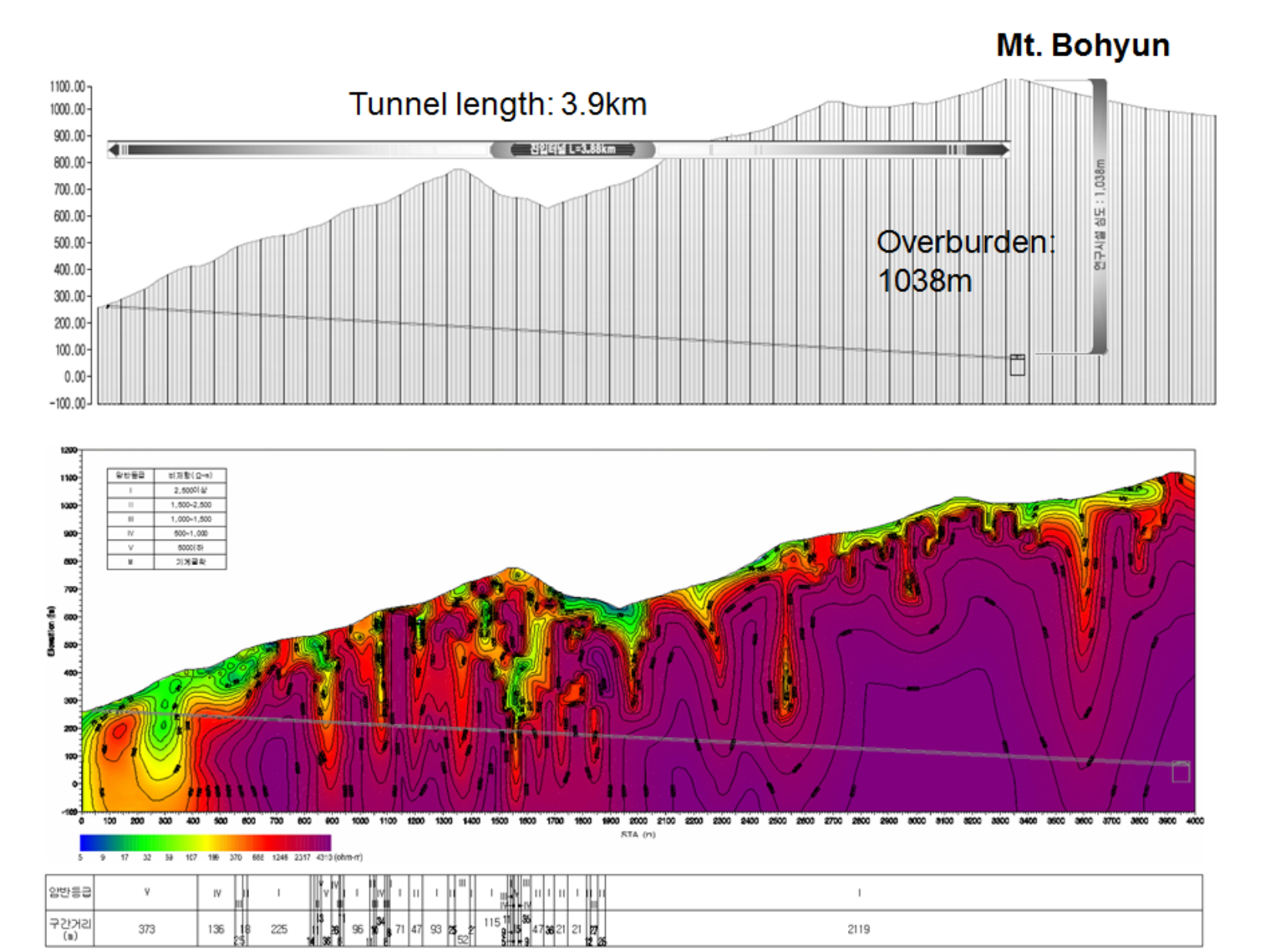}
\caption{
Mt. Bohyun as a candidate site for the second Hyper-K detector in Korea. It is 1124~m high 
and provides $\sim$1000~m overburden with a slightly inclined access-tunnel of 3.9~km long. 
A geological survey shows an excellent bedrock of the candidate site, suitable for a large cavern.
}
\label{bohyun-site}
\end{center}
\end{figure}

The Mt. Bohyun is located at Youngcheon and is also the site of the Bohyunsan Optical Astronomy Observatory, 
as shown in Fig.~\ref{bohyun-site}.
The mountain is 1124~m high and made of fairly hard rocks: granite, volcanic rocks and volcanic breccia.
It is an excellent candidate site for a large cavern.
A detector at Bohyun is expected to have $\sim$1000~m overburden and to be exposed to a 2.3 degree off-axis neutrino 
beam. The site coordinates are 
$\mathrm{N}36^\circ\,09^\prime\,47\vphantom{.105}^{\prime\prime}$ in latitude and 
$\mathrm{E}128^\circ\,58^\prime\,26\vphantom{.105}^{\prime\prime}$ in longitude. 
The baseline from J-PARC is 1040~km. 
The geological survey also shows an excellent bedrock of the candidate site.
Based on nearby rivers, sufficient underground water is expected in the site.
The survey result obtains an estimate of $\sim 3.7\, \textrm{m}^3 /\textrm{km}/\textrm{min}$ underground-water flow into the access tunnel. 
Its accessibility is reasonably good.

In summary, both candidate sites seem to be suitable for building a second Hyper-K detector, and can provide roughly 
1000~m overburden for low-energy neutrino measurements. Investigation of the suitability of the sites for a larger cavern 
is already advanced, and initial estimates suggest the excavation cost in Korea would be comparatively low.

%\input{detector/det_sens_t.tex}

%======================================================================
\section{Physics Motivation for the T2HKK Configuration}
\label{sec:moti}
In this section we explain in more detail the expected benefits of a
detector sited in Korea, and why these are sufficient to make the
proposal attractive despite the inevitable loss of statistics from
using a longer baseline.

To do so, it is first worth restating the goals of the next generation
of neutrino oscillation experiments. A general goal is to test the
3-neutrino PMNS model.  It is almost tautological that by using a
different range of $L/E$ we subject the model to a broader range of
checks than existing long-baseline experiments.  An example of these
kind of checks is the study of non-standard interactions (between
neutrinos and normal matter), which is described in
Section~\ref{sec:nonSTD}.  Even without resorting to any particular
new model, an expanded range of $L/E$ values provides a check of the
complete model used to analyse neutrino experiments. In the broadest
sense, this includes the models used for systematic uncertainties,
such as the modelling of neutrino fluxes and cross sections.

\begin {figure}[htbp]
\captionsetup{justification=raggedright,singlelinecheck=false}
  \begin{center}
  \includegraphics[width=0.8\textwidth]{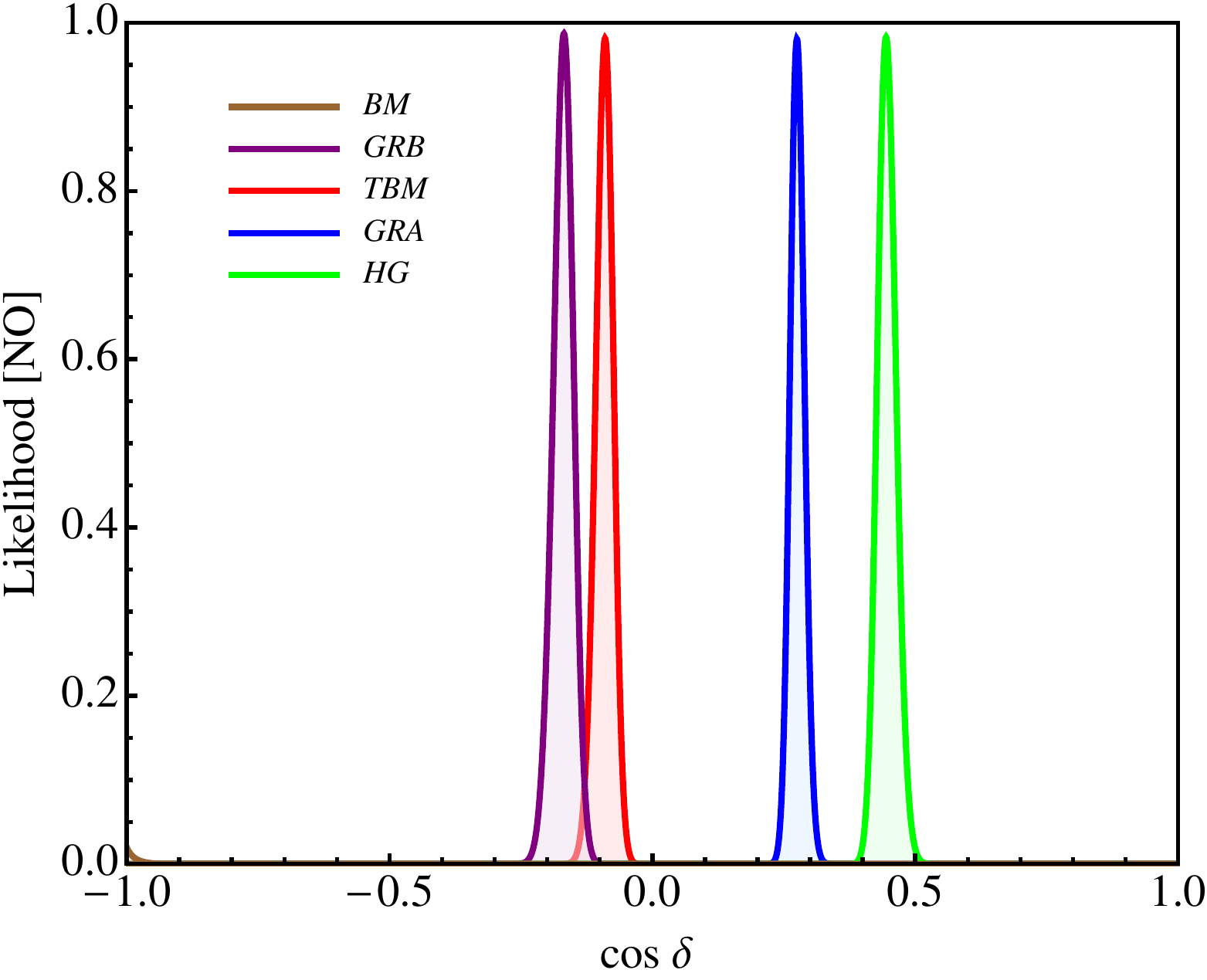}
  \caption{The likelihood function versus $\cos{\delta_{CP}}$ for
  normal ordering neutrino mass spectrum for different types of flavor
  symmetries assuming the prospective 1$\sigma$ uncertainties in the
  determination of the neutrino mixing angles~\cite{Petkov2015}.}
\label{fig:GUT-CP}
  \end{center}
\end {figure}

Assuming that the 3-neutrino oscillation model is correct and the
uncertainties we assign to the broader model are sensible, the primary
goals of the next generation of neutrino oscillation experiments are:
\begin{itemize}
\item Establish the ordering of neutrino masses.
That is, whether or not $m^2_3 > m^2_1$.
\item Establish the existence (or not) of CP violation in the neutrino sector.
That is, whether or not $\delta_{CP} = {0,\pi}$.
\item Make precision measurements of all oscillation parameters.
Most important is to make a precision measurement of
$\delta_{CP}$, (as opposed to simply determining that it is not zero
or $\pi$).
\end{itemize}
The first two of these are well-known, and need not be elaborated.
As an example of the third goal, Fig.~\ref{fig:GUT-CP} shows the
prediction of GUTs with different flavour symmetries for the value of
$\cos\delta$. Four of the five models predict a sizeable CP violation effect
(CP conservation occurs for $\cos\delta = \pm 1$), and are only
separated by precision measurement.  %Such measurements are roughly
%analogous to the unitarity triangle measurements made in hadronic
%flavour physics.
%

At the second maximum, the effect on the observed spectrum from both
the CP-phase and mass hierarchy are enhanced in comparison to the
Kamioka site, but the exact behaviour is dependent on the site. At the
Bisul site (or any hypothetical site with an off-axis angle below
about 1.5 degrees, shown in Fig.~\ref{fig:osc_probs}.)  the beam
spectrum peaks at around 1\,GeV with a wider energy band, making it
possible to sample events from both the first and second oscillation
maxima. In addition, because of the energy dependence of
matter-enhancement effects, the maximum sensitivity for determining
the neutrino mass ordering is achieved by going to higher neutrino
energies, so the Bisul site will provide better discrimination between
normal and inverted orderings.

Alternatively, the off-axis angle can be chosen in the region of 2.5
degrees, to provide a similar flux (without oscillations) to that seen
at Kamioka.  In that case, the J-PARC neutrino beam spectrum peaks at
$E_{\nu} = 0.6$\,GeV with a narrower energy band in the vicinity of
the second oscillation maximum.  With identical off-axis angles for HK
and HKK, the combined analysis is essentially a ratio measurement
between the two sites, which can be expected to greatly reduce the
uncertainties from fluxes and cross-sections.

Both alternatives sample a wider range of $L/E$ than the Kamioka site,
and this enables improved measurement resolution of the CP-phase,
especially near the maximally CP-violating values.

In the following section we examine the oscillation probabilities to
explain the enhancements seen at various Korean sites, and how
measurements at HKK would differ from the baseline HK. Following this
sensitivity studies are presented using both generic sites at a
typical baseline of 1100\,km and the Bisul and Bohyun sites identified
above.

\subsection{Neutrino oscillation probabilities}
\label{sec:osc_prob}
The sensitivity enhancement of a second detector in Korea can be
understood by first examining the $P(\nu_{\mu}\rightarrow\nu_{e})$ and
$P(\overline{\nu}_{\mu}\rightarrow\overline{\nu}_{e})$ probabilities. We do this
first studying the probabilities in vacuum and then the probabilities
with the matter effect included.  The approximate oscillation
probability in vacuum is: \\
%\small
%\footnotesize
\begin{align}
P(\nupbar_\mu \rightarrow \nupbar_e) \approx & \sin^{2}\!\theta_{23} \sin^{2}\!2\theta_{13} \sin^{2}(\Delta_{31}) \nonumber \\
& + \sin\!2\theta_{23} \sin\!2\theta_{13} \sin\!2\theta_{12} \cos\theta_{13} \sin(\Delta_{31}) \sin(\Delta_{21}) \cos(\Delta_{32}) \cos\delta \nonumber \\
& \mp \sin\!2\theta_{23} \sin\!2\theta_{13} \sin\!2\theta_{12} \cos\theta_{13} \sin(\Delta_{31}) \sin(\Delta_{21}) \sin(\Delta_{32}) \sin\delta \nonumber \\ 
& + \cos^{2}\!\theta_{13}\cos^2\!\theta_{23} \sin^{2}\!2\theta_{12} \sin^{2}(\Delta_{21}).
\end{align}
%\normalsize
\noindent
%%% TODO {mention the cos-Delta cos-delta term for (0-pi) degeneracy}

%Here, $J_{CP} = \frac{1}{2} \sin(2\theta_{23}) \sin(2\theta_{12}) \sin(2\theta_{13})\cos\theta_{13}$
%is the Jarlskog invariant, and the minus sign applies for neutrinos.
Here, we use the shorthand $\Delta_{ji} = \tfrac{\Delta
m^2_{ji}L}{4E}$.% and the $\Delta m^{2}_{32}$ and $\Delta m^{2}_{31}$
%mass splittings are treated as equal.
The first line represents the oscillations at the atmospheric mass
splitting, and this term dominates for $L/E$ values of
$\sim$(500~km)/(1~GeV) typical of accelerator based long-baseline
oscillation experiments.  The fourth line gives the oscillations
driven by the solar mass splitting, which are small for the $L/E$ values
of interest.  The second and third lines are the CP conserving and CP
violating parts respectively of the interference term.  The sign of
the third line flips to positive when considering antineutrinos,
introducing the CP violation effect.

\begin {figure}
  \captionsetup{justification=raggedright,singlelinecheck=false}
  \centering        
  \includegraphics[width=0.8\textwidth]{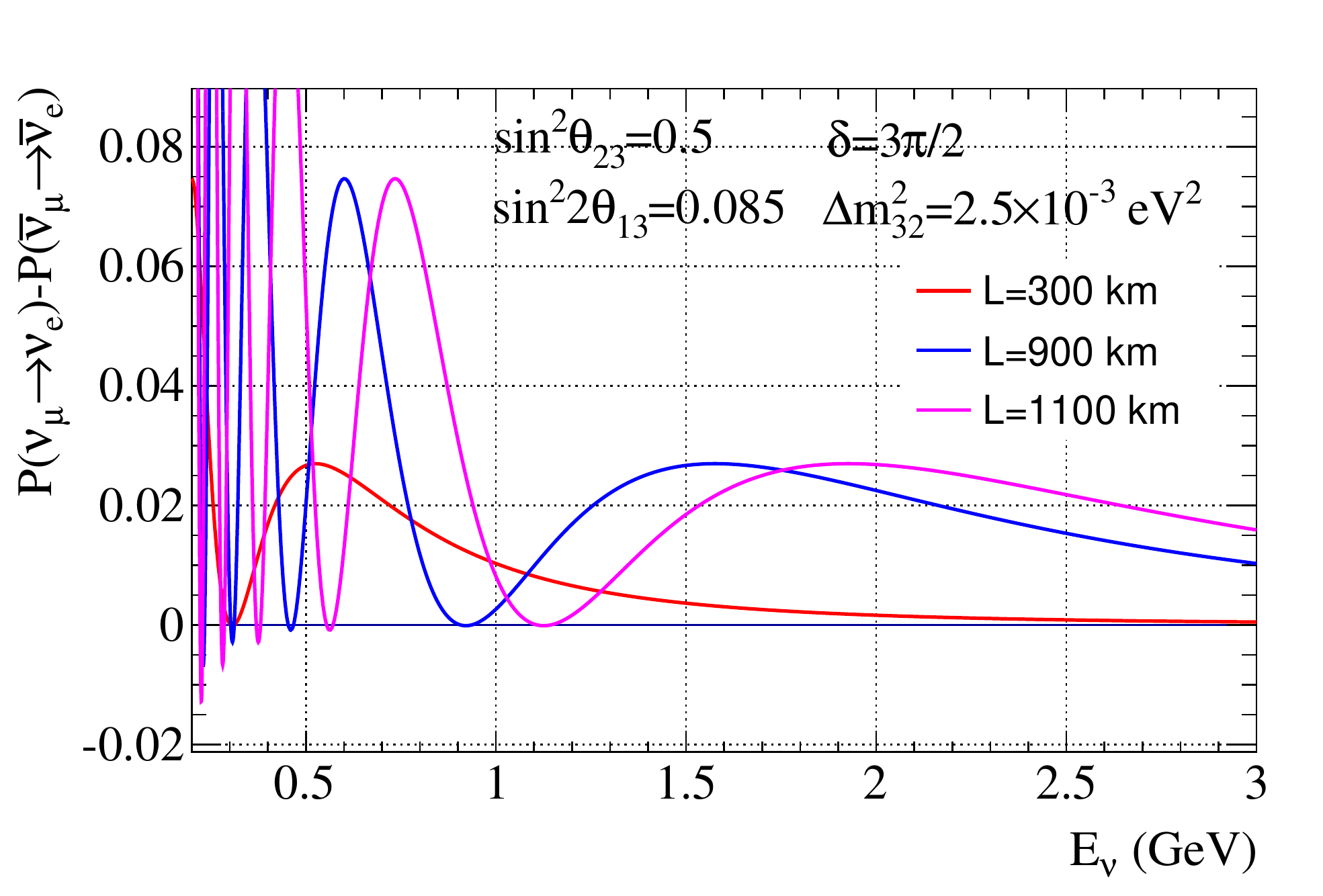}
  \caption{The neutrino-antineutrino probability difference for
  $\delta=3\pi/2$ at 300, 900, and 1100\,km baselines for oscillations
  in vacuum.  As the effects of matter are ignored here, this
  indicates the empirical size of `true' CP-violation, driven only by
  $\delta_{CP}$.} \label{fig:cp_asymm_vacuum}
\end {figure}

To understand the benefit of a second-maximum experiment, note that
the CP violating interference term depends on $\sin(\frac{\Delta
  m^{2}_{21}L}{4E})$.  Since the argument is small for the $L/E$
values of interest, this dependence is approximately linear.  Thus,
for a given energy, a larger CP effect will be observed at longer
baselines. This is illustrated in Fig.~\ref{fig:cp_asymm_vacuum} where
the intrinsic CP-driven neutrino--antineutrino probability difference
is shown for baselines of 300, 900 and 1100\,km.

We identify the ``oscillation maxima'' as points where $\frac{\Delta
m^{2}_{32}L}{4E}=n\pi/2$ and $n$ is an odd integer.  For a fixed
baseline, the second oscillation maximum will be located at $1/3$ the
energy of the first oscillation maximum.  Alternatively, for a fixed
energy, the necessary baseline to observe the second oscillation
maximum will be 3 times larger than the baseline needed to observe the
first oscillation maximum.  The neutrino flux will decrease by the
ratio of the baselines squared---a factor of 1/9 in this case---which
means the statistical sensitivity decreases by a factor of three. At
the same time the CP effect is around 3 times larger at the second
oscillation maximum. The result is that CP violation measurements made
at the second oscillation maximum have similar sensitivity to the
first maximum, despite the reduction in event rate that comes with a
3-times larger baseline.

The significant benefit for second-maxima experiments is that this
near-indifference of the statistical sensitivity to the baseline does
not apply in relation to systematic uncertainties.  Most important
systematic errors are not significantly constrained by data from the
oscillated neutrino beam.  This can be because parameters are better
constrained by a near detector in the same beamline, by other
independent experiments, or perhaps by using data taken with local
calibration sources.  Such uncertainties do \emph{not} grow with the
baseline and as such are reduced relative to the larger CP violation
effect.

Another uncertainty that potentially impacts the CP measurement is
that associated with the leading atmospheric term.  This dominates the
overall appearance probability at the first maximum, so in a realistic
case where we can't do a perfect neutrino-antineutrino comparison,
there will be some contribution to the overall uncertainty arising
from our imperfect measurement of the parameters that appear in this
term.  But at the second maximum, where the size of the interference
term is roughly as large as the leading term, the finite precision to
which the latter is known has less impact.  In essence there is less
`background' due to contributions that are independent of
$\delta_{CP}$. The picture is not that simple in practice, but this
effectively means correlations between oscillation parameter
measurements at Kamioka and Korean detector can be different, leading
to potential synergies by having both.
\\

When neutrinos propagate in matter, the matter potential is added to
the Hamiltonian of the system, modifying the neutrino oscillation
probabilities.  
There are few expressions of the approximate probability in matter~\cite{Nunokawa:2007qh,DMP:2016,DMP:2018,IP:2018}
and one of them can be written as~\cite{Nunokawa:2007qh}: 
%\small
%\footnotesize
\begin{align}
P(\nupbar_\mu\rightarrow \nupbar_e)\nonumber \approx
%P(\nu_{\mu}(\bar{\nu}_{\mu})\rightarrow\nu_{e}(\bar{\nu}_{e})) \approx
& \sin^{2}\!\theta_{23} \sin^{2}\!2\theta_{13} \frac{\sin^{2}(\Delta_{31}\mp aL)}{(\Delta_{31} \mp aL)^2} \Delta_{31}^{2} \nonumber \\
& + \sin\!2\theta_{23} \sin\!2\theta_{13} \sin\!2\theta_{12} \cos\theta_{13} \frac{\sin(\Delta_{31} \mp aL)}{(\Delta_{31} \mp aL)} \Delta_{31} \frac{\sin(aL)}{aL} \Delta_{21} \cos(\Delta_{32}) \cos\!\delta \nonumber \\
& \mp \sin\!2\theta_{23} \sin\!2\theta_{13} \sin\!2\theta_{12} \cos\theta_{13} \frac{\sin(\Delta_{31}\mp aL)}{(\Delta_{31}\mp aL)} \Delta_{31} \frac{\sin(aL)}{aL} \Delta_{21} \sin(\Delta_{32}) \sin\!\delta \nonumber \\
& + \cos^{2}\!\theta_{13} \cos^2\!\theta_{23}\;\sin^{2}\!2\theta_{12} \frac{\sin^{2}(aL)}{(aL)^{2}} \Delta_{21}^{2}.
\label{eqn:osc_prob_matter}
\end{align}
%\normalsize
\noindent
%Here, $\Delta_{21}=\frac{\Delta m^{2}_{21}L}{4E}$ and
%$\Delta_{31}=\frac{\Delta m^{2}_{31}L}{4E}$.
The matter effect depends on $a=G_{F}N_{e}/\sqrt{2}$, where $G_{F}$ is
Fermi's constant and $N_{e}$ is the number density of electrons in the
matter.  The sign of the $aL$ terms flip for antineutrinos,
introducing an effect that can mimic CP violation for some
experimental configurations.

From here, note that the amplitude of the first, second and third terms
are dependent on the ratio:
\begin{equation}
\frac{\Delta_{31}}{\Delta_{31} \mp aL} = \big(1 \mp A\big)^{-1}
%\textrm{,\quad where\quad} A = aL/\Delta_{31} = 2\sqrt{2}G_F N_e E/\Delta m^2_{31}
\end{equation}
where $A= aL/\Delta_{31}= {2\sqrt{2} G_F N_e E}/{\Delta m^2_{31}}$ is
directly proportional to $E$ and inversely proportional to the signed
value of $\Delta m^2_{31}$.  Similarly, the vacuum oscillation phase
$\Delta_{31}$ in $\sin(\Delta_{31})$ is scaled by a factor of
$(1 \mp A)$.  Since $A$ is signed, this factor is less than unity for
the combination of normal ordering with neutrinos or inverted ordering
with anti-neutrinos, but greater than unity for the opposite
combinations.  As a result, both the amplitude of the appearance
probability and the location of the oscillation maxima shift depend
on the mass ordering, and can be used to determine it experimentally.

The shift in the amplitude of the probability depends only on $E$, and
this is the reason the sensitivity of first-maxima experiments
increases with $L$ and $E$ at fixed $L/E$.  Among the Korean sites,
it also means that a detector at Mt. Bisul will have greater
sensitivity to the mass ordering than other locations.

The shift in the phase is more interesting.  It also depends on energy,
but when considering a fixed value of $E$, the physical effect will
grow linearly as the baseline increases.  Thus even Korean sites at a
$2.5^\circ$ off-axis angle are more sensitive to the effect compared
to Kamioka, even though the amplitude difference is not larger.

\begin {figure}[htbp]
  \captionsetup{justification=raggedright,singlelinecheck=false}
  \centering      
  \includegraphics[width=0.8\textwidth]{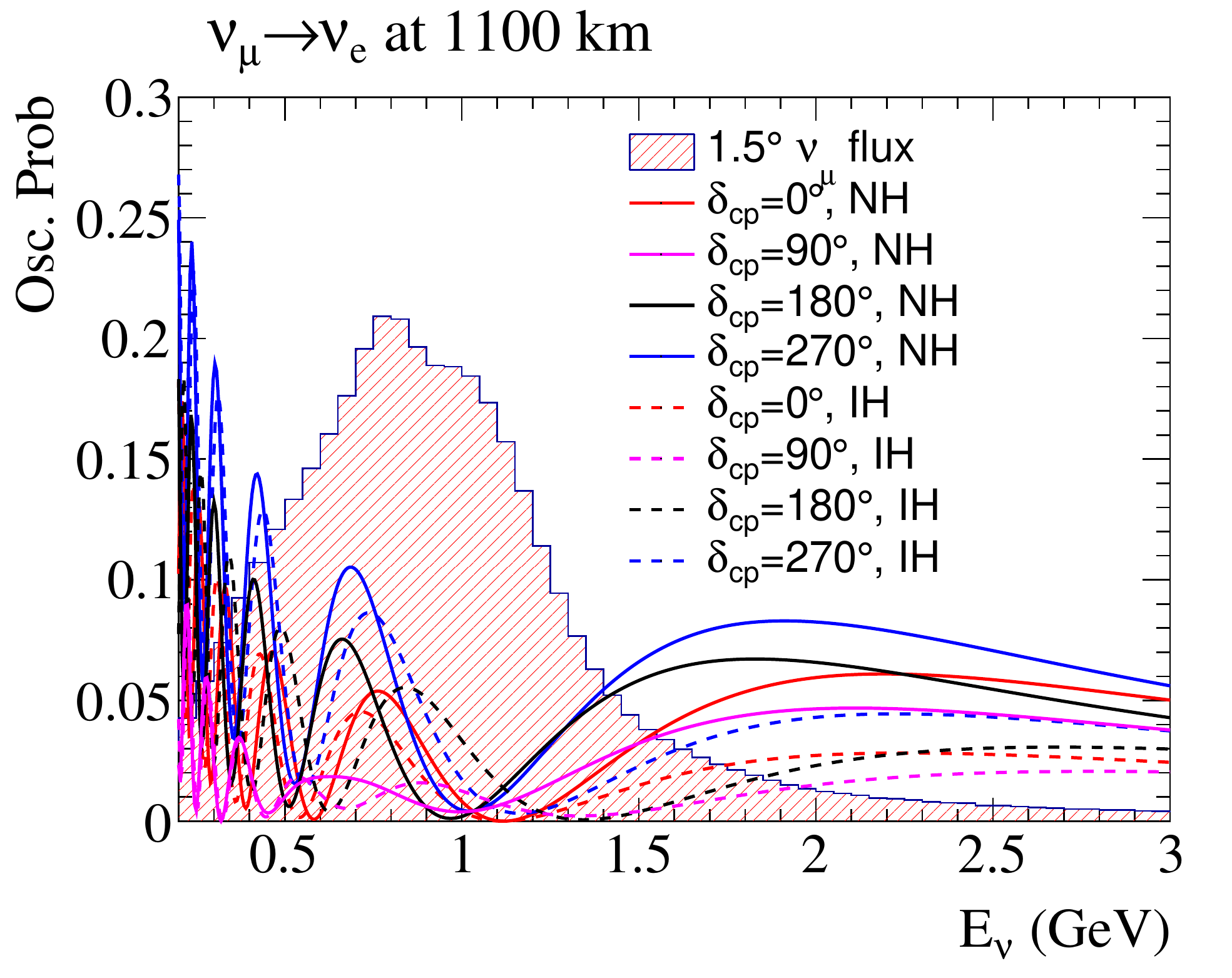}\\
  \includegraphics[width=0.8\textwidth]{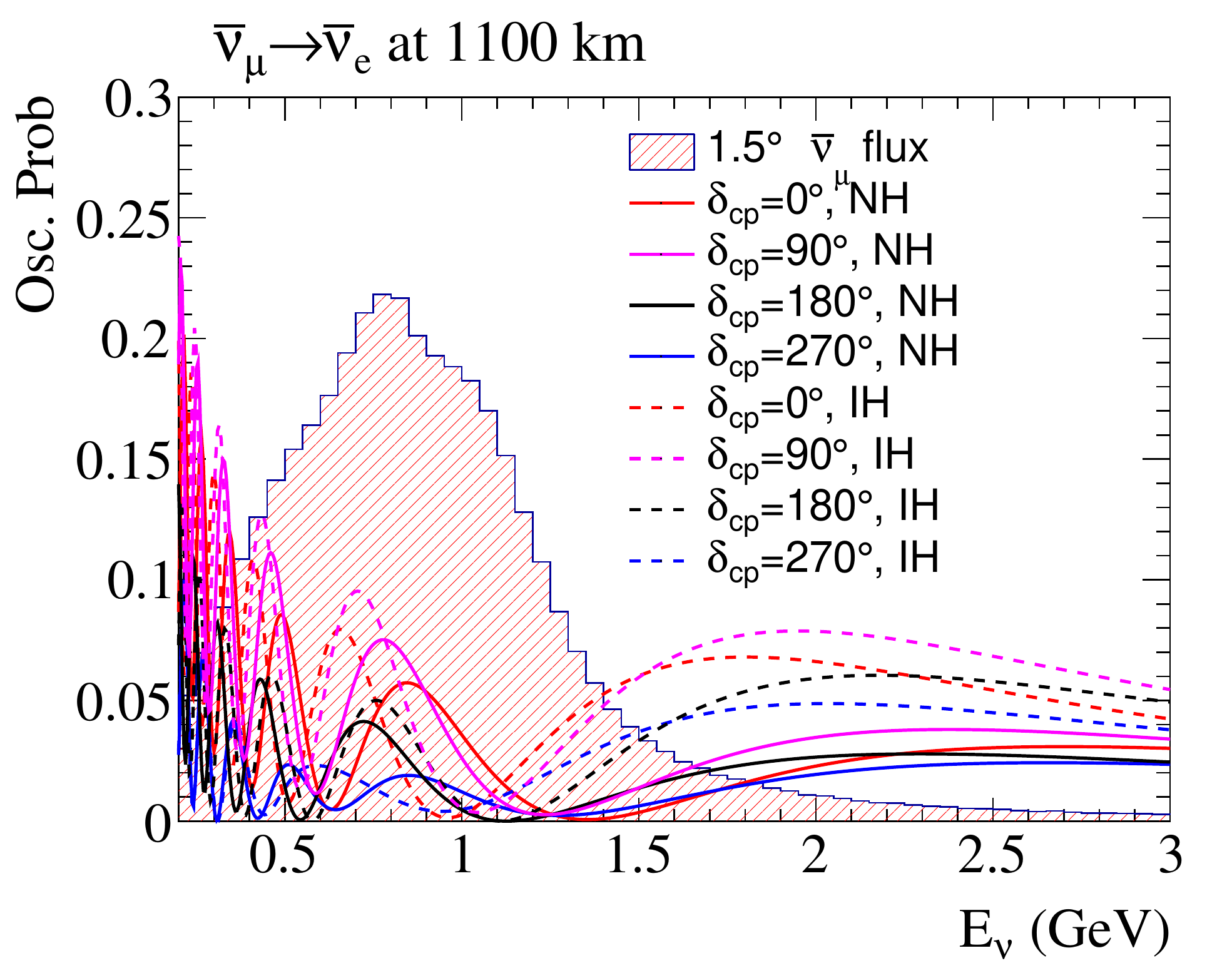}
  \caption{The oscillation probabilities for
  $\delta=0,\pi/2,\pi,3\pi/2$ and normal and inverted mass ordering
  are shown for neutrinos (top) and antineutrinos (bottom). Expected
  muon (anti)neutrino spectra at $1.5^{\circ}$ off-axis with arbitrary
  normalization are shown for comparison.}
\label{fig:osc_probs}
\end{figure}

This enhanced sensitivity to the mass ordering and $\delta_{CP}$ are
illustrated first in Fig.~\ref{fig:osc_probs} and further in the
following section.  Figure~\ref{fig:osc_probs} shows the oscillation
probabilities as a function of energy, at a baseline of $L=1100$\,km.
In the region of the first oscillation maximum above 1.2\,GeV, the
matter effect has separated the oscillation probabilities for normal
and inverted ordering for all values of the CP phase.  In the region
of the second oscillation maximum, $0.5\sim1.2$~GeV, the CP
probability differences are significant, while the matter effect also
affects the height and position of the oscillation maximum.  The
spectrum of neutrino interactions (ignoring oscillations) for a
1.5$^{\circ}$ off-axis beam is also shown for comparison.
%These suggest that with such a beam, it would
%be possible to measure the mass ordering with the high energy part of
%the neutrino spectrum at the first oscillation maximum, while
%measuring the CP phase with the second and even third oscillation
%maxima.  Other configurations are not shown, but enhance the event
%rate below 0.5\,GeV, at the cost of a reduced rate overall,
%particularly above 1.2\,GeV.

\subsubsection{Bi-probability plots}
\label{sec:biProb}
While it is relatively straightforward for the eye to interpret the
way the oscillation probabilities depend on $\delta_{CP}$ and the mass
ordering near the first oscillation maximum, e.g. above 1.2\,GeV in
Fig~\ref{fig:osc_probs}, these plots become complicated around the
second maximum, and it becomes difficult to understand the overall
pattern.  A popular alternative is the ``bi-probability'' plot~\cite{Minakata:2001qm}: 
in which the $\nu_\mu \rightarrow \nu_e$ appearance probability is
plotted along the horizontal axis, and the
$\overline{\nu}_\mu \rightarrow \overline{\nu}_e$ probability along
the vertical.  Such plots in principle show CP violation in a very
direct way---the diagonal is CP-conserving, and everywhere in the
space is CP-violating, although that includes induced CP-violation due
to matter effects.

Assuming a specific baseline and energy, the probabilities approximated
by~\eqref{eqn:osc_prob_matter}, can be calculated for various values
of the oscillation parameters.  In particular, by allowing $\delta$ to
vary with other parameters held constant, a closed ellipse is traced
out on the bi-probability plot.  The binary choice of mass orderings
can then also be displayed as a second ellipse. The center point of
the ellipse will depend on the first and fourth lines of the
probability, and the binary choice of mass orderings will (if the
matter density is non-zero) appear as a separation into two separated
ellipses, one for each ordering.

Such plots have one particular weakness when considering
long-baseline experiments: the ellipses are plotted for a fixed
energy.  An experiment which measures oscillations
over a range of energies is summarised by a single value.  This
single-energy simplification is reasonable when comparing two
experiments at the first oscillation maximum, where the
oscillation is typically rather broad in comparison to the flux. But
it does not represent the situation in T2HKK well, as this
configuration samples a much wider range of $L/E$. 

In order to use these plots for showing the more complex oscillation patterns
at the T2HKK it is necessary to show multiple neutrino energies.  It
is not sufficient to average the probability over the interaction
spectrum; this is equivalent to doing a rate-only measurement, and
suppresses all the shape information that a real analysis would use.
Instead in the following plots three curves are shown: one (in blue)
for the peak interaction rate, as normal, and two more from the upper
(red) and lower (green) tails of the spectrum, chosen so that 50\% of
the (potentially) interacting neutrinos lie between these two
ellipses.  (The exact method by which these points are calculated is
described in Appendix~\ref{sec:biProb_appendix}.) This results in a
set of six representative ellipses instead of two.
\begin{figure}[ht]
  \centering
  \includegraphics[width=0.49\textwidth]{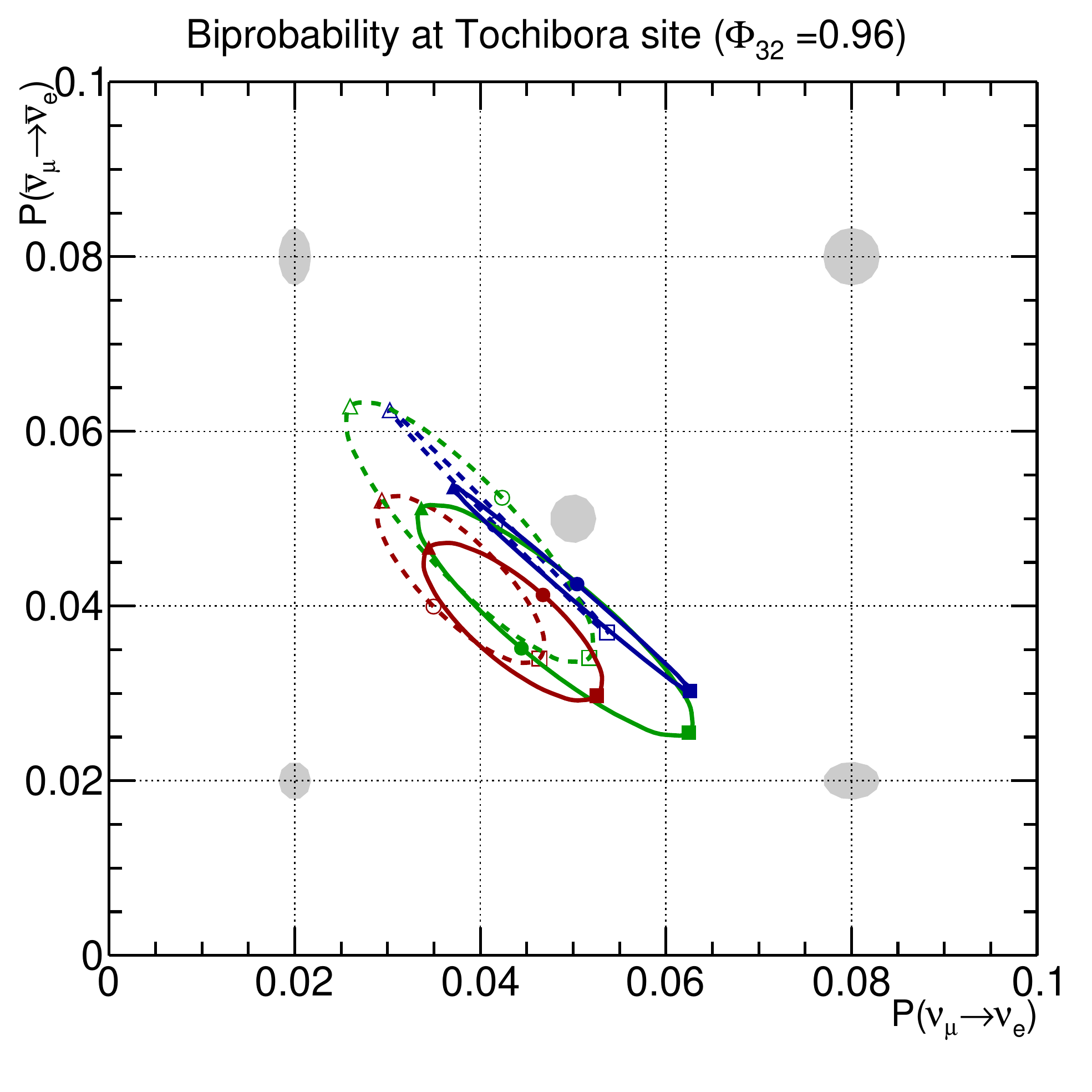}
  \includegraphics[width=0.49\textwidth]{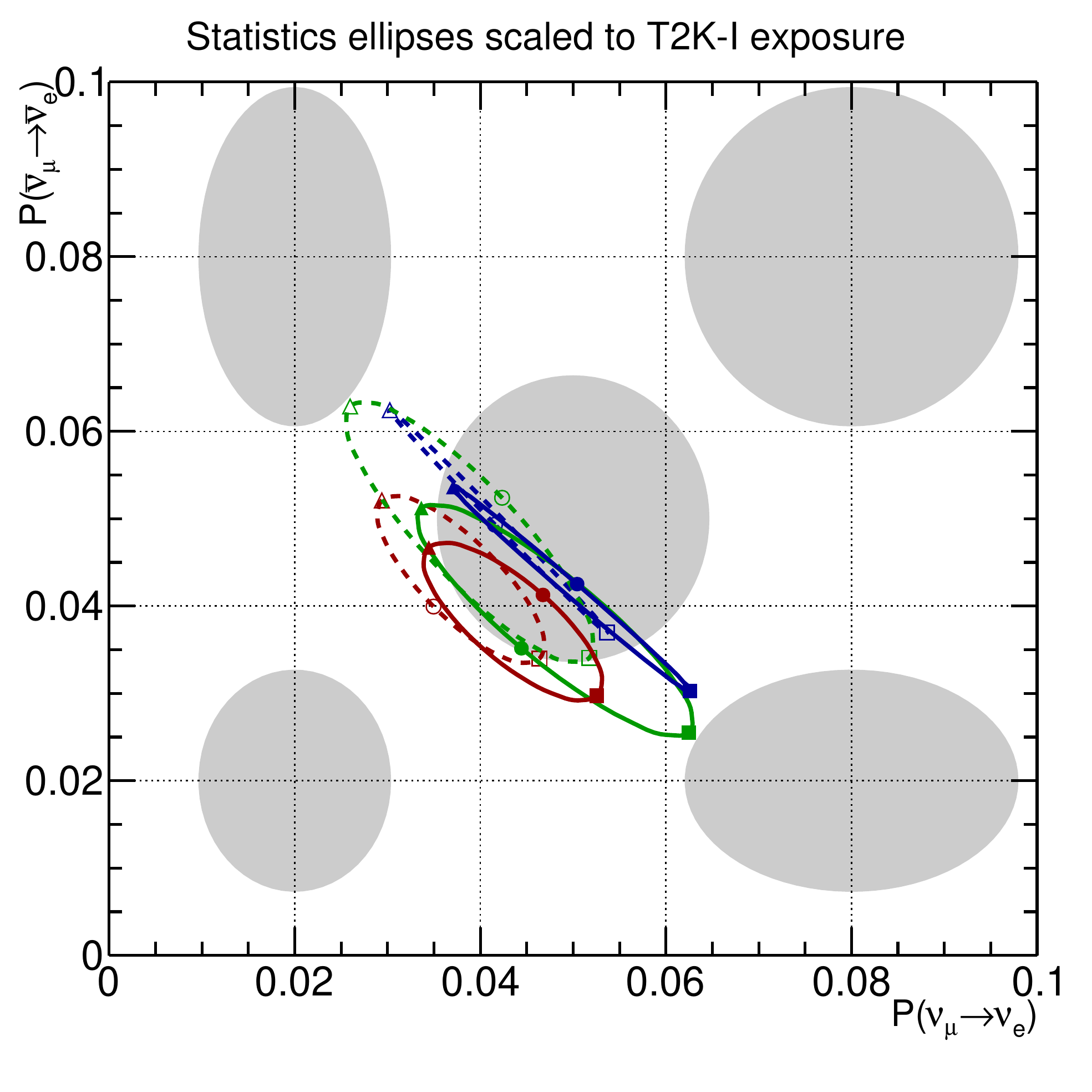}
  \begin{minipage}{0.98\textwidth}
  \centering
  %Legend\\
  \begin{tabular}{c@{\quad}l@{\qquad}c@{\quad}l@{\qquad}c@{\quad}l}
  \textcolor{bp_green}{\rule[.5ex]{3em}{1.5pt}} & 520\,MeV   
    & \rule[.5ex]{3em}{1.5pt} &  $\Delta m^2_{31} > 0$
    &\scalebox{1.5}{$\bullet$}\quad\scalebox{1.55}{$\circ$} & $\delta = 0$\\

  \textcolor{bp_blue}{\rule[.5ex]{3em}{1.5pt}} & 620\,MeV
    &\rule[.5ex]{0.6em}{1.5pt}\rule[.5ex]{0.6em}{0pt}\rule[.5ex]{0.6em}{1.5pt}\rule[.5ex]{0.6em}{0pt}\rule[.5ex]{0.6em}{1.5pt} & $\Delta m^2_{31} < 0$
    &\scalebox{1.2}{$\blacktriangle$}\quad\scalebox{1.2}{$\vartriangle$}& $\delta = \pi/2$\\

  \textcolor{bp_red}{\rule[.5ex]{3em}{1.5pt}} & 770\,MeV   
    &&
    & $\blacksquare\quad\square$ & $\delta = -\pi/2$\\%[0.4ex]

\end{tabular}
\end{minipage}
\caption{Appearance bi-probabilities at the Hyper-K site in Kamioka. Grey ellipses show the relative sensitivity (left) for a ten year exposure of one Hyper-K tank and (right) for the original T2K design goal. Full explanation is given in the text. 
  %The green, blue, and red ellipses are representative energies of 520, 620, and 770\,MeV, respectively.
}
\label{f:bp_tochibora}
\end{figure}

Figure~\ref{f:bp_tochibora}, shows the resulting ellipses for the
Kamioka site. The three ellipses differ in size and eccentricity, but
the separation between the two mass orderings, and the dependence of
the appearance probabilities on the value of $sin(\delta)$ is similar
for all three energies. The similarity of the three pairs of ellipses
explains why the narrow-band, first-maximum configuration can be
reasonably approximated as a single (`rate-only') measurement.

Two more details are added to the plots. The first is a summary of the
$L/E$ value probed (at the peak energy) at the given site.  The quantity
\begin{equation}
  \Phi_{32} = \frac{2}{\pi}\frac{\left|\Delta m^2_{32}\right|L}{4E}
\end{equation}
is defined, which has (near) odd-integer values for oscillation
maxima, and (near) even-integer values for oscillation minima.  For
calculation of $\Phi_{32}$ and drawing all bi-probability plots, a
value of $2.5\times10^{-3}$\,eV$\vphantom{V}^2$ is assumed for
$\left|\Delta m^2_{32}\right|$.

The second detail is the grey ellipses, which provide a comparative
scale for the statistical power of the measurement: this depends on
the number of events seen with energy similar to each colored ellipse.
Since this will depend on the $\nu_e$ appearance probabilities and the
relative size of the background contamination, this varies across the
measurement space. Since this is essentially unrelated to how the
analysis is actually performed, it's not possible to accurately
indicate the absolute sensitivity, but does give an indication of the
relative sensitivity at different sites.  Full details are in
Appendix~\ref{sec:biProb_appendix}.

Since the flux at the Hyper-K site near Kamioka should be the same as
the existing T2K experiment, the benefit of a larger detector and
upgraded beam is entirely due to the higher statistics.  In terms of
the bi-probability plots, the probability ellipses remain the same,
but the statistical error ellipses shrink.  This is illustrated by the
comparison of left and right panels of Fig~\ref{f:bp_tochibora}
showing that Hyper-K will be much more sensitive than the design goal
of T2K-I ($7.8\times10^{21}\,\mathrm{POT}$) by virtue of much smaller
statistical errors.  In contrast, at the Korean sites the statistical
uncertainties are about three times larger than the default Hyper-K
configuration, but they gain sensitivity due to the fact the
probability ellipses change drastically.

\begin{figure}[ht]
  \centering
  \includegraphics[width=0.59\textwidth]{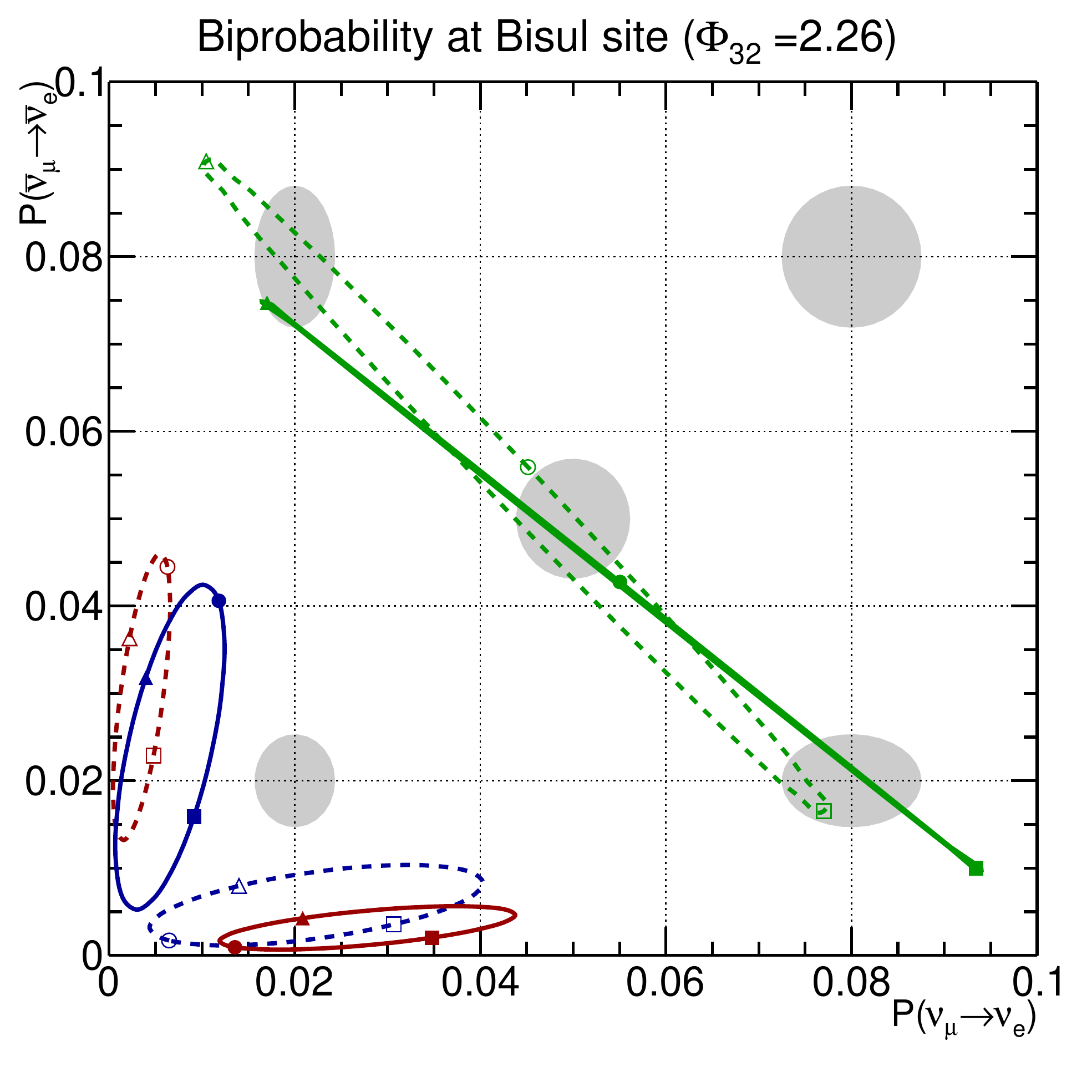}
  \begin{minipage}{0.39\textwidth}
  \centering
  \vspace{-1.59\textwidth}
  ~\\ %Legend\\
  \begin{tabular}{c@{\quad}l}
  \textcolor{bp_green}{\rule[.5ex]{3em}{1.5pt}} & 740\,MeV \\   
  \textcolor{bp_blue}{\rule[.5ex]{3em}{1.5pt}} & 970\,MeV \\   
  \textcolor{bp_red}{\rule[.5ex]{3em}{1.5pt}} & 1300\,MeV \\[3.0ex]   
  \rule[.5ex]{3em}{1.5pt} &  $\Delta m^2_{31} > 0$ \\%[0.4ex]
  \rule[.5ex]{0.6em}{1.5pt}\rule[.5ex]{0.6em}{0pt}\rule[.5ex]{0.6em}{1.5pt}\rule[.5ex]{0.6em}{0pt}\rule[.5ex]{0.6em}{1.5pt} & $\Delta m^2_{31} < 0$\\[3.0ex]
  \scalebox{1.5}{$\bullet$}\quad\scalebox{1.55}{$\circ$} & $\delta = 0$ \\%[0.4ex]
  \scalebox{1.2}{$\blacktriangle$}\quad\scalebox{1.2}{$\vartriangle$}& $\delta = \pi/2$\\%[0.4ex]
  $\blacksquare\quad\square$ & $\delta = -\pi/2$\\%[0.4ex]
  \end{tabular}
  \end{minipage}
  \caption{Appearance bi-probabilities at the Mt. Bisul site at 1088\,km.
  %The green, blue, and red ellipses are representative energies of 740, 970, and 1300\,MeV, respectively.
}
\label{f:bp_bisul}
\end{figure}

The first example shown is for the most on-axis site under
consideration---Mt. Bisul---in Fig.~\ref{f:bp_bisul}. A detector at
Bisul would see completely separated ellipses depending on the mass
ordering, i.e. the measurement is completely non-degenerate with CP
violation for events at the peak, and in the high-energy tail.  This
reflects the benefit of high energy neutrinos for resolving the mass
ordering.  At the peak energy however, the normal ordering enhances
the \emph{antineutrino} probabilities relative to neutrinos, which is
opposite to the amplitude-dominated effect seen at the first maximum.
This is due to the matter effect also changing the effective
oscillation length for neutrinos and antineutrinos, so that the
neutrinos are closer to an appearance minimum than are the
antineutrinos of the same energy.  Unlike a first maximum experiment,
the experimental signature determining the mass ordering uses shape as
well as rate information, and the impact of systematic uncertainties
(most obviously on neutrino vs antineutrino cross sections and
detection efficiencies) is correspondingly different. Although most
pronounced at the higher beam energies of the Bisul site, this
characteristic is seen for some energies at any Korean site.

\begin{figure}[ht]
\centering
\includegraphics[width=0.59\textwidth]{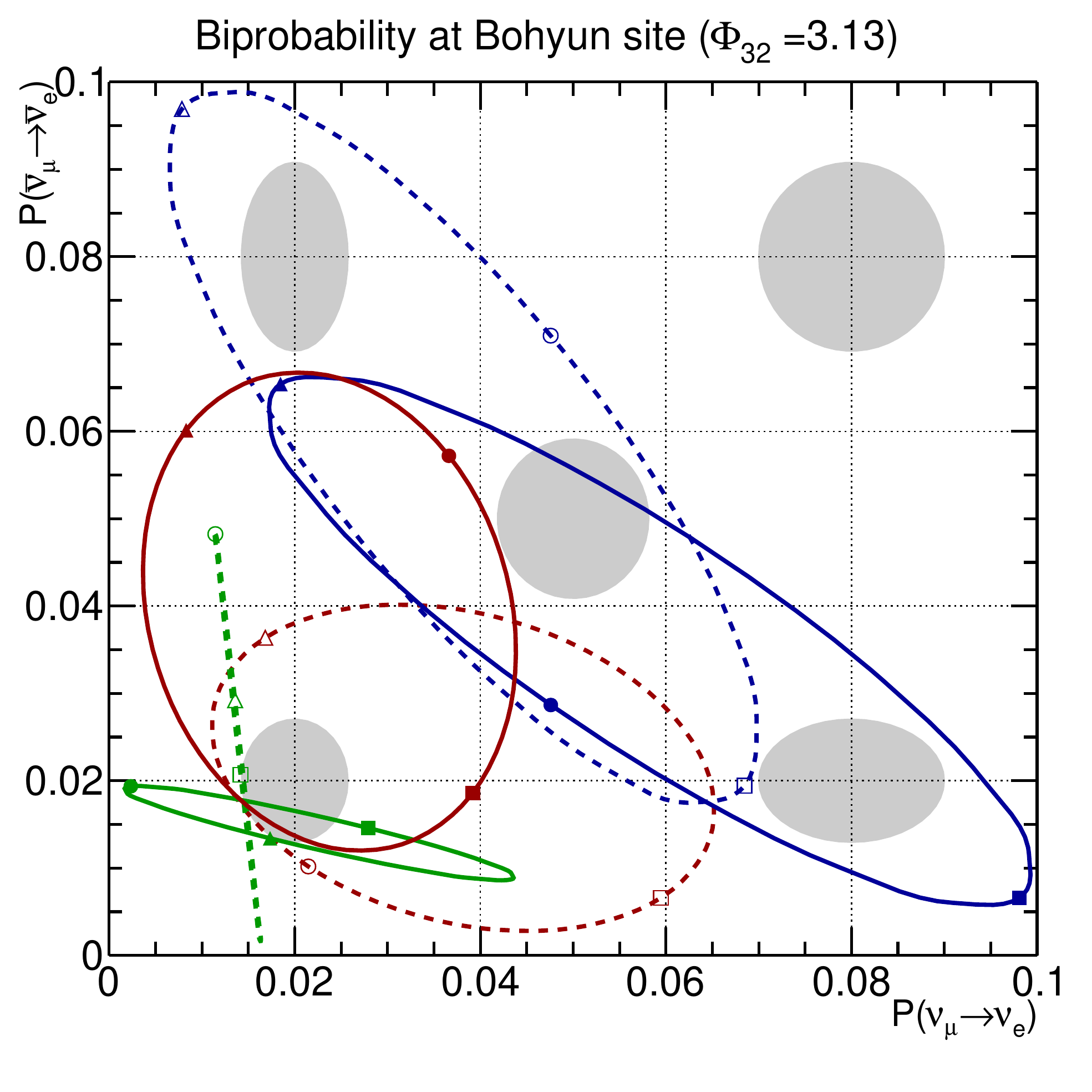}
\begin{minipage}{0.39\textwidth}
\centering
\vspace{-1.59\textwidth}
 ~\\ %Legend\\
\begin{tabular}{c@{\quad}l}
\textcolor{bp_green}{\rule[.5ex]{3em}{1.5pt}} & 550\,MeV \\   
\textcolor{bp_blue}{\rule[.5ex]{3em}{1.5pt}} & 670\,MeV \\   
\textcolor{bp_red}{\rule[.5ex]{3em}{1.5pt}} & 830\,MeV \\[3.0ex]   
\rule[.5ex]{3em}{1.5pt} &  $\Delta m^2_{31} > 0$ \\%[0.4ex]
\rule[.5ex]{0.6em}{1.5pt}\rule[.5ex]{0.6em}{0pt}\rule[.5ex]{0.6em}{1.5pt}\rule[.5ex]{0.6em}{0pt}\rule[.5ex]{0.6em}{1.5pt} & $\Delta m^2_{31} < 0$\\[3.0ex]
\scalebox{1.5}{$\bullet$}\quad\scalebox{1.55}{$\circ$} & $\delta = 0$ \\%[0.4ex]
\scalebox{1.2}{$\blacktriangle$}\quad\scalebox{1.2}{$\vartriangle$}& $\delta = \pi/2$\\%[0.4ex]
$\blacksquare\quad\square$ & $\delta = -\pi/2$\\%[0.4ex]
\end{tabular}
\end{minipage}
\caption{Appearance bi-probabilities at the Mt. Bohyun site at 1043\,km.
  %The green, blue, and red ellipses are representative energies of 550, 670, and 830\,MeV, respectively.
}
\label{f:bp_bohyun}
\end{figure}

The low energy tail of the Bisul site corresponds to an $(L,E)$ regime
more similar to the other Korean sites. Figure~ \ref{f:bp_bohyun}
shows the Bohyun site, which has the shortest baseline of the sites
under consideration, and a flux more similar to that at Kamioka.  It
is immediately apparent (as it was for the lower-energy ellipse at
Bisul) that the ellipses are much larger, demonstrating a greatly
enhanced CP-violation signal.  For the energy peak at Bohyun (which is
close to the oscillation maximum), the overlap between normal and
inverted ordering ellipses is still large, but the effect for the
higher and lower energy ellipses is quite different, so information
provided by the spectrum of observed events is less degenerate, and
correspondingly more interesting, than it is for first-maxima
experiments.  The larger effect size almost entirely compensates for
the lower statistics compared to T2HK at Kamioka, but the absolute
scale of detector systematic errors does \emph{not} grow for a more
distant site, so systematic errors are expected to be much less
important.

\begin{figure}[ht]
\centering
\includegraphics[width=0.59\textwidth]{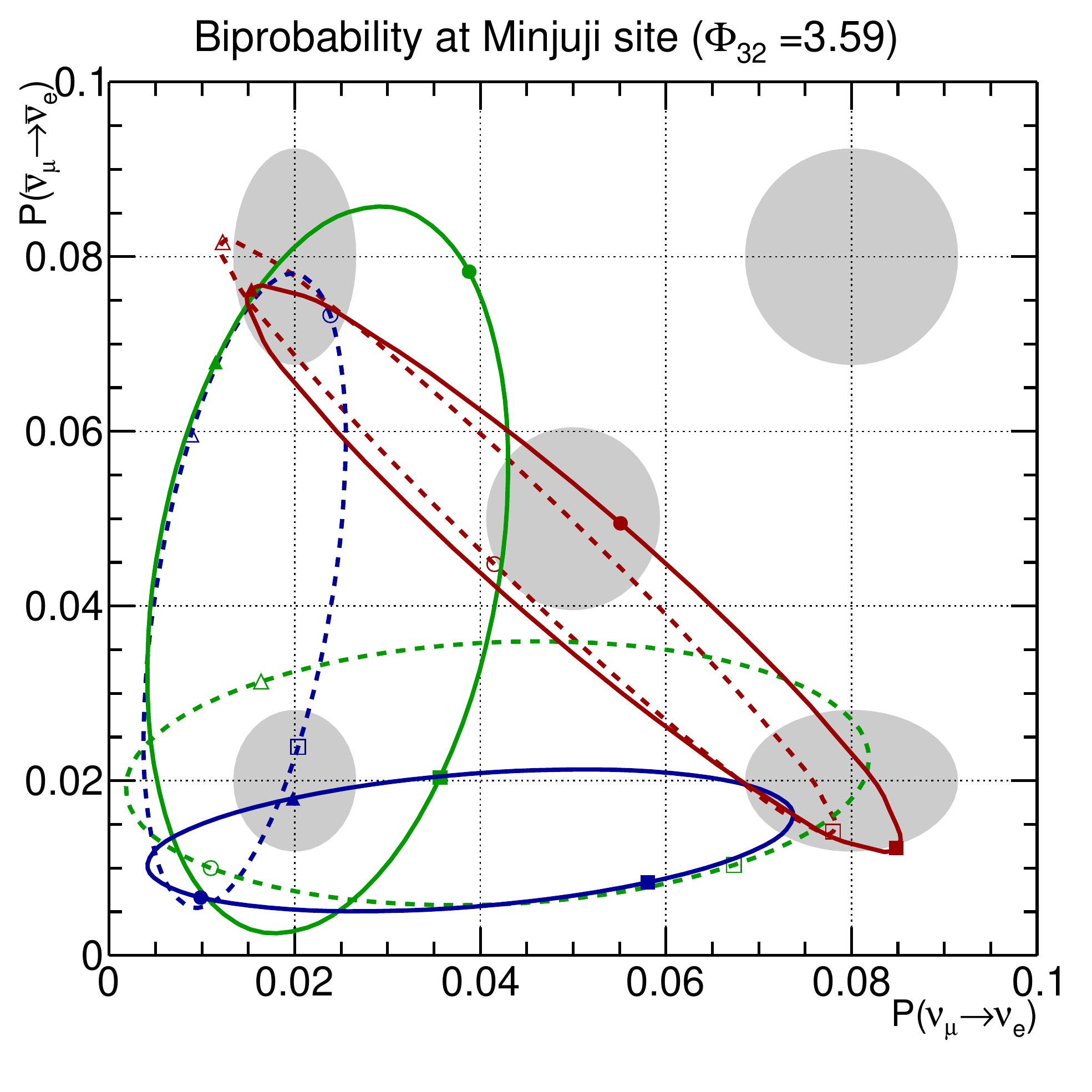}
\begin{minipage}{0.39\textwidth}
\centering
\vspace{-1.59\textwidth}
 ~\\ %Legend\\
\begin{tabular}{c@{\quad}l}
\textcolor{bp_green}{\rule[.5ex]{3em}{1.5pt}} & 530\,MeV \\   
\textcolor{bp_blue}{\rule[.5ex]{3em}{1.5pt}} & 640\,MeV \\   
\textcolor{bp_red}{\rule[.5ex]{3em}{1.5pt}} & 800\,MeV \\[3.0ex]   
\rule[.5ex]{3em}{1.5pt} &  $\Delta m^2_{31} > 0$ \\%[0.4ex]
\rule[.5ex]{0.6em}{1.5pt}\rule[.5ex]{0.6em}{0pt}\rule[.5ex]{0.6em}{1.5pt}\rule[.5ex]{0.6em}{0pt}\rule[.5ex]{0.6em}{1.5pt} & $\Delta m^2_{31} < 0$\\[3.0ex]
\scalebox{1.5}{$\bullet$}\quad\scalebox{1.55}{$\circ$} & $\delta = 0$ \\%[0.4ex]
\scalebox{1.2}{$\blacktriangle$}\quad\scalebox{1.2}{$\vartriangle$}& $\delta = \pi/2$\\%[0.4ex]
$\blacksquare\quad\square$ & $\delta = -\pi/2$\\%[0.4ex]
\end{tabular}
\end{minipage}
\caption{Appearance bi-probabilities at the Mt. Minjuji site at 1145\,km.
  %The green, blue, and red ellipses are representative energies of 530, 640, and 800\,MeV, respectively.
}
\label{f:bp_minjuji}
\end{figure}

Finally, Fig.~\ref{f:bp_minjuji} shows the Minjuji site, which probes
the largest values of $L/E$.  For Minjuji, the low-energy tail is
actually probing the region near the third oscillation maximum. The
independence of information available at different energies is again
clearly apparent or the peak energy and below the ellipses are
noticeably `fatter'. More precisely, we can say that $dP/d{\delta}$ is
larger in the vicinity of $\delta = \pm\pi/2$.  Narrow ellipses are
characteristic of a energy band near the oscillation maximum, where
the dependence on $\delta$ is dominated by a $\sin\delta$ dependence.
This is optimal for making a measurement of $\delta$ near ${0, \pi}$, and
therefore for establishing (or not) CP violation.  However, if the
true value of $\delta$ is in the vicinity of $\pm\pi/2$, this provides
the least precision on the value.  The wider $L/E$ band of a Korean
detector improves the sensitivity to $\cos\delta$ and the measurement
precision on the CP phase in large CP-violation scenarios.

However, we reiterate that although these kind of plots are useful for understanding the merits of a Korean detector, they are largely unrelated to how the actual analysis is done: using binned energy spectra; multiple backgrounds and finite energy resolution.  To determine real sensitivities, more detailed studies must be done with full simulations.  Such studies are presented in the following sections.

%======================================================================
%%%=======================================================================================
\graphicspath{{sensitivity/plots}}
%======================================================================
\section{\label{sec:sense} Improved Neutrino Mass Ordering and CP Sensitivities }
This section describes the sensitivity to measure the neutrino mass
ordering and discover CP violation using a configuration of Hyper-K
with one tank in Japan and the second tank in Korea.  Expected
reconstructed event spectra for the Korean detector are presented and
the effect of the oscillation parameters on these spectra are
considered.  A model of systematic uncertainties is added, and
studies of the resulting sensitivity for the mass ordering
measurement, CP violation discovery, and precision of the CP phase
measurement are presented.

\subsection{Event rates at Korean detectors} \label{sec:event_rates}
For the purpose of the sensitivity studies presented here, we consider generic detector locations in South Korea at a baseline of 1100~km and an off-axis angle of 1.5$^{\circ}$,  2.0$^{\circ}$
or 2.5$^{\circ}$.  The expected event rates are estimated by using the NEUT~\cite{Hayato:2009zz} 5.3.2 neutrino interaction generator and a GEANT3-based simulation of the Super-K detector, 
where the fiducial mass has been scaled from 22.5~kton to 187~kton.  The
simulated events are scaled to give good agreement with NEUT 5.1.4.2, which has been tuned against T2K near detector data.  Following the running plan of Hyper-K, an exposure of 
(1.3~MW)$\times$($10\times10^{7}$~sec) is assumed with a 3:1 ratio of antineutrino mode to neutrino mode operation. This corresponds to 10 years of Hyper-K operation, or $27\times10^{21}$ protons on target.
 Oscillation probabilities are calculated using Prob3++~\cite{prob3++}, and a constant matter density of 
3.0~g/cm$^{3}$ is assumed for the 1100 km baseline~\cite{Hagiwara:2011kw}. Unless otherwise specified, simulated event rates are calculated with the oscillation parameters shown in Table~\ref{tab:osc_param}.  
When fitting, the parameters sin$^{2}\theta_{13}$, sin$^{2}\theta_{12}$ and $\Delta m^{2}_{21}$ are constrained by Gaussian constraint terms with the 1$\sigma$ uncertainties shown in Table~\ref{tab:osc_param}, which 
are extracted from the Particle Data Group (PDG) 2016 Review of Particle Physics~\cite{Patrignani:2016xqp}.
For each detector configuration, reconstructed events are classified in 4 categories:
\begin{itemize}
\item Neutrino mode, 1R$e$: Single electron-like ring candidates collected in the neutrino mode operation of the beam.
\item Antineutrino mode, 1R$e$: Single electron-like ring candidates collected in the antineutrino mode operation of the beam.
\item Neutrino mode, 1R$\mu$: Single muon-like ring candidates collected in the neutrino mode operation of the beam.
\item Antineutrino mode, 1R$\mu$: Single muon-like ring candidates collected in the antineutrino mode operation of the beam.
\end{itemize}
The selection cuts for these candidate samples are identical to the selection cuts used in recent T2K oscillation measurements~\cite{Abe:2015awa},
except for the reconstructed energy, $E_{rec}<$1.25~GeV cut on the 1R$e$ samples.  This cut has been removed since the matter effect which constrains the 
mass ordering is most strongly manifested in events with reconstructed energy greater than 1.25~GeV.

\begin{table}[tbp]
\captionsetup{justification=raggedright,singlelinecheck=false}
\caption{\label{tab:osc_param}
The default oscillation parameter values and prior uncertainties used in the studies presented in this section.
}
\begin{center}%
\scalebox{1.0}{
\begin{tabular}{l|c|c} \hline \hline
Parameter & Value & Prior Error \\ \hline
$\delta_{cp}$ & 0 & No Prior Constraint \\
$\Delta m^{2}_{32}$ & $2.5\times10^{-3}$ eV & No Prior Constraint \\
sin$^{2}\theta_{23}$ & 0.5 & No Prior Constraint \\
$\Delta m^{2}_{21}$ & $7.53\times10^{-5}$ eV & $0.18\times10^{-5}$ eV \\
sin$^{2}\theta_{12}$ & 0.304 & 0.041 \\
sin$^{2}\theta_{13}$ & 0.0219 & 0.0012 \\
\hline \hline
\end{tabular}%
}
\end{center}
\end{table}%

\begin {figure}[htbp]
\captionsetup{justification=raggedright,singlelinecheck=false}
  \begin{center}
    \includegraphics[width=0.49\textwidth]{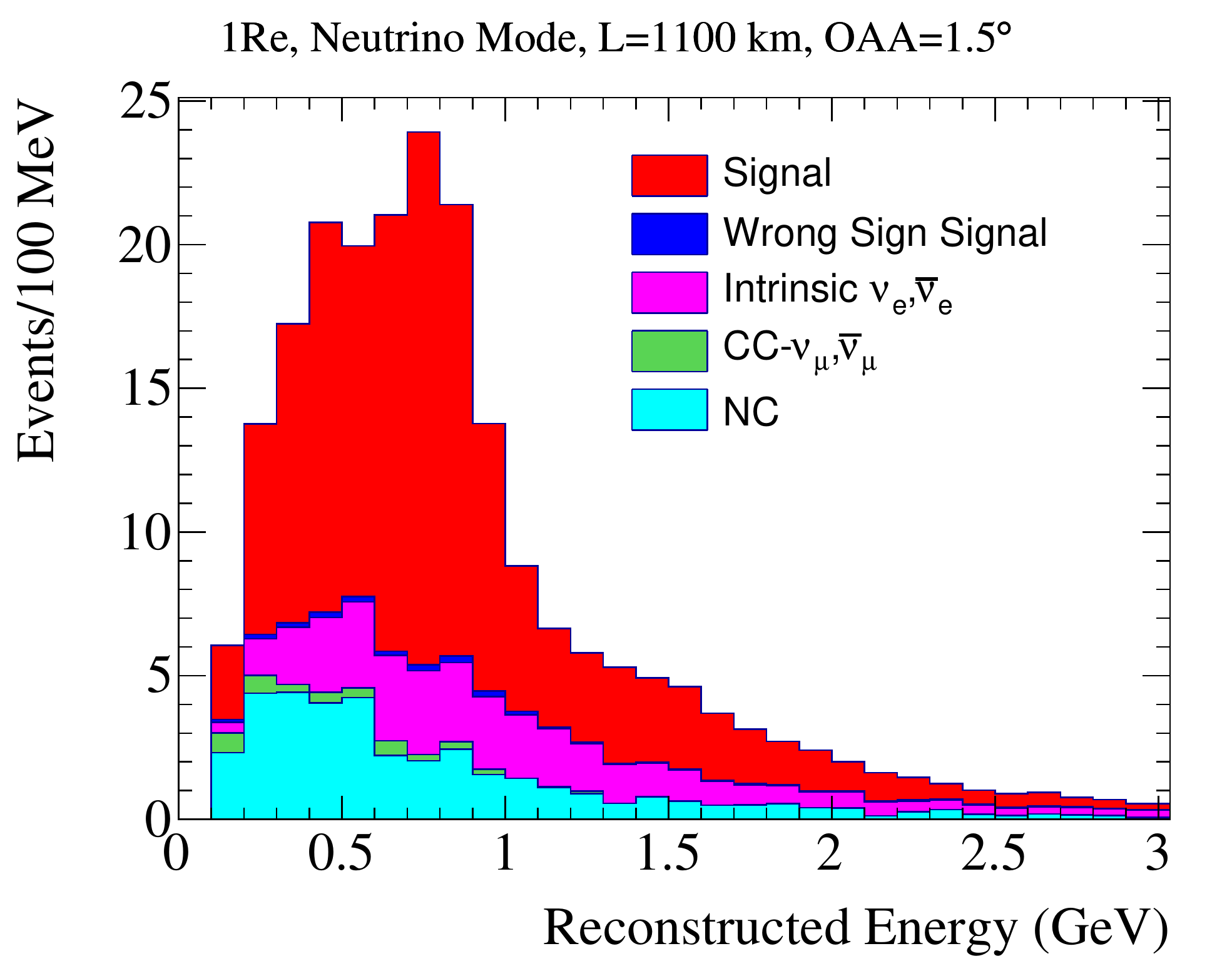}
    \includegraphics[width=0.49\textwidth]{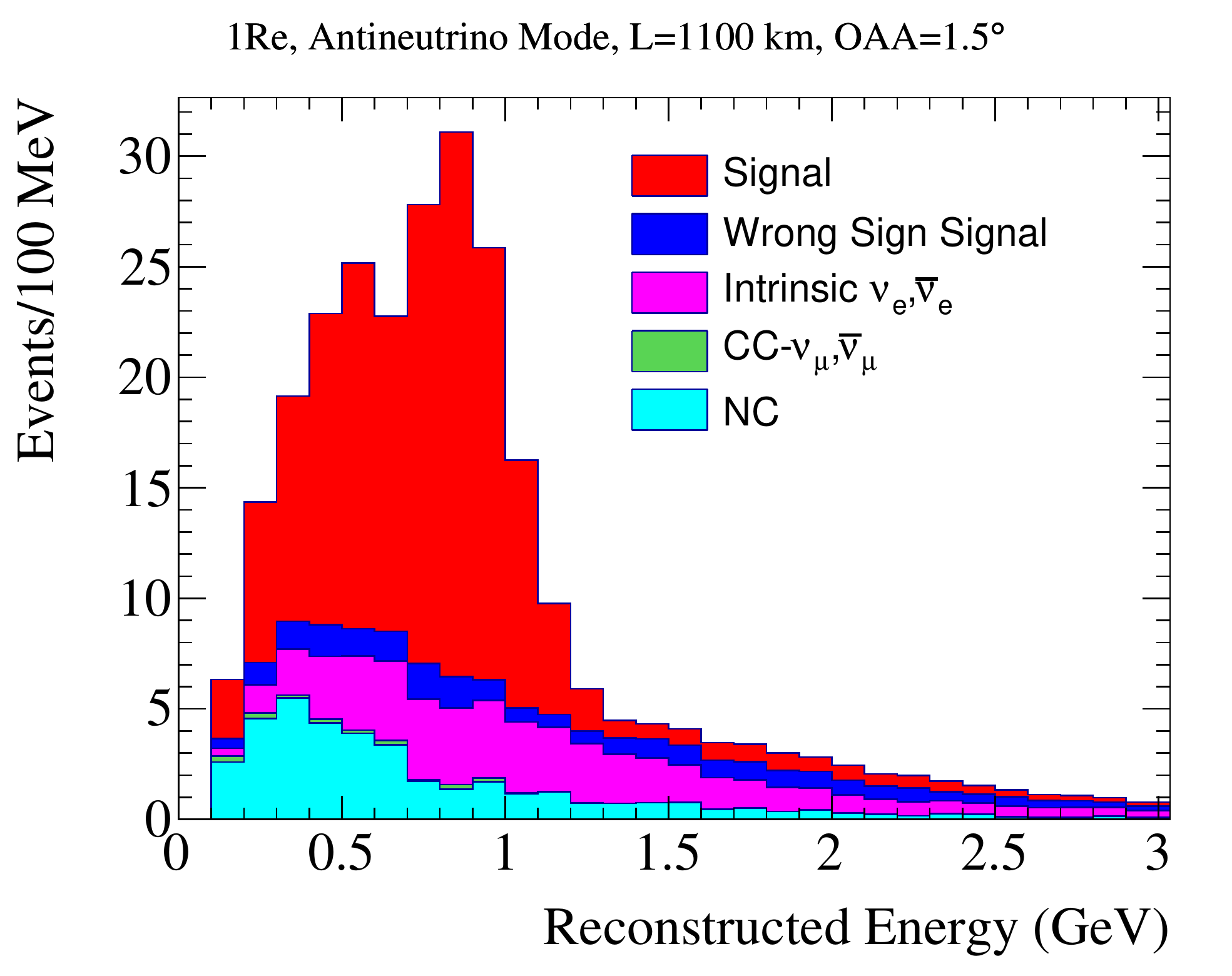}\\
    \includegraphics[width=0.49\textwidth]{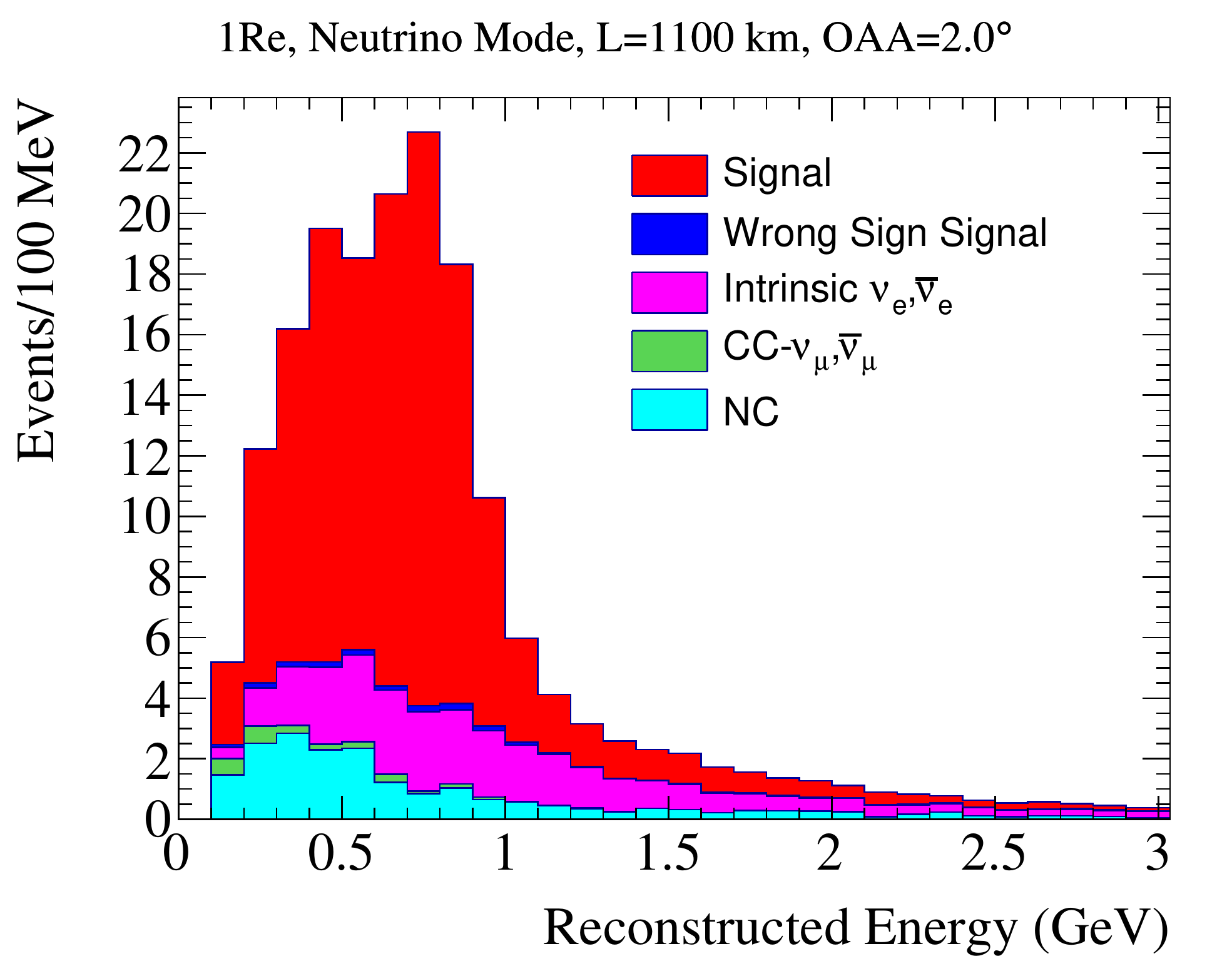}
    \includegraphics[width=0.49\textwidth]{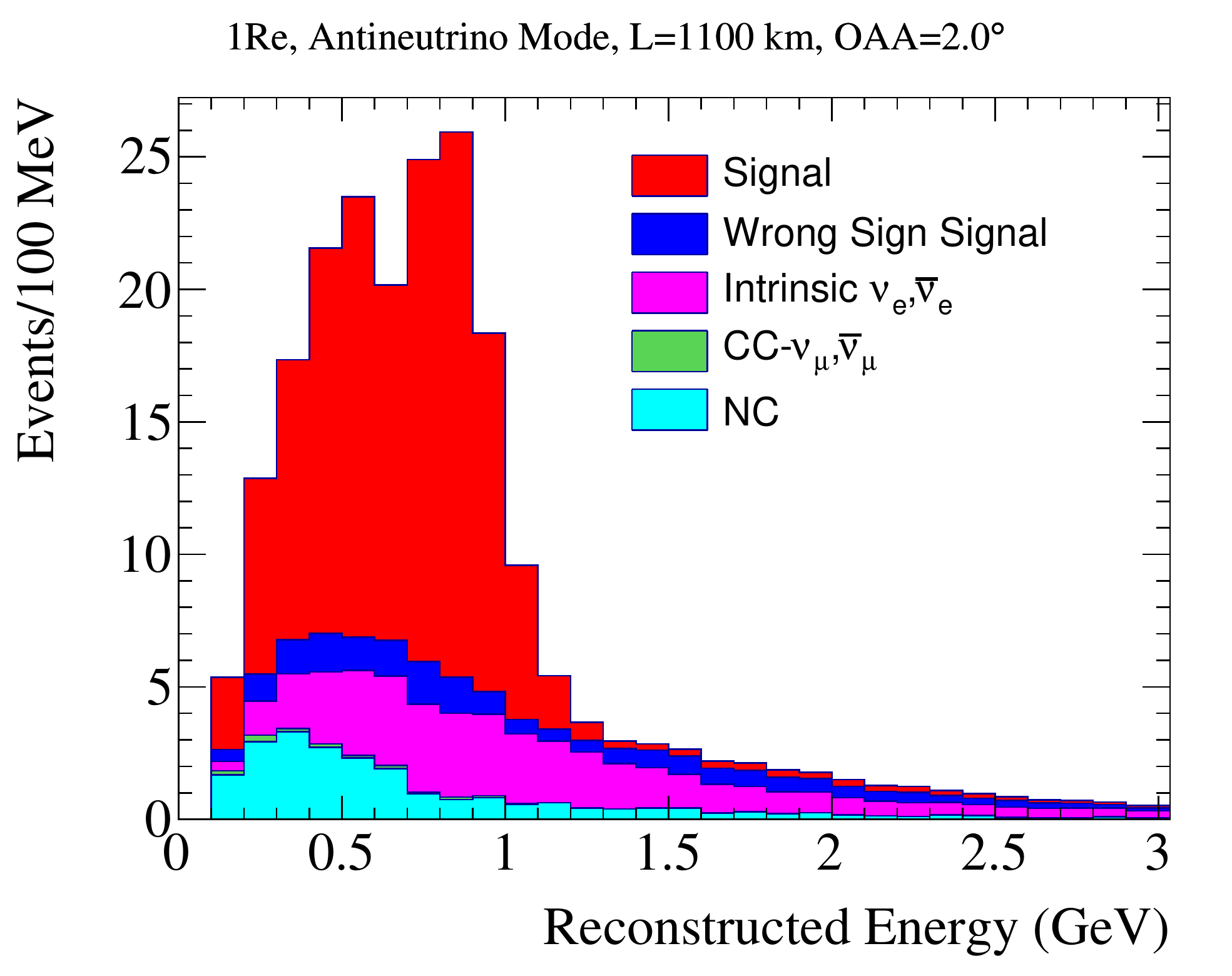}\\
    \includegraphics[width=0.49\textwidth]{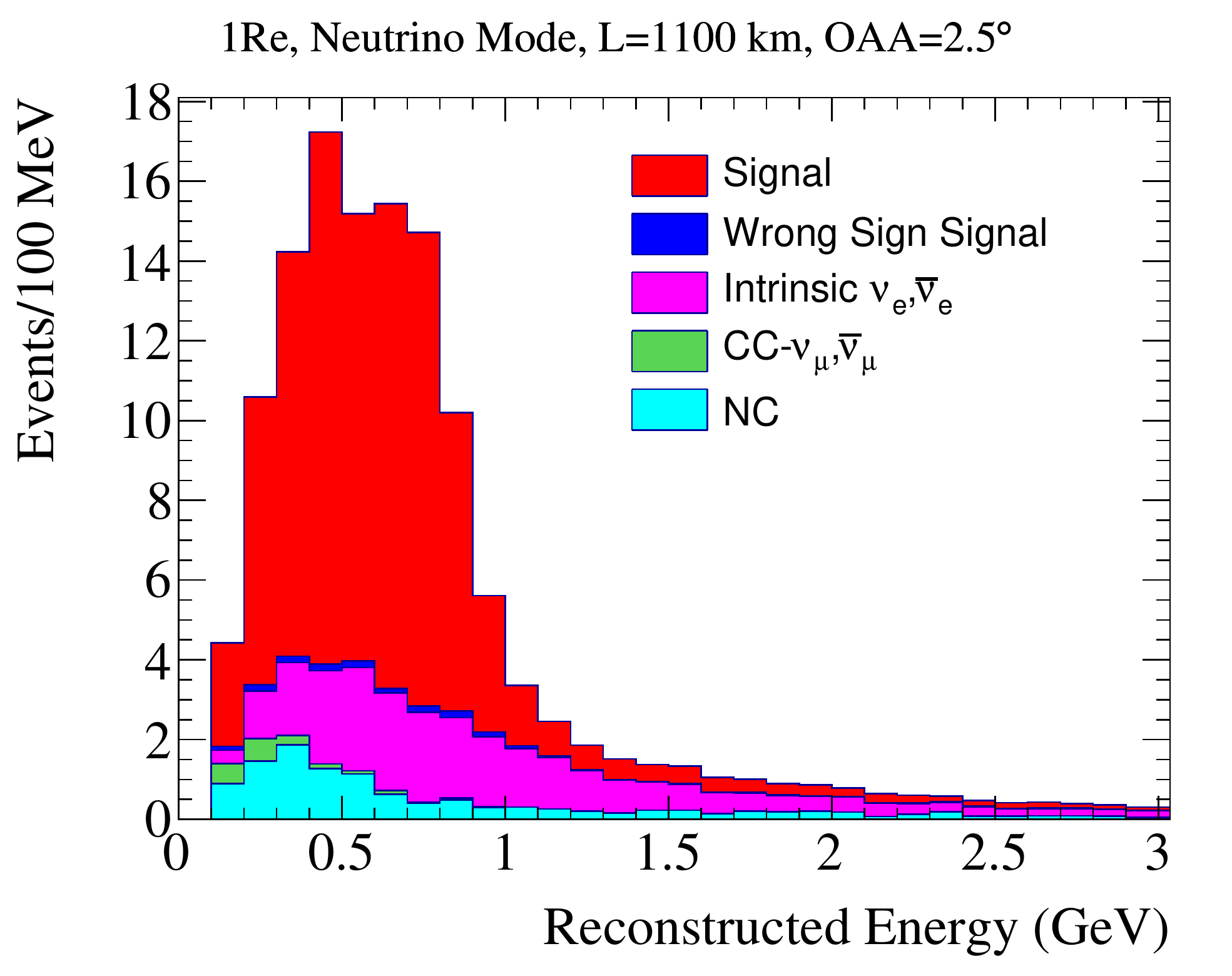}
    \includegraphics[width=0.49\textwidth]{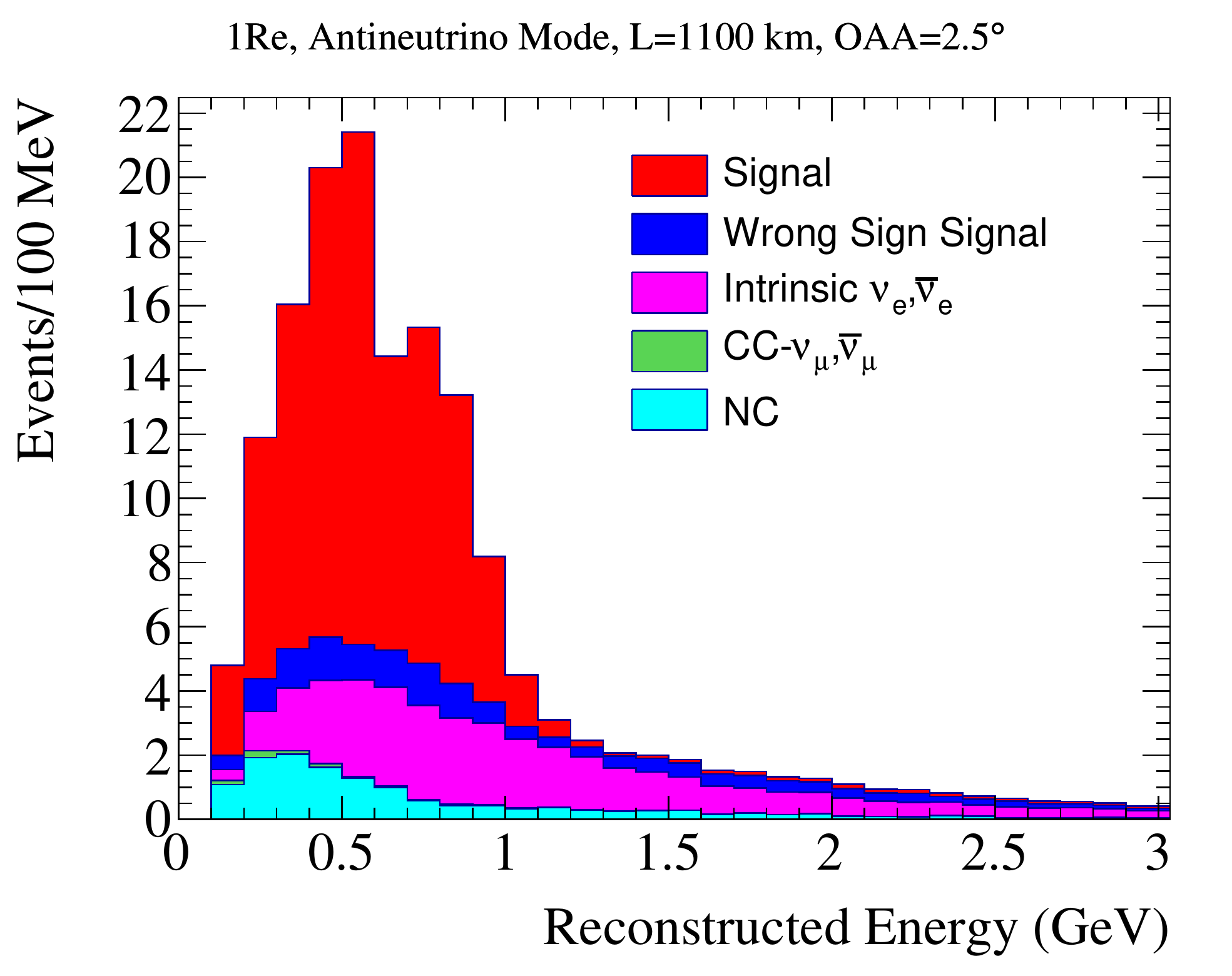}
    \caption{Predicted 1R$e$ candidate rates for neutrino mode (left) and antineutrino mode (right) with the detector at a 1.5$^{\circ}$ (top), 2.0$^{\circ}$ (middle) or
             2.5$^{\circ}$ (bottom) off-axis angle.  The oscillation parameters are set to $\delta_{cp}$=0, $\Delta m^{2}_{32}=2.5\times10^{-3}$~eV$^{2}$ (normal mass ordering), sin$^{2}\theta_{23}$=0.5,
              sin$^{2}\theta_{13}$=0.0219.}
    \label{fig:evt_rates_1Re}
  \end{center}
\end {figure}

\begin {figure}[htbp]
\captionsetup{justification=raggedright,singlelinecheck=false}
  \begin{center}
    \includegraphics[width=0.49\textwidth]{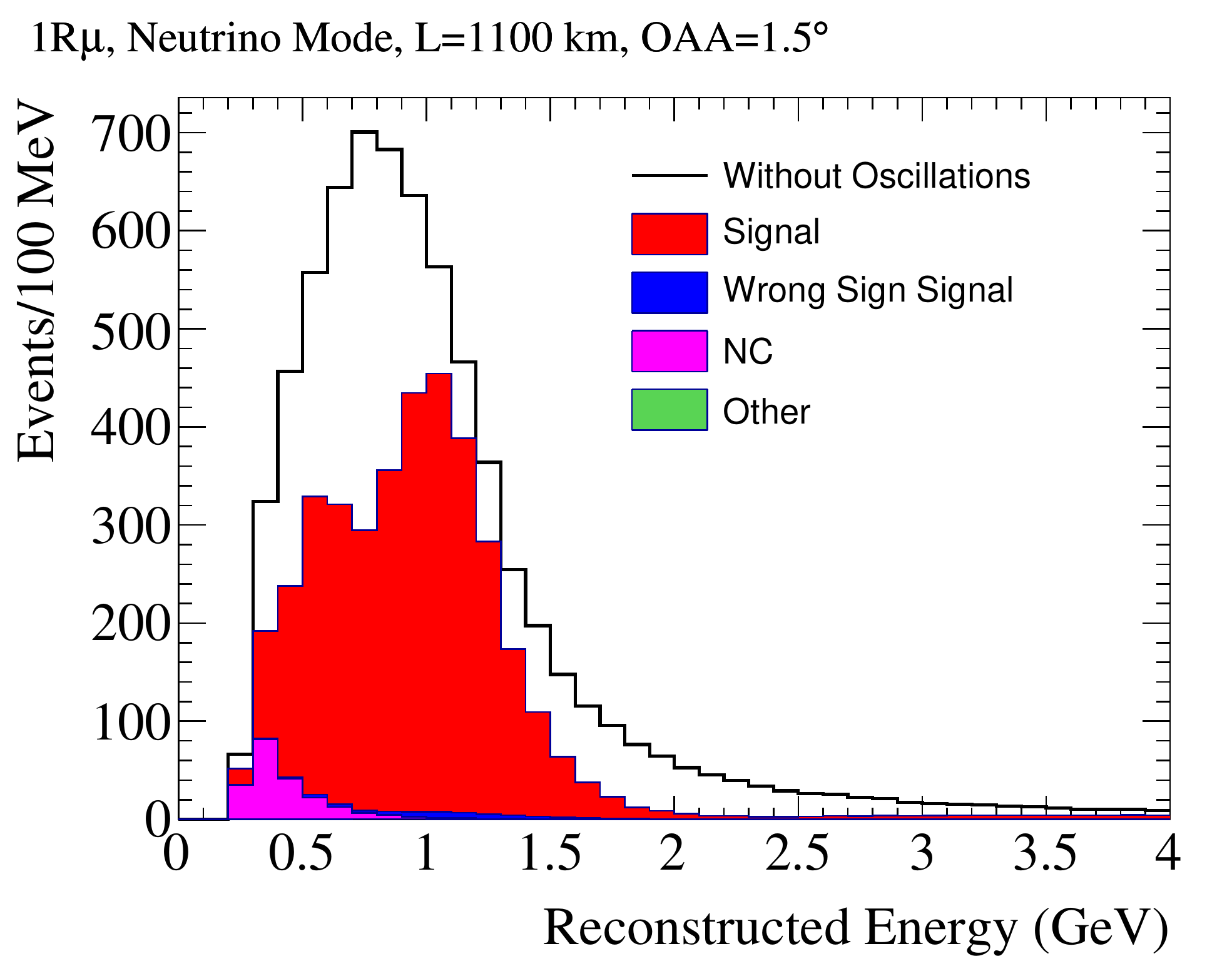}
    \includegraphics[width=0.49\textwidth]{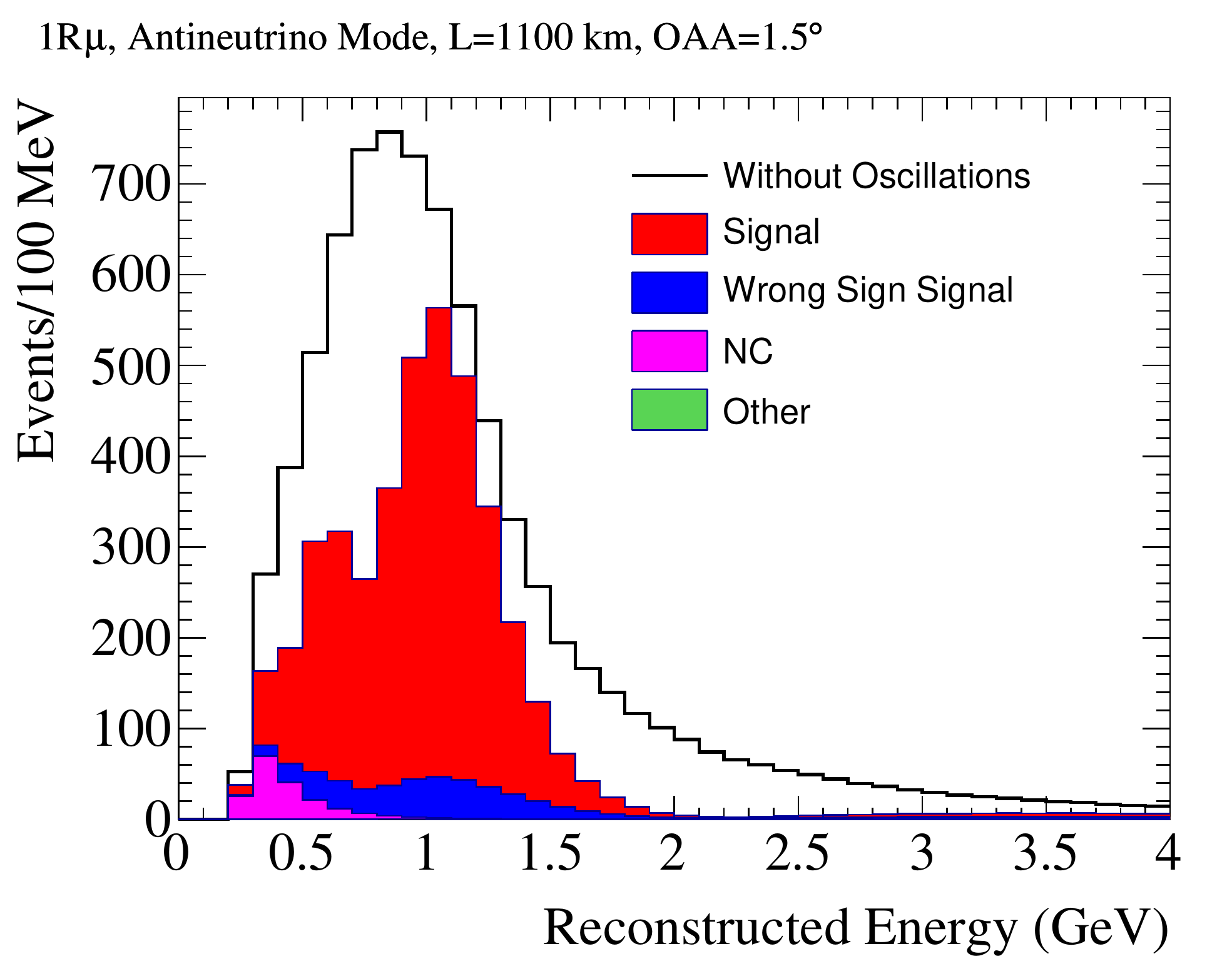}\\
    \includegraphics[width=0.49\textwidth]{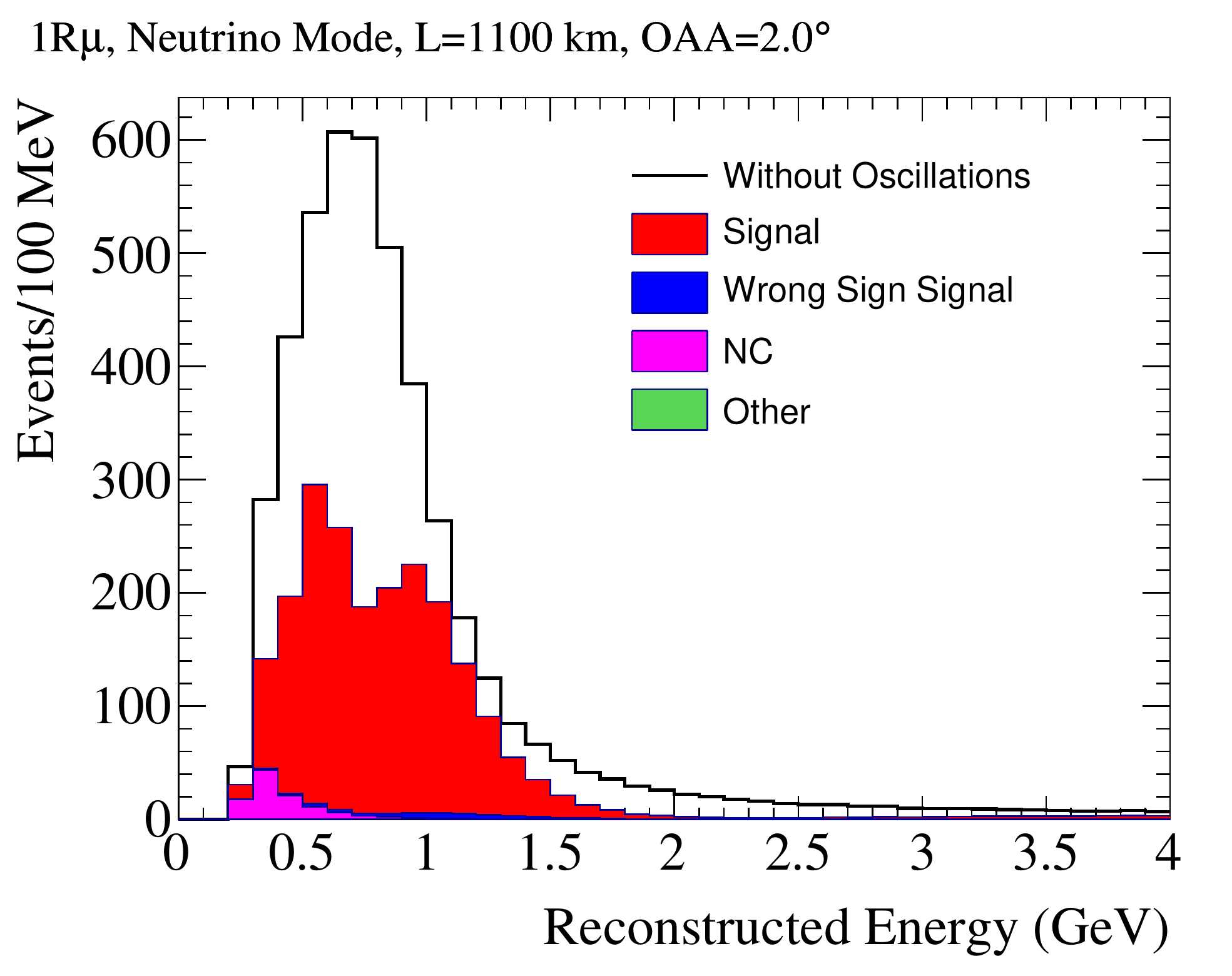}
    \includegraphics[width=0.49\textwidth]{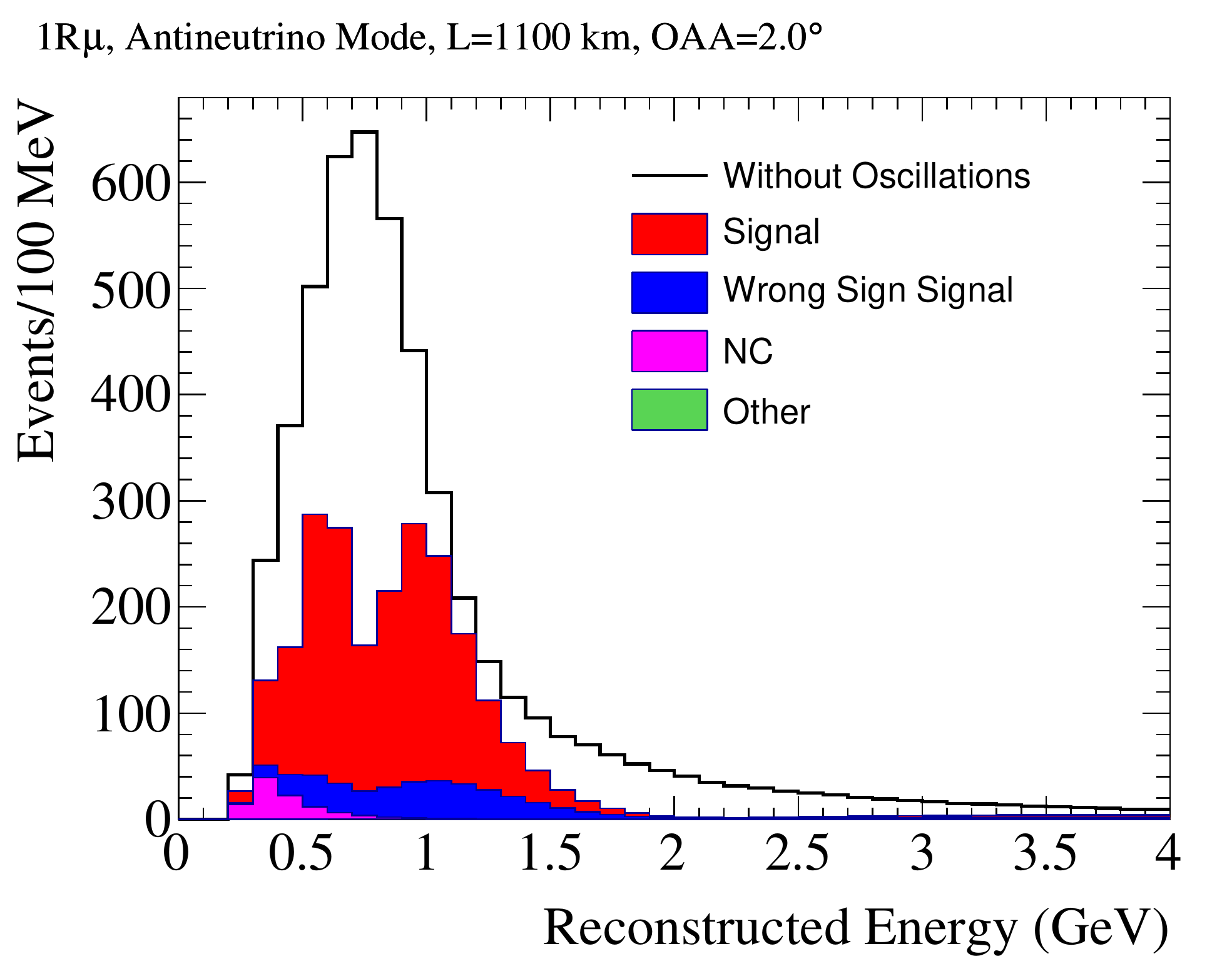}\\
    \includegraphics[width=0.49\textwidth]{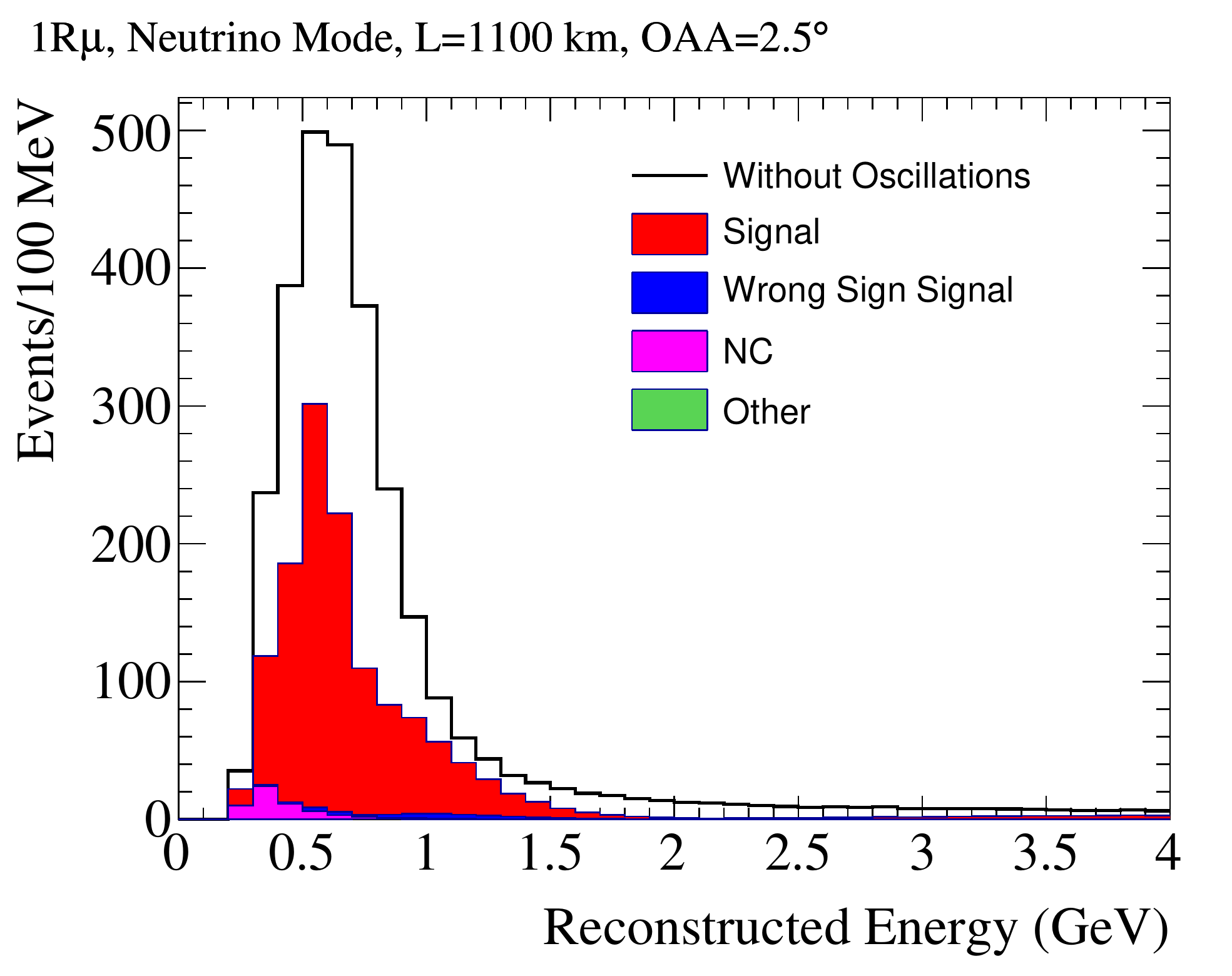}
    \includegraphics[width=0.49\textwidth]{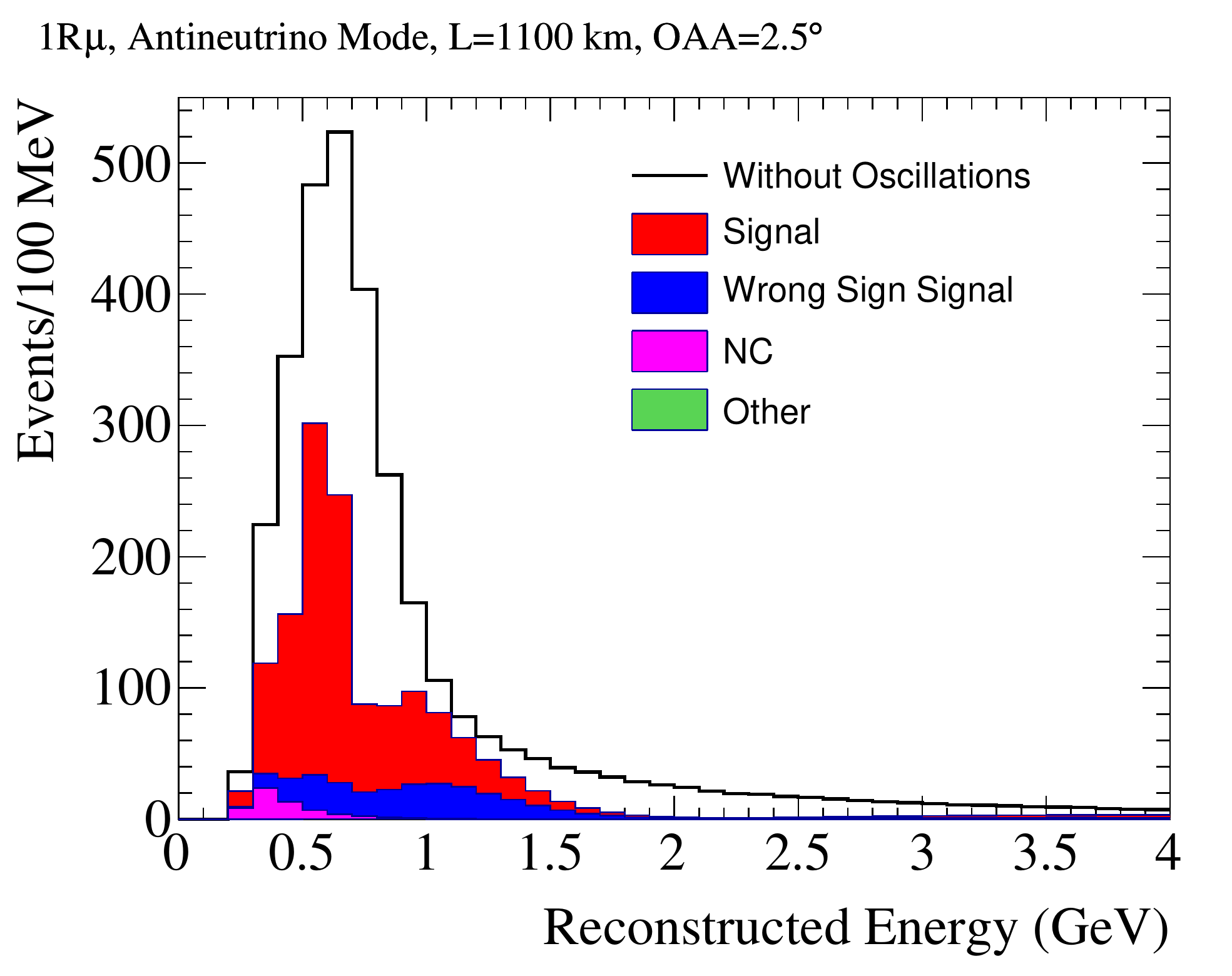}
    \caption{Predicted 1R$\mu$ candidate rates for neutrino mode (left) and antineutrino mode (right) with the detector at a 1.5$^{\circ}$ (top), 2.0$^{\circ}$ (middle) or
             2.5$^{\circ}$ (bottom) off-axis angle.  The oscillation parameters are set to $\delta_{cp}$=0, $\Delta m^{2}_{32}=2.5\times10^{-3}$~eV$^{2}$ (normal mass ordering), sin$^{2}\theta_{23}$=0.5,
              sin$^{2}\theta_{13}$=0.0219.}
    \label{fig:evt_rates_1Rmu}
  \end{center}
\end {figure}

Predicted event rates for normal mass ordering and $\delta_{cp}$=0 are shown for 1Re and 1R$\mu$ samples in Fig.~\ref{fig:evt_rates_1Re}/Table~\ref{tab:evts_1Re} and 
Fig.~\ref{fig:evt_rates_1Rmu}/Table~\ref{tab:evts_1Rmu} respectively.  In Tables~\ref{tab:evts_1Re},~\ref{tab:evts_1Rmu}, the predicted event rates for the
nominal Hyper-K tank location are shown for comparison.  
%These differ from those presented in the Hyper-K Design Report since the value for $\sin^{2}\theta_{13}$ 
%has been updated to the 2015 PDG value, a new version of NEUT is used for the neutrino interaction generation, and a 10 year exposure with one tank is presented here,
%while the Design Report assumes a 10 year exposure with one tank for the first 6 years and a second tank for the remaining 4 years.   
The 1R$e$ candidate rates in 
Korea are $\sim1/10$ the rates at the 295~km baseline due to the $1/L^{2}$ dependence 
of the flux.  In the 1R$\mu$ samples, the first and second oscillation maxima can be observed at 2~GeV and 0.7~GeV respectively. 

\begin{table}[tbp]
\captionsetup{justification=raggedright,singlelinecheck=false}
\caption{\label{tab:evts_1Re}
The expected number of $\nu_{e}$ and $\bar{\nu}_{e}$ 1R$e$ candidate events.
Normal mass ordering with $\sin^2\theta_{13}=0.0219$ and $\delta_{cp}=0$ are assumed.  Background is
categorized by the flavor before oscillation. Signal are 1R$e$ candidates produced by $\nu_{e}$ in 
neutrino mode operation and $\bar{\nu}_{e}$ in antineutrino mode operation.  Wrong-sign Signal are
1R$e$ candidates produce by $\bar{\nu}_{e}$ in neutrino mode operation and $\nu_{e}$ in antineutrino
mode operation. }
\begin{center}%
\scalebox{1.0}{
\begin{tabular}{l|c|c|c|c|c|c} \hline \hline
Detector Location &   Signal   & Wrong-sign Signal & Intrinsic $\nu_{e}$, $\bar{\nu}_{e}$ & NC & CC $\nu_{\mu}$,$\bar{\nu}_{\mu}$ & Total \\ \hline
OAA,  L &\multicolumn{5}{c}{Neutrino Mode} \\ \hline
2.5$^{\circ}$, $295$~km & 1426.1 & 15.4 & 269.3 & 125.0 & 7.1 & 1842.9 \\ 
2.5$^{\circ}$, $1100$~km & 87.9 & 1.7 & 28.3 & 12.5 & 1.7 & 132.2 \\ 
2.0$^{\circ}$, $1100$~km & 122.6 & 2.0 & 33.8 & 21.4 & 2.4 & 182.3 \\ 
1.5$^{\circ}$, $1100$~km & 140.6 & 2.4 & 39.1 & 39.1 & 3.7 & 224.8 \\ \hline
OAA,  L &\multicolumn{5}{c}{Antineutrino Mode} \\ \hline
2.5$^{\circ}$, $295$~km & 1053.1 & 164.3 & 338.3 & 153.5 & 4.2 & 1713.4 \\ 
2.5$^{\circ}$, $1100$~km & 89.8 & 15.5 & 39.4 & 14.3 & 0.8 & 159.8 \\ 
2.0$^{\circ}$, $1100$~km & 131.5 & 19.8 & 46.3 & 23.4 & 1.1 & 222.1 \\ 
1.5$^{\circ}$, $1100$~km & 159.1 & 23.9 & 54.3 & 39.5 & 1.7 & 278.5 \\ 
\hline \hline
\end{tabular}%
}
\end{center}
\end{table}%

\begin{table}[tbp]
\captionsetup{justification=raggedright,singlelinecheck=false}
\caption{\label{tab:evts_1Rmu}
The expected number of $\nu_{\mu}$ and $\bar{\nu}_{\mu}$ 1R$\mu$ candidate events.
Normal mass ordering with
$\sin^2\theta_{23}=0.5$, $\Delta m^{2}_{32}=2.5\times10^{-3}$~eV$^{2}$ and sin$^2\theta_{13}$=0.0219 are assumed.}
\begin{center}%
\scalebox{1.0}{
\begin{tabular}{l|c|c|c|c|c} \hline \hline
Detector Location &   Signal   & Wrong-sign Signal & NC & CC-$\nu_{e}$,$\bar{\nu}_{e}$ & Total \\ \hline
OAA,  L &\multicolumn{5}{c}{Neutrino Mode} \\ \hline
2.5$^{\circ}$, $295$~km & 9062.5 & 571.2 & 813.6 & 29.5 & 10476.9 \\ 
2.5$^{\circ}$, $1100$~km & 1275.0 & 32.7 & 58.5 & 1.9 & 1368.1 \\ 
2.0$^{\circ}$, $1100$~km & 2047.2 & 42.8 & 107.7 & 2.5 & 2200.2 \\ 
1.5$^{\circ}$, $1100$~km & 3652.0 & 55.4 & 210.4 & 2.9 & 3920.7 \\ \hline
OAA,  L &\multicolumn{5}{c}{Antineutrino Mode} \\ \hline
2.5$^{\circ}$, $295$~km & 8636.1 & 4905.9 & 860.8 & 23.6 & 14426.5 \\ 
2.5$^{\circ}$, $1100$~km & 1119.5 & 300.6 & 61.9 & 2.0 & 1484.0 \\ 
2.0$^{\circ}$, $1100$~km & 1888.5 & 390.0 & 102.6 & 2.4 & 2384.4 \\ 
1.5$^{\circ}$, $1100$~km & 3579.2 & 490.8 & 185.1 & 2.8 & 4257.9 \\ 
\hline \hline
\end{tabular}%
}
\end{center}
\end{table}%

\begin {figure}[htbp]
\captionsetup{justification=raggedright,singlelinecheck=false}
  \begin{center}
    \includegraphics[width=0.42\textwidth]{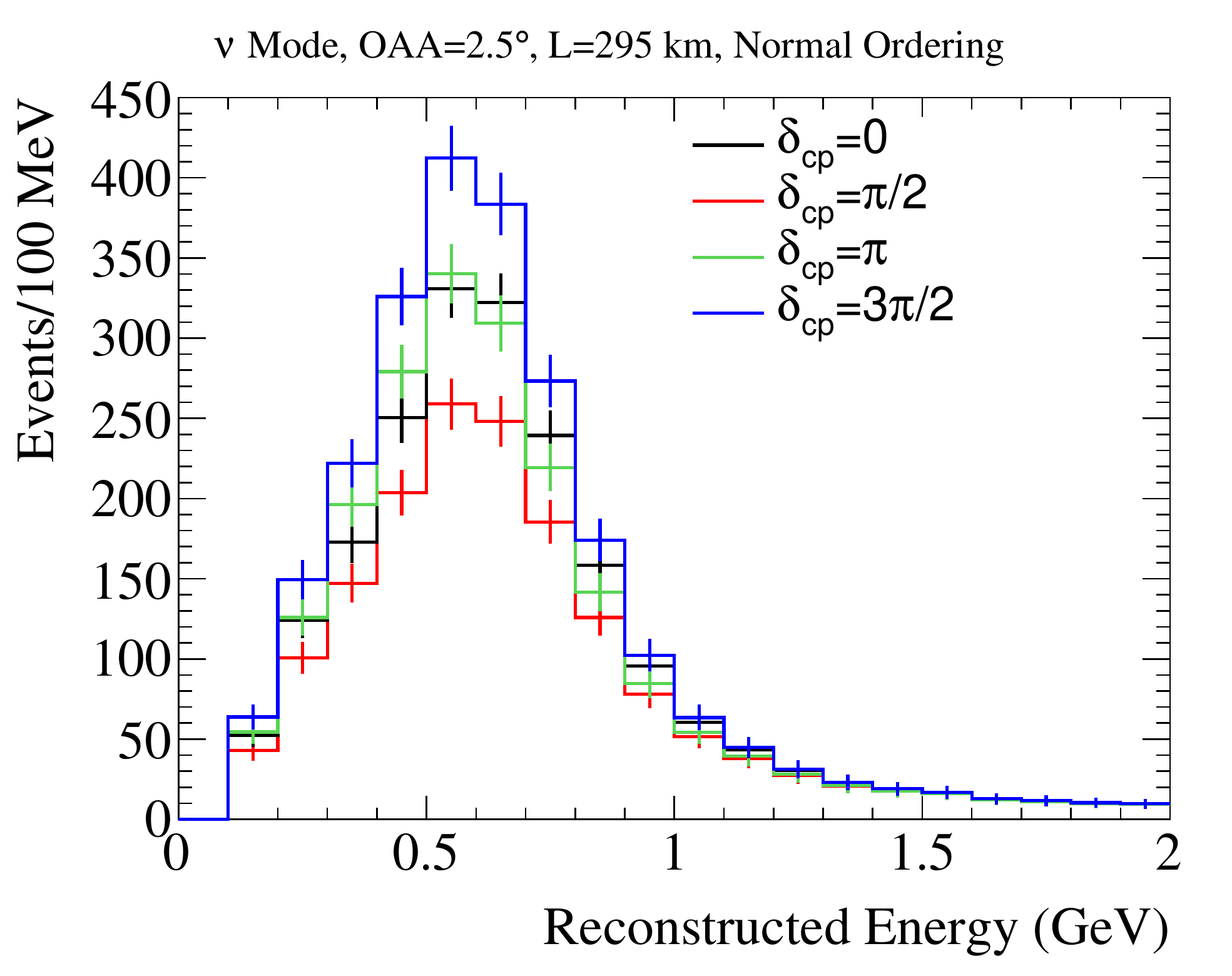}
    \includegraphics[width=0.42\textwidth]{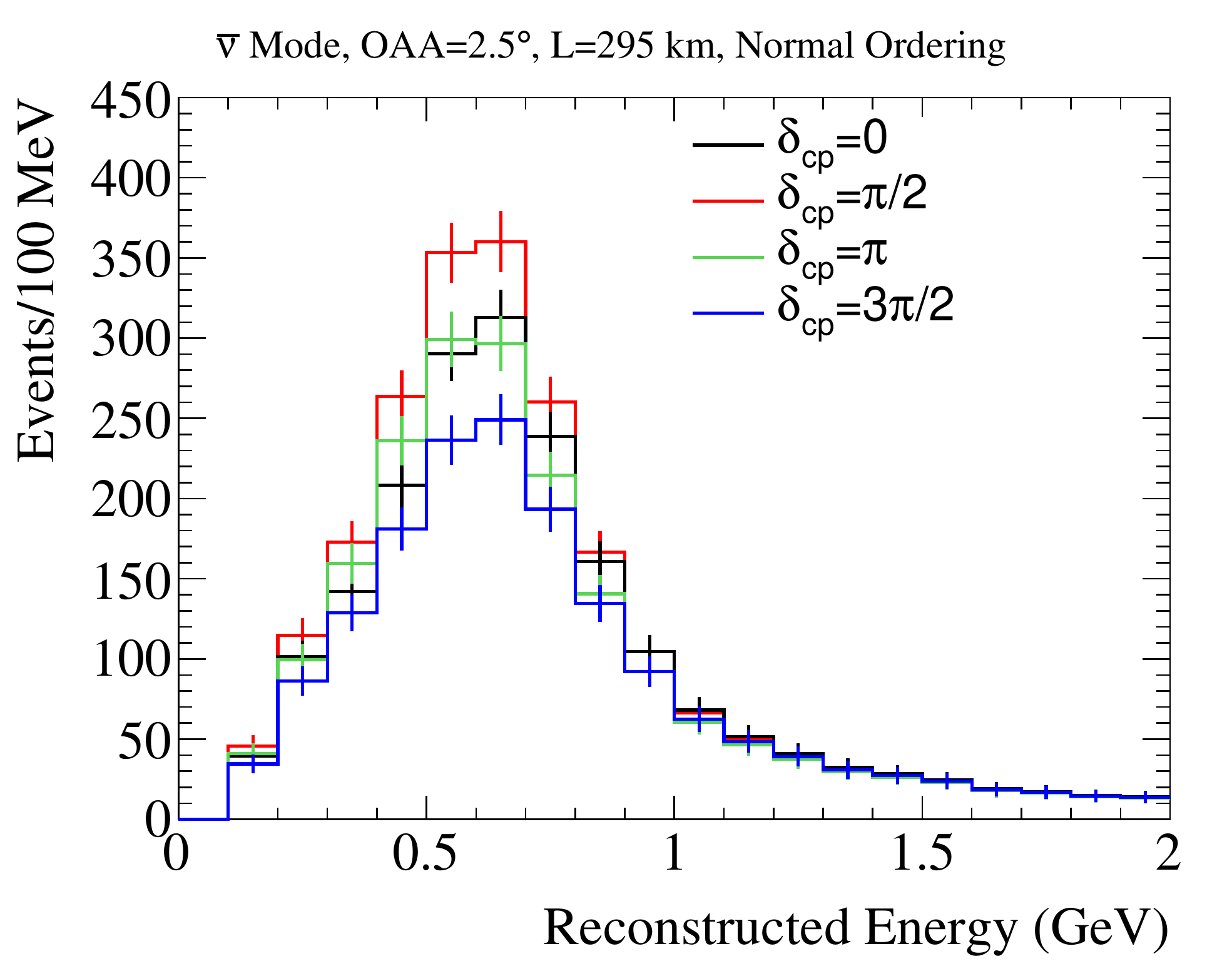} \\
    \includegraphics[width=0.42\textwidth]{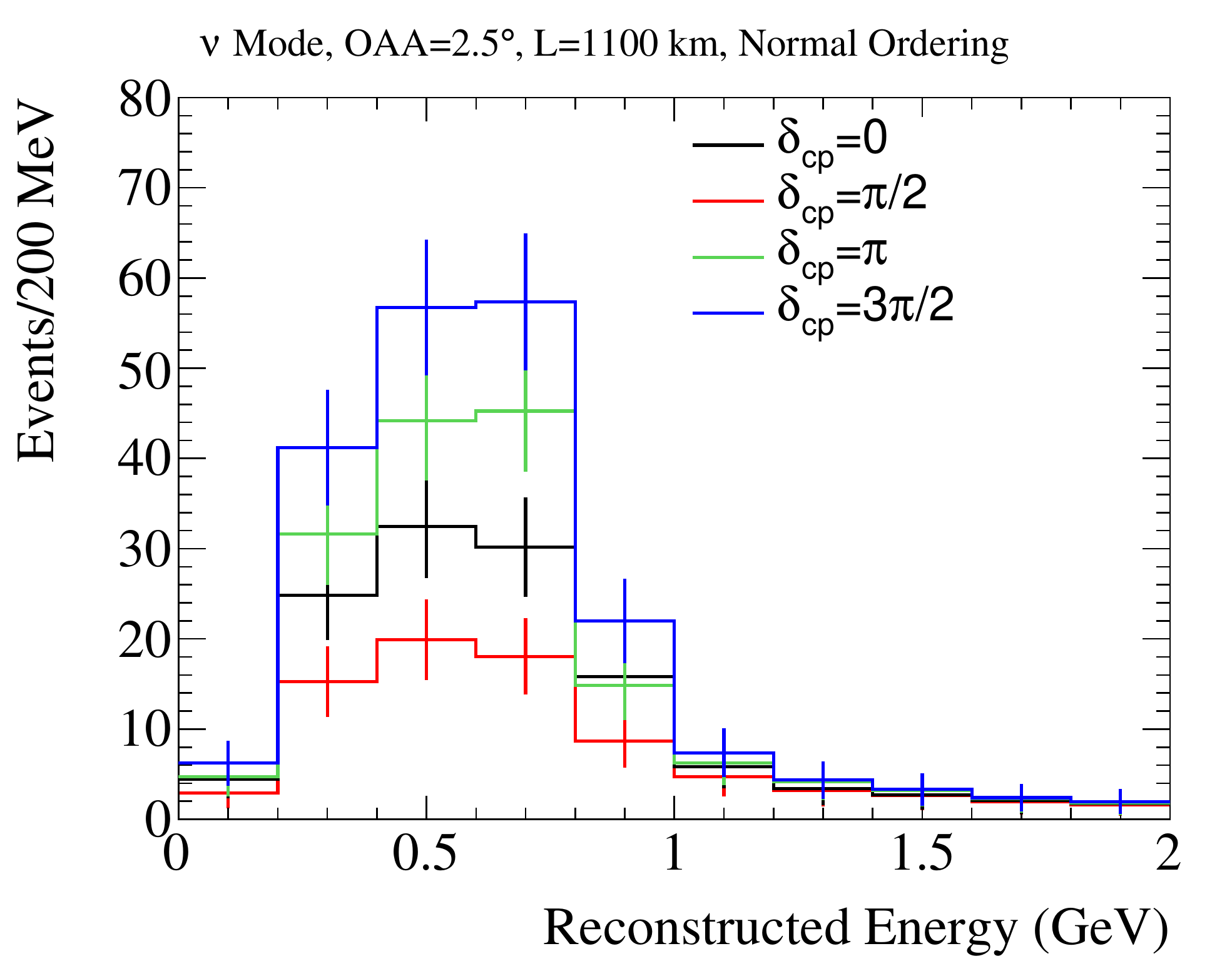}
    \includegraphics[width=0.42\textwidth]{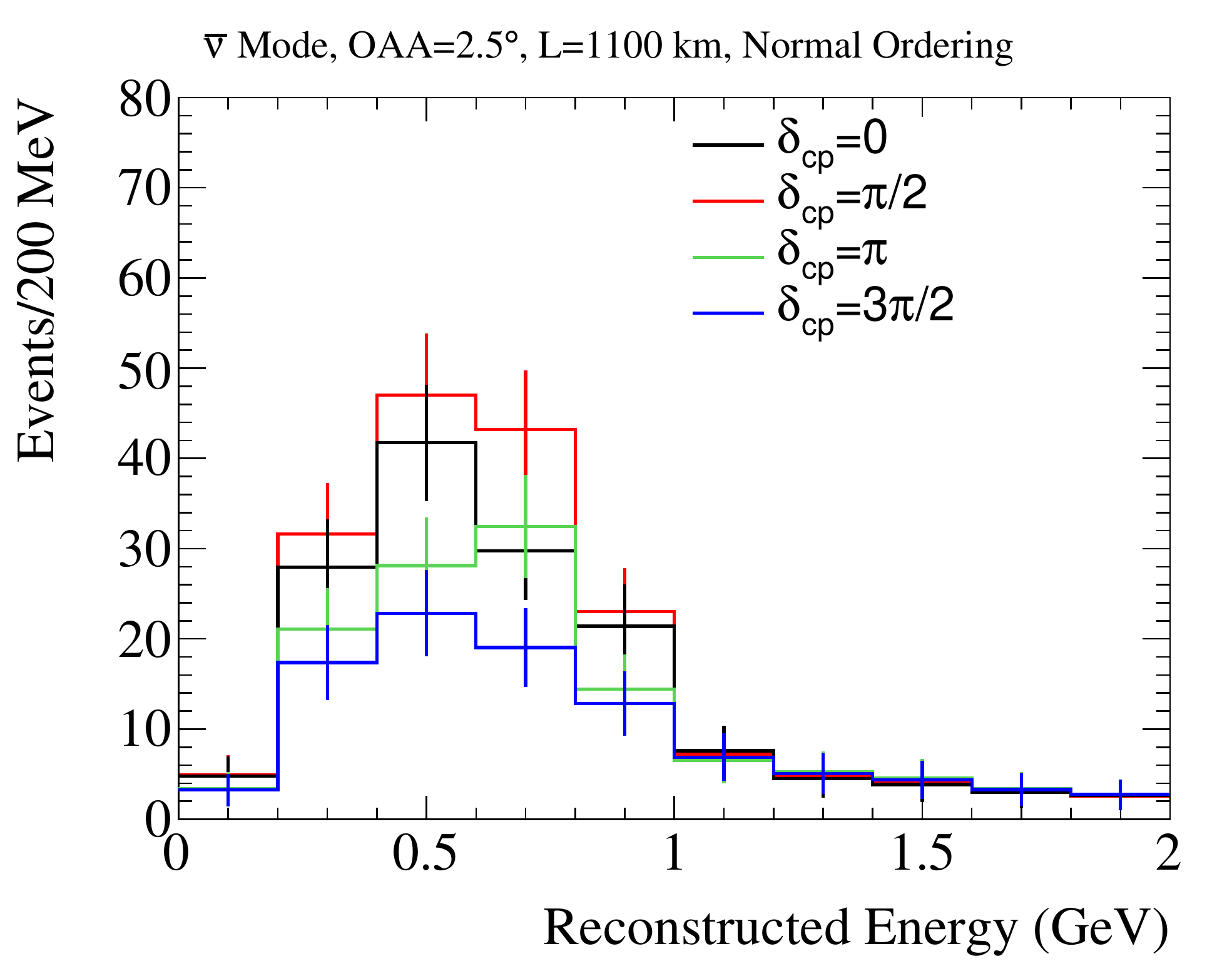} \\
    \includegraphics[width=0.42\textwidth]{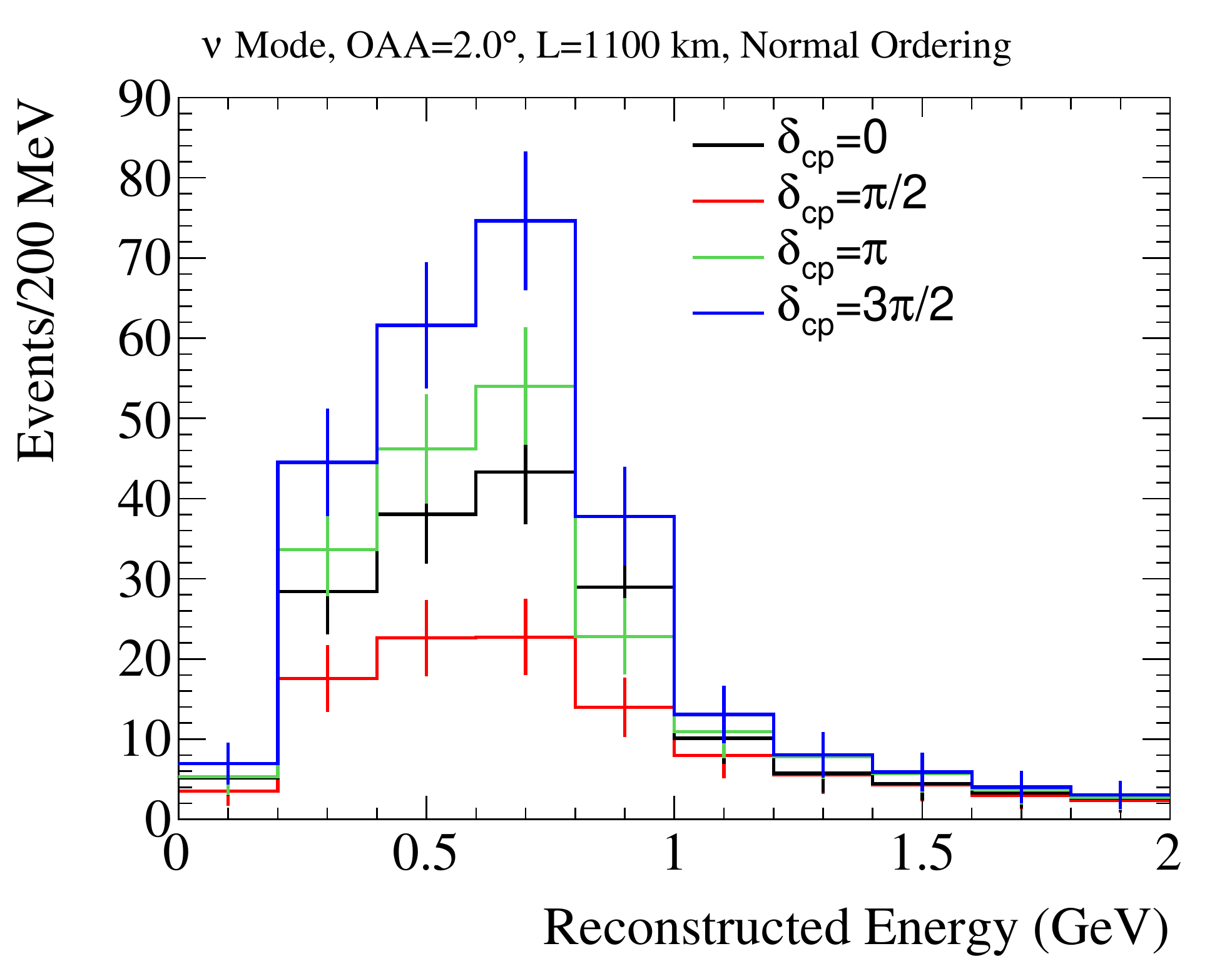}
    \includegraphics[width=0.42\textwidth]{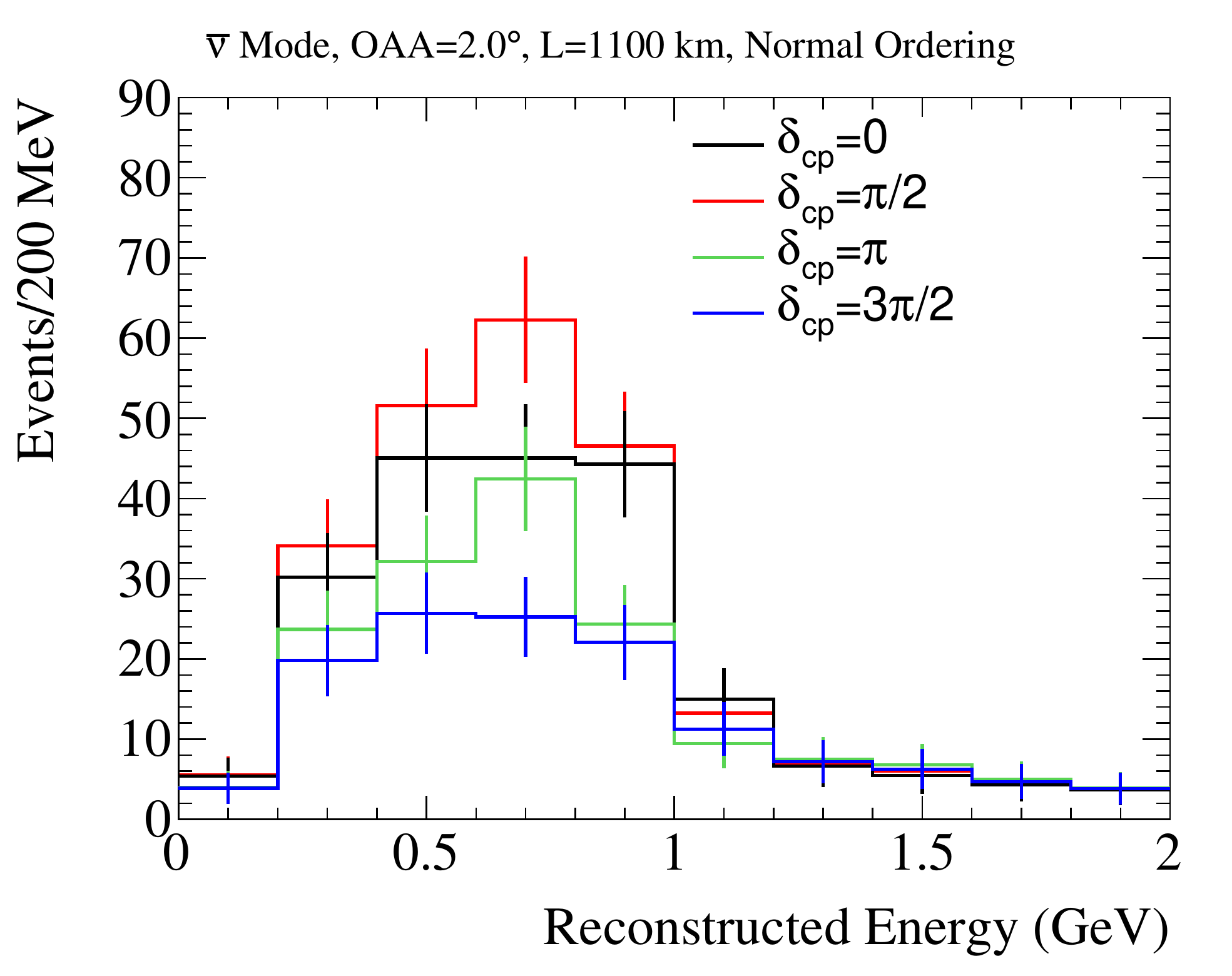}\\
    \includegraphics[width=0.42\textwidth]{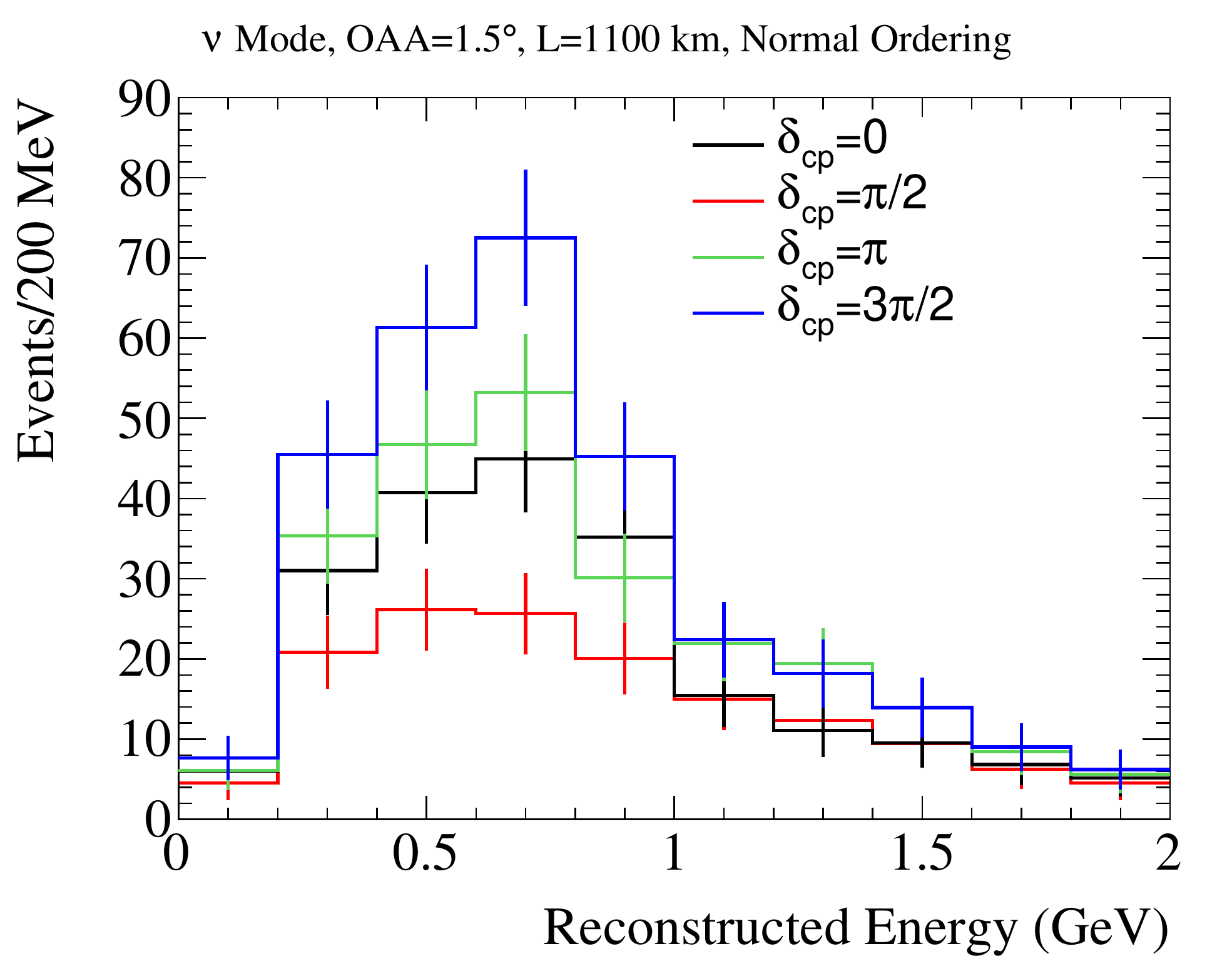}
    \includegraphics[width=0.42\textwidth]{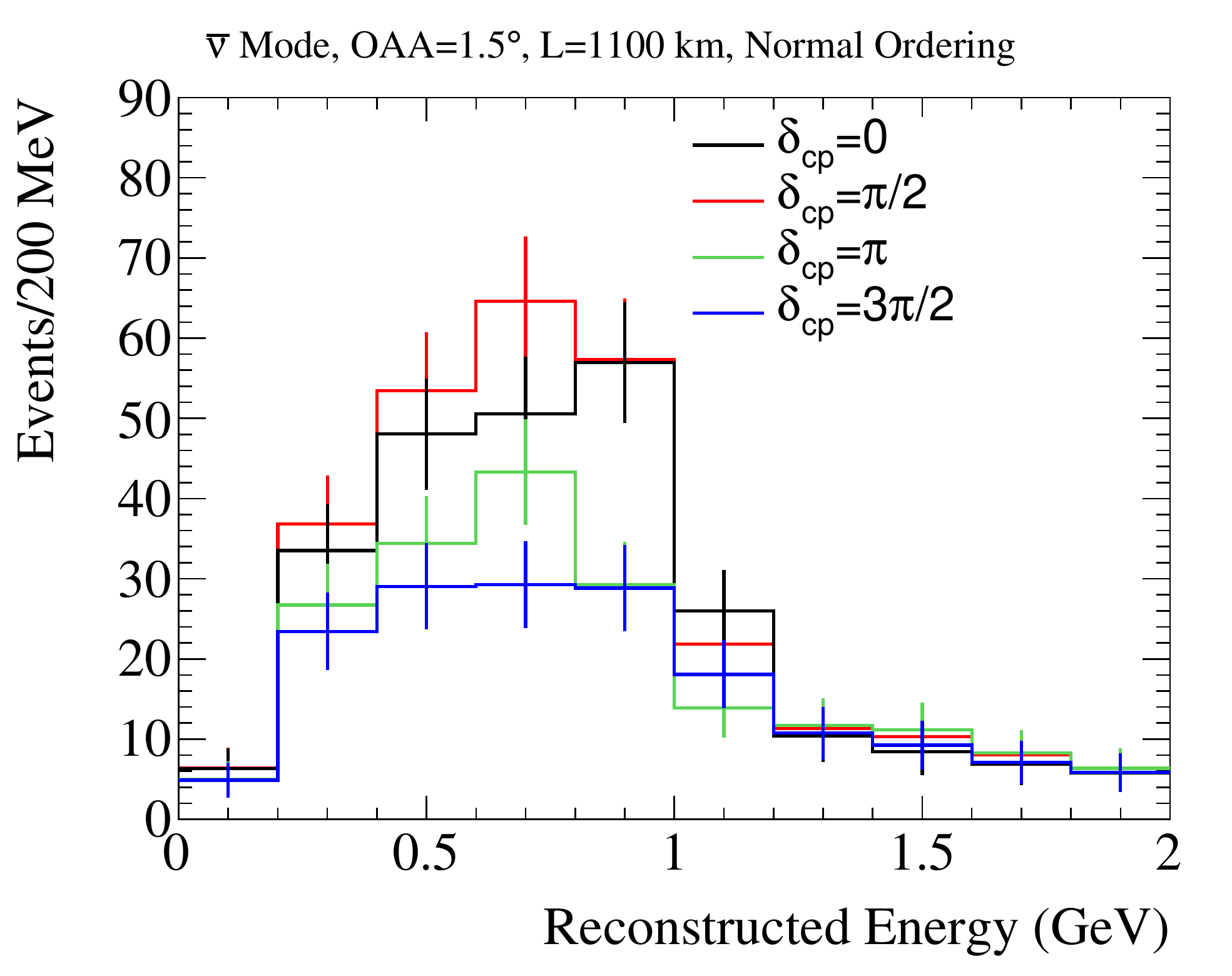}
 \caption{The predicted 1R$e$ spectra in neutrino mode (left) and antineutrino mode (right) for different values of $\delta_{cp}$.  Error bars represent statistical errors only.}
\label{fig:cp_effect_spectra}
  \end{center}
\end {figure}

\begin {figure}[htbp]
\captionsetup{justification=raggedright,singlelinecheck=false}
  \begin{center}
    \includegraphics[width=0.49\textwidth]{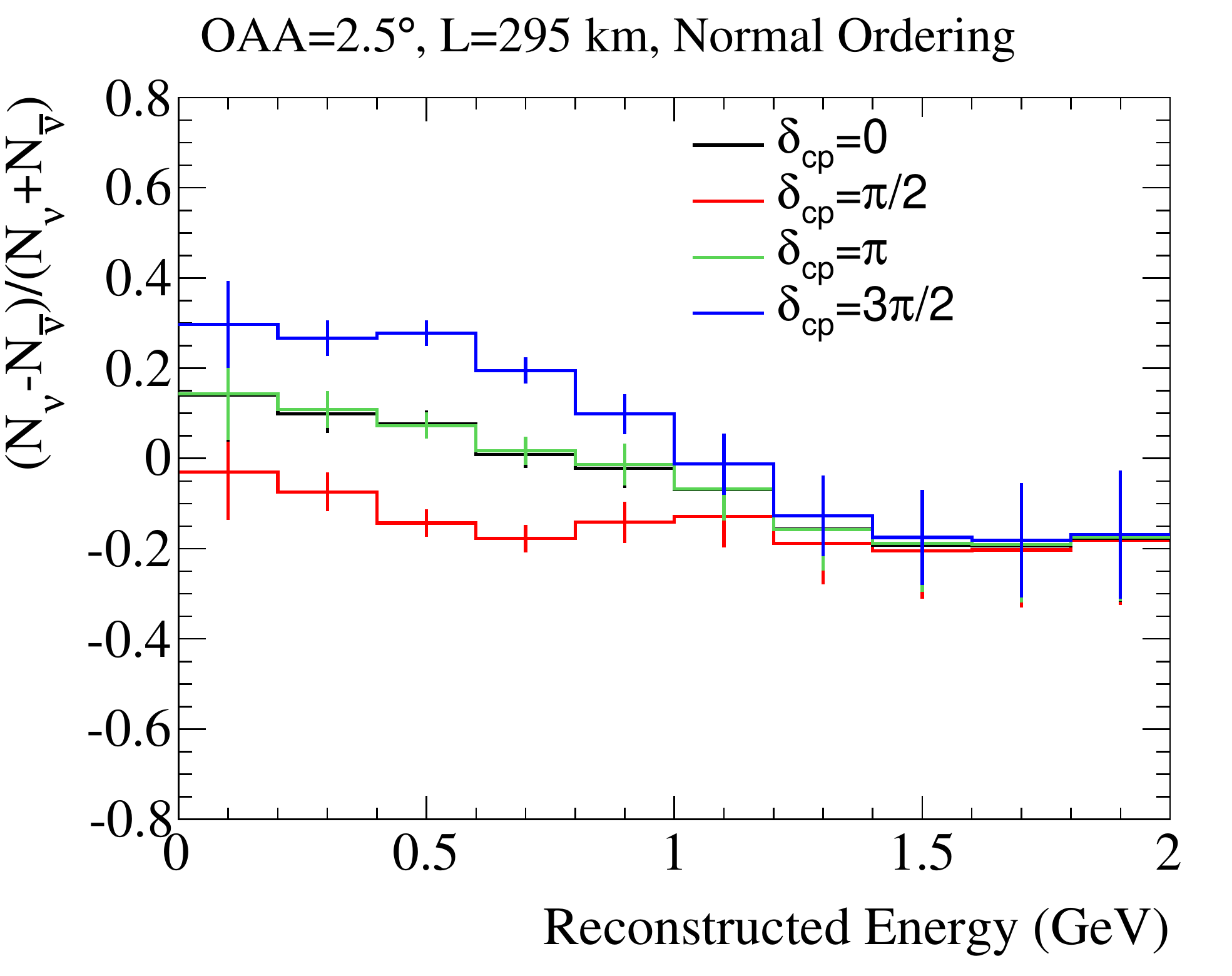}
    \includegraphics[width=0.49\textwidth]{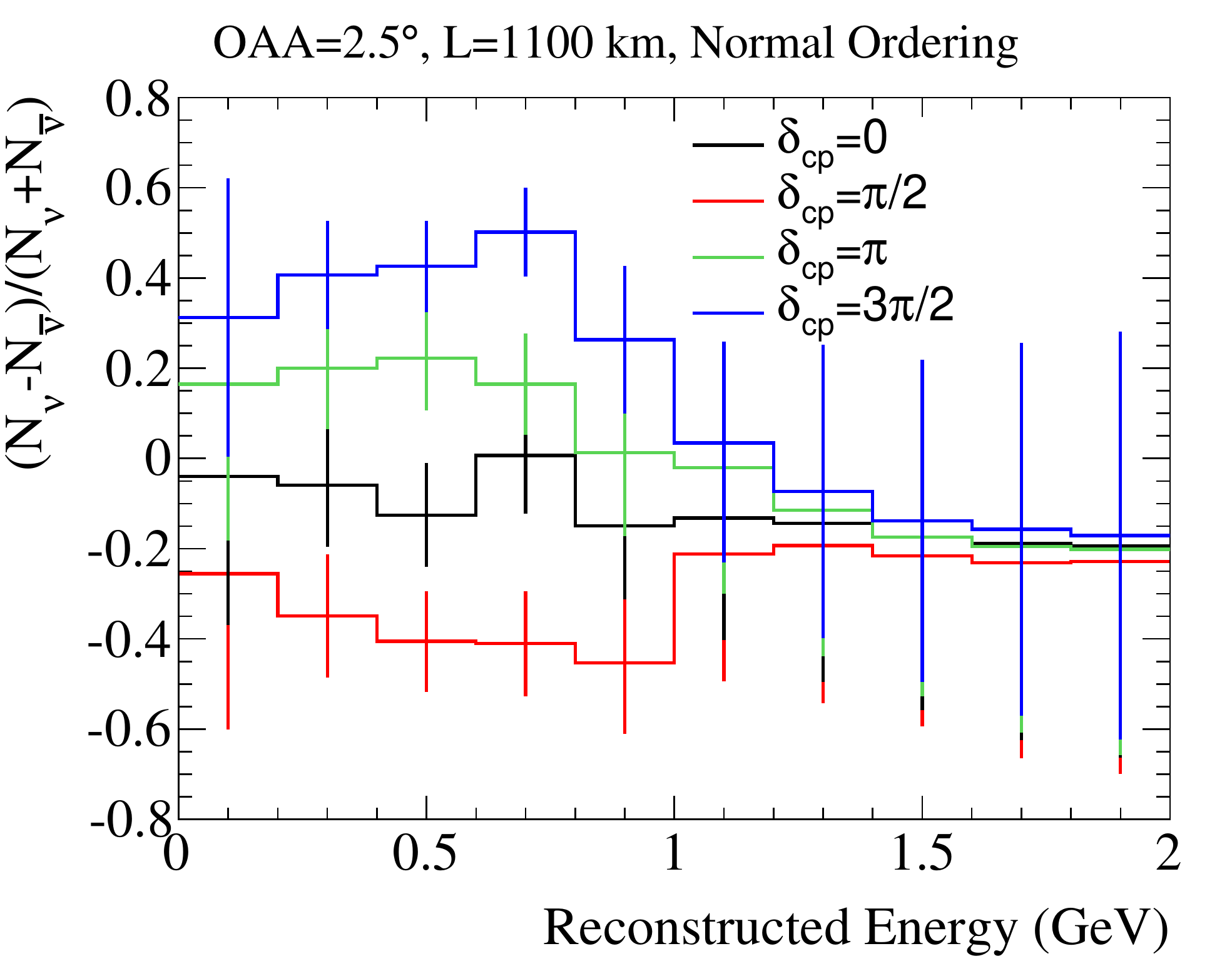}\\
    \includegraphics[width=0.49\textwidth]{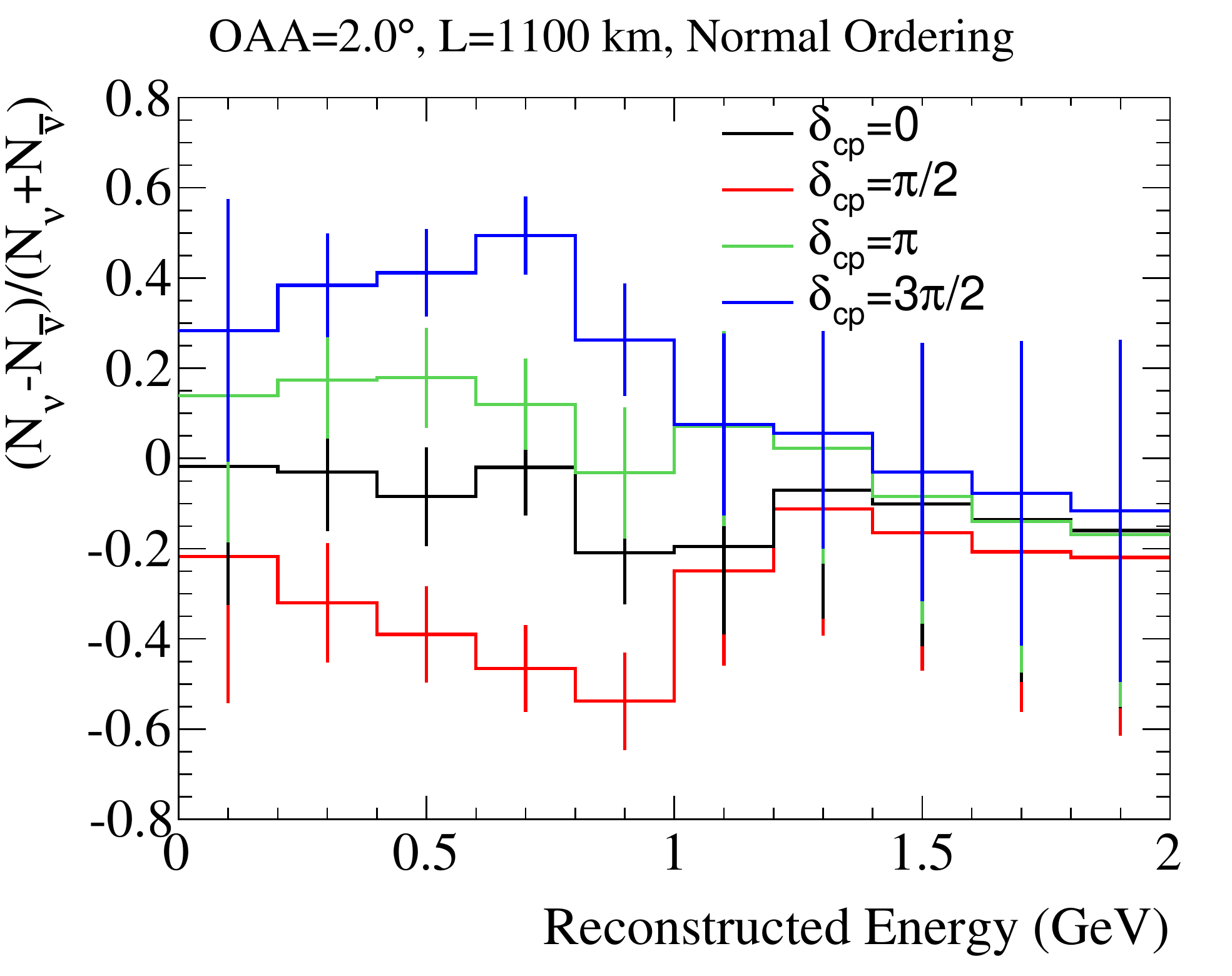}
    \includegraphics[width=0.49\textwidth]{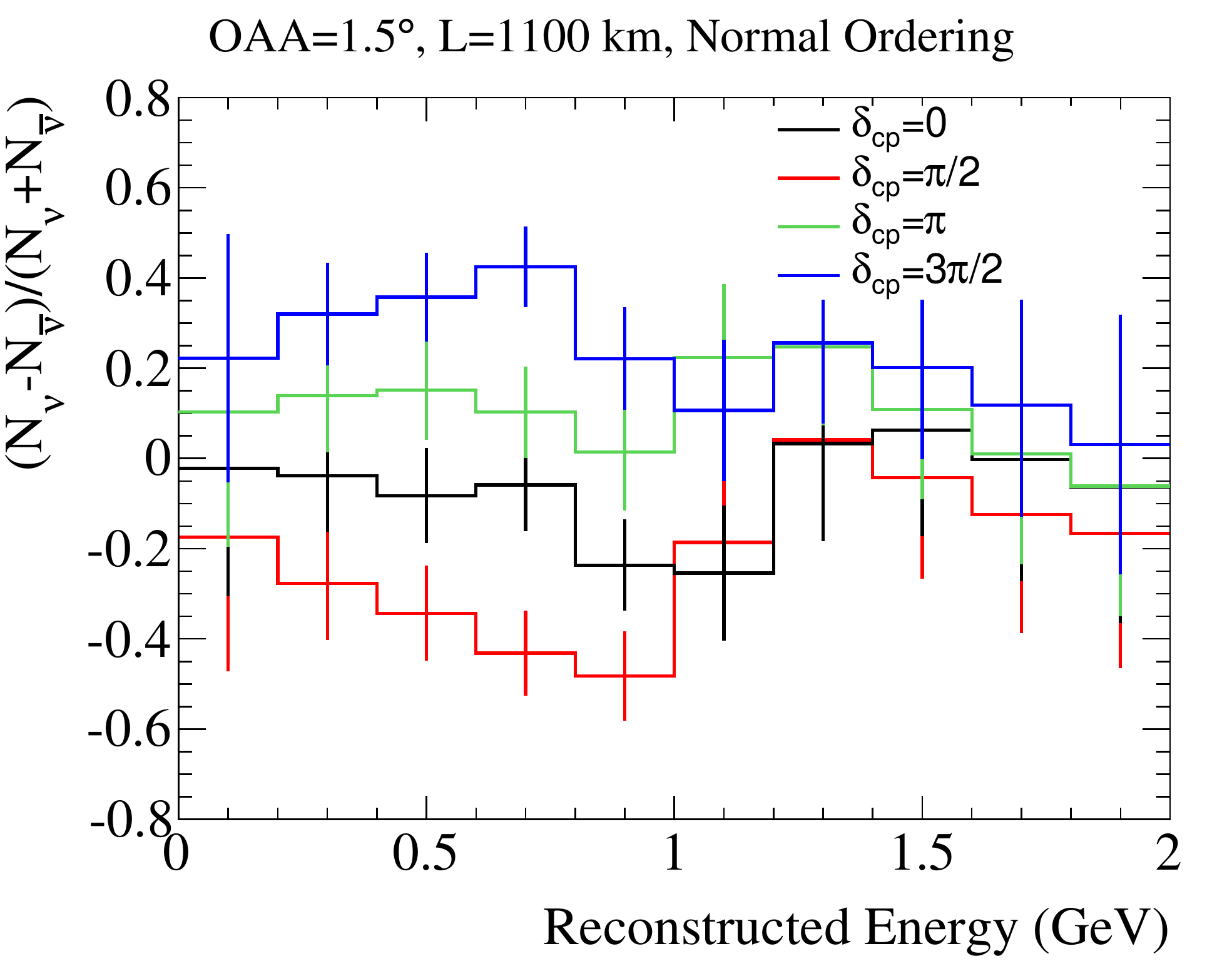}
 \caption{The event rate asymmetry between neutrino mode and antineutrino mode for variations of $\delta_{cp}$ at different detector site locations. Error bars represent statistical errors only.}
\label{fig:cp_effect_asymm}
  \end{center}
\end {figure}

The variations of the 1R$e$ spectra in neutrino mode and antineutrino mode for different $\delta_{cp}$ values at different detector locations are
shown in Fig.~\ref{fig:cp_effect_spectra}.
Similarly, the asymmetries of predicted 1R$e$ spectra between neutrino mode and antineutrino mode as a function of $\delta_{cp}$
are shown in Fig.~\ref{fig:cp_effect_asymm}.  For the detectors in Korea, the magnitude of the potential neutrino/antineutrino asymmetry is larger and 
this effect can partially compensate for the larger statistical uncertainties at the 3.7 times longer baseline.
The purely statistical separations between
the maximally CP violating and  CP conserving hypotheses are listed in Table~\ref{tab:cp_statistical}, where it is assumed that the mass ordering is known.  The $2.0^{\circ}$
off-axis slice has the strongest statistical separation between CP violating and CP conserving hypotheses.  Here, the detectors at 1100~km do not match the significance for 
CP violation discovery of the 295~km detectors as suggested in Section~\ref{sec:osc_prob}.  The lower significance is due to a few factors. First, the 1100~km baseline
means that the second oscillation maximum is at a higher energy than the first oscillation maximum at the 295~km baseline, and the neutrino flux is decreased by a factor of 14
at 1100~km compared 295~km.  The width of the second oscillation maximum is also narrower than the first oscillation maximum, so for a 2.0$^{\circ}$ off-axis beam, a larger fraction
of events don't have energies very close to the second oscillation maximum.  The introduction of the matter effect also decreases the CP violation significance as it introduces degeneracies 
between the mass ordering and $\delta_{CP}$.  Despite the lower statistical significance at the 1100~km baseline, the overall significance with a second detector in Korea is
higher when systematic uncertainties are accounted for, as is shown in the sensitivity studies presented later in this section.

\begin{table}[tbp]
\captionsetup{justification=raggedright,singlelinecheck=false}
\caption{\label{tab:cp_statistical}  The statistical separation of the predicted maximally CP violating spectra from the 
predicted CP conserving spectrum.  Here the significance is calculated for both CP conserving hypotheses and the smallest
significance is shown.  The mass ordering is assumed to be known.}
\begin{center}%
\scalebox{1.0}{
\begin{tabular}{l|c|c|c|c} \hline \hline
Detector Location     &   \multicolumn{4}{c}{Significance ($\sigma$)} \\ \hline
                      & \multicolumn{2}{c|}{NH} &  \multicolumn{2}{c}{IH} \\
OAA, L &   $\delta_{cp}=\pi/2$ & $\delta_{cp}=3\pi/2$ & $\delta_{cp}=\pi/2$ & $\delta_{cp}=3\pi/2$  \\ \hline
2.5$^{\circ}$, $295$~km & 11.6 & 11.0 & 11.8 & 10.9 \\ \hline
2.5$^{\circ}$, $1100$~km & 6.1 & 4.9 & 6.5 & 4.9 \\ \hline
2.0$^{\circ}$, $1100$~km & 7.9 & 5.9 & 7.1 & 6.3 \\ \hline
1.5$^{\circ}$, $1100$~km & 6.9 & 5.3 & 5.9 & 5.7 \\ 
\hline \hline
\end{tabular}%
}
\end{center}
\end{table}%

\begin {figure}[htbp]
\captionsetup{justification=raggedright,singlelinecheck=false}
  \begin{center}
    \includegraphics[width=0.49\textwidth]{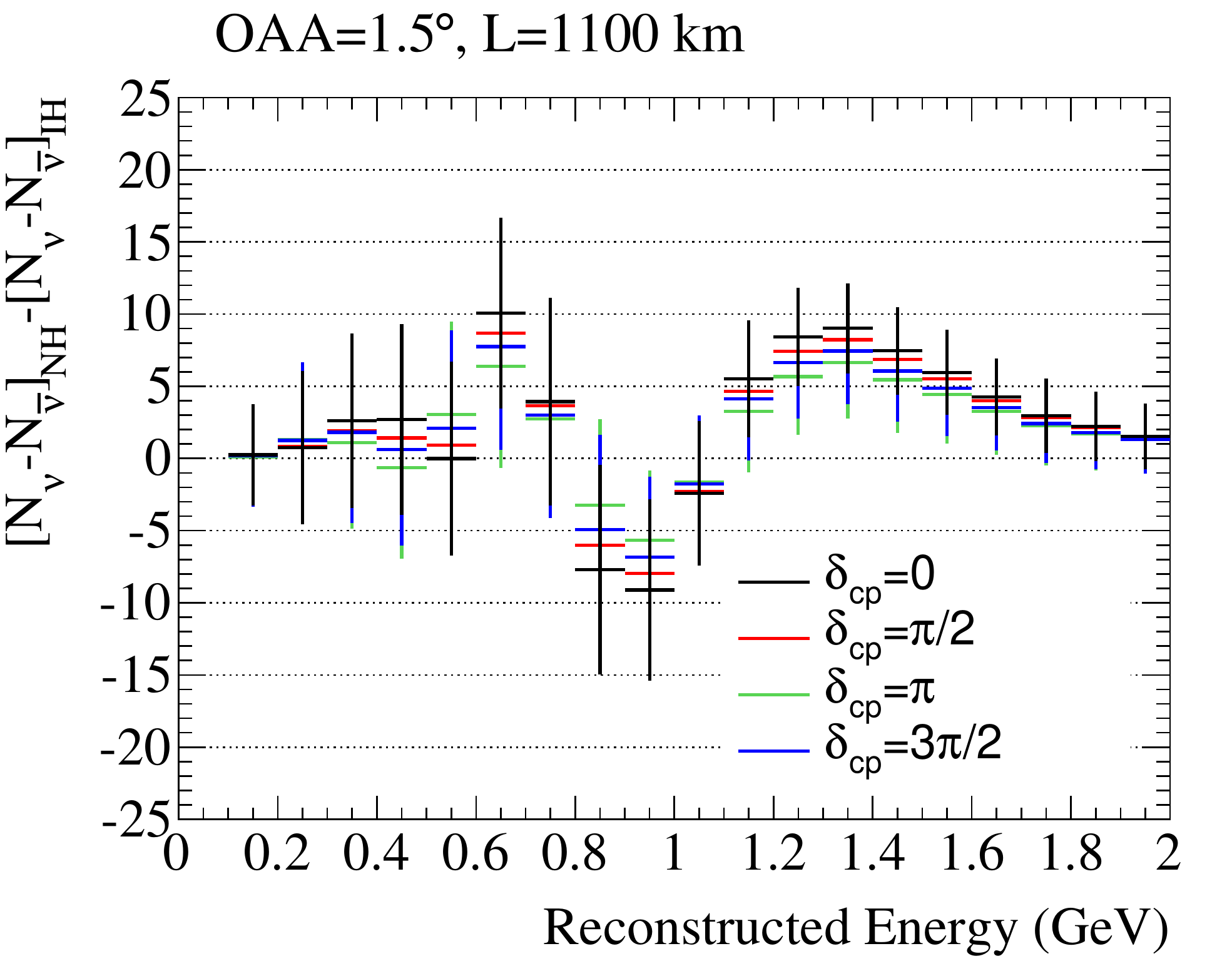}\\
    \includegraphics[width=0.49\textwidth]{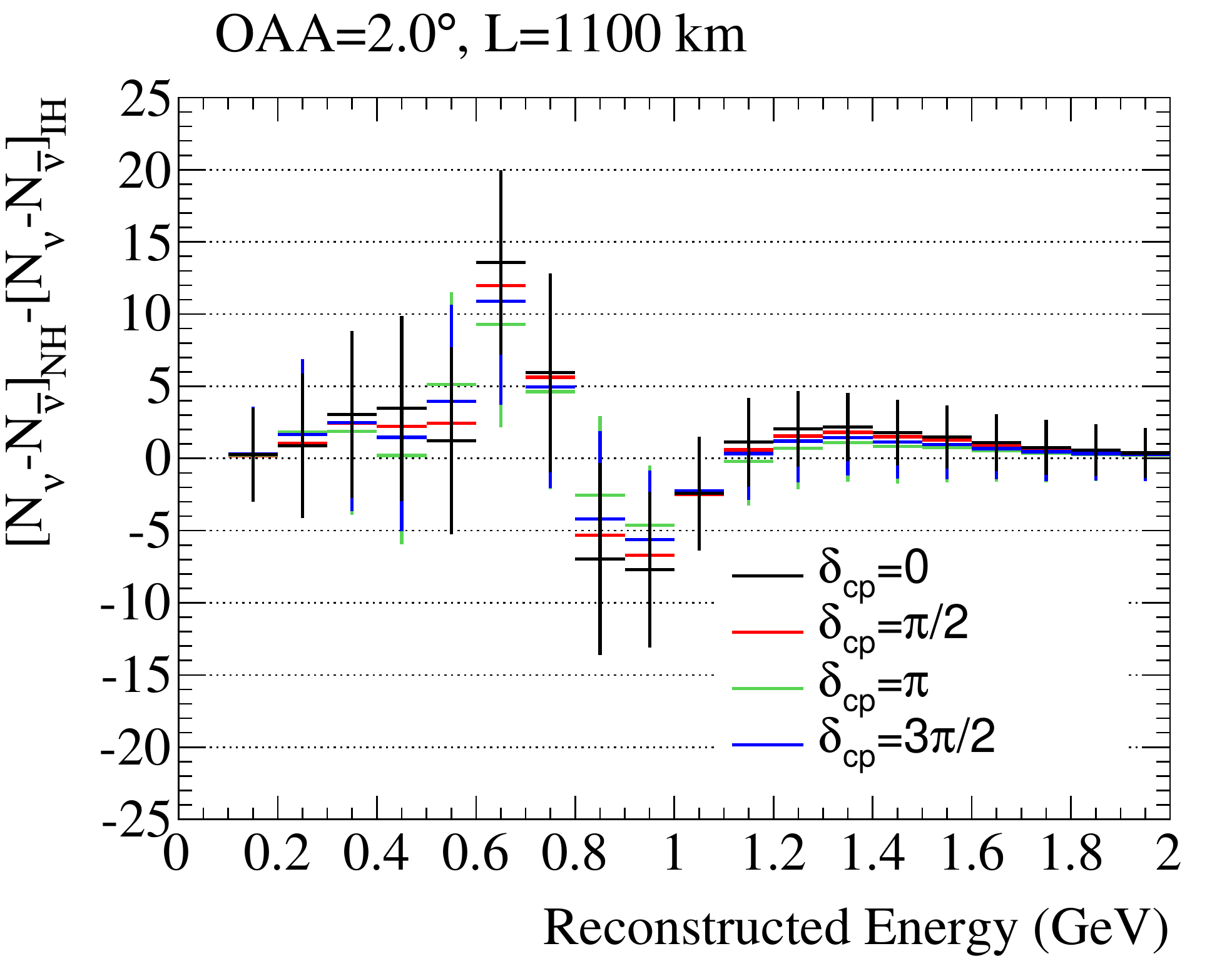}
    \includegraphics[width=0.49\textwidth]{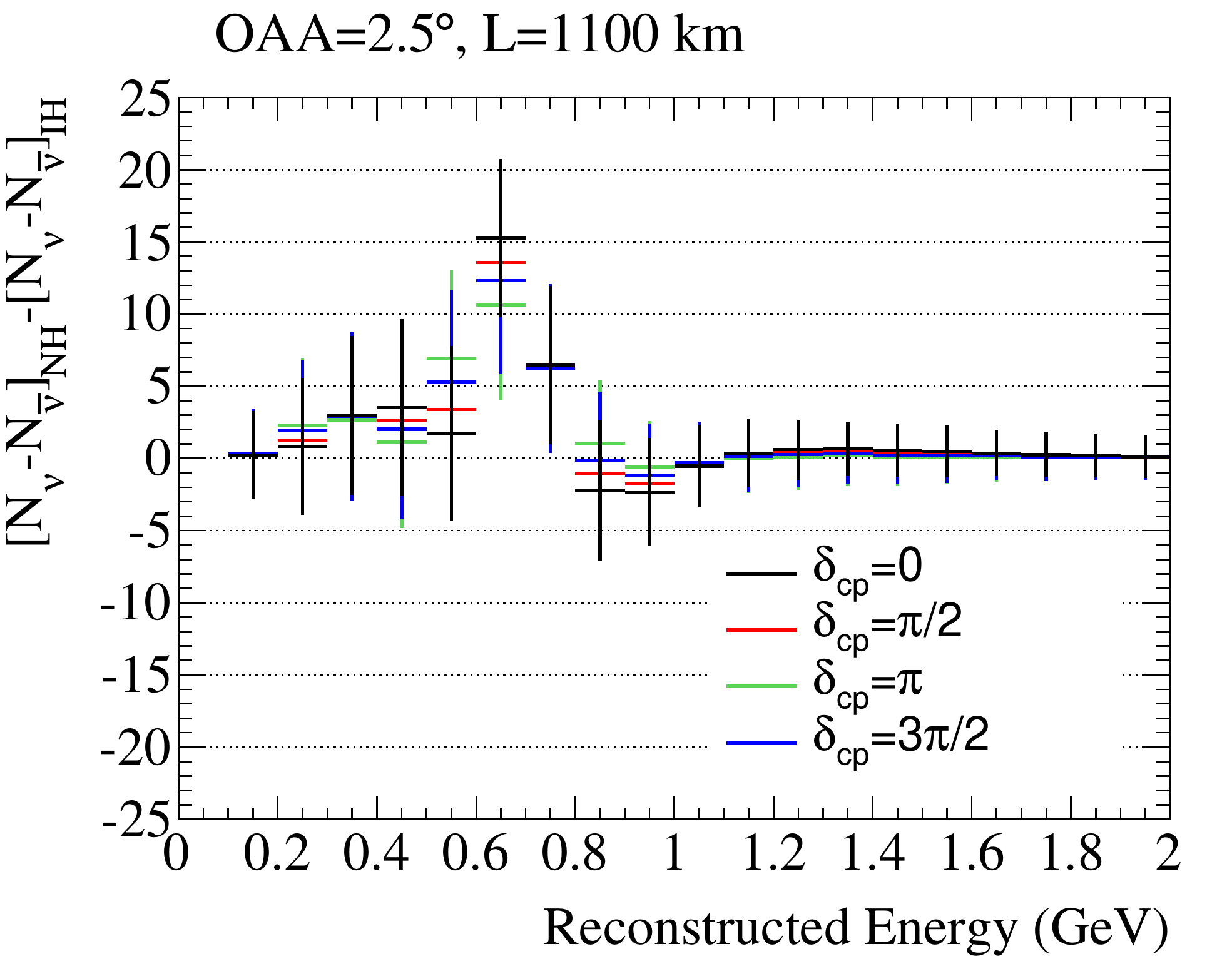}
    \caption{The difference of the observed neutrino-antineutrino difference in the 1R$e$ samples for normal mass ordering relative to the expected differences for inverted mass ordering.  Error
bars are the propagated statistical errors for the neutrino mode and antineutrino mode 1R$e$ samples.}
    \label{fig:matter_effect}
  \end{center}
\end {figure}

The impact of the matter effect and sensitivity to mass ordering is illustrated in Fig.~\ref{fig:matter_effect}.  Here, a double difference is presented.  First the 
difference in observed neutrino mode and antineutrino mode 1R$e$ candidates is calculated as a function of reconstructed energy.  This difference is calculated
for both the normal and inverted hierarchies and the difference between hierarchies is taken.  It can be seen that the neutrino-antineutrino difference varies 
differently with reconstructed energy for normal and inverted hierarchies.  For the normal mass ordering, the neutrinos are enhanced in the $<0.8$~GeV and $>1.1$~GeV 
regions and diminished in the 0.8-1.0~GeV region relative to the inverted mass ordering.  This relative difference is nearly independent of the true value of $\delta_{cp}$,
as illustrated in Fig.~\ref{fig:matter_effect}.
 The 1.5$^{\circ}$ off-axis angle configuration allows for a significant
observation of this spectral dependence of the asymmetry in the 0.8-1.0~GeV and  $>1.1$~GeV 
regions.  The 2.0$^{\circ}$ off-axis angle configuration has little sensitivity to the
$>1.1$~GeV region, and the 2.5$^{\circ}$ off-axis configuration is only sensitive to the $<0.8$~GeV region.  Fig.~\ref{fig:matter_effect} illustrates the mass 
ordering sensitivity for the neutrino/antineutrino mode difference for fixed bins of reconstructed energy, but additional sensitivity arises from the energy dependent 
enhancements and deficits observed in each mode independently.

\begin {figure}[htbp]
\captionsetup{justification=raggedright,singlelinecheck=false}
  \begin{center}
    \includegraphics[width=0.49\textwidth]{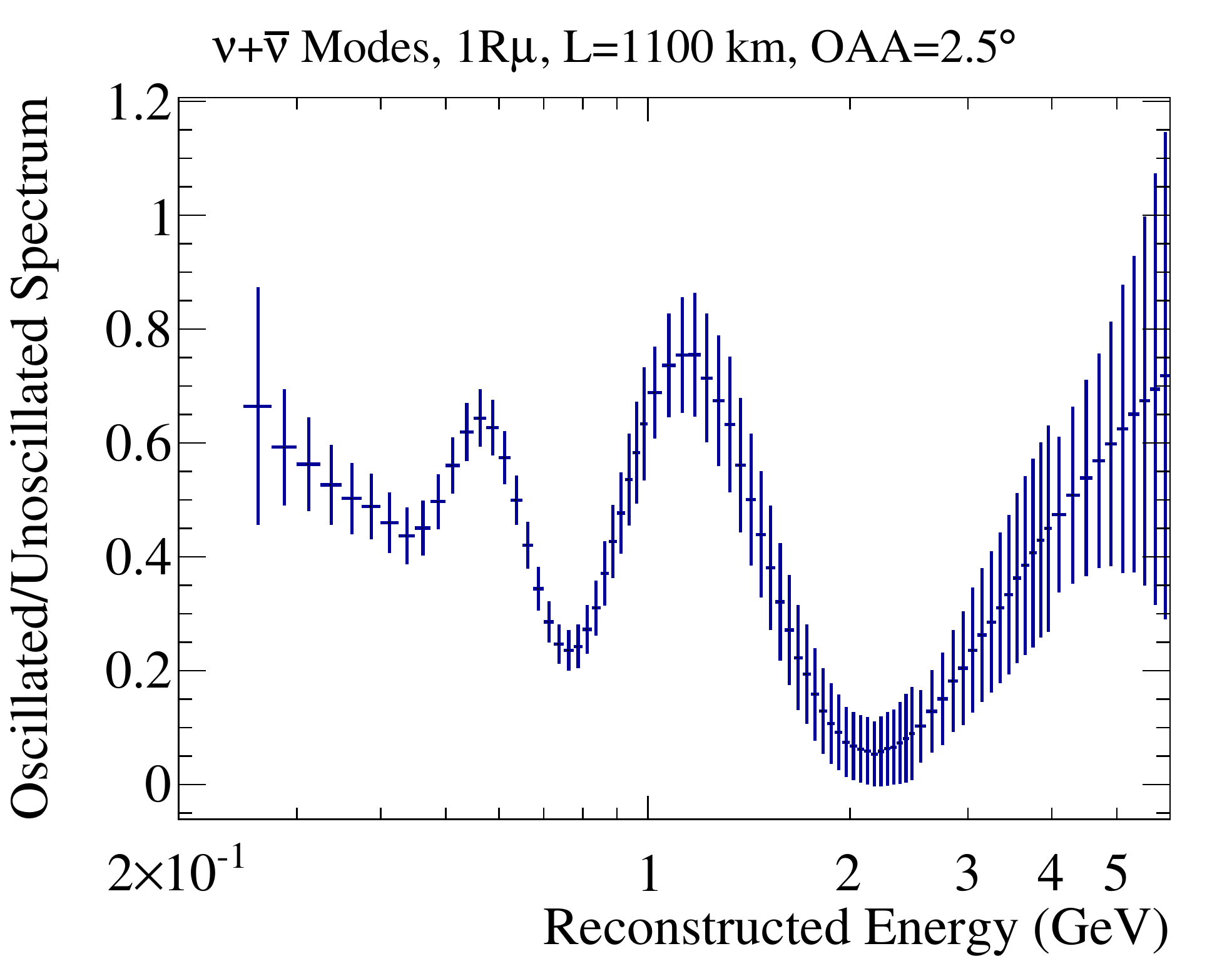}
    \includegraphics[width=0.49\textwidth]{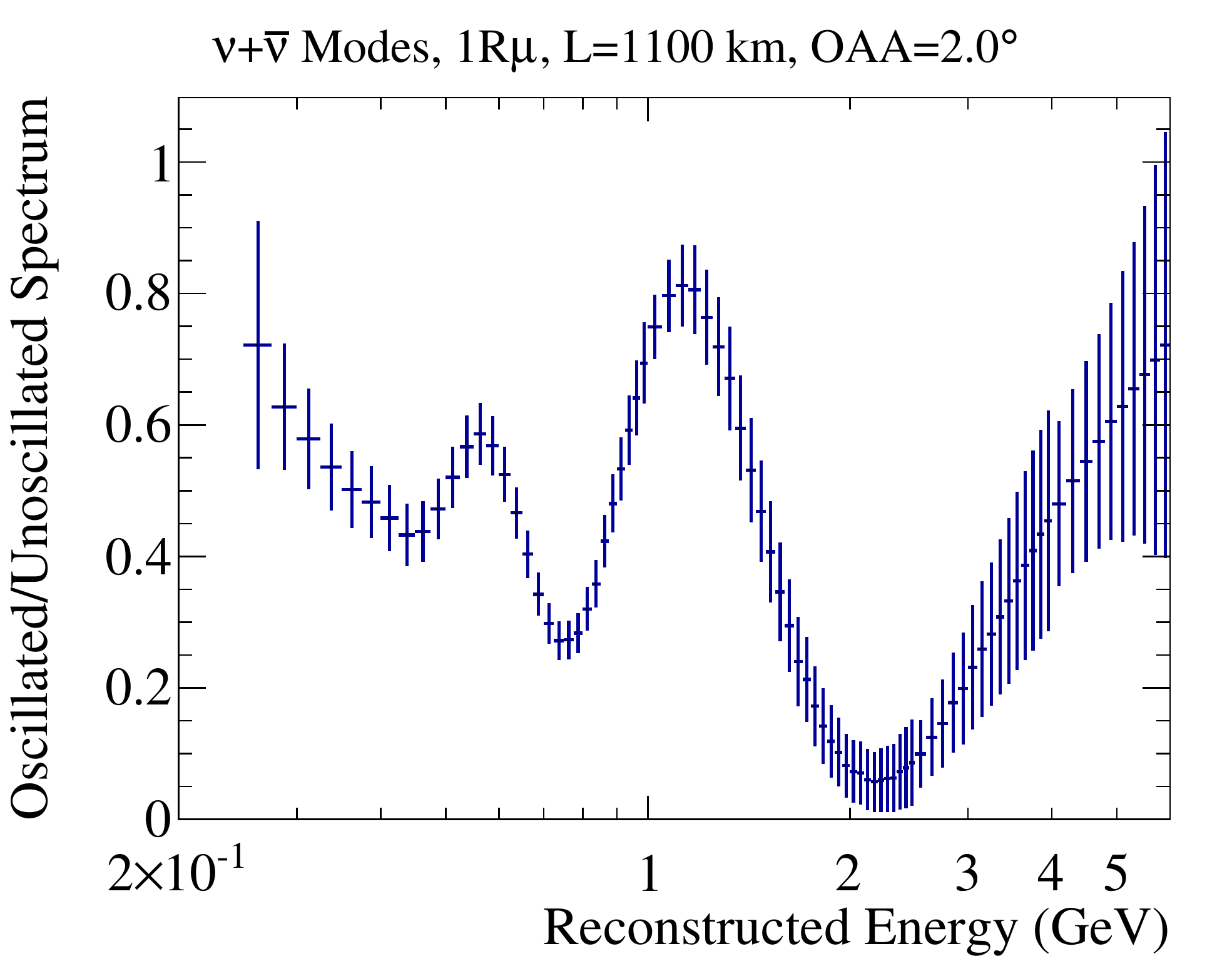}\\
    \includegraphics[width=0.49\textwidth]{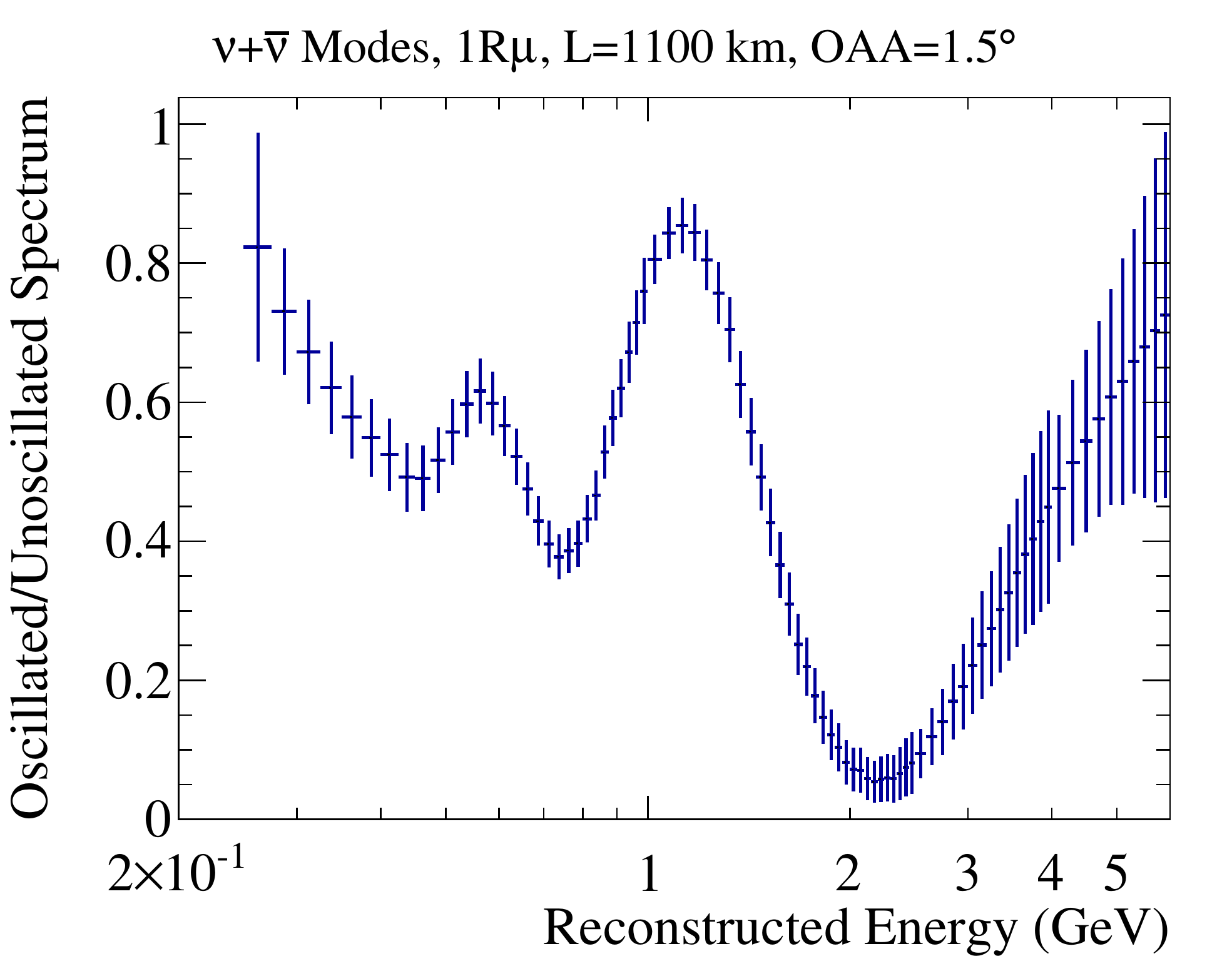}
    \includegraphics[width=0.49\textwidth]{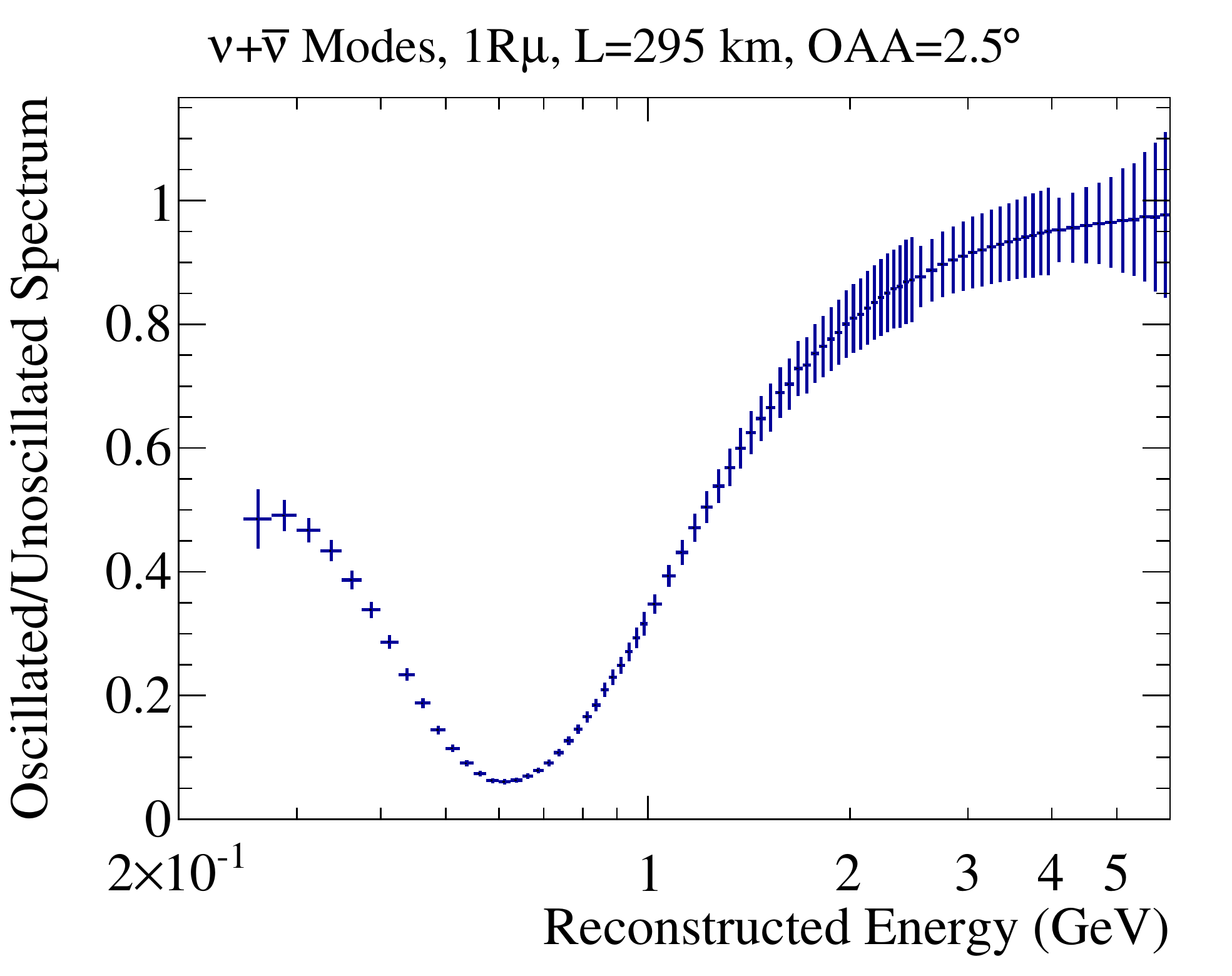}
    \caption{The ratio of the predicted 1R$\mu$ spectrum with oscillations to the predicted 1R$\mu$ spectrum without oscillations.  Here, the neutrino mode and antineutrino mode
             data have been summed.  The bin width varies from 25 MeV at low energy to 100 MeV at high energy, and the errors on each bin represent the statistical error for that bin.}
    \label{fig:muon_disappearance}
  \end{center}
\end {figure}

While the CP-even and CP-odd interference terms in the electron (anti)neutrino appearance probability are enhanced at the 1100~km baseline due to the $\Delta_{21}$ dependence,
no such enhancement is present in the muon (anti)neutrino survival probability.  Hence, the statistical constraint from the 1R$\mu$ samples on $\Delta m^{2}_{32}$ and
$\sin^{2}2\theta_{23}$ will be stronger for the detector at $L=295$~km due to the larger statistics.  The Korean detector, however, has the unique feature of measuring the 
oscillation pattern over two periods, confirming the oscillatory behavior of the neutrino transitions.  Fig.~\ref{fig:muon_disappearance} shows the ratio of the expected spectrum after oscillations
to the expected spectrum without oscillations. For all three Korean detector locations, the oscillation pattern over two periods may be observed.  While the measurement
in Hyper-K provides higher statistics, only one period of oscillations can be observed.

\subsection{Systematic errors}
Due to the statistically large samples available in the Hyper-K experiment, systematic errors are likely to represent the ultimate limit on oscillation parameter
measurement precision.  An advantage of a Korean detector is to enhance the contribution of the $\delta_{cp}$ dependent interference terms at the cost of fewer statistics, 
achieving similar sensitivity in a statistics limited measurement.  To evaluate the impact of the Korean detector on the Hyper-K sensitivities, it is necessary to 
implement a systematic error model that takes into account what are expected to be the dominant systematic errors for Hyper-K.  The systematic error model should also 
account for any new systematic errors introduced by having a detector in Korea. The systematic errors considered for the sensitivity studies presented in this 
paper are:
\begin{itemize}
\item { {\bf $\sigma_{\nu_{e}}/\sigma_{\nu_{\mu}}$ and  $\sigma_{\bar{\nu}_{e}}/\sigma_{\bar{\nu}_{\mu}}$} - The interaction cross sections for $\nu_{e}$ and $\bar{\nu}_{e}$
are not currently precisely measured with near detector data, although they may be more precisely measured in the Hyper-K era.  
When extrapolating the measured $\nu_{\mu}$ and $\bar{\nu}_{\mu}$ rates from the near detectors to predict the $\nu_{e}$ and $\bar{\nu}_{e}$ appearance rates at the far detector, it is necessary to 
assign an uncertainty on the interaction cross section ratios $\sigma_{\nu_{e}}/\sigma_{\nu_{\mu}}$ and  $\sigma_{\bar{\nu}_{e}}/\sigma_{\bar{\nu}_{\mu}}$.  Here the T2K
approach based on the work of Day \& McFarland~\cite{Day:2012gb} is taken.  Separate normalization parameters are assigned to vary  $\sigma_{\nu_{e}}$ and $\sigma_{\bar{\nu}_{e}}$.
The correlation between these parameters is assigned assuming there is a 2\% systematic effect that is uncorrelated between neutrinos and antineutrinos and an additional 
2\% systematic effect with anticorrelation between neutrinos and antineutrinos.}
\item { {\bf Energy scale at the far detectors} - The energy scale at Super-K is calibrated using samples of Michel electrons, $\pi^{0}$s and stopping cosmic muons.  In 
T2K oscillation analyses, the energy scale error is found to be 2.4\%~\cite{Abe:2015awa}.  Here a 2.4\% energy scale uncertainty is applied to the reconstructed energy
for events in Hyper-K and the Korean detector.  Independent parameters with no correlation are used for Hyper-K and the Korean detector.  100\% correlation between the 1R$\mu$ and 
1R$e$ samples is assumed.}
\item{ {\bf Matter density} - For results presented here, a constant matter density of 3.0~g/cm$^3$ is assumed for the path to the Korean detector.  An uncertainty of
6\% is assigned based on previous estimates~\cite{Hagiwara:2011kw}.}
\item{ {\bf The NC$\pi^+$ background} - NC$\pi^+$ interactions are a significant background in the 1R$\mu$ samples.  Based on the approach taken by T2K~\cite{Abe:2015awa}, a 30\% error is applied here.}
\item{ {\bf The intrinsic $\nu_{e}(\bar{\nu}_{e})$ and NC$\pi^{0}$ backgrounds} - The backgrounds for the 1R$e$ samples are the intrinsic $\nu_{e}(\bar{\nu}_{e})$ in the beam and NC$\pi^{0}$ interactions mistaken for 
an electron.  It is expected that these backgrounds will be measured by an intermediate water Cherenkov detector with similar $\nu_{e}(\bar{\nu}_{e})$ and total fluxes to the
far detector fluxes. The fluxes are similar since the oscillation effect on  $\nu_{e}(\bar{\nu}_{e})$ is $\sim$5\%, and oscillations don't affect the neutral current event rate.
 Studies of this measurement with the NuPRISM detector show an expected statistical error of 3\%.  A total error of 5\% is considered to account for uncertainties in the
different efficiency and fluxes between the near and far detectors.  100\% correlation is assumed between Hyper-K and the Korean detector, but no correlation is assumed between the neutrino and 
antineutrino beam modes.}
\item{ {\bf The CC non-quasielastic fraction} - The fraction of non-quasielastic interactions in the candidates samples affects the predicted normalization and reconstructed energy distribution.  In 
T2K near detector fits, the normalization of the non-quasielastic 2p-2h component of the cross section is fitted with a 20\% error~\cite{Abe:2017vif}.  The 2p-2h interactions are sometimes called multi-nucleon
interactions, and they consist of interactions on correlated pairs of nucleons rather than a single nucleon.  T2K models these interactions based on the work of Nieves {\it et al.}~\cite{Nieves:2011pp,Gran:2013kda}.
Here a 20\% error is applied to the normalization of the non-quasielastic
interactions, which includes 2p-2h events as well as events where a pion is produced, but is absorbed before exiting the nucleus.  
An anticorrelated parameter is applied to the quasielastic interactions, and its error is chosen such that the normalization of the unoscillated event rate is conserved for variations of 
these parameters.  This approach models the effect of the near detector constraint.}
\item{ {\bf Near to far extrapolation} - The T2K oscillation analysis~\cite{Abe:2015awa} accounts for an uncertainty from the flux and cross section model parameters that are constrained by the near detector data.  This
error includes the near detector measurement error and extrapolation uncertainties in the flux and cross section models that arise due to different neutrino fluxes and detector acceptances at the near and far detectors.  
To model this uncertainty, the T2K errors are applied as an overall uncertainty
on the charged current event rate. In
principle, the extrapolation error includes the effect of the previously described uncertainty on the non-quasielastic fraction.  To avoid double counting the error on the non-quasielastic fraction, the T2K errors are corrected by subtracting in quadrature the normalization uncertainty 
that is explicitly calculated from the non-quasielastic uncertainty.} 
\item{ {\bf Far detector modeling} - In addition to the energy scale uncertainty, there are uncertainties related to the modeling of efficiencies in the far detector.  This uncertainty is estimated based
on the uncertainty evaluated for T2K.  Since the far detector efficiency model is tuned using atmospheric neutrino control samples, it is assumed that the uncertainty will be reduced with the larger
sample of atmospheric neutrinos available in Hyper-K.  For the studies presented here, the assumption is that 50\% of the error is reduced by a factor of $1/\sqrt{8.3}$, where 8.3 is the fiducial mass ratio
between Hyper-K and Super-K.  The remaining 50\% of the error remains unchanged under the assumption the perfect agreement between the detector model and control samples may not be achieved and systematic
errors may be applied to cover any disagreement.  For this error source, there are no correlations between Hyper-K and the Korean detector.}
\end{itemize}
For the purpose of this document, the above systematic error model is used in place of the model adopted for the Hyper-K Design Report.  This is done because the systematic
errors used in the Hyper-K design report are based on the T2K systematic error estimate for a 2.5$^{\circ}$ off-axis angle flux and a 1R$e$ sample with a
$E_{rec}<1.25$~GeV cut applied.  The T2K systematic error model has not yet been applied to the other off-axis angle positions and 1R$e$ samples with the 
reconstructed energy cut removed.

\begin {figure}[ht]
\captionsetup{justification=raggedright,singlelinecheck=false}
  \begin{center}
    \includegraphics[width=0.49\textwidth]{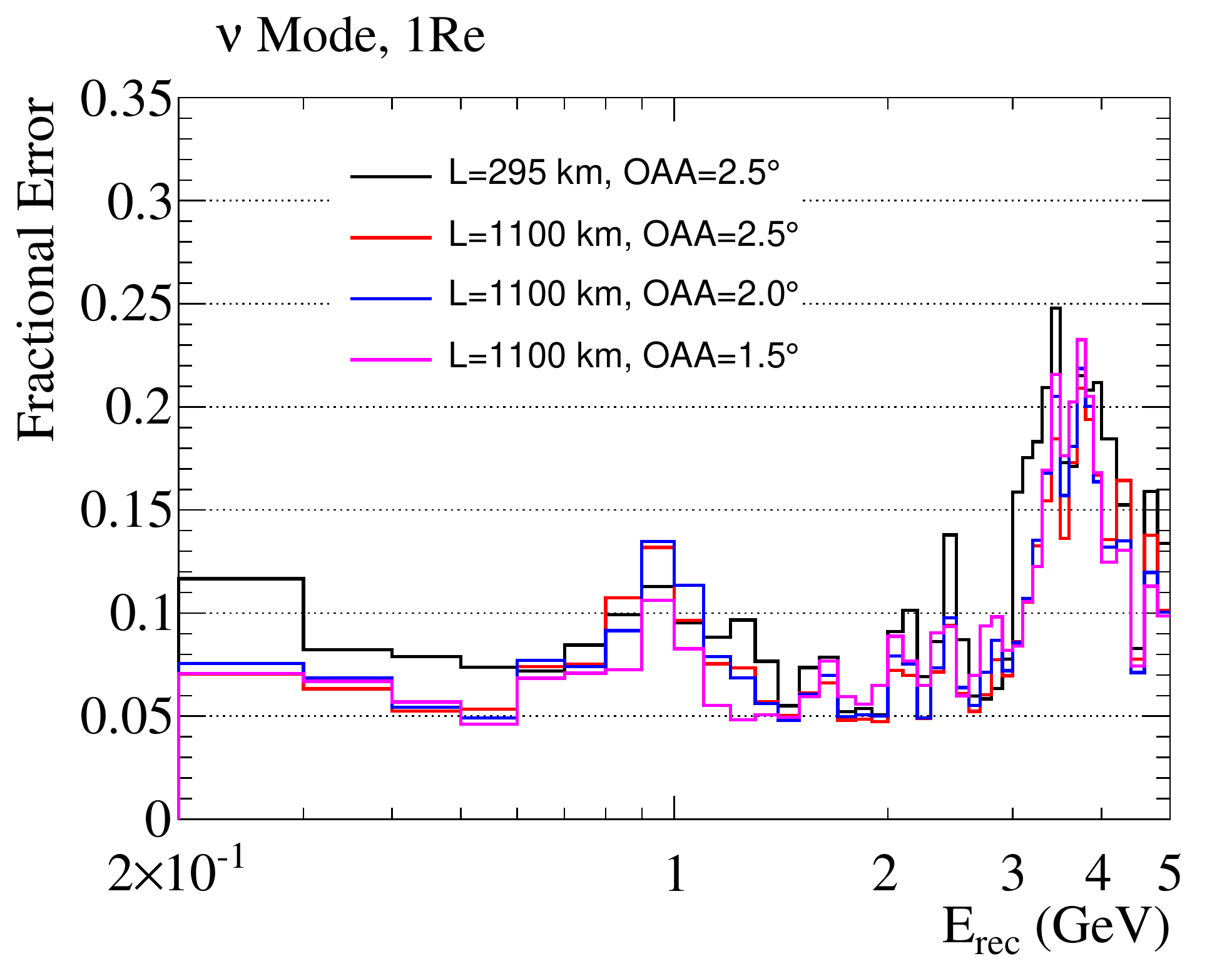}
    \includegraphics[width=0.49\textwidth]{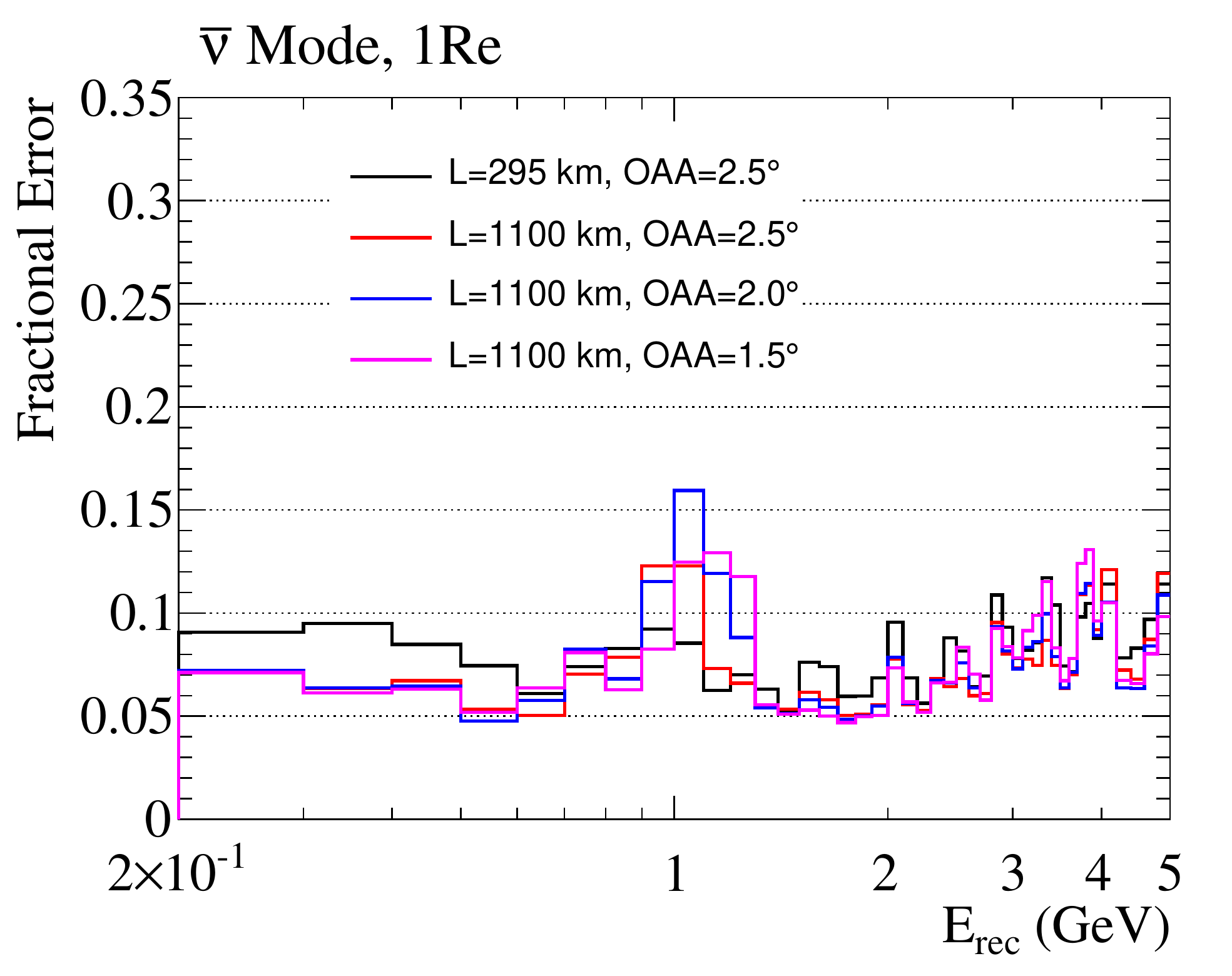}\\
    \includegraphics[width=0.49\textwidth]{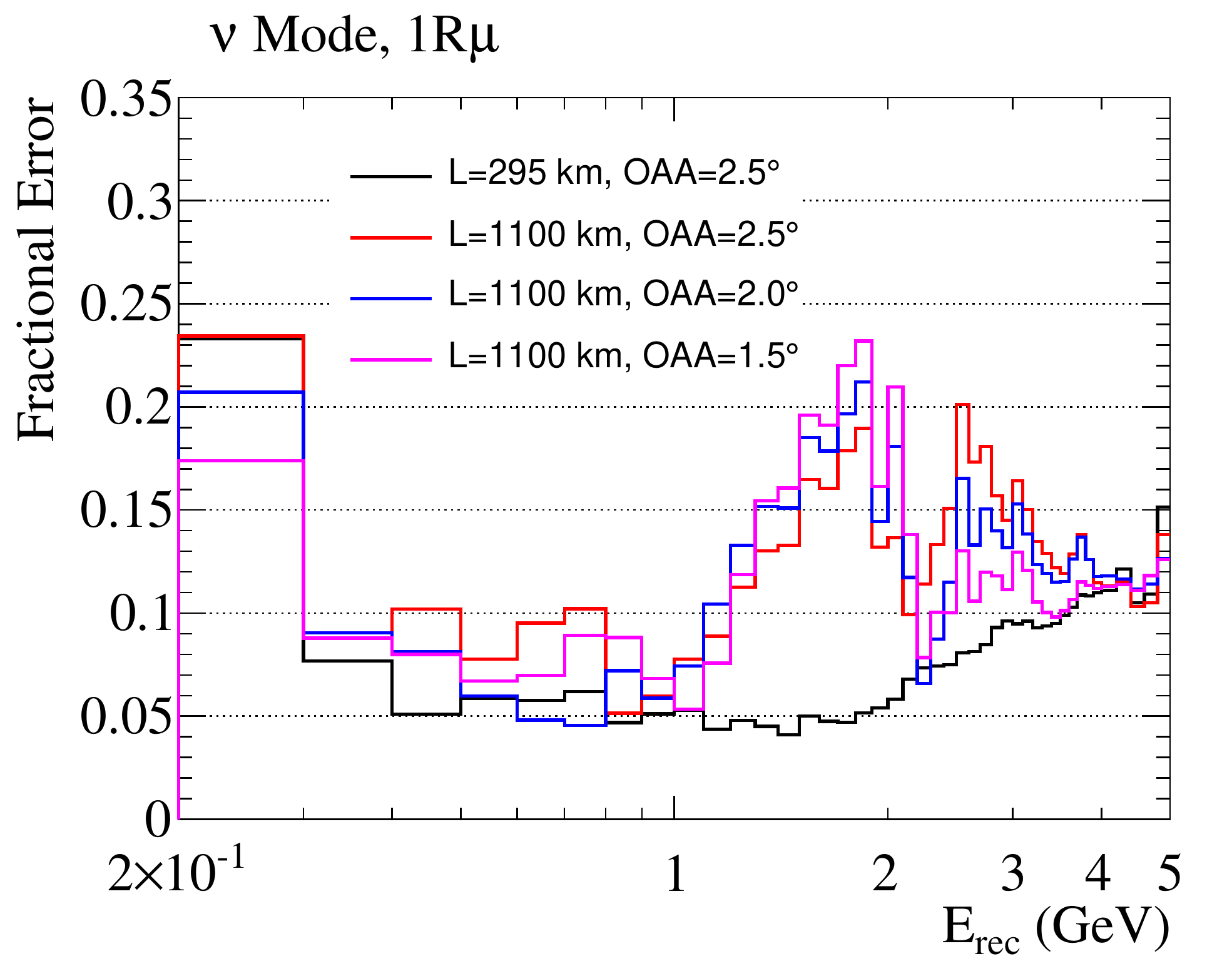}
    \includegraphics[width=0.49\textwidth]{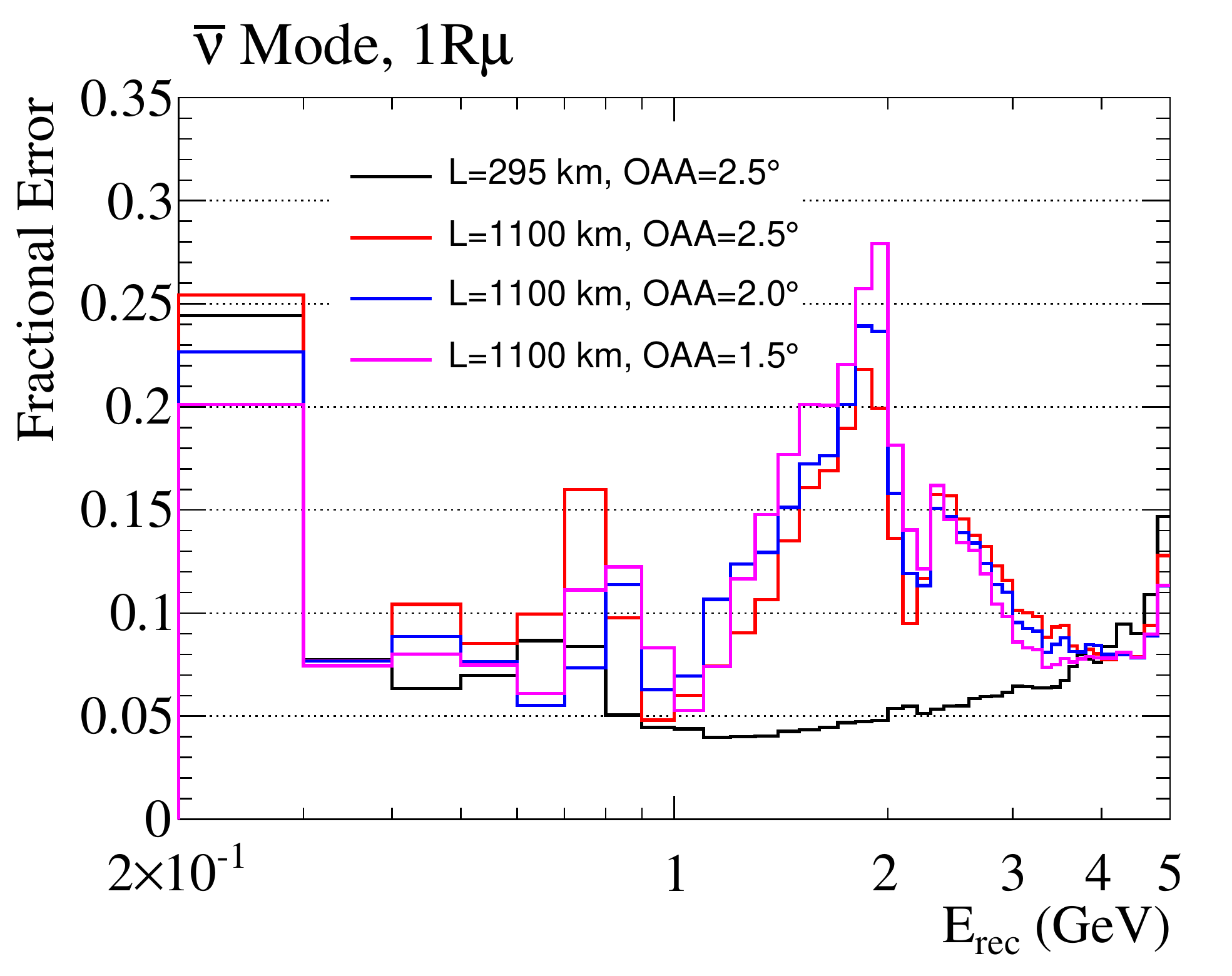}
    \caption{The fractional systematic errors per bin on the predicted spectra binned in reconstructed energy.}
    \label{fig:erec_error}
  \end{center}
\end {figure}

The effect of systematic errors propagated to the normalization uncertainties on the 1R$\mu$ and 1R$e$ samples are summarized in Table~\ref{tab:sys_error}.  The 
normalization uncertainties for individual samples are in the 4-5\% range.  These uncertainties are slightly more conservative than those presented in the Hyper-K
design report, which included a total systematic error between 3\% and 4\% depending on the sample.  The uncertainties for the more on-axis detector locations
appear marginally smaller because the broader spectrum tends to average over shape uncertainties more.  The uncertainties as a fraction of the total predicted event rate and as a function
of reconstructed energy are shown in Fig.~\ref{fig:erec_error}.  Here, the most prominent feature is the large uncertainty in the 1-3~GeV region of the 
1R$\mu$ samples for the detector at $L=1100$~km.  This energy range is the location of the first oscillation maximum and the large uncertainty arises from 
energy scale and non-quasielastic fraction uncertainties that can cause feed-down or feed-up (in the case of energy scale) into the region of the oscillation
maximum. 

\begin {figure}[htbp]
\captionsetup{justification=raggedright,singlelinecheck=false}
  \begin{center}
    \includegraphics[width=0.80\textwidth]{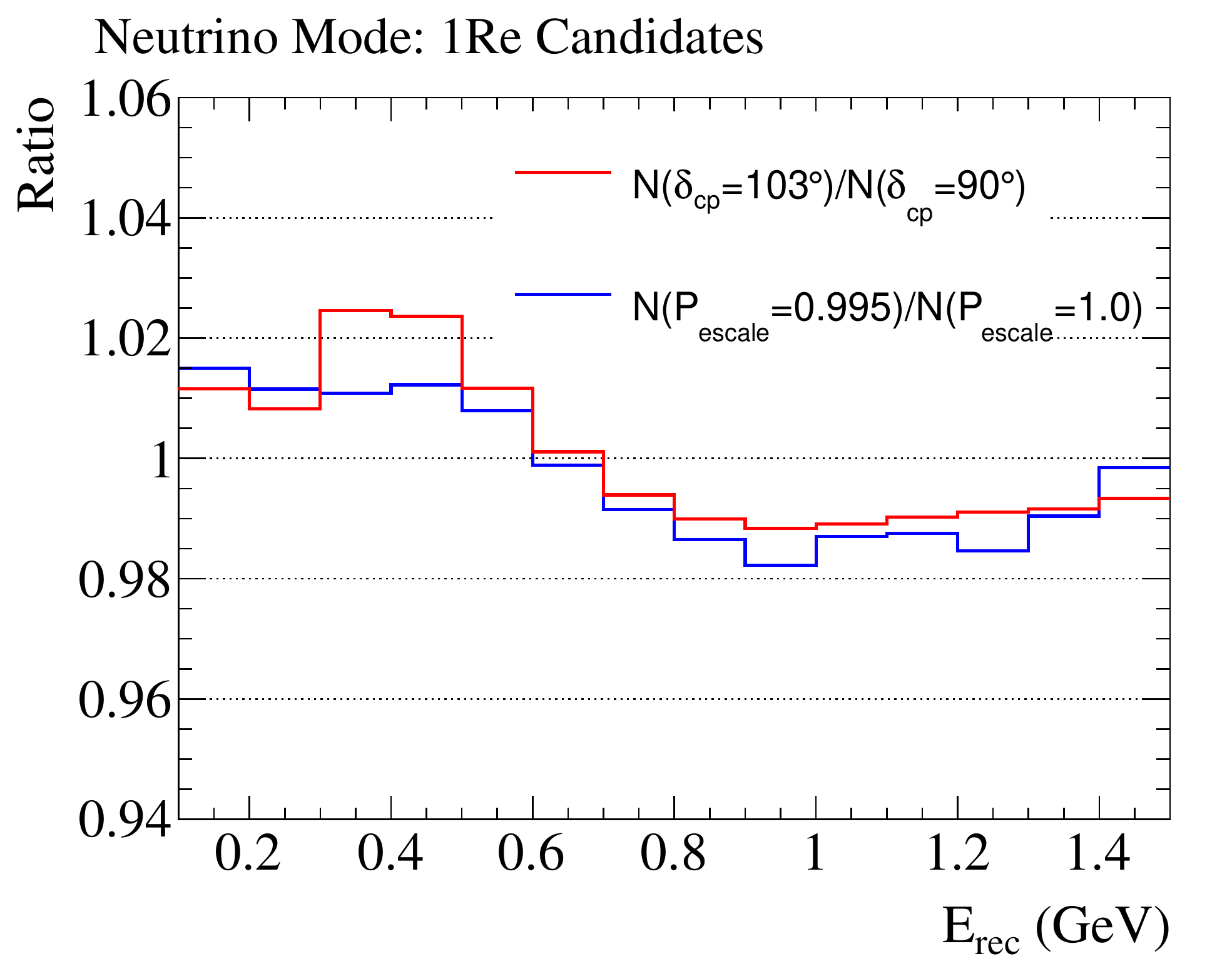} \\
    \includegraphics[width=0.80\textwidth]{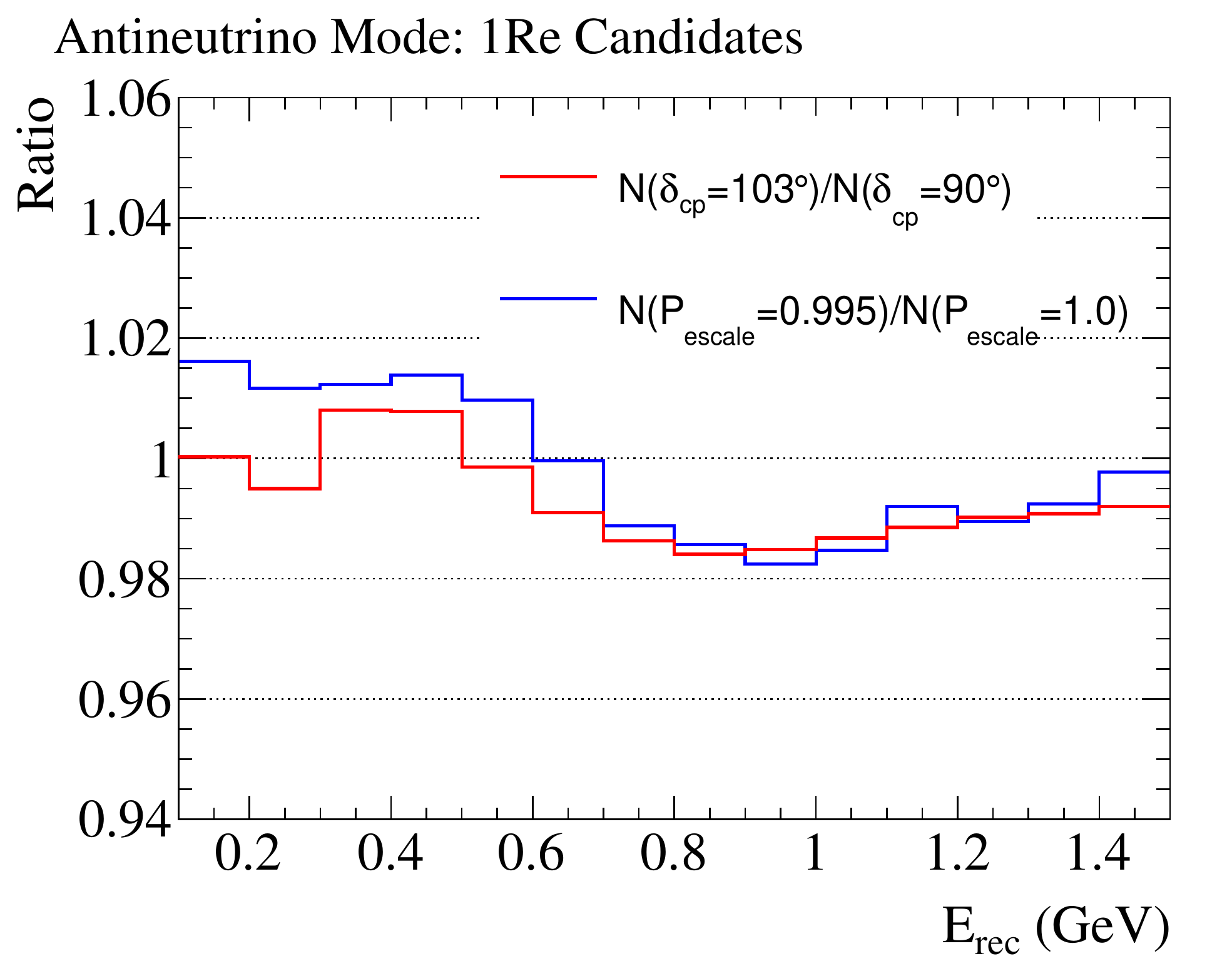}
    \caption{The ratios to nominal predicted spectra ($\delta_{cp}=\pi/2$) for a $\delta_{cp}$ shift of $+13^{\circ}$ and an 
             energy scale shift of $-0.5\%$.  The ratios are shown for the 1R$e$ samples in neutrino mode (top) and antineutrino mode (bottom).
             The ratios are calculated for the Hyper-K detector at 295~km and 2.5$^{\circ}$ off-axis.}
    \label{fig:hk_dcp_escale}
  \end{center}
\end {figure}

\begin {figure}[htbp]
\captionsetup{justification=raggedright,singlelinecheck=false}
  \begin{center}
    \includegraphics[width=0.80\textwidth]{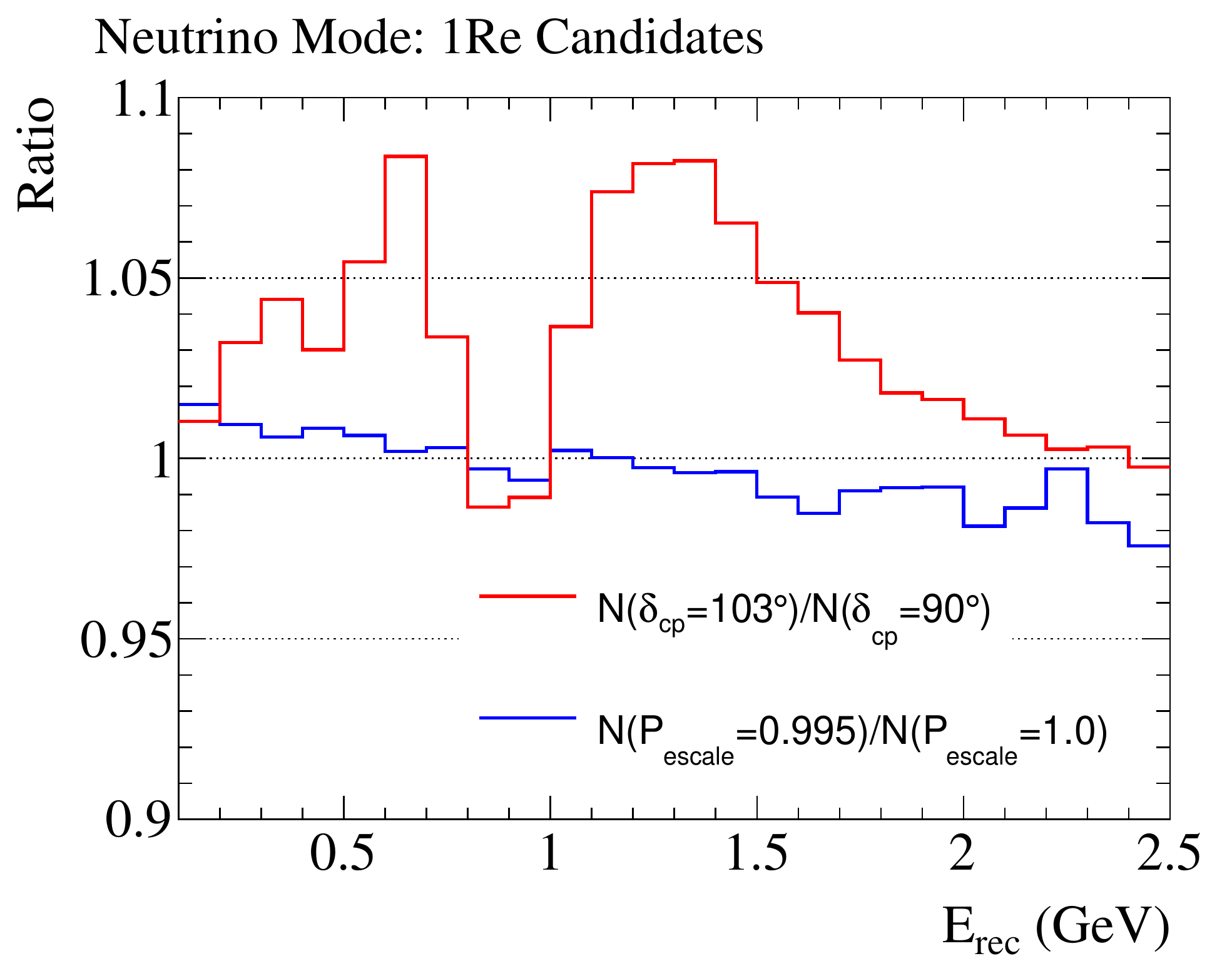} \\
    \includegraphics[width=0.80\textwidth]{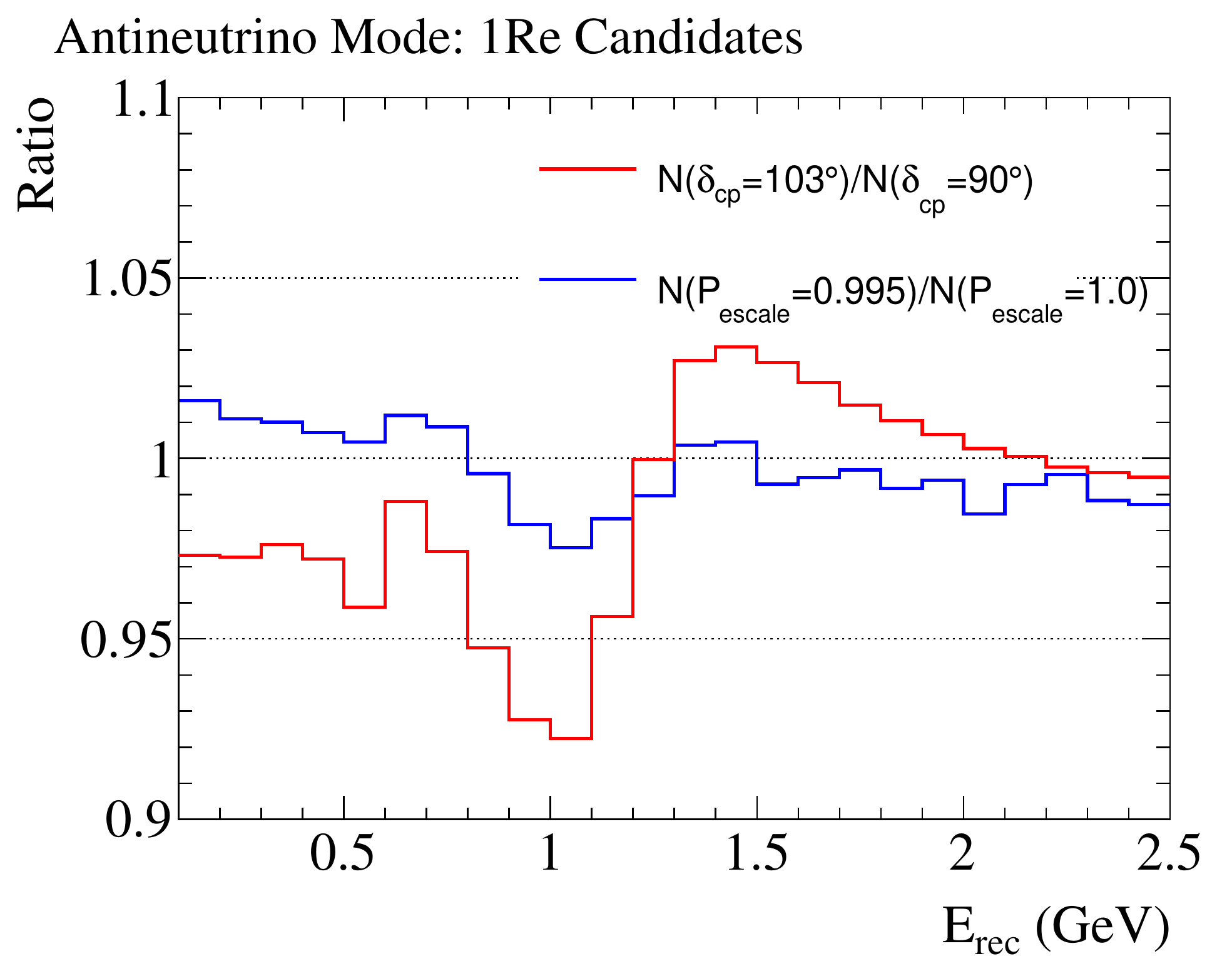}
    \caption{The ratios to nominal predicted spectra ($\delta_{cp}=\pi/2$) for a $\delta_{cp}$ shift of $+13^{\circ}$ and an 
             energy scale shift of $-0.5\%$.  The ratios are shown for the 1R$e$ samples in neutrino mode (top) and antineutrino mode (bottom).
             The ratios are calculated for the Korean detector at 1100~km and 1.5$^{\circ}$ off-axis.}
    \label{fig:kd_1p5_dcp_escale}
  \end{center}
\end {figure}

The relationship between systematic uncertainties and the physics sensitivity with a Korean detector can be better understood by
investigating a specific measurement, the precision measurement of $\delta_{cp}$ when $\delta_{cp}$ is near a maximally CP violating
value of $\pi/2$ or $3\pi/2$.  Near these values, the derivative of sin$(\delta_{cp})$ approaches zero, degrading the sensitivity to
the CP odd term in the oscillation probability.  Here, the CP even term, which depends on cos$(\delta_{cp})$ may contribute to the
precision measurement of the phase.  Fig.~\ref{fig:hk_dcp_escale} shows the changes to the spectra for a change in a $\delta_{cp}$ by $+13^{\circ}$ from
an initial value of $\pi/2$ for the Hyper-K detector. Here, $13^{\circ}$ is chosen since it is expected to be the ultimate precision of Hyper-K after
a 10~year$\times$1.3~MW exposure with 2 tanks. It can be seen that the change to $\delta_{cp}$ by $13^{\circ}$ largely effects the spectrum
through the cos$(\delta_{cp})$ term, causing a downward shift in energy with little change to the overall normalization.  Fig.~\ref{fig:hk_dcp_escale}
also shows the effect of an energy scale shift by $-0.5\%$ for comparison.  The energy scale shift has a similar effect on the spectrum, indicating
that even a 0.5\% uncertainty on the energy scale can degrade the $\delta_{cp}$ precision near maximally CP violating values.
%\clearpage
%\vspace{0.5cm}
%\captionsetup{width=16cm}
%\captionsetup{justification=raggedright,singlelinecheck=false}
%\begin{longtable}{@{\extracolsep{\fill}} l|c|c|c|c|c @{}}
%\lipsum[1]
%\begin{longtabu}{l|c|c|c|c|c}
%\begin{longtable}{@{\extracolsep{\fill}} l|c|c|c|c|c @{}}
\begin{table}[tbp]
\caption{\label{tab:sys_error} Percent error on the normalization of the predicted 1R$\mu$ and 1R$e$ samples
in neutrino and antineutrino mode for each systematic error source.  The error on the ratio of neutrino mode to
antineutrino mode is also shown for 1R$e$ since this uncertainty is relevant for the detection of a CP asymmetry.}
%\endfirsthead
\scalebox{0.83}{
%\footnotesize{
\renewcommand{\arraystretch}{0.6}% Tighter
\begin{tabular}{l|c|c|c|c|c} 
\hline \hline
                 & \multicolumn{5}{c}{Percent Error (\%)} \\ \hline
Error Source     & $\nu$ 1R$\mu$ &  $\bar{\nu}$ 1R$\mu$ &  $\nu$ 1R$e$ &  $\bar{\nu}$ 1R$e$ &  ($\nu$ 1R$e$)/($\bar{\nu}$ 1R$e$) \\ \hline
\multicolumn{6}{c}{OAA=2.5$^{\circ}$, $L=1100$~km} \\ \hline
$\sigma_{\nu_{e}}/\sigma_{\nu_{\mu}}$, $\sigma_{\bar{\nu}_{e}}/\sigma_{\bar{\nu}_{\mu}}$ & 0.00 & 0.00 & 2.10 & 1.68 & 3.12 \\ \hline
Energy Scale & 0.02 & 0.02 & 0.01 & 0.01 & 0.01 \\ \hline
Matter Density & 0.04 & 0.08 & 0.43 & 0.09 & 0.53 \\ \hline
NC$\pi^{+}$ Bgnd. & 1.28 & 1.25 & 0.00 & 0.00 & 0.00 \\ \hline
$\nu_{e}$ \& NC$\pi^{0}$ Bgnd. & 0.00 & 0.00 & 1.32 & 1.41 & 1.88 \\ \hline
CC non-QE Fraction & 2.76 & 1.88 & 1.98 & 1.29 & 2.35 \\ \hline
Extrapolation & 2.70 & 2.60 & 2.44 & 3.06 & 1.95 \\ \hline
Far Detector Model & 2.64 & 2.64 & 2.08 & 2.08 & 0.00 \\ \hline
Total & 4.69 & 4.16 & 4.54 & 4.47 & 4.86 \\ \hline
\multicolumn{6}{c}{OAA=2.0$^{\circ}$, $L=1100$~km} \\ \hline
$\sigma_{\nu_{e}}/\sigma_{\nu_{\mu}}$, $\sigma_{\bar{\nu}_{e}}/\sigma_{\bar{\nu}_{\mu}}$ & 0.00 & 0.00 & 2.01 & 1.67 & 3.07 \\ \hline
Energy Scale & 0.02 & 0.01 & 0.01 & 0.01 & 0.01 \\ \hline
Matter Density & 0.02 & 0.06 & 0.55 & 0.12 & 0.67 \\ \hline
NC$\pi^{+}$ Bgnd. & 1.47 & 1.29 & 0.00 & 0.00 & 0.00 \\ \hline
$\nu_{e}$ \& NC$\pi^{0}$ Bgnd. & 0.00 & 0.00 & 1.26 & 1.29 & 1.76 \\ \hline
CC non-QE Fraction & 0.87 & 0.82 & 1.24 & 0.76 & 1.51 \\ \hline
Extrapolation & 2.68 & 2.68 & 2.38 & 3.00 & 1.92 \\ \hline
Far Detector Model & 2.64 & 2.64 & 2.08 & 2.08 & 0.00 \\ \hline
Total & 3.89 & 3.83 & 4.18 & 4.27 & 4.39 \\ \hline
\multicolumn{6}{c}{OAA=1.5$^{\circ}$, $L=1100$~km} \\ \hline
$\sigma_{\nu_{e}}/\sigma_{\nu_{\mu}}$, $\sigma_{\bar{\nu}_{e}}/\sigma_{\bar{\nu}_{\mu}}$ & 0.00 & 0.00 & 1.72 & 1.41 & 2.67 \\ \hline
Energy Scale & 0.01 & 0.01 & 0.01 & 0.01 & 0.01 \\ \hline
Matter Density & 0.01 & 0.06 & 0.24 & 0.28 & 0.53 \\ \hline
NC$\pi^{+}$ Bgnd. & 1.61 & 1.30 & 0.00 & 0.00 & 0.00 \\ \hline
$\nu_{e}$ \& NC$\pi^{0}$ Bgnd. & 0.00 & 0.00 & 1.42 & 1.37 & 1.93 \\ \hline
CC non-QE Fraction & 0.44 & 0.30 & 0.52 & 0.37 & 0.75 \\ \hline
Extrapolation & 2.67 & 2.60 & 2.23 & 2.88 & 1.84 \\ \hline
Far Detector Model & 2.64 & 2.64 & 2.08 & 2.08 & 0.00 \\ \hline
Total & 3.83 & 3.81 & 3.84 & 4.11 & 3.91 \\ \hline
\multicolumn{6}{c}{OAA=2.5$^{\circ}$, $L=295$~km} \\ \hline
$\sigma_{\nu_{e}}/\sigma_{\nu_{\mu}}$, $\sigma_{\bar{\nu}_{e}}/\sigma_{\bar{\nu}_{\mu}}$ & 0.01 & 0.00 & 2.44 & 1.82 & 3.53 \\ \hline
Energy Scale & 0.04 & 0.03 & 0.42 & 0.63 & 0.21 \\ \hline
Matter Density & -- & -- & -- & -- & -- \\ \hline
NC$\pi^{+}$ Bgnd. & 2.33 & 1.79 & 0.00 & 0.00 & 0.00 \\ \hline
$\nu_{e}$ \& NC$\pi^{0}$ Bgnd. & 0.00 & 0.00 & 0.94 & 1.22 & 1.51 \\ \hline
CC non-QE Fraction & 1.68 & 1.72 & 2.07 & 1.00 & 2.25 \\ \hline
Extrapolation & 2.60 & 2.56 & 2.51 & 3.05 & 1.96 \\ \hline
Far Detector Model & 2.64 & 2.64 & 2.08 & 2.08 & 0.00 \\ \hline
Total & 4.13 & 4.15 & 4.71 & 4.47 & 4.90 \\ \hline
\multicolumn{6}{c}{OAA=2.5$^{\circ}$, $L=295$~km (Hyper-K Design Report)} \\ \hline
Total & 3.6 & 3.6 & 3.2 & 3.9 & -- \\ \hline
\hline 
%\end{longtable}
\end{tabular}%
}
\end{table}%
%\vspace{1.0cm}
%\begin {figure}[htbp]
The Korean detector is constraining $\delta_{cp}$ with a significant number of events at the second and third oscillation maximum.  Near the second oscillation maximum, the
effect of the CP-odd interference term in the oscillation probability is 3 times larger, and for the same shift in $\delta_{cp}$, the CP-odd
effect may be observable.  Fig~\ref{fig:kd_1p5_dcp_escale} shows the spectrum ratios for the Korean detector at 1100~km baseline and 
1.5$^{\circ}$ off-axis.  Here, the effect of both the CP-even term can be seen in the increased rate from 1.3~GeV and above for both neutrino
and antineutrino mode.  The CP-odd term causes an asymmetry in the normalization of the neutrino mode and antineutrino mode samples 
below 1~GeV.  These effects can not be reproduced with a small variation of the energy scale parameter, as is the case for Hyper-K.
This study shows that the constraint on $\delta_{cp}$ near  $\delta_{cp}=\pi/2,3\pi/2$ is sensitive to different systematic errors for
Hyper-K and the Korean detector.  It also shows that the fractional change to spectrum from the $\delta_{cp}$ variation is larger 
for the detector at a longer baseline, suggesting that the measurement is less likely to be systematics limited.  The full impact of the
Korean detector on the  $\delta_{cp}$ precision will be shown in the following section where the physics sensitivities are presented.

\subsection{Impact of the Korean detector on physics results}
For the physics sensitivity studies presented here, it is assumed that one or two 187~kton detectors will be operated for 10~years ($10\times10^{7}$~sec) at 1.3~MW beam power, corresponding to
 $27\times10^{21}$ protons on target.  
For the initial studies, five configurations are considered:
\begin{itemize}
\item{{\bf JD$\times$1} - A single detector is located in Japan at the Tochibora site with a baseline of 295~km and an off-axis angle of 2.5$^{\circ}$.}
\item{{\bf JD$\times$2} - Two detectors are located in Japan at the Tochibora site with a baseline of 295~km and an off-axis angle of 2.5$^{\circ}$.}
\item{{\bf JD$+$KD at 2.5$^{\circ}$} - One detector is located in Japan at a baseline of 295~km and an off-axis angle of 2.5$^{\circ}$, while
the second is located in Korea at a baseline of 1100~km and an off-axis angle of  2.5$^{\circ}$. }
\item{{\bf JD$+$KD at 2.0$^{\circ}$} - One detector is located in Japan at a baseline of 295~km and an off-axis angle of 2.5$^{\circ}$, while
the second is located in Korea at a baseline of 1100~km and an off-axis angle of  2.0$^{\circ}$. }
\item{{\bf JD$+$KD at 1.5$^{\circ}$} - One detector is located in Japan at a baseline of 295~km and an off-axis angle of 2.5$^{\circ}$, while
the second is located in Korea at a baseline of 1100~km and an off-axis angle of  1.5$^{\circ}$. }
\end{itemize}
Later in this section, the sensitivities for the Mt. Bisul site ($L=1084$~km and OAA=1.3$^{\circ}$) and the Mt. Bohyun site ($L=1043$~km and OAA=2.3$^{\circ}$) will also be presented.

The initial physics sensitivity studies focus on 3 measurements: the determination of the mass ordering, the discovery of CP violation
through the exclusion of the $\sin(\delta_{cp})=0$ hypothesis, and the precision measurement of $\delta_{cp}$.  In all cases, the
sensitivities are evaluated on pseudo-data generated with the following true oscillation parameter values:
\begin{itemize}
\item{$|\Delta m^{2}_{32}|=2.5\times10^{-3}$ eV$^{2}$}
\item{$\sin^{2}\theta_{23}=0.5$}
\item{$\sin^{2}\theta_{13}=0.0219$}
\item{$\Delta m^{2}_{21}=7.53\times10^{-5}$ eV$^{2}$}
\item{$\sin^{2}\theta_{12}=0.304$}
\end{itemize}
The pseudo-data are also generated for multiple values of $\delta_{cp}$ and both mass orderings, and the sensitivities are presented
as a function of the true value of $\delta_{cp}$ and the mass ordering.  In the fits to the pseudo-data, $\Delta m^{2}_{32}$, $\sin^{2}\theta_{23}$
and $\delta_{cp}$ are free parameters with no prior constraints.  $\sin^{2}\theta_{13}$, $\sin^{2}\theta_{12}$ and $\Delta m^{2}_{21}$ also vary
in the fits, but they have prior Gaussian constraints with 1$\sigma$ uncertainties of 0.0012, 0.041 and $0.18\times10^{-5}$~eV$^{2}$ respectively.   The prior uncertainties
on these parameters are taken from the 2015 edition of the PDG Review of Particle Physics~\cite{Patrignani:2016xqp}.  The systematic parameters described in the 
previous section are also allowed to vary as nuisance parameters in the fit within their prior constraints.  In all cases, the sensitivities
are evaluated on the fit to the so-called Asimov set, where the prediction for each sample is made for the nominal values of the oscillation parameters and
systematic parameters, and no statistical variations are applied.  All four samples (neutrino mode 1R$e$, 1R$\mu$ and antineutrino mode 1R$e$, 1R$\mu$) are used to construct a binned likelihood
and the product of the pseudo-data likelihood is taken with the Gaussian priors for constrained oscillation parameters and systematic parameters to construct
the full likelihood, $L$. To simplify the notation, we write $-2\tlog(L)$ as $\Delta\chi^2$.  

The test statistic used for the mass ordering determination is:
\begin{equation} 
\sqrt{\Delta\chi^{2}_{MH}} = \sqrt{\chi^2_{WH}-\chi^2_{CH}}
\label{eq:mh_test}
\end{equation}
Here,  $\chi^2_{WH}$ and $\chi^2_{CH}$ are the best-fit $-2\tlog(L)$ for the wrong and correct mass hierarchies respectively.  In the Gaussian limit,
the test parameter can be interpreted as the significance of the mass ordering determination.  Here sensitivities are shown for the Hyper-K accelerator neutrinos only and
do not account for the additional constraint from Hyper-K atmospheric neutrinos.

The test statistic used for the CP violation discovery potential is:
\begin{equation} 
\sqrt{\Delta\chi^{2}_{CPV}} = \sqrt{\tMIN[\chi^2_{BF}(\delta_{cp}=0),\chi^2_{BF}(\delta_{cp}=\pi)]-\chi^2_{BF}}
\label{eq:cpv_test}
\end{equation}
Here, $\chi^2_{BF}(\delta_{cp}=0)$ and $\chi^2_{BF}(\delta_{cp}=\pi)$ are the best-fit $-2\tlog(L)$ where $\delta_{cp}$ is fixed to one of the CP conserving values.  
The minimum of these two is used for the test statistic.  $\chi^2_{BF}$ is the best-fit minimum of  $-2\tlog(L)$  where $\delta_{cp}$ is allowed to 
vary.  Two cases are treated for the CP violation studies.  In the first case, the mass ordering is assumed to be known based on external measurements and the measurement
using the Hyper-K atmospheric neutrinos. In the second case, the constraints from external measurements and Hyper-K atmospheric neutrinos are not used, in order to estimate
the sensitivity from the accelerator neutrinos alone.  
When the mass ordering is determined with Hyper-K accelerator neutrinos alone, the sign of $\Delta m^{2}_{32}$ is allowed to vary in the minimization procedure.
The test parameter can be interpreted as the significance to exclude the CP conserving hypotheses.

For the evaluation of the $\delta_{cp}$ measurement precision the fitted value of $\delta_{cp}$ is scanned and the $-2\tlog(L)$ is minimized at each
value of $\delta_{cp}$, {\it i.e.} the profiling method.  The $\delta_{cp}$ values that correspond to a 1 unit change in $-2\tlog(L)$ relative to the minimum 
are taken as the bounds for the 68\% confidence interval.  The plotted 1$\sigma$ error is the width of the  68\% confidence interval divided by two. 

\begin {figure}[htbp]
\captionsetup{justification=raggedright,singlelinecheck=false}
  \begin{center}
    \includegraphics[width=0.80\textwidth]{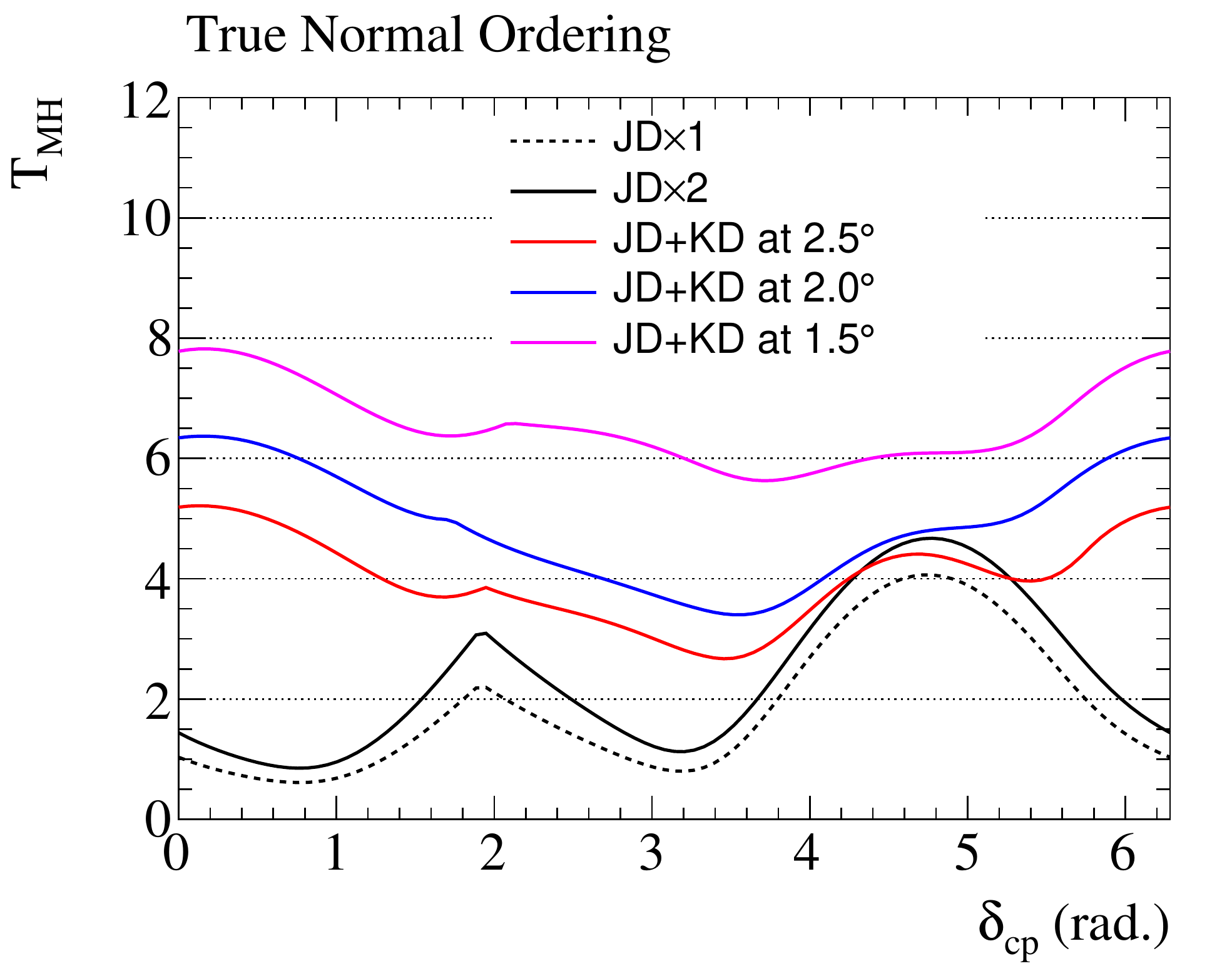}\\
    \includegraphics[width=0.80\textwidth]{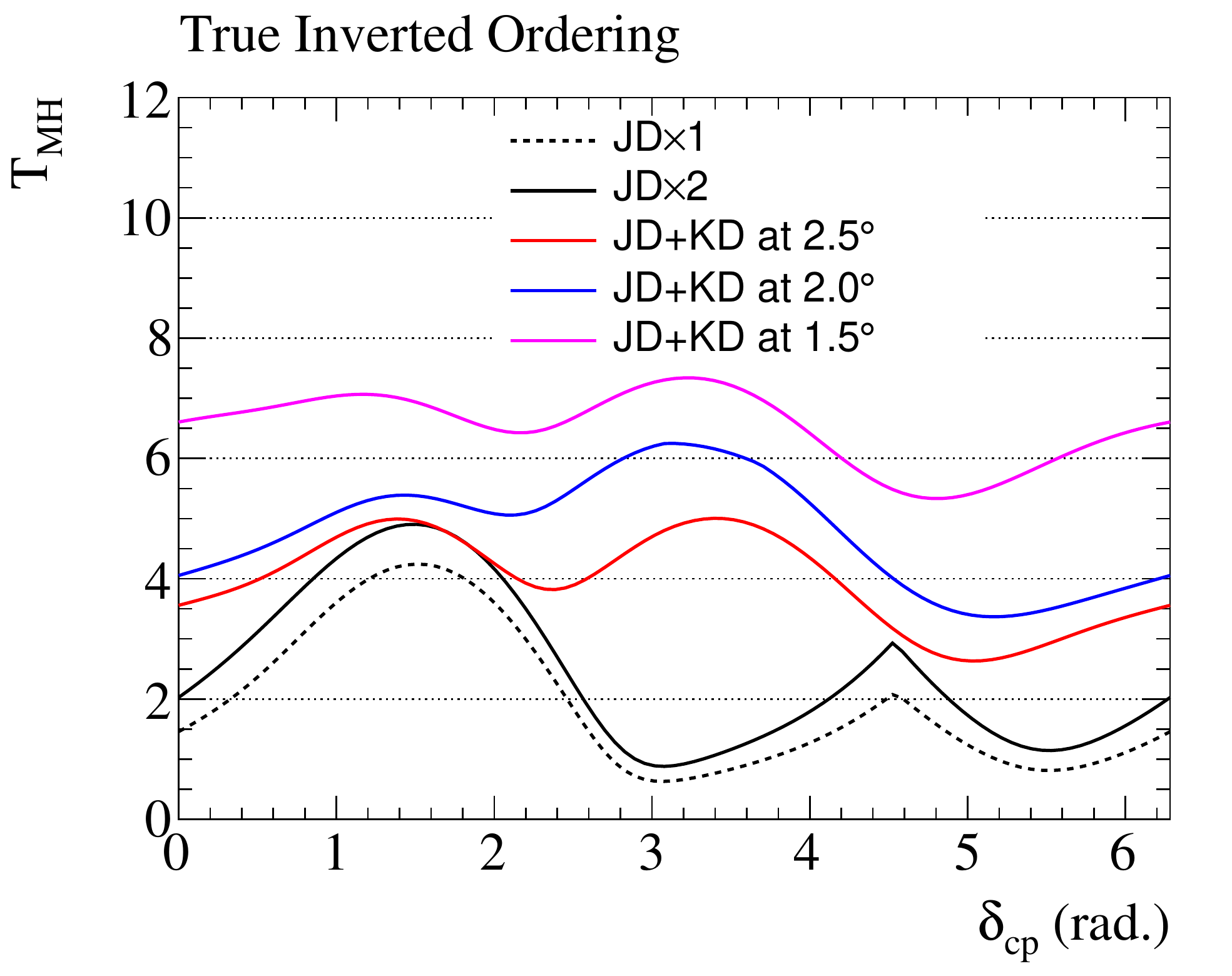}
    \caption{The significance for the wrong mass ordering rejection as a function of the true value of $\delta_{cp}$ and the true mass ordering (top=normal, bottom=inverted).}
    \label{fig:mh_sensitivity}
  \end{center}
\end {figure}

\begin {figure}[htbp]
\captionsetup{justification=raggedright,singlelinecheck=false}
  \begin{center}
    \includegraphics[width=0.8\textwidth]{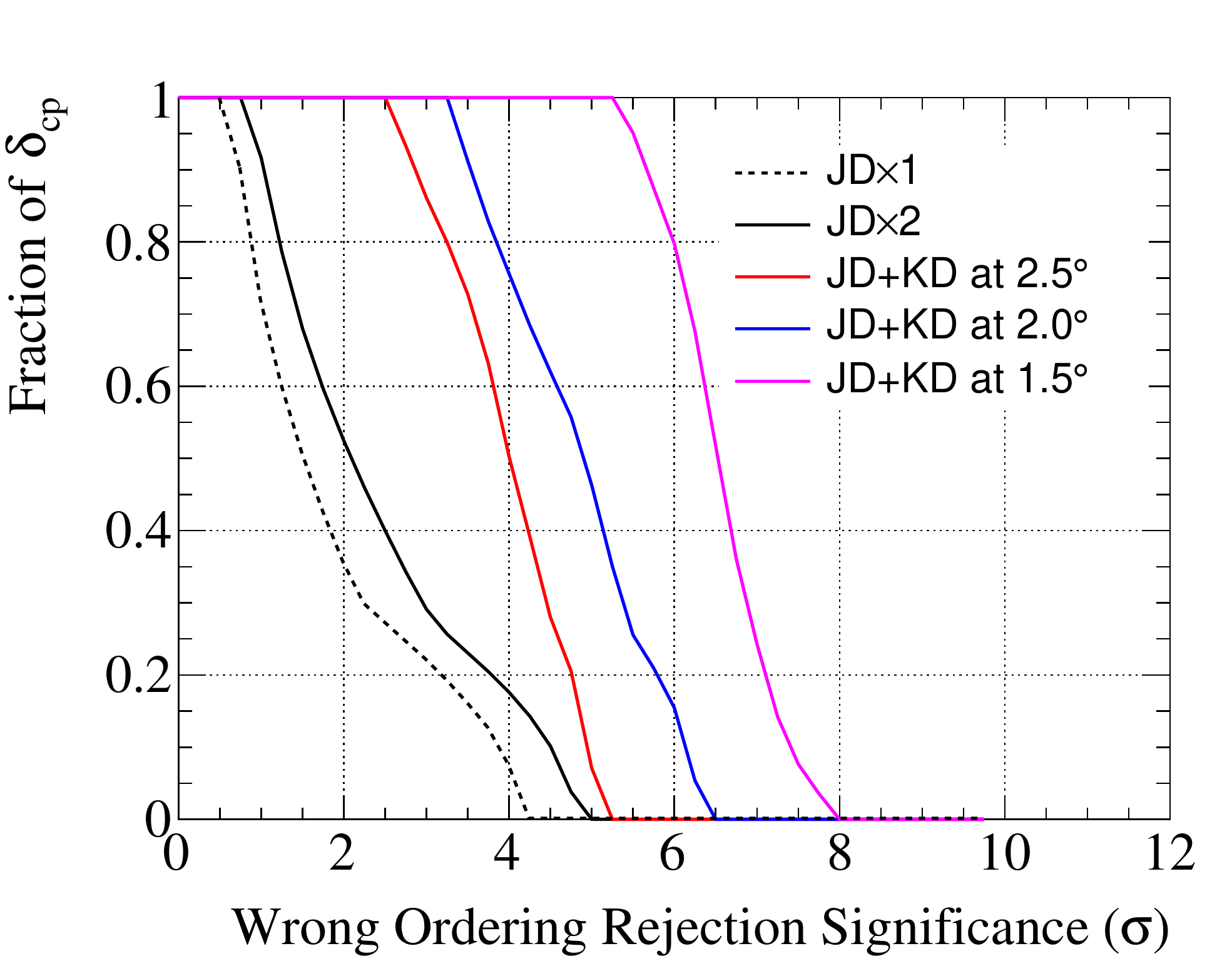}
    \caption{The fraction of $\delta_{cp}$ values (averaging over the true mass ordering) for which the wrong hierarchy can be rejected with a given significance or greater.}
    \label{fig:mh_cp_frac}
  \end{center}
\end {figure}

The significances to reject the wrong mass ordering are shown in Fig.~\ref{fig:mh_sensitivity}, and the fraction of $\delta_{cp}$ values for which a given
significance is achieved is shown in Fig.~\ref{fig:mh_cp_frac}.  As is expected based on Fig.~\ref{fig:matter_effect}, the significance
is largest for the configuration with the Korean detector at 1.5$^{\circ}$ off-axis since the more on-axis position gives more events in the 1-2~GeV range where the 
matter effect is large.  For this configuration, the significance to reject the wrong mass ordering is greater than 6$\sigma$ for most values of $\delta_{cp}$ and 
greater than 5$\sigma$ for all values of $\delta_{cp}$.  The significance of the wrong mass ordering rejection degrades as the Korean detector is moved to more
off-axis locations.  However, even the configuration with 2.5$^{\circ}$ off-axis Korean detector has 3$\sigma$ rejection sensitivity for most values of $\delta_{cp}$ and 
improved sensitivity over the configuration with one (both) detector(s) in Japan for all (most) values of $\delta_{cp}$.  Based on this study, it is clear that the sensitivity may
be improved further by adding events above 1~GeV in reconstructed energy.  This may be achieved by moving to a more on-axis position (see Mt. Bisul) or by including
multi-ring event reconstruction that allows the inclusion of higher energy events with one or more detected pions.  The multi-ring event reconstruction will be the
topic of a future study.  Based on this study of the configurations with a detector in Korea, the accelerator neutrinos can provide an alternative measurement of the mass ordering
that is complimentary to the measurement using atmospheric neutrinos.  By combining the two measurements, an even stronger constraint can be obtained, and better sensitivity can be achieved 
earlier in the lifetime of the Hyper-K.  

\begin {figure}[htbp]
\captionsetup{justification=raggedright,singlelinecheck=false}
  \begin{center}
    \includegraphics[width=0.49\textwidth]{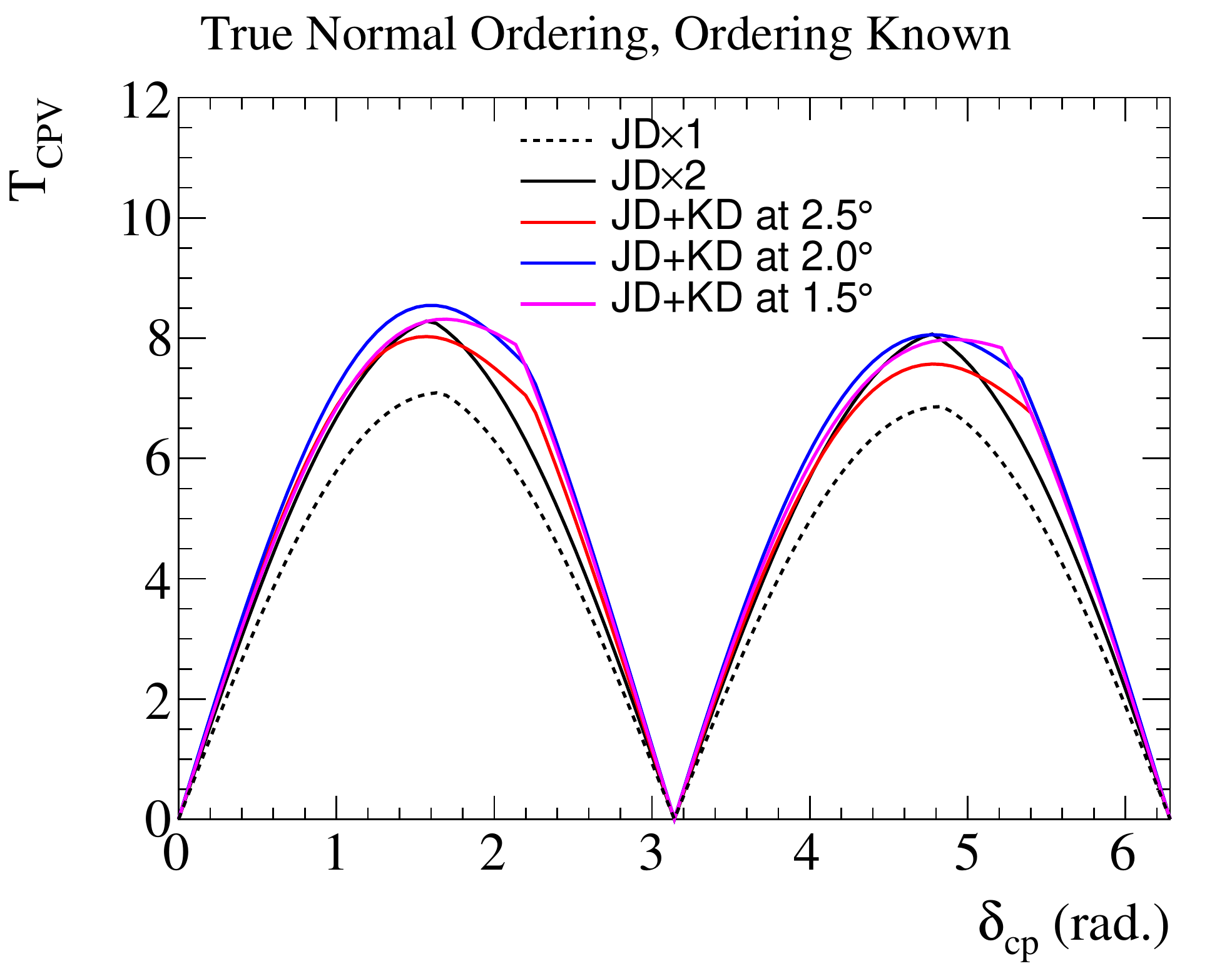}
    \includegraphics[width=0.49\textwidth]{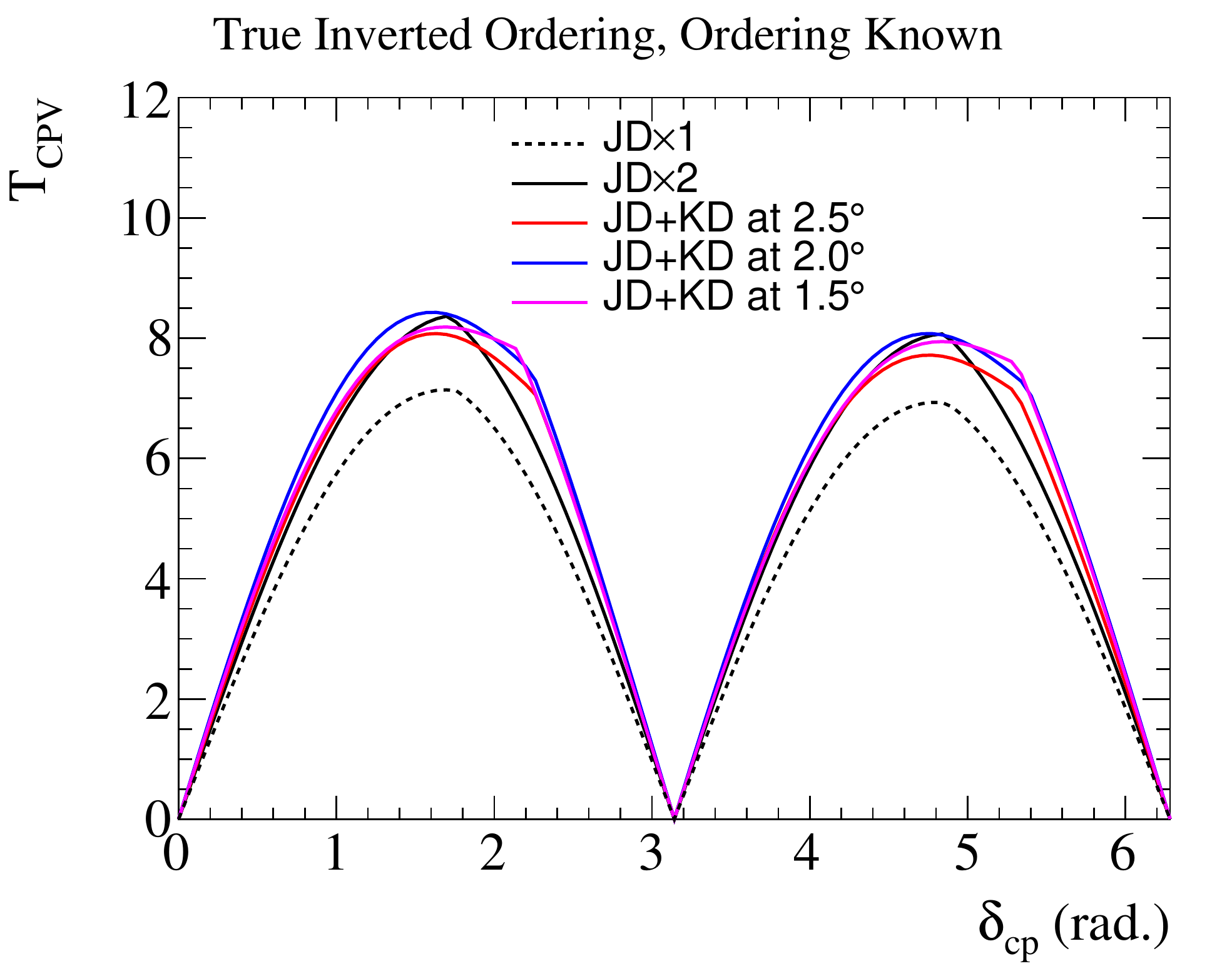}\\
    \includegraphics[width=0.49\textwidth]{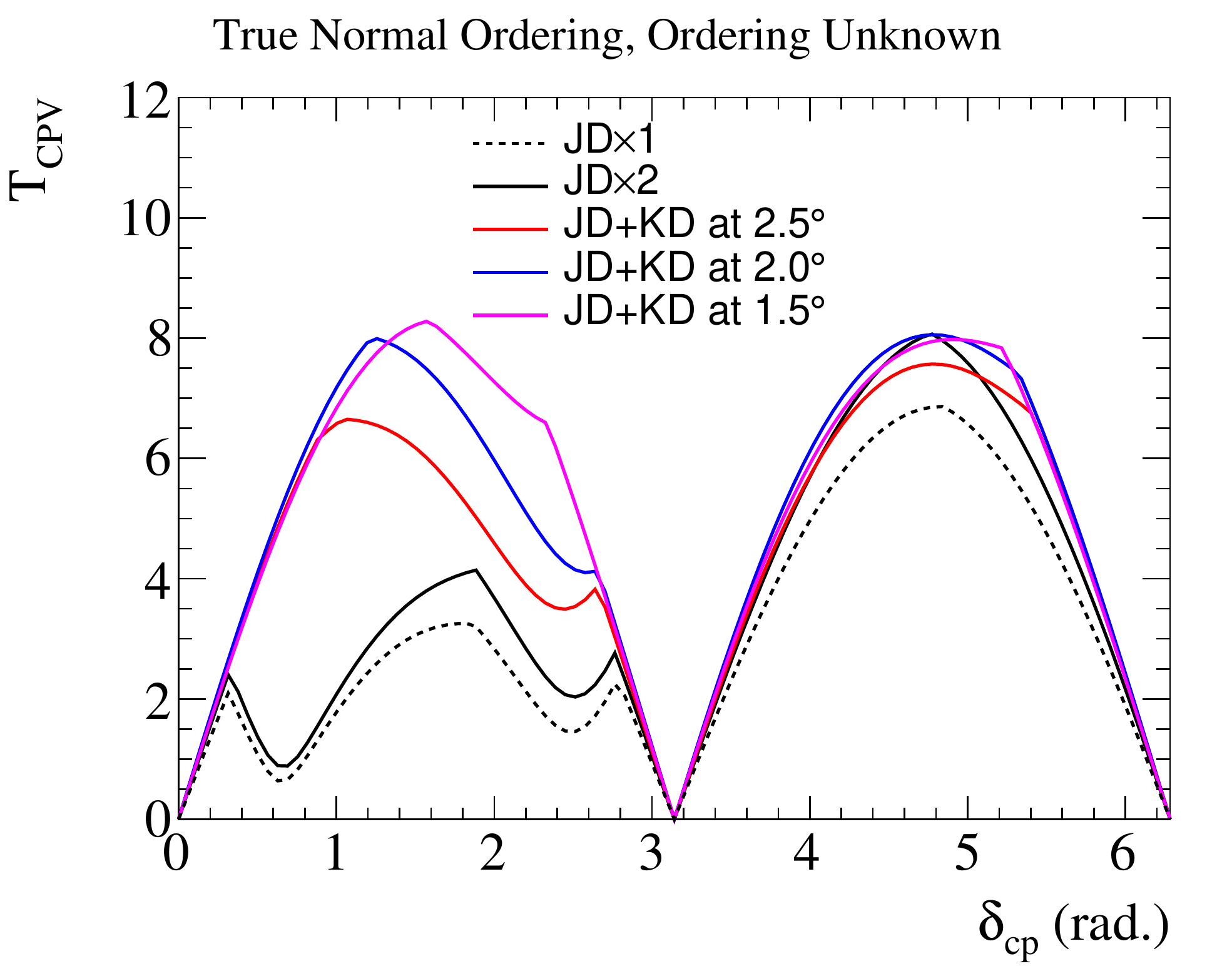}
    \includegraphics[width=0.49\textwidth]{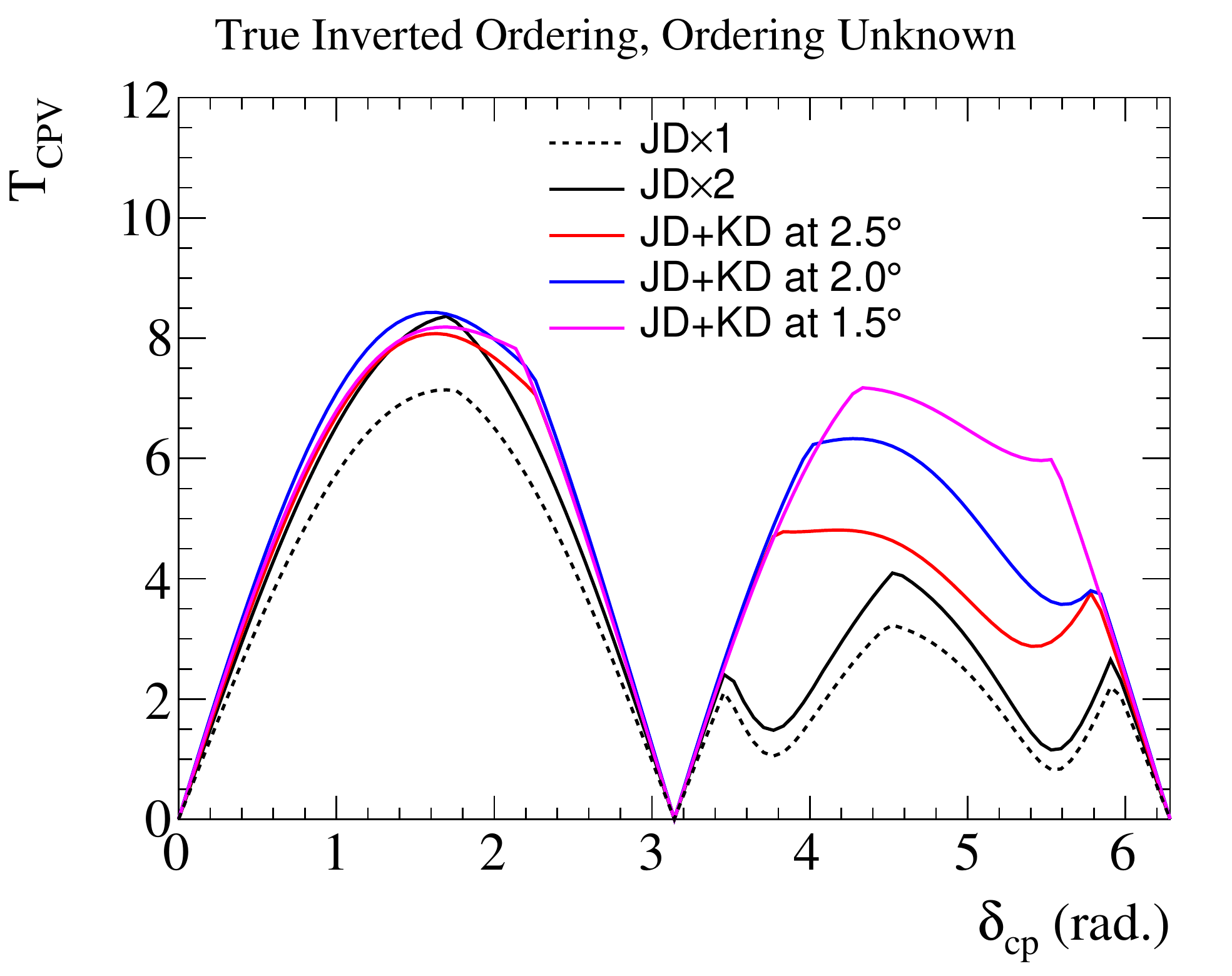}
    \caption{The significance for CP conservation rejection as a function of the true value of $\delta_{cp}$ and the true mass ordering (left=normal, right=inverted).  The top row
    shows the significance when the mass ordering is determined independently from the accelerator neutrinos, while the bottom 
    row shows the significance when the mass ordering is determined only by accelerator neutrinos observed in the Hyper-K detectors.}
    \label{fig:cpv_sensitivity}
  \end{center}
\end {figure}

\begin {figure}[htbp]
\captionsetup{justification=raggedright,singlelinecheck=false}
  \begin{center}
    \includegraphics[width=0.70\textwidth]{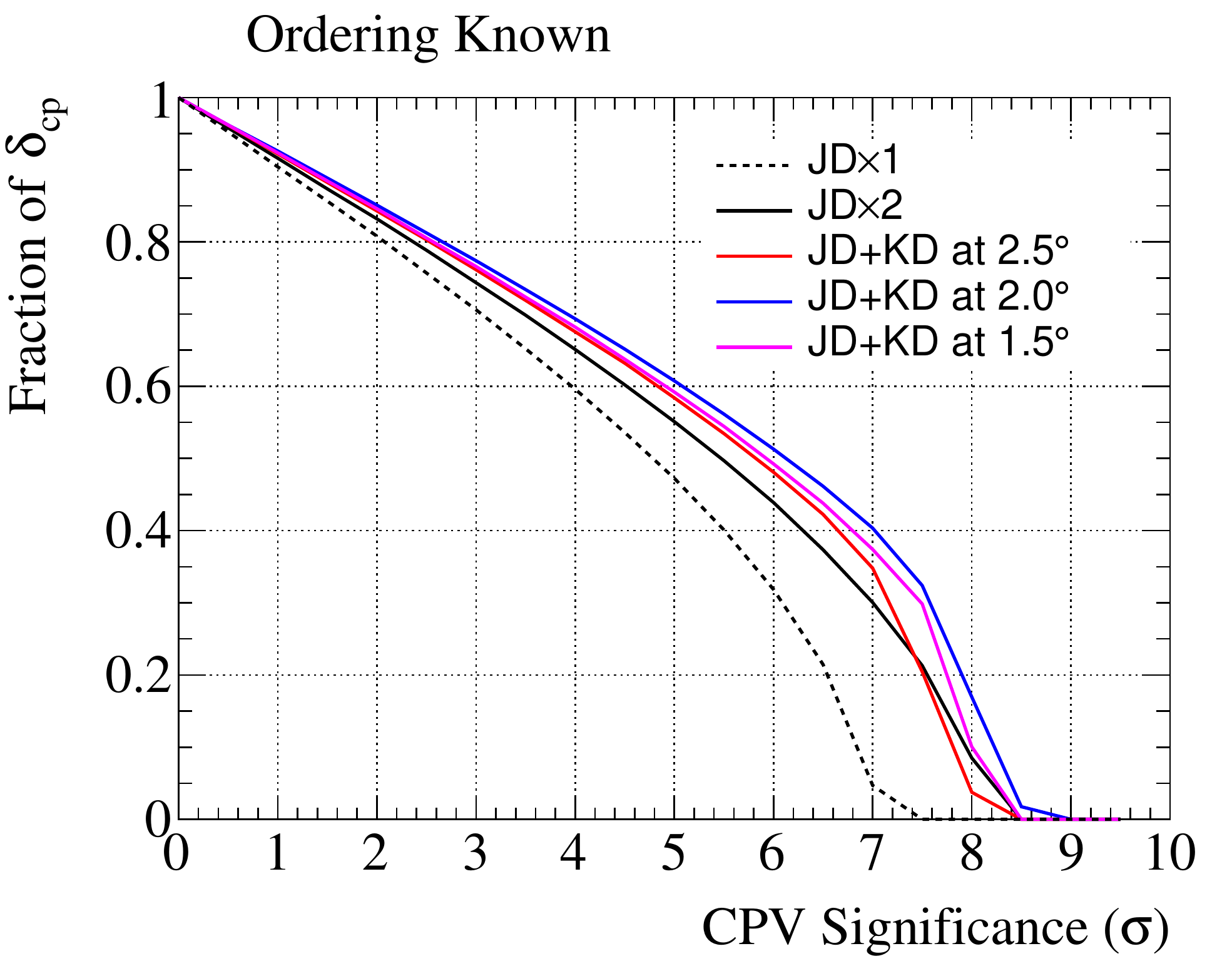}\\
    \includegraphics[width=0.70\textwidth]{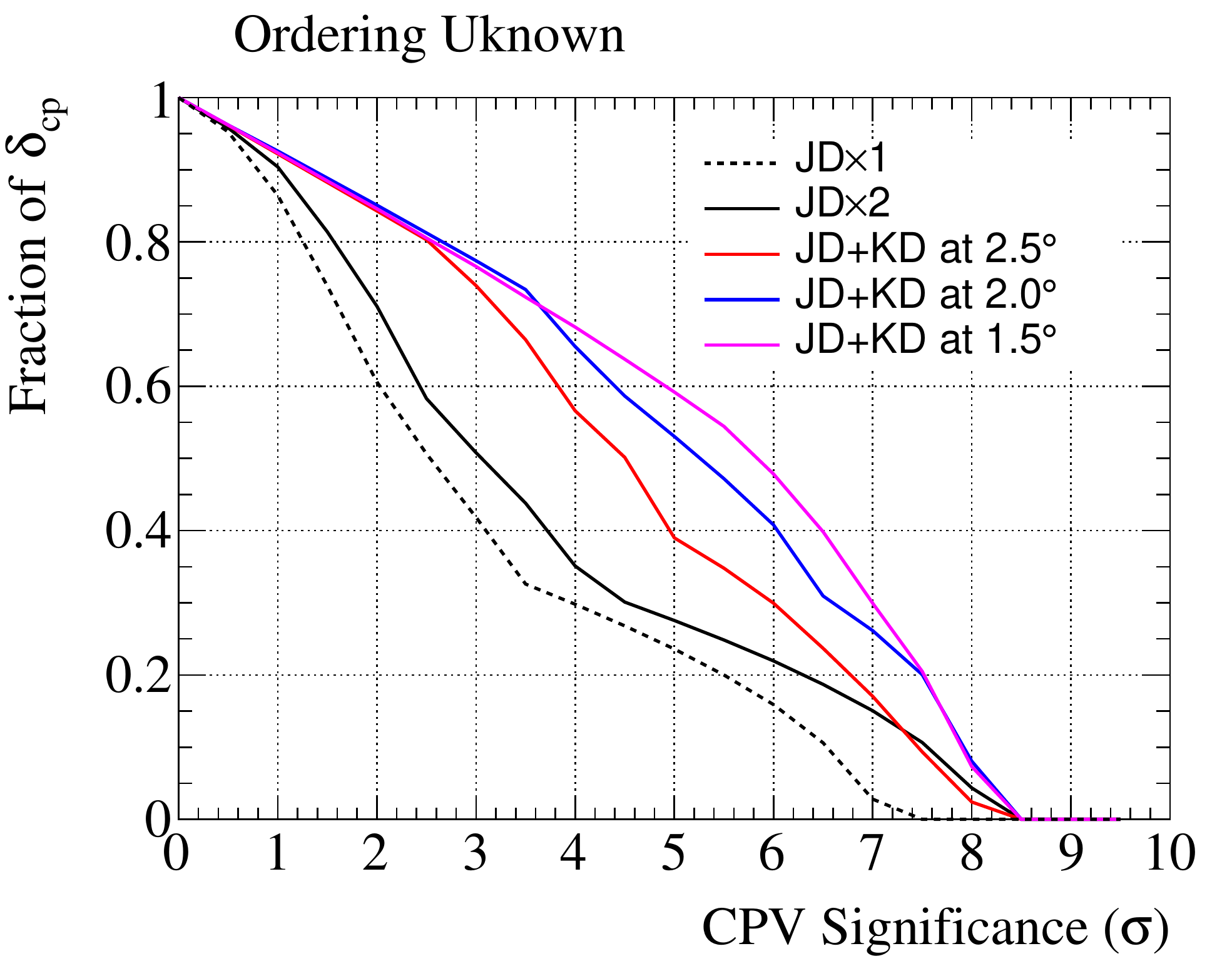}
    \caption{The fraction of $\delta_{cp}$ values (averaging over the true mass ordering) for which the CP conserving values can be rejected with a given significance or greater.
    The top figure shows the significance when the mass ordering is determined independently from the accelerator neutrinos, while the bottom 
    figure shows the significance when the mass ordering is determined only by accelerator neutrinos observed in the Hyper-K detectors. }
    \label{fig:cpv_cp_frac}
  \end{center}
\end {figure}

The plots showing the significance to reject the CP conserving hypotheses are in Fig.~\ref{fig:cpv_sensitivity}, and Fig.~\ref{fig:cpv_cp_frac} shows
the fraction of $\delta_{cp}$ values for which a given significance can be achieved.  The fractions of true $\delta_{cp}$ values for which
$3\sigma$ and $5\sigma$ sensitivity are achieved are listed in Table~\ref{tab:cpv_fraction}.
When the mass ordering is already known, all four two-detector configurations have similar sensitivity, but the best sensitivity is available when the Korean
detector is placed at 2.0$^{\circ}$ off-axis. It should be mentioned that in this study, it is assumed that the mass ordering is determined by external experiments and 
Hyper-K atmospheric neutrinos with
a significance greater than the CP conservation rejection significance being studied.  Compared to a single detector in Japan, the configuration with a second detector in Korea at 2.0$^{\circ}$ off-axis
has 3$\sigma$(5$\sigma$) sensitivity for an additional 7-8\%(13-14\%) of $\delta_{cp}$ values depending on the mass ordering.
When the mass ordering is only determined by the accelerator neutrinos, the configuration with the Korean detector at 1.5$^{\circ}$ off-axis gives the largest fraction of true $\delta_{cp}$ values
for which a $5\sigma$ discovery is possible.  This is true because this configuration has the best sensitivity to determine the mass ordering, breaking the 
mass ordering-$\delta_{cp}$ degeneracy.  For the case with the 1.5$^{\circ}$ off-axis configuration, the dependence of the CP violation discovery sensitivity on
the relative fraction of antineutrino mode to neutrino mode operation has been evaluate for ratios ranging from 3:1 (default) to 1:3.  The fraction of $\delta_{CP}$ values
with a $5\sigma$ discovery changes by only 0.01 depending on the relative fraction of antineutrino mode and neutrino mode operation.

\begin {figure}[htbp]
\captionsetup{justification=raggedright,singlelinecheck=false}
  \begin{center}
    \includegraphics[width=0.70\textwidth]{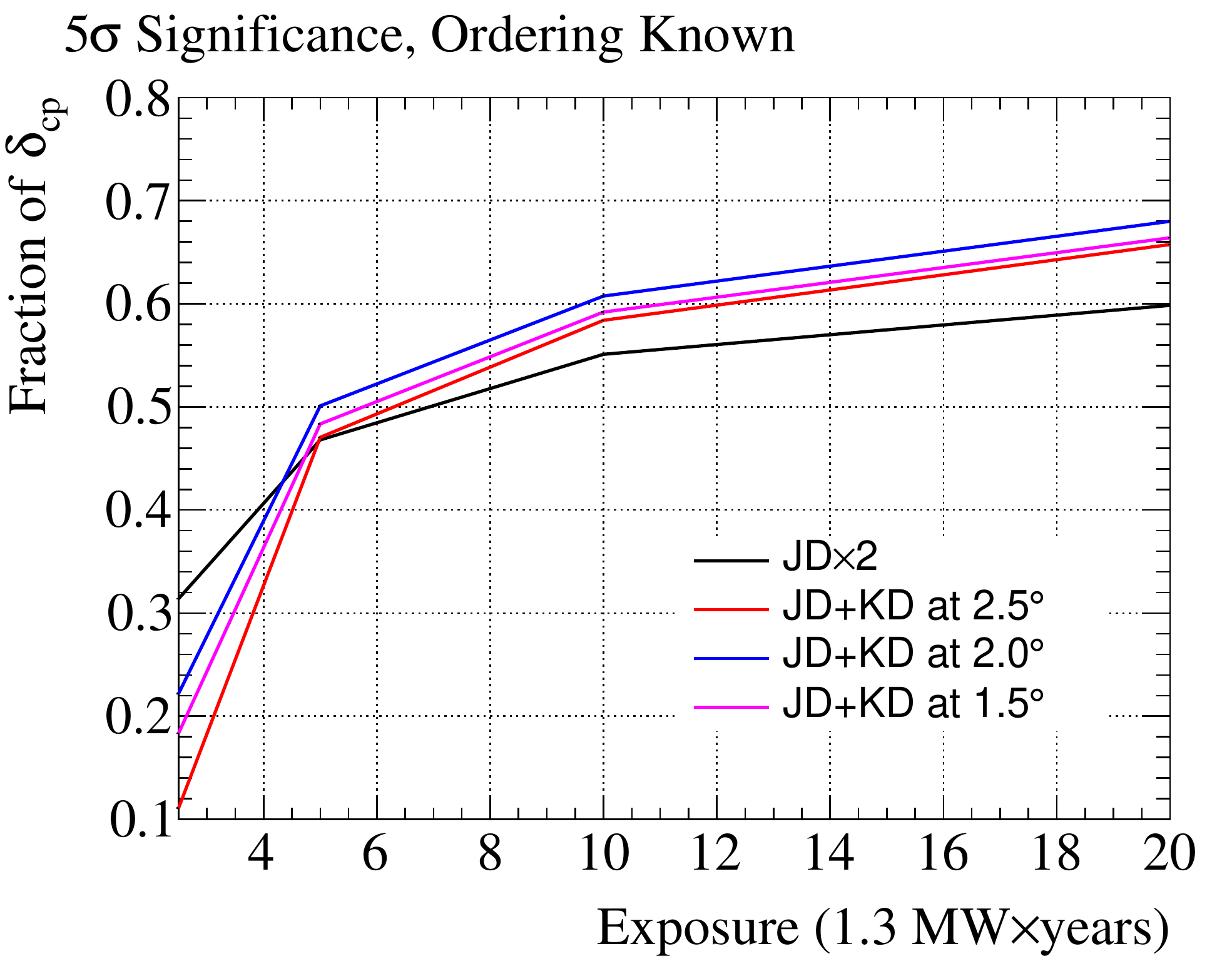}\\
    \includegraphics[width=0.70\textwidth]{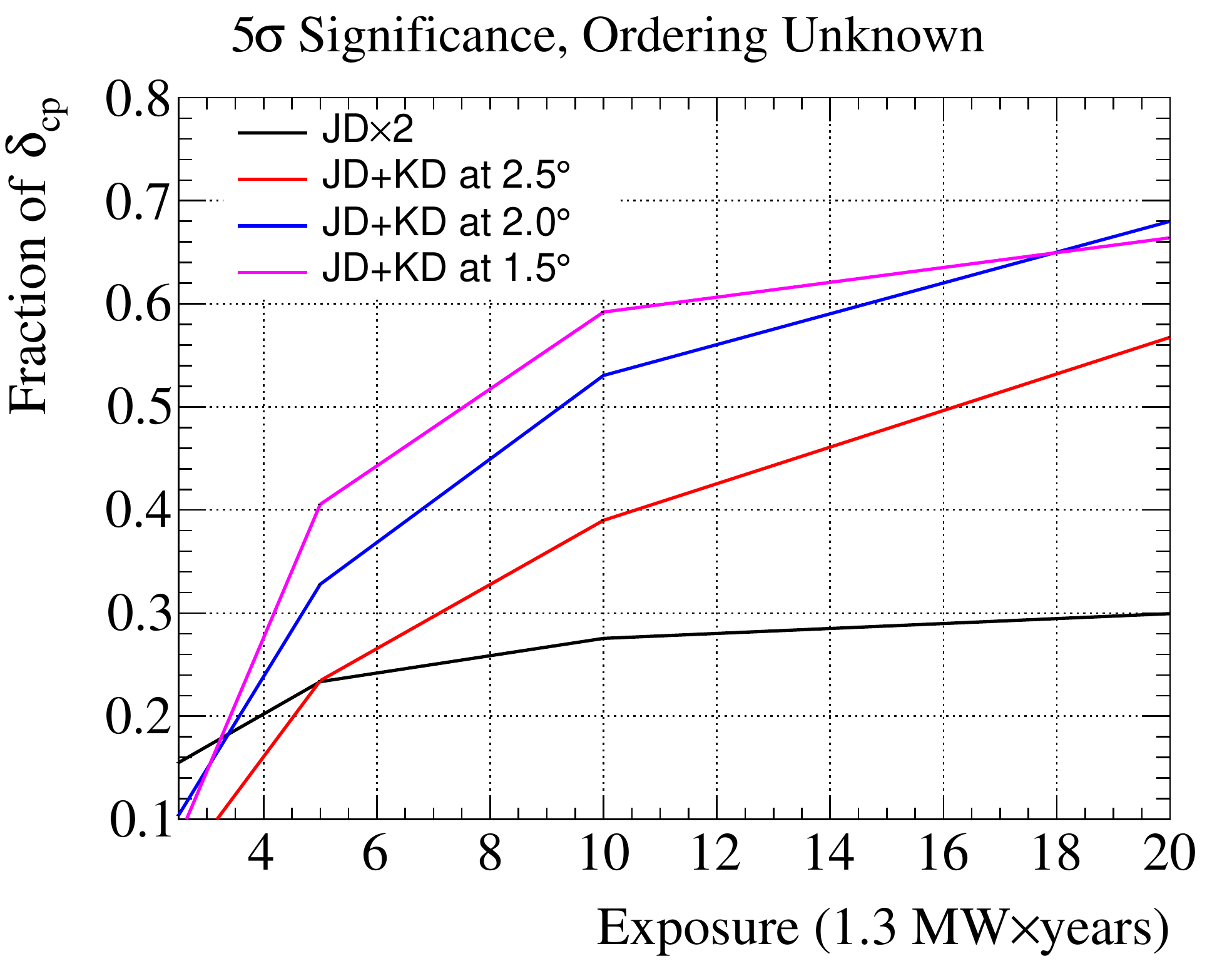}
    \caption{The fraction of $\delta_{cp}$ values (averaging over the true mass ordering) with at least a 5$\sigma$ significance to reject the CP conserving values of $\delta_{cp}$ .
    The top figure shows the significance when the mass ordering is determined independently from the accelerator neutrinos, while the bottom 
    figure shows the significance when the mass ordering is determined only by accelerator neutrinos observed in the Hyper-K detectors. }
    \label{fig:cpv_exp}
  \end{center}
\end {figure}

The evolution of the CP violation discovery potential with exposure is summarized in Fig.~\ref{fig:cpv_exp}.  At a 20~year$\times$1.3~MW exposure, the 
presence of the Korean detector can increase the fraction of $\delta_{cp}$ values for which a 5$\sigma$ discovery is possible by up to 8\%.  This is a
27\% reduction in the number of  $\delta_{cp}$ values for which a 5$\sigma$ discovery of CP violation would not be possible.

\begin{table}[tbp]
\captionsetup{justification=raggedright,singlelinecheck=false}
\caption{\label{tab:cpv_fraction}  The fraction of true $\delta_{cp}$ values for which CP violation can be discovered at 
3$\sigma$ or 5$\sigma$.}
\begin{center}%
\scalebox{1.0}{
\begin{tabular}{l|c|c|c|c|c|c|c|c} \hline \hline
 & \multicolumn{2}{|c}{True NH, Known} &  \multicolumn{2}{|c}{True IH, Known} &  \multicolumn{2}{|c}{True NH, Unknown} &  \multicolumn{2}{|c}{True IH, Unknown} \\ \hline
                         & $3\sigma$ & $5\sigma$ & $3\sigma$ & $5\sigma$ & $3\sigma$ & $5\sigma$ & $3\sigma$ & $5\sigma$ \\ \hline 
JD$\times$1              & 0.70      & 0.47      & 0.71      & 0.48      & 0.43      & 0.23      &  0.41     & 0.24 \\ \hline
JD$\times$2              & 0.74      & 0.55      & 0.74      & 0.55      & 0.52      & 0.27      &  0.50     & 0.28 \\ \hline
JD$+$KD at 2.5$^{\circ}$ & 0.76      & 0.58      & 0.76      & 0.59      & 0.76      & 0.48      &  0.72     & 0.30 \\ \hline
JD$+$KD at 2.0$^{\circ}$ & 0.78      & 0.61      & 0.78      & 0.61      & 0.77      & 0.55      &  0.79     & 0.51 \\ \hline
JD$+$KD at 1.5$^{\circ}$ & 0.77      & 0.59      & 0.77      & 0.59      & 0.77      & 0.59      &  0.77     & 0.59 \\ \hline

\hline \hline
\end{tabular}%
}
\end{center}
\end{table}%

\begin {figure}[htbp]
\captionsetup{justification=raggedright,singlelinecheck=false}
  \begin{center}
    \includegraphics[width=0.80\textwidth]{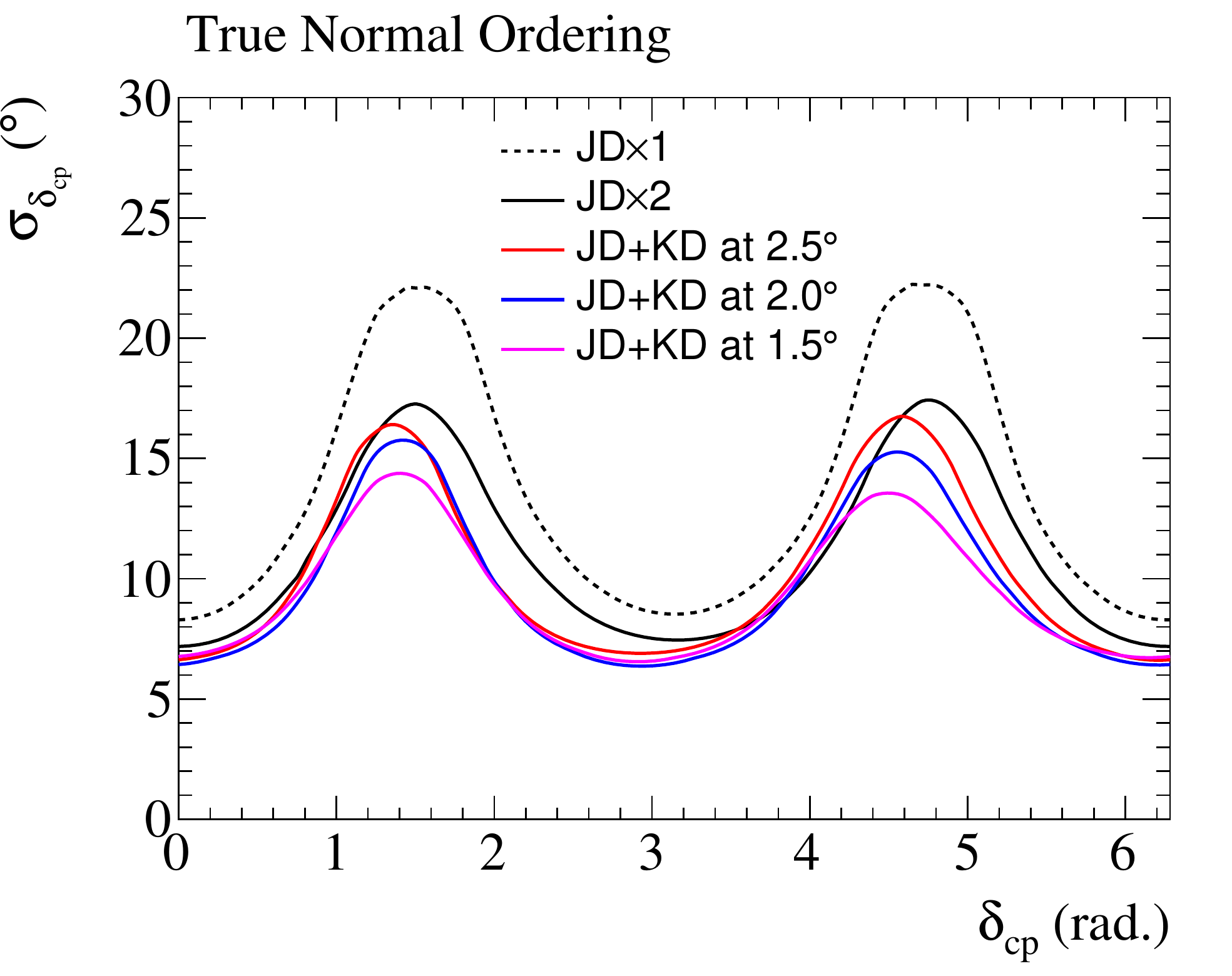}\\
    \includegraphics[width=0.80\textwidth]{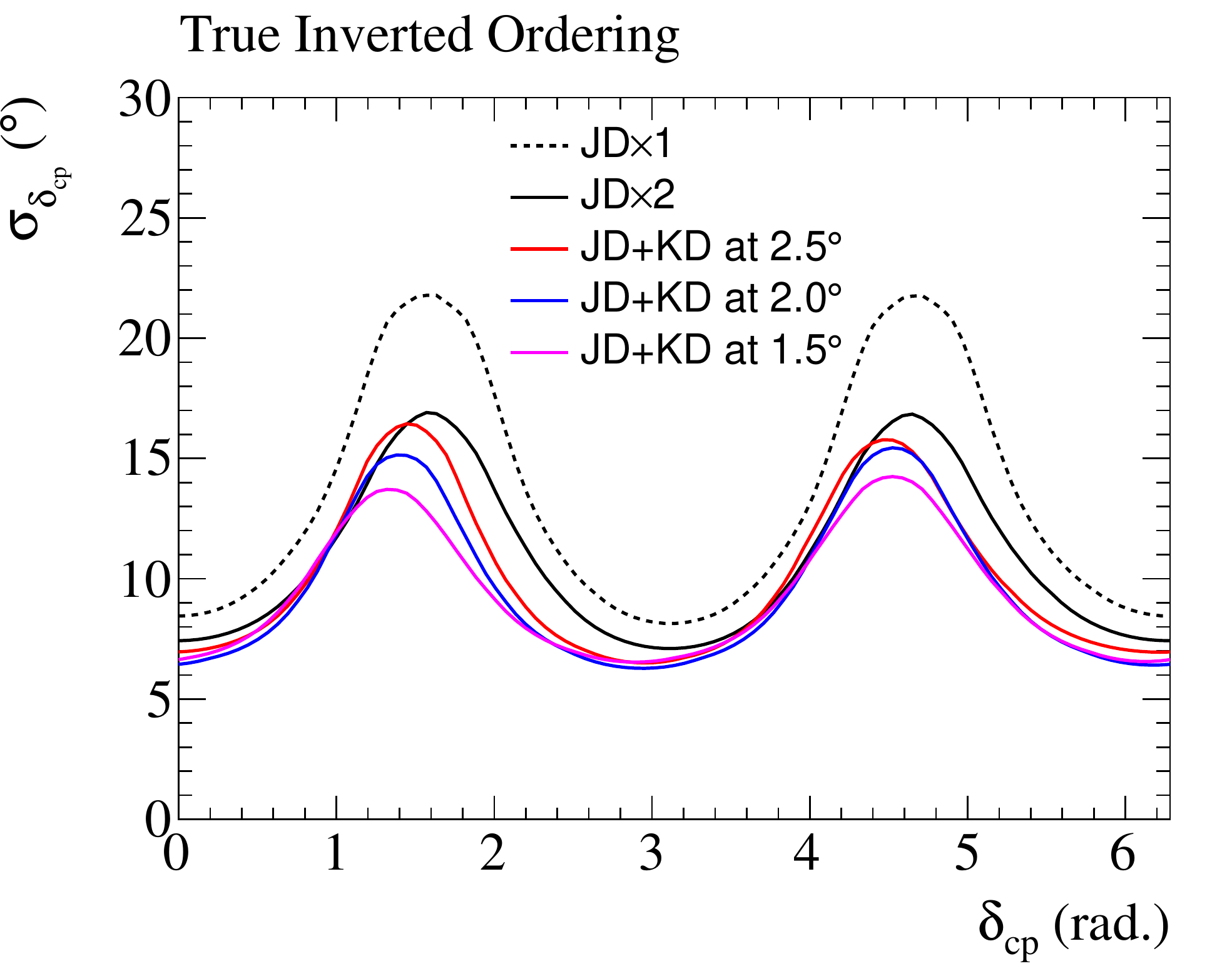}
    \caption{The 1$\sigma$ precision of the $\delta_{cp}$ measurement as a function of the true $\delta_{cp}$ value.  Here, it is assumed there is no prior knowledge of the
mass ordering.}
    \label{fig:cp_precision}
  \end{center}
\end {figure}

\begin {figure}[htbp]
\captionsetup{justification=raggedright,singlelinecheck=false}
  \begin{center}
    \includegraphics[width=0.8\textwidth]{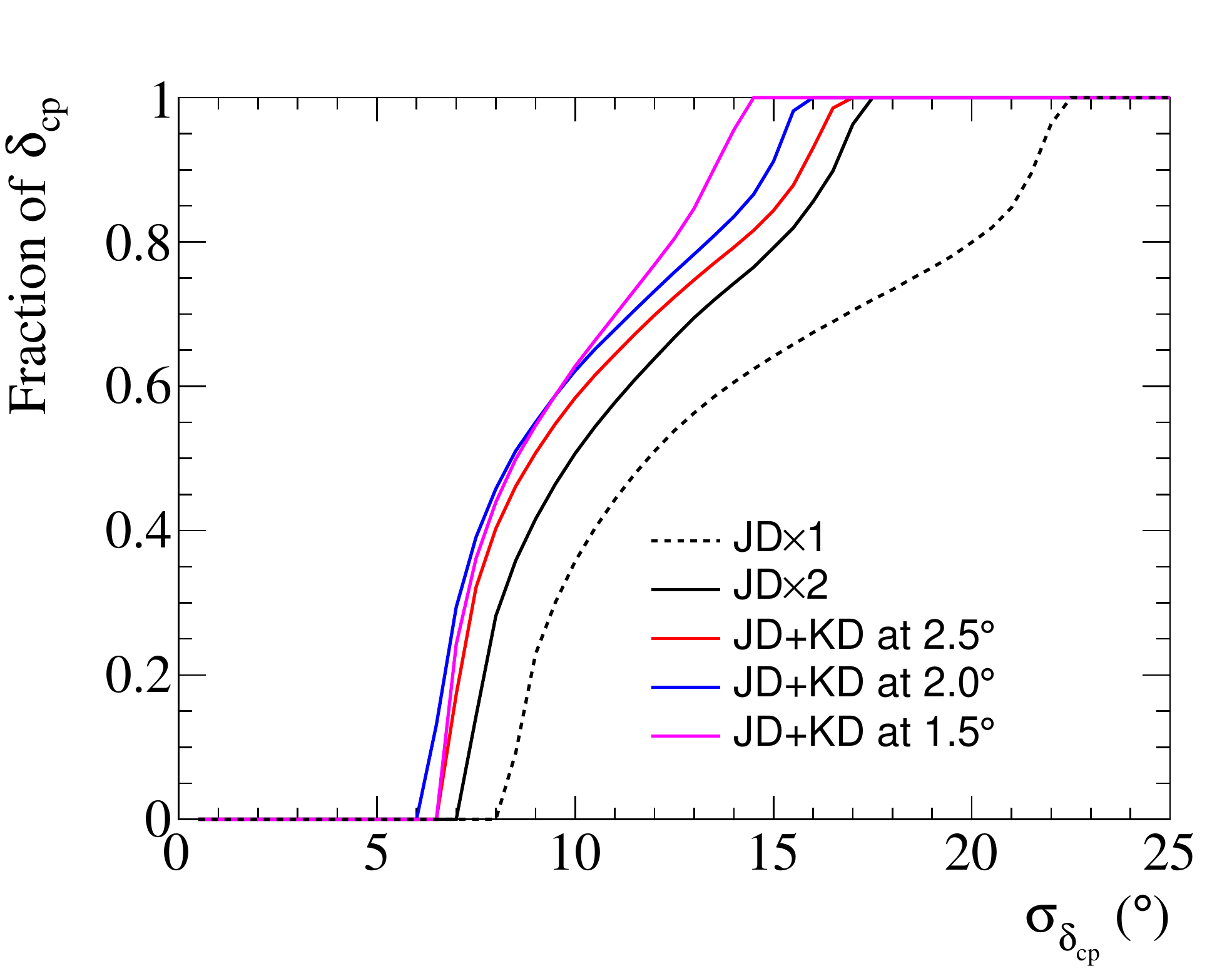}
    \caption{The fraction of $\delta_{cp}$ values (averaging over the true mass ordering) for which a given precision or better on $\delta_{cp}$ can be achieved.}
    \label{fig:cp_prec_cp_frac}
  \end{center}
\end {figure}

The $\delta_{cp}$ measurement precision is shown in Fig.~\ref{fig:cp_precision}, and Fig.~\ref{fig:cp_prec_cp_frac} shows the fraction of 
$\delta_{cp}$ values for which a given level of precision can be achieved.  The configurations with the Korean detector give the best $\delta_{cp}$ precision on 
average.  Near the CP conserving values, the configurations with the 2.0$^{\circ}$ and   1.5$^{\circ}$ off-axis Korean detectors have similar precision.  However,
near the maximally CP violating values of $\delta_{cp}$ the  1.5$^{\circ}$ off-axis configuration has 1.5$^{\circ}$ better precision for $\delta_{cp}$ than the 2.0$^{\circ}$ off-axis configuration.
The configuration with the 1.5$^{\circ}$ off-axis Korean detector also improves on the precision of the configuration with 2 detectors in Japan by 3$^{\circ}$ near the 
maximally CP violating values of $\delta_{cp}$.  The precision for the configuration with 2 detectors in Japan is $3^{\circ}$ better than what is presented in the
Hyper-K design report.  An improved sensitivity is expected since the Hyper-K design report assumes a staged approach with the second detector starting operation after 
6~years, while these studies assume that both detectors start operation simultaneously.

%  Further studies are necessary to determine if this difference arises due to differences in the systematic error model.  However, it is likely
%that any additional systematic errors will more strongly impact the measurement with 2 detectors in Japan since the measurements at 295~km baseline are systematics limited, 
%while the measurements at the 1100~km baseline are statistics limited.  

\begin {figure}[htbp]
\captionsetup{justification=raggedright,singlelinecheck=false}
  \begin{center}
    \includegraphics[width=0.8\textwidth]{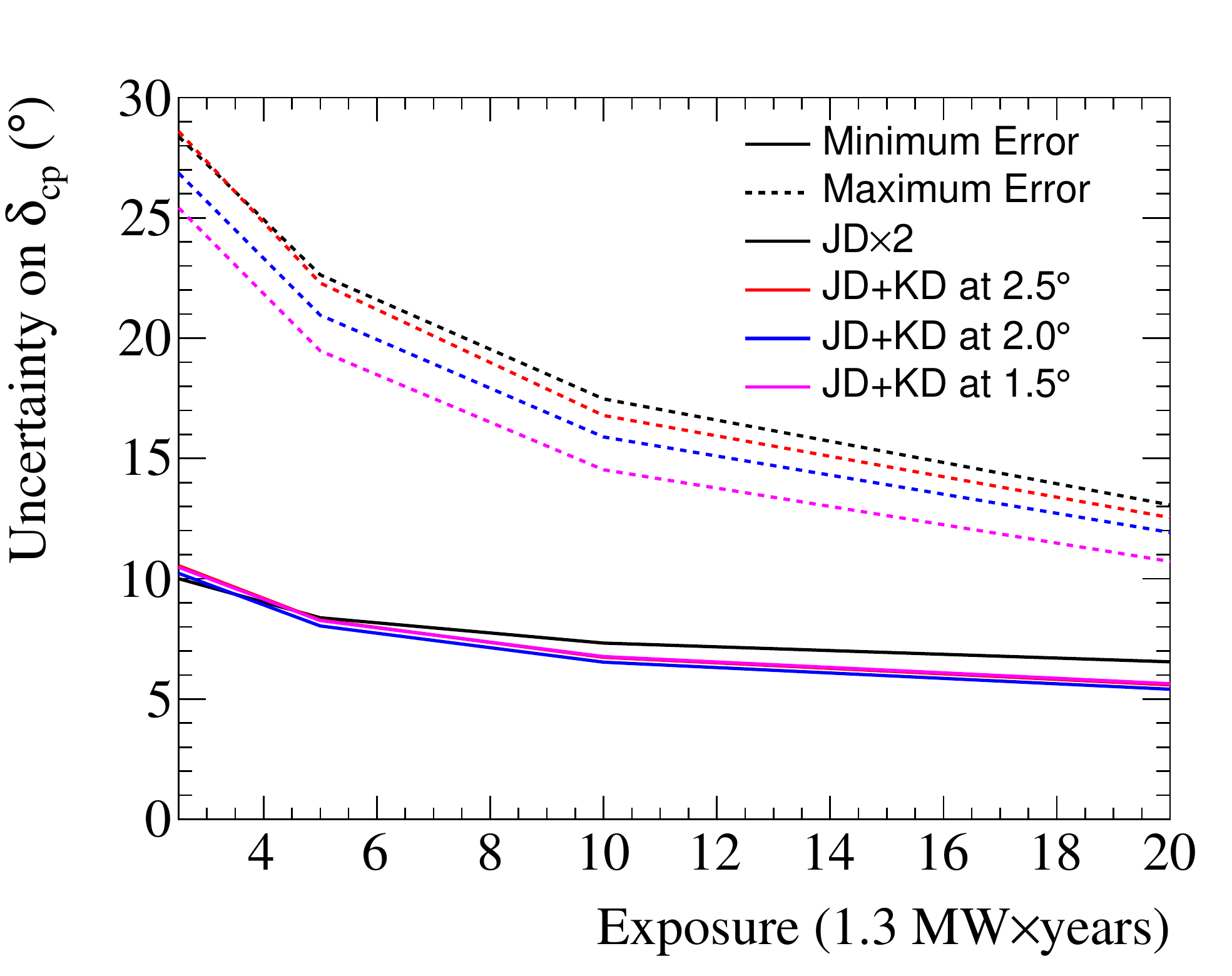}
    \caption{The evolution of the $\delta_{cp}$ measurement precision with exposure.  The ``Minimum" and ``Maximum" errors are the uncertainties at the true $\delta_{cp}$ and 
mass ordering values with the best and worst measurement resolution respectively. }
    \label{fig:cp_prec_exp}
  \end{center}
\end {figure}

The evolution of the  $\delta_{cp}$ precision with exposure is summarized in Fig.~\ref{fig:cp_prec_exp}. For the worst-case uncertainty, when $\delta_{cp}$  
is near the maximally CP violation values, the relative advantage of the detector in Korea remains constant with exposure. It should be noted that these 
conclusions depend on the systematic errors that are assumed, and may change if different levels of systematic uncertainty can be achieved.

% It should be noted, however, 
%that this may be an artifact of the systematic error model used in these studies, which likely underestimates the uncertainties on the shape of the observed
%spectra.  For a more realistic systematic error model, the $\delta_{cp}$ resolution may be degraded, particularly for the detector in Japan which is 
%more systematics limited.  

\subsubsection{Sensitivity studies for the Mt. Bisul and Mt. Bohyun site}
The potential Mt. Bisul site is located at a baseline of 1084~km and an off-axis angle of 1.3$^{\circ}$.  The primary effect of the off-axis angle change from 1.5$^{\circ}$ to 
1.3$^{\circ}$ is to decrease the (anti)neutrino flux at 700~MeV by $\sim10\%$ while increasing flux above 1.2~GeV by $\sim50\%$.  With these flux changes, it is expected
that the Mt. Bisul location should provide better sensitivity to determine the mass ordering, while the CP violation discovery potential may be slightly degraded. The Mt. Bohyun
site is located at a baseline of 1043~km and an off-axis angle of 2.3$^{\circ}$.  With a lower energy flux that more directly probes the second oscillation maximum, it is expected
that Mt. Bohyun should have a slightly improved CP violation discovery sensitivity compared to the Mt. Bisul site, but will have less sensitivity to the mass ordering.

For the sensitivities presented here, the combinations of the detector in Japan with a detector at Mt. Bisul or Mt. Bohyun are shown.  For comparison, the sensitivity with two 
detectors in Japan is shown.  Here the sensitivities are shown as a function of true $\delta_{CP}$ and the band of sensitivities shows the variation of the sensitivity in for
the range of true sin$^2\theta_{23}$ values between 0.4 and 0.6.

\begin {figure}[htbp]
\captionsetup{justification=raggedright,singlelinecheck=false}
  \begin{center}
    \includegraphics[width=0.70\textwidth]{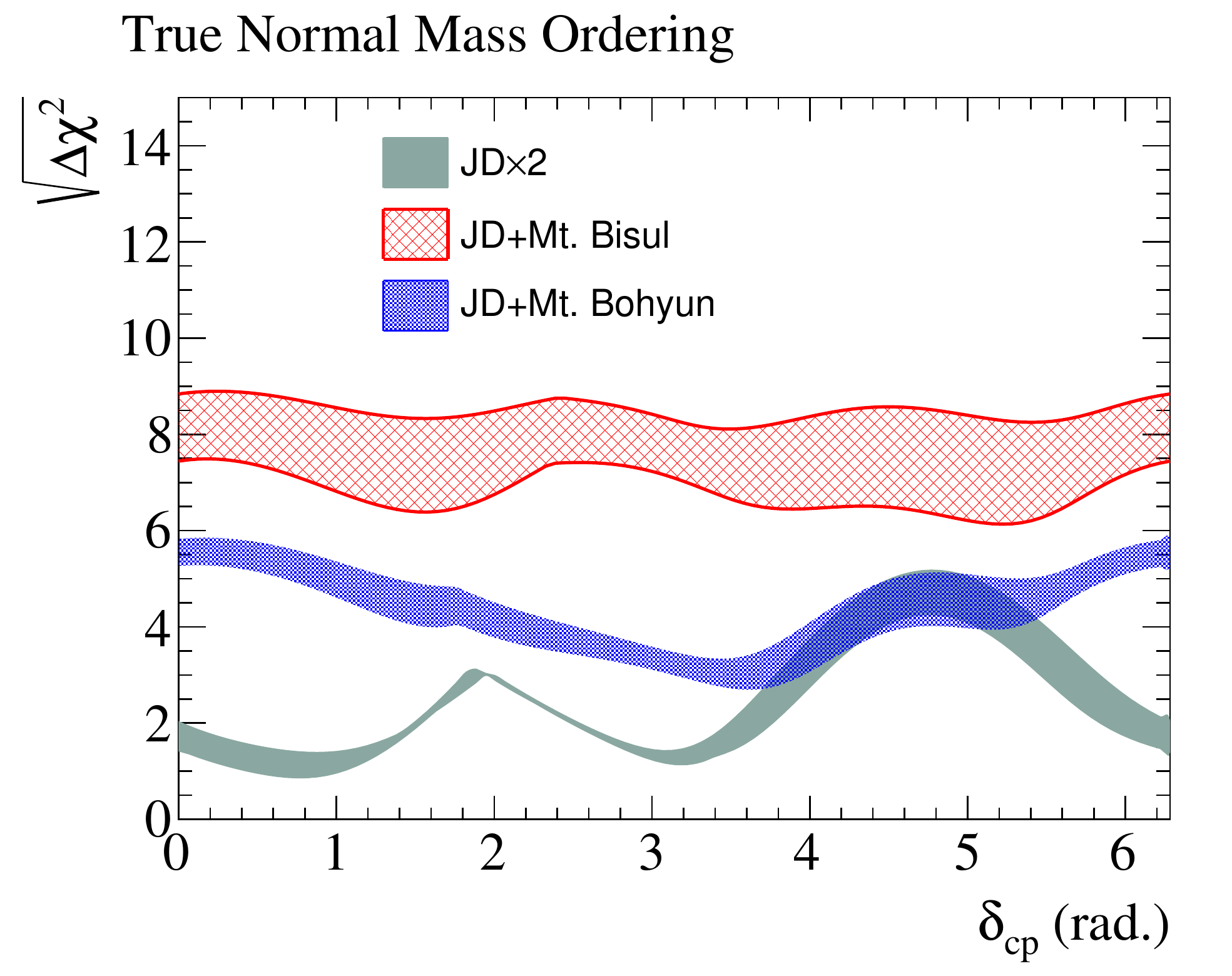}\\
    \includegraphics[width=0.70\textwidth]{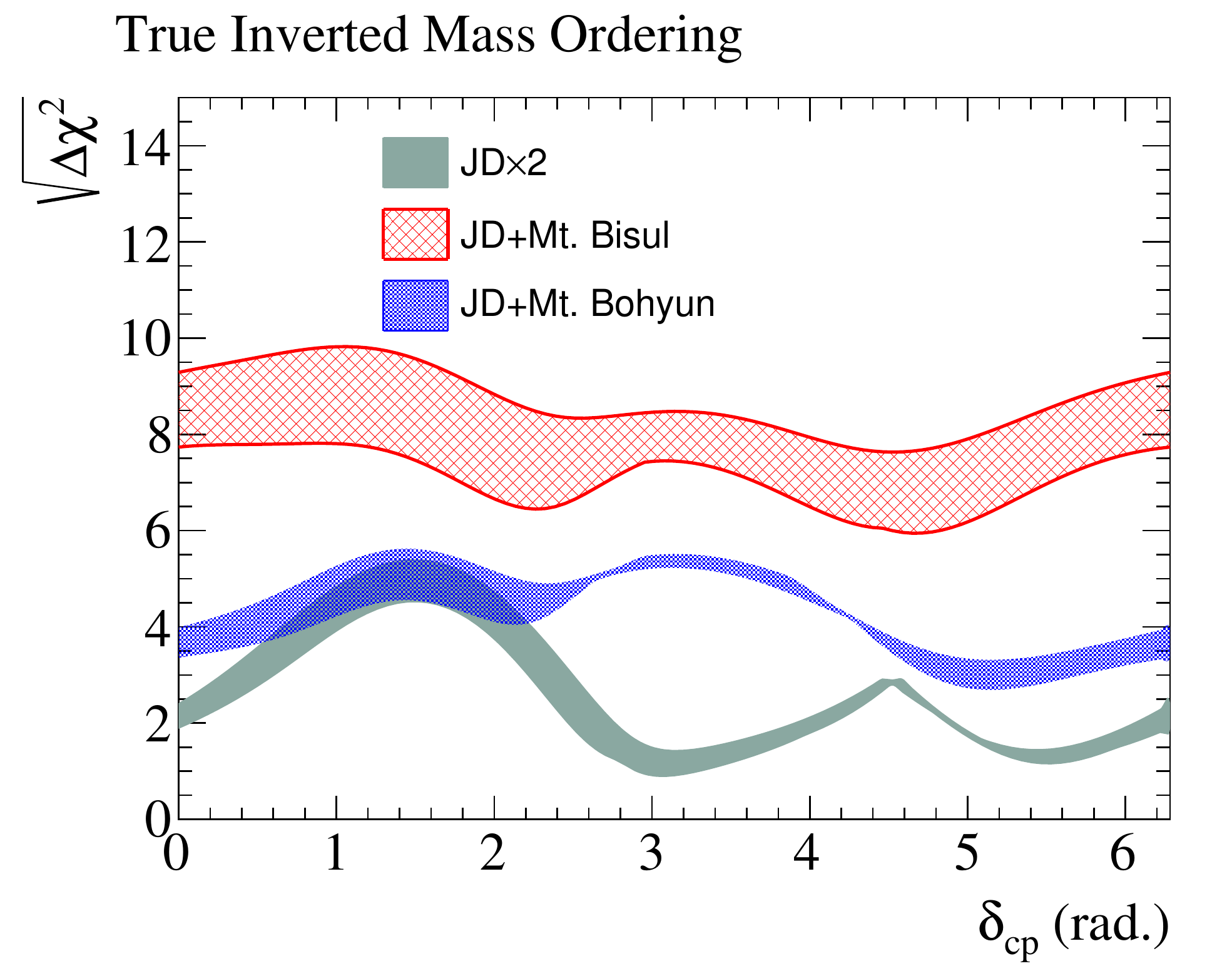}
    \caption{The significance for the wrong mass ordering rejection as a function of the true value of $\delta_{cp}$ and the true mass ordering (top=normal, bottom=inverted).
Results are shown for Mt. Bisul and Mt. Bohyun sites.  The bands represent the dependence of the sensitivity on the true value sin$^2\theta_{23}$ in the range $0.4<$sin$^2\theta_{23}<0.6$.  }
    \label{fig:mh_sensitivity_mt_bisul}
  \end{center}
\end {figure}

The wrong mass ordering rejection significances including the Mt. Bisul and Mt. Bohyun configurations are shown in Fig.~\ref{fig:mh_sensitivity_mt_bisul}.  The wrong mass ordering rejection significance
is largest for the Mt. Bisul configuration for all true values of the mass ordering and $\delta_{cp}$, and is above 6$\sigma$ for almost all true values of the oscillation parameters.  The mass 
ordering rejection sensitivity with the Mt. Bohyun site is above 3$\sigma$ for almost all true values of the oscillation parameters and higher than the configuration with two detectors in Japan
for most values.

\begin {figure}[htbp]
\captionsetup{justification=raggedright,singlelinecheck=false}
  \begin{center}
    \includegraphics[width=0.49\textwidth]{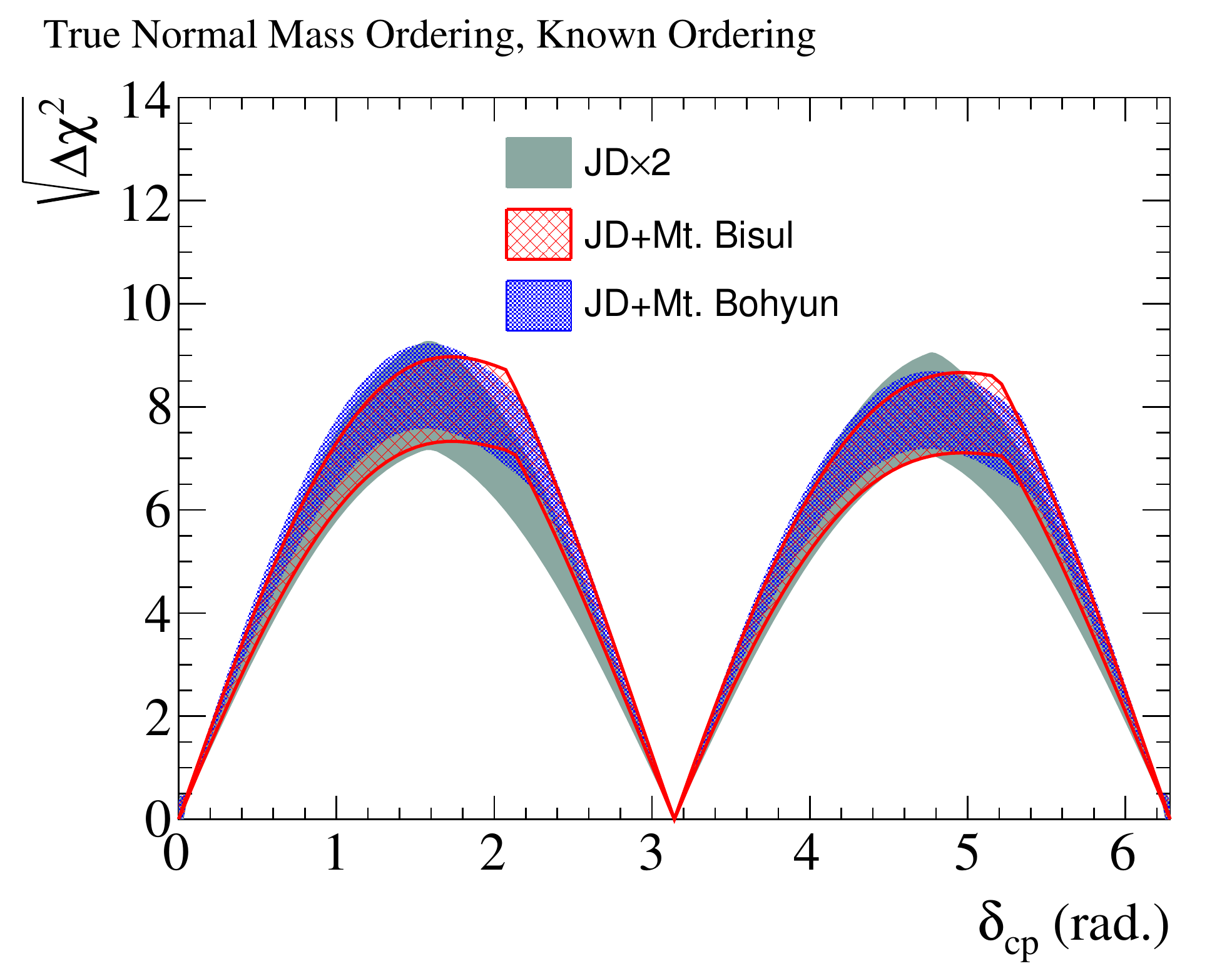}
    \includegraphics[width=0.49\textwidth]{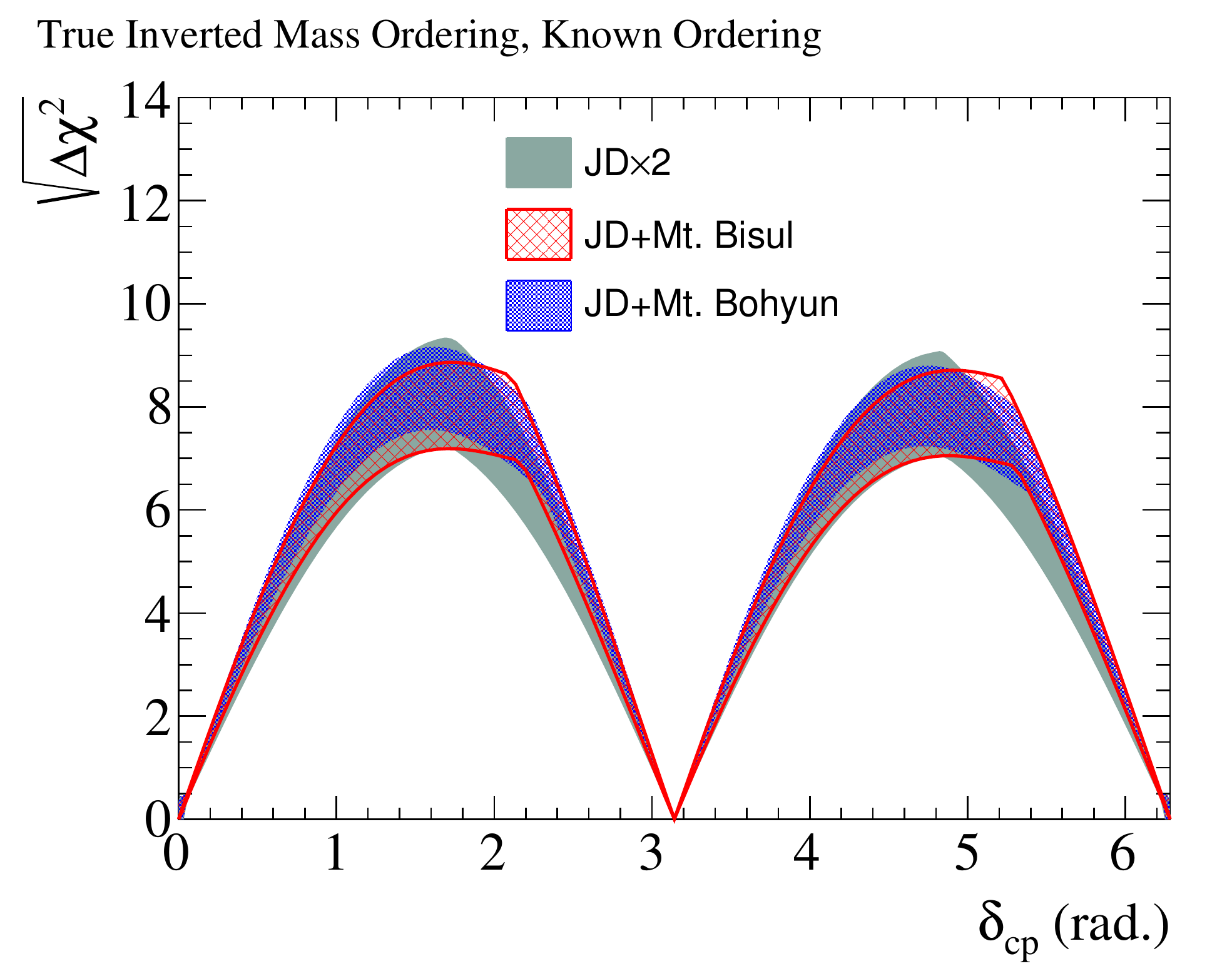}\\
    \includegraphics[width=0.49\textwidth]{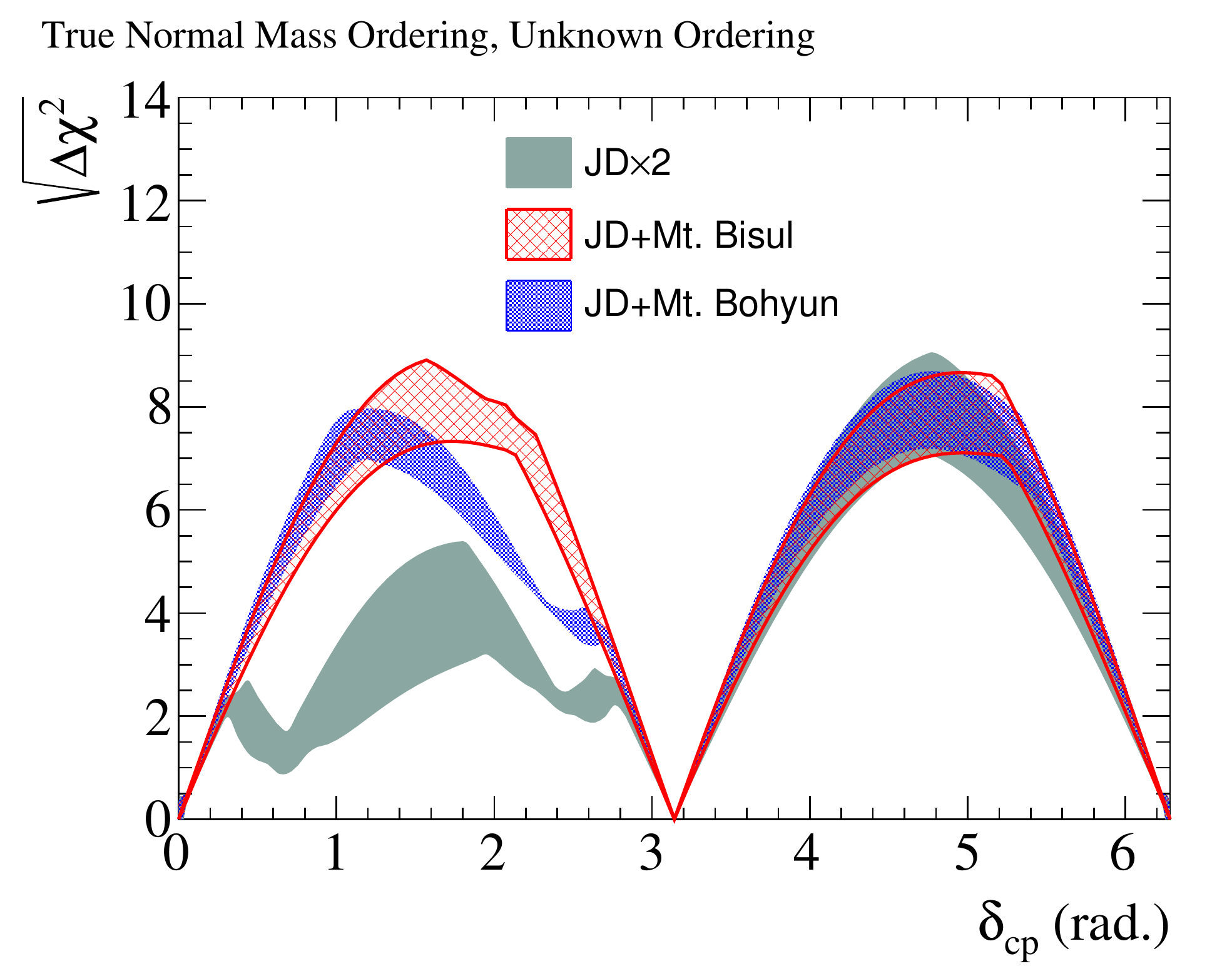}
    \includegraphics[width=0.49\textwidth]{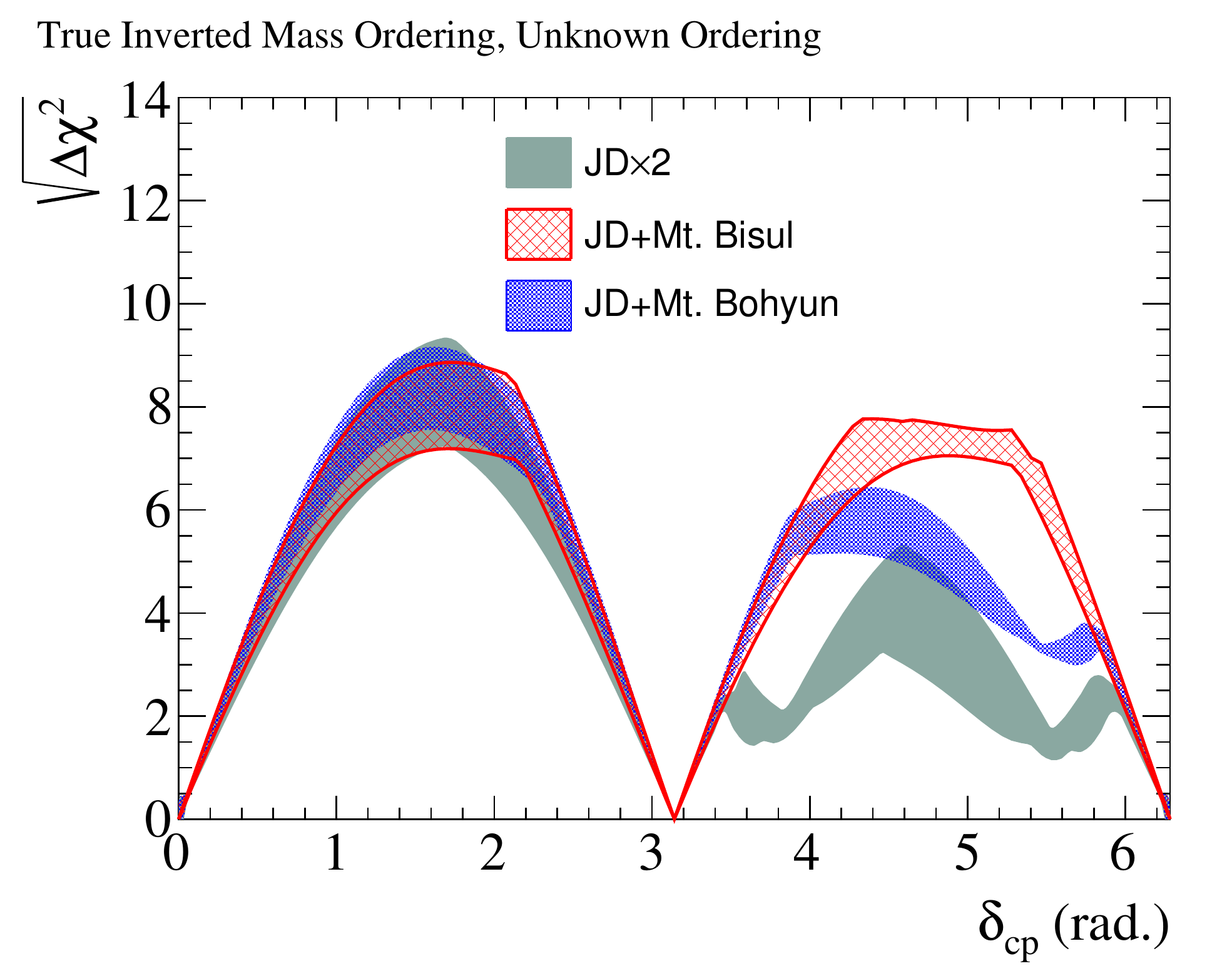}
    \caption{The significance for CP conservation rejection as a function of the true value of $\delta_{cp}$ and the true mass ordering (left=normal, right=inverted).  The top row
    shows the significance when the mass ordering is determined using external data and Hyper-K atmospheric neutrinos, while the bottom 
    row shows the significance when the mass ordering is determined only by accelerator neutrinos observed in the Hyper-K detectors . Results are shown for the Mt. Bisul and Mt. Bohyun sites.
    The bands represent the dependence of the sensitivity on the true value of sin$^2\theta_{23}$ in the range $0.4<$sin$^2\theta_{23}<0.6$.}
    \label{fig:cpv_sensitivity_mt_bisul}
  \end{center}
\end {figure}

The CP conservation rejection significances including the Mt. Bisul and Mt. Bohyun configurations are shown in Fig.~\ref{fig:cpv_sensitivity_mt_bisul}.  There is little change to the fraction
of $\delta_{cp}$ values with 3$\sigma$ or 5$\sigma$ rejection compared to the configuration with the Korean detector at 1.5$^{\circ}$ and 2.5$^{\circ}$ off-axis. For the scenarios where the 
mass ordering is determined by the accelerator neutrinos only, better CP conservation rejection with Mt. Bisul is achieved for some values
of $\delta_{cp}$ where the improved wrong mass ordering rejection impacts the CP violation measurement.

\begin {figure}[htbp]
\captionsetup{justification=raggedright,singlelinecheck=false}
  \begin{center}
    \includegraphics[width=0.70\textwidth]{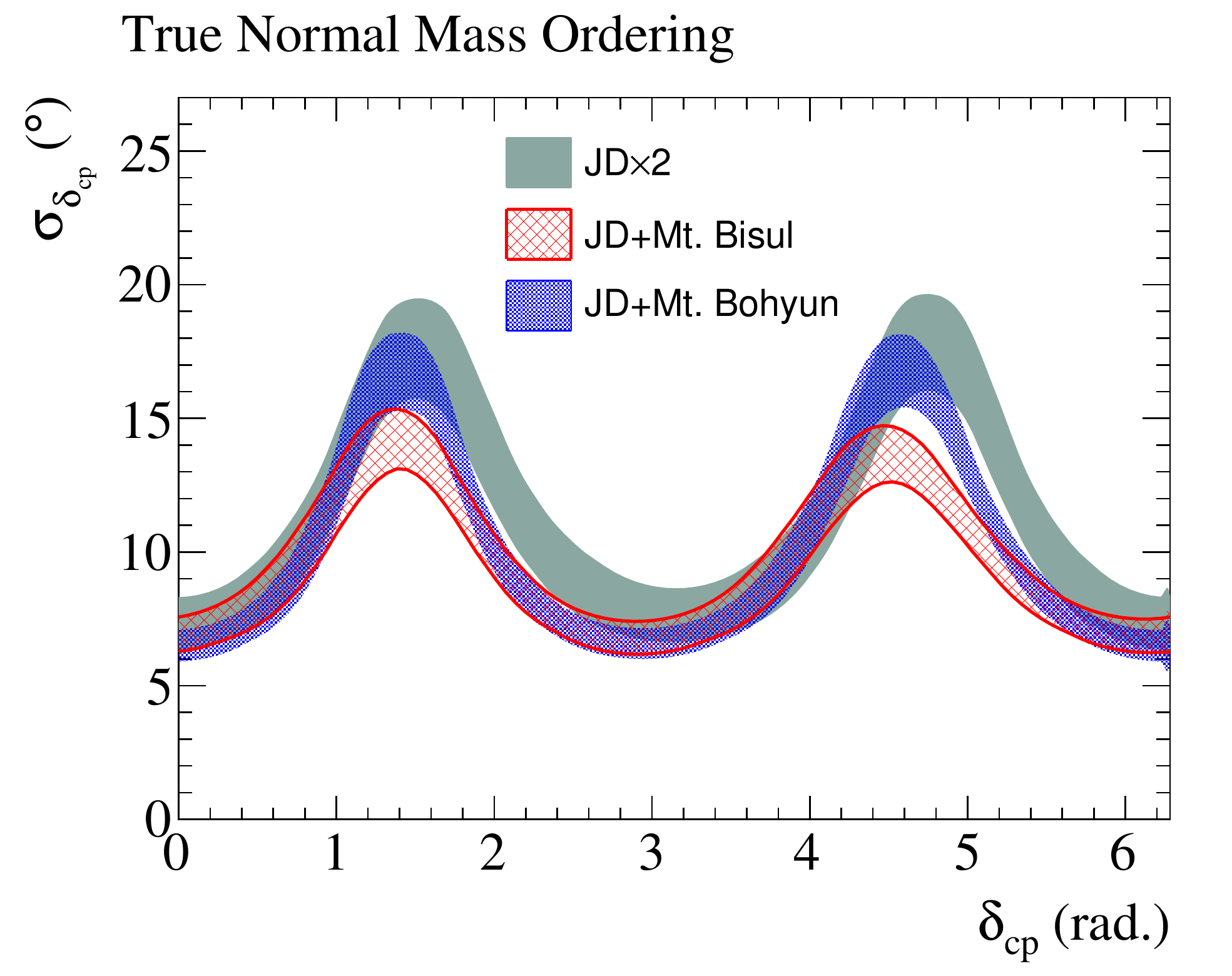}
    \includegraphics[width=0.70\textwidth]{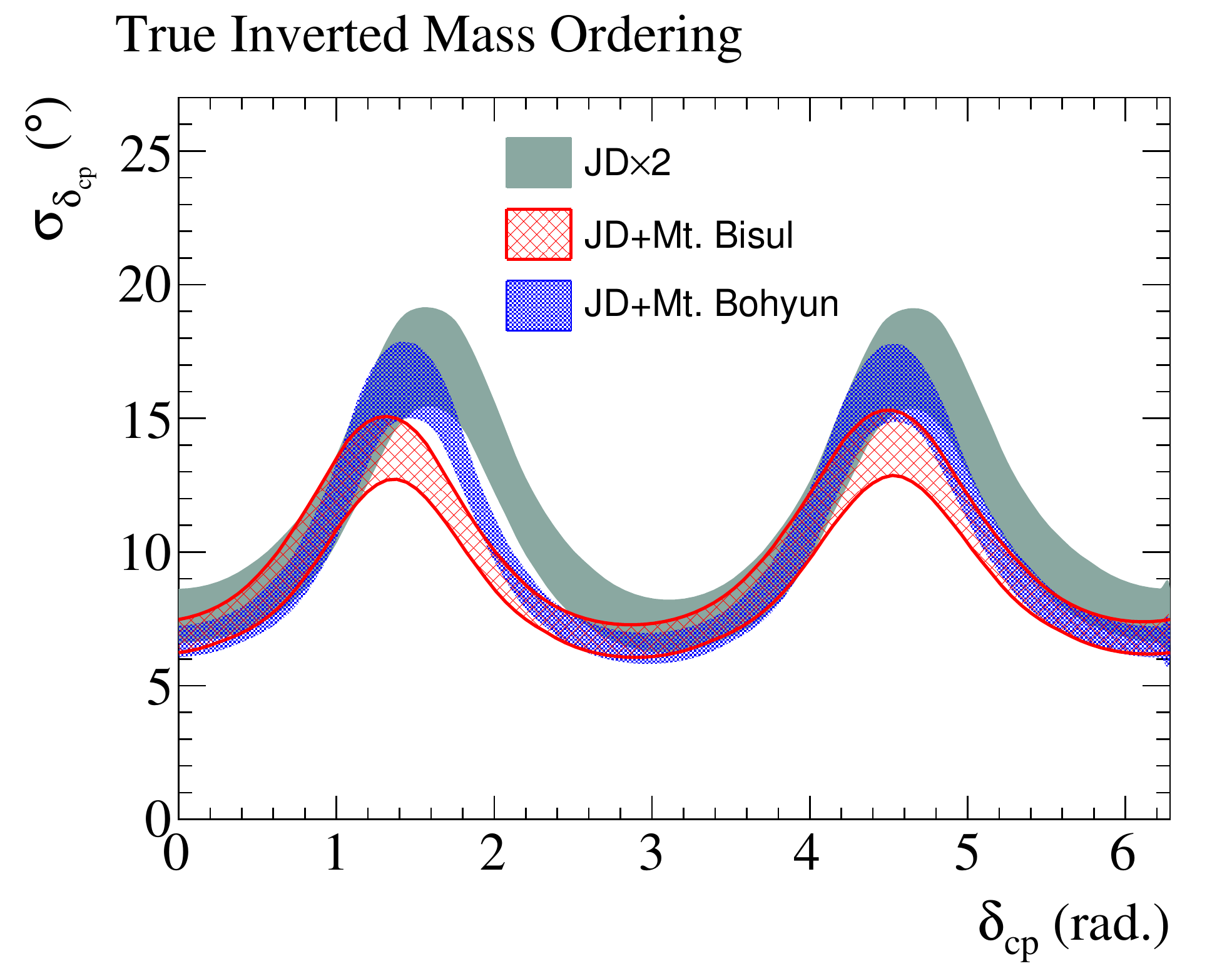}
    \caption{The 1$\sigma$ precision of the $\delta_{cp}$ measurement as a function of the true $\delta_{cp}$ value. Results are shown for the Mt. Bisul and Mt. Bohyun sites.
    The bands represent the dependence of the sensitivity on the true value of sin$^2\theta_{23}$ in the range $0.4<$sin$^2\theta_{23}<0.6$.}
    \label{fig:cp_precision_mt_bisul}
  \end{center}
\end {figure}

The $\delta_{cp}$ precision is shown in Fig.~\ref{fig:cp_precision_mt_bisul}.  Near the maximally CP violation values, the Mt. Bisul site has the best precision, indicating that
the measurement is in part due to the spectrum distortion in the $>1$~GeV region arising from the $\cos(\delta_{cp})$ dependent term.  Near the CP conserving values, the 
configurations with the Mt. Bisul and Mt. Bohyun sites gives nearly identical precision.  Both configurations with a detector in Korea show improved precision compared to the 
configuration with two detectors in Japan for almost all true values of the oscillation parameters.

\subsubsection{Atmospheric parameters and octant sensitivity}
The Korean detector has enhanced sensitivity for the CP violation and CP phase measurements due to the $L/E$ dependence in the $\delta_{CP}$ dependent interference terms of the 
electron (anti)neutrino appearance probability.  No such enhancement is present in the leading terms of the muon (anti)neutrino survival probability or the
electron (anti)neutrino appearance probability.  Since the leading terms in these probabilities provide the constraints on $\Delta m^{2}_{32}$ and sin$^{2}\theta_{23}$ we may
expect no advantage for a configuration with one detector in Japan and one in Korea over a configuration with two detectors in Japan.  In fact, there may be a reduced sensitivity
with one detector in Korea due to the lower statistics at the longer baseline.  We have studied the sensitivity to  $\Delta m^{2}_{32}$ and sin$^{2}\theta_{23}$ as well as the
$\theta_{23}$ octant determination in configurations that include detectors in Japan and Korea or detectors only in Japan.  

\begin {figure}[htbp]
\captionsetup{justification=raggedright,singlelinecheck=false}
  \begin{center}
    \includegraphics[width=0.49\textwidth]{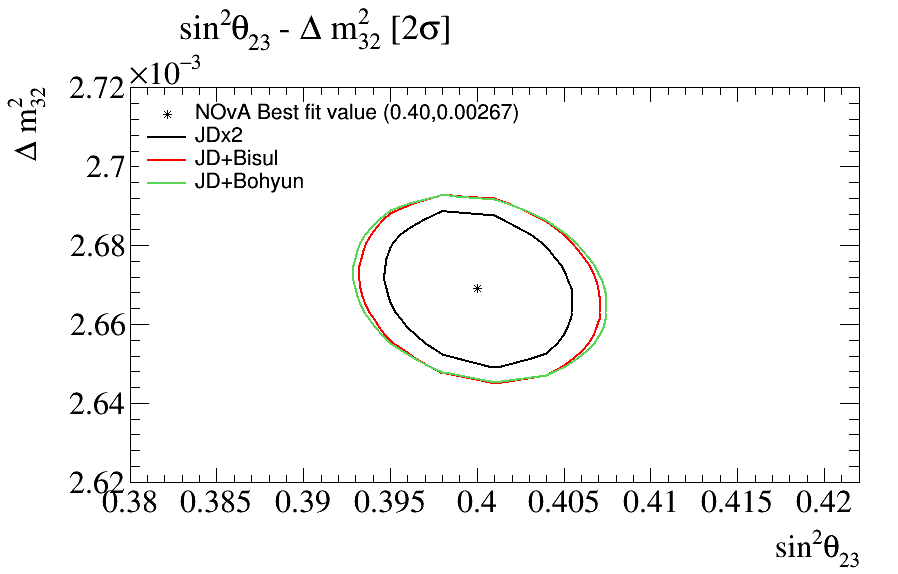}
    \includegraphics[width=0.49\textwidth]{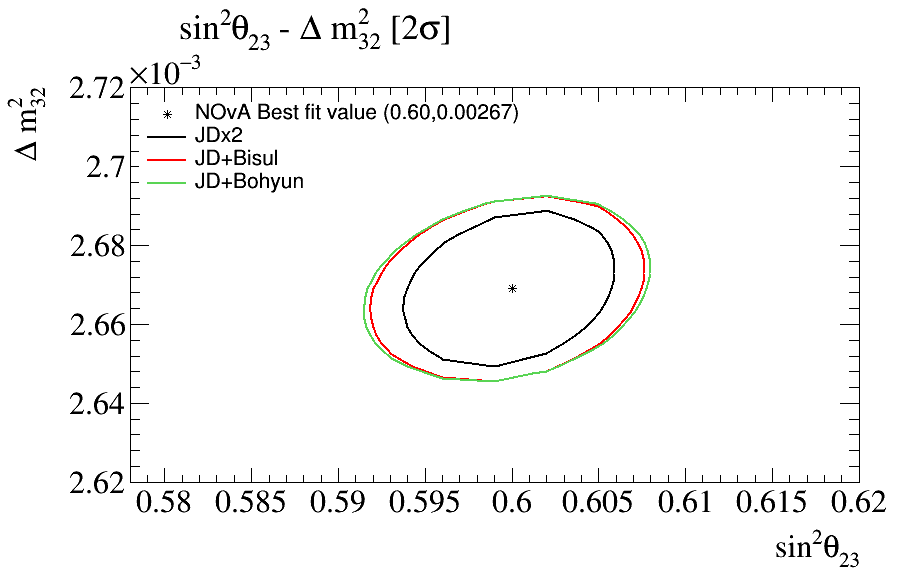}
    \includegraphics[width=0.49\textwidth]{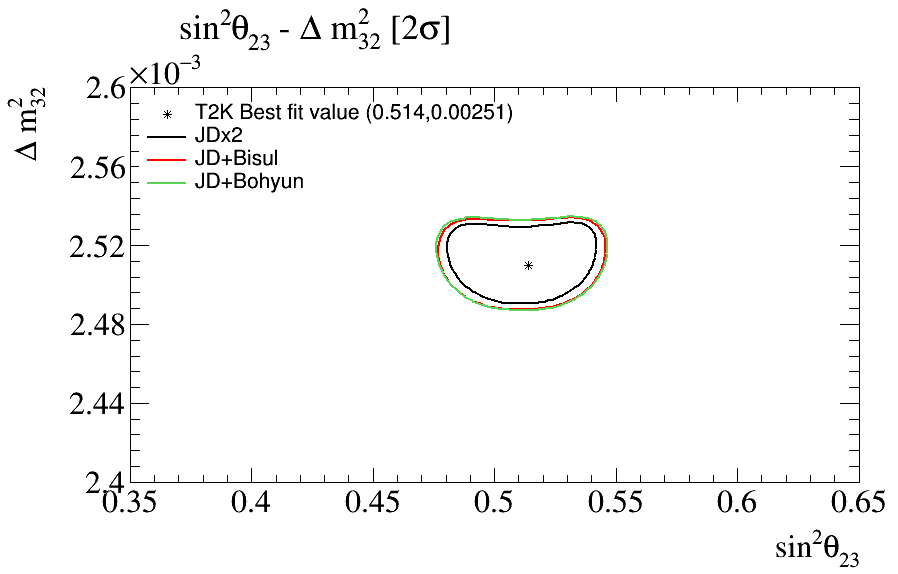}
    \caption{2$\sigma$ sensitivity curves for the atmospheric parameter determination at true values of $\Delta m^{2}_{32}=2.67\times10^{-3}$~eV$^{2}$ and sin$^{2}\theta_{23}$=0.4 (top left),
 $\Delta m^{2}_{32}=2.67\times10^{-3}$~eV$^{2}$ and sin$^{2}\theta_{23}$=0.6 (top right),  $\Delta m^{2}_{32}=2.51\times10^{-3}$~eV$^{2}$ and sin$^{2}\theta_{23}$=0.514 (bottom).  }
    \label{fig:atm_parameter_prec}
  \end{center}
\end {figure}

\begin {figure}[htbp]
\captionsetup{justification=raggedright,singlelinecheck=false}
  \begin{center}
    \includegraphics[width=0.80\textwidth]{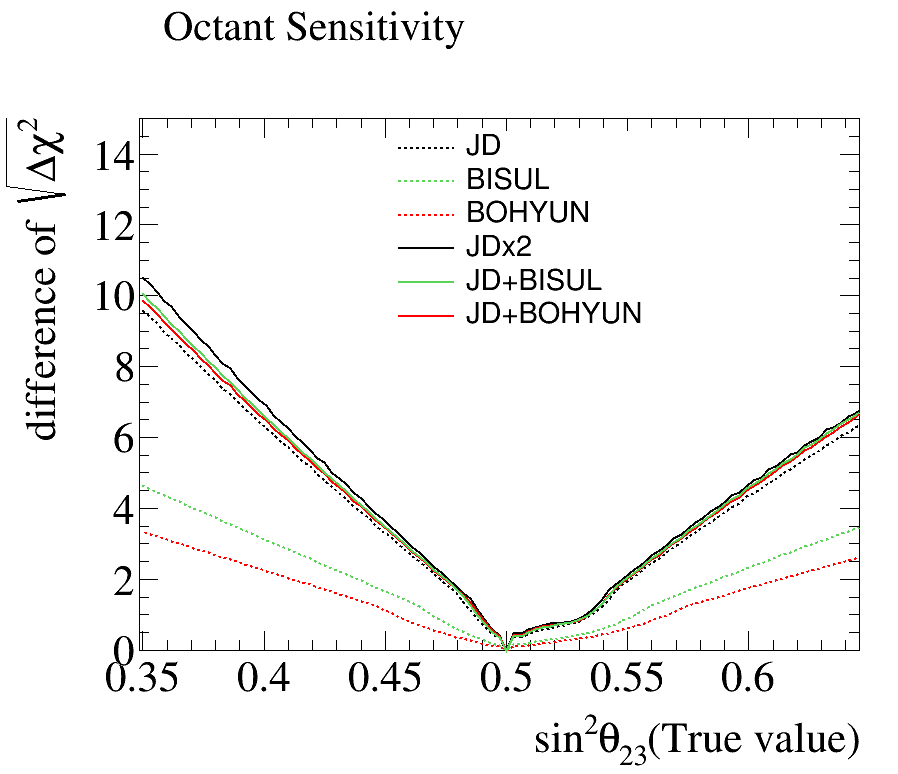}
    \caption{The significance of the octant determination as a function of the true value of sin$^{2}\theta_{23}$ for different detector configurations.}
    \label{fig:octant_sensitivity}
  \end{center}
\end {figure}

Fig.~\ref{fig:atm_parameter_prec} shows the 2$\sigma$ sensitivities for the $\Delta m^{2}_{32}$ and sin$^{2}\theta_{23}$ parameter determination for different true values of these
parameters.  There is a reduction in the sensitivity when a configuration with a Japanese and Korean detector is used relative to a configuration with two Japanese detectors, but the
reduction in sensitivity is not large. Fig.~\ref{fig:octant_sensitivity} shows the significance of the octant determination as a function of the true value of sin$^{2}\theta_{23}$
for different detector configurations.  The addition of a second detector in Japan or Korea does little to improve the octant sensitivity.  

From the sensitivity studies presented here, we conclude that a configuration with one detector in Japan and the second in Korea has similar but slightly worse sensitivity for 
atmospheric parameter determination than a configuration with two detectors in Japan.  The measurement program for these parameters will not be significantly degraded in a configuration
where the second detector is in Korea.

%\clearpage

%======================================================================
%\input{cc1pi/cc1pi_t}

%======================================================================
\graphicspath{{lbl_atm/plots}}
%======================================================================
\section{Sensitivities with LBL + Atmospheric Neutrinos}
\label{sec:lbl_atm}
In addition to neutrino data from the J-PARC beam, Hyper-K will collect atmospheric 
neutrino data, which will add sensitivity to its oscillation measurements.
These neutrinos are produced in the decays of particles emerging from 
the interaction of primary cosmic rays with nuclei in the atmosphere. 
Indeed the production mechanism is identical to that of beam neutrinos 
with the notable exception that the lack of an absorber results in a 
significant fraction of $\nu_{e}$ and $\bar{\nu}_{e}$ in the flux. 
Until $\sim 10$~GeV the ratio of muon to electron type neutrinos is 
roughly 2:1 with the fraction of muon neutrinos increasing at higher energies.
Since the primary cosmic ray flux spans several orders of magnitude 
and is roughly isotropic about the Earth, the resulting atmospheric neutrino 
spectrum covers an equally wide range of energies and a given detector 
can expect to observe neutrinos with a variety of pathlengths from O(10)~km,
for neutrinos produced overhead, to $\sim 10,000$~km, for those produced 
on the opposite side of the planet.
Importantly, these events will be accumulated at both detectors in 
Japan and Korea in similar proportion, modulo differences in the atmospheric densities 
and the local geomagnetic fields.

\begin {figure}[htbp]
\begin{center}
    \includegraphics[width=0.45\textwidth]{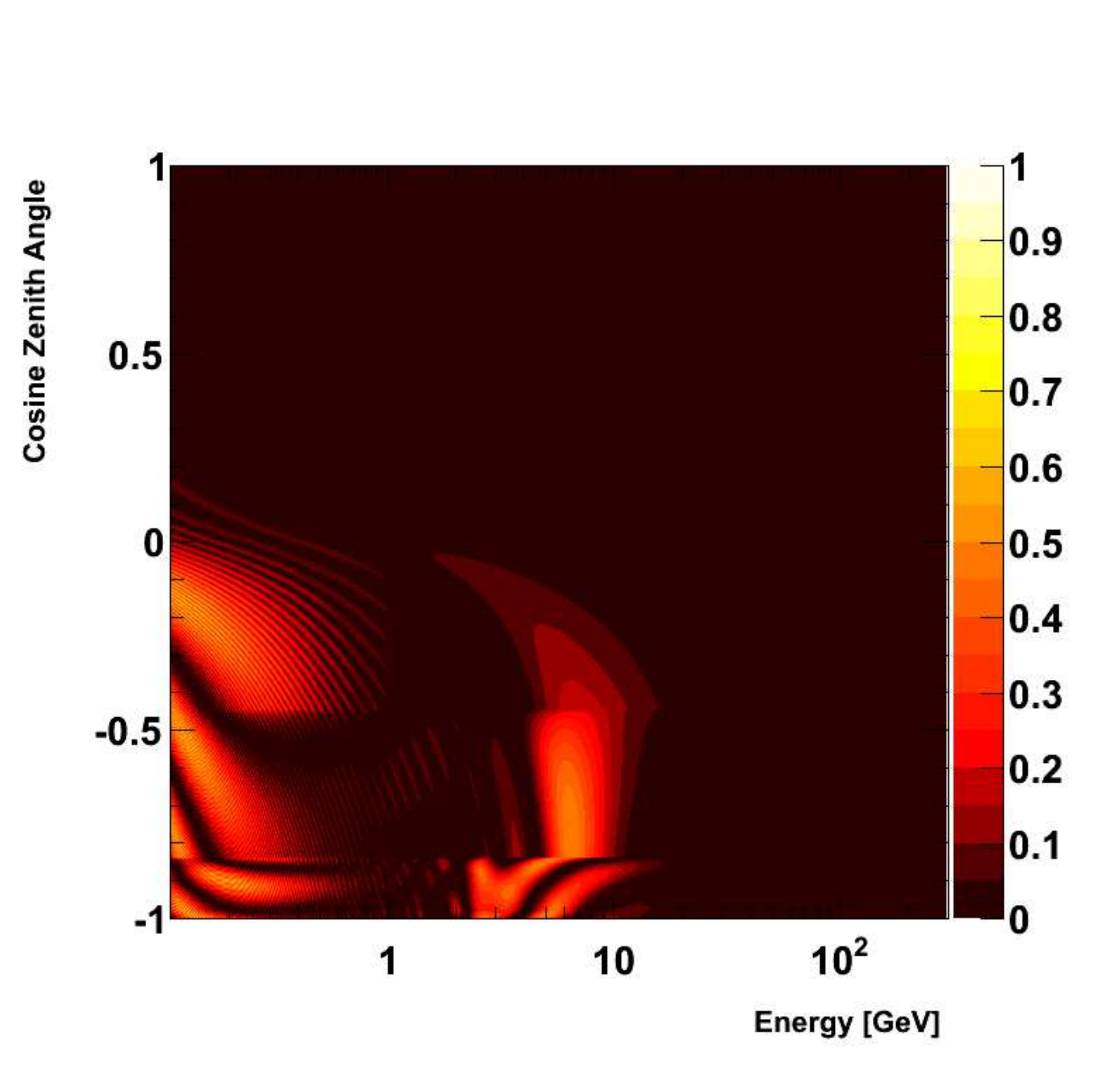}
    \includegraphics[width=0.45\textwidth]{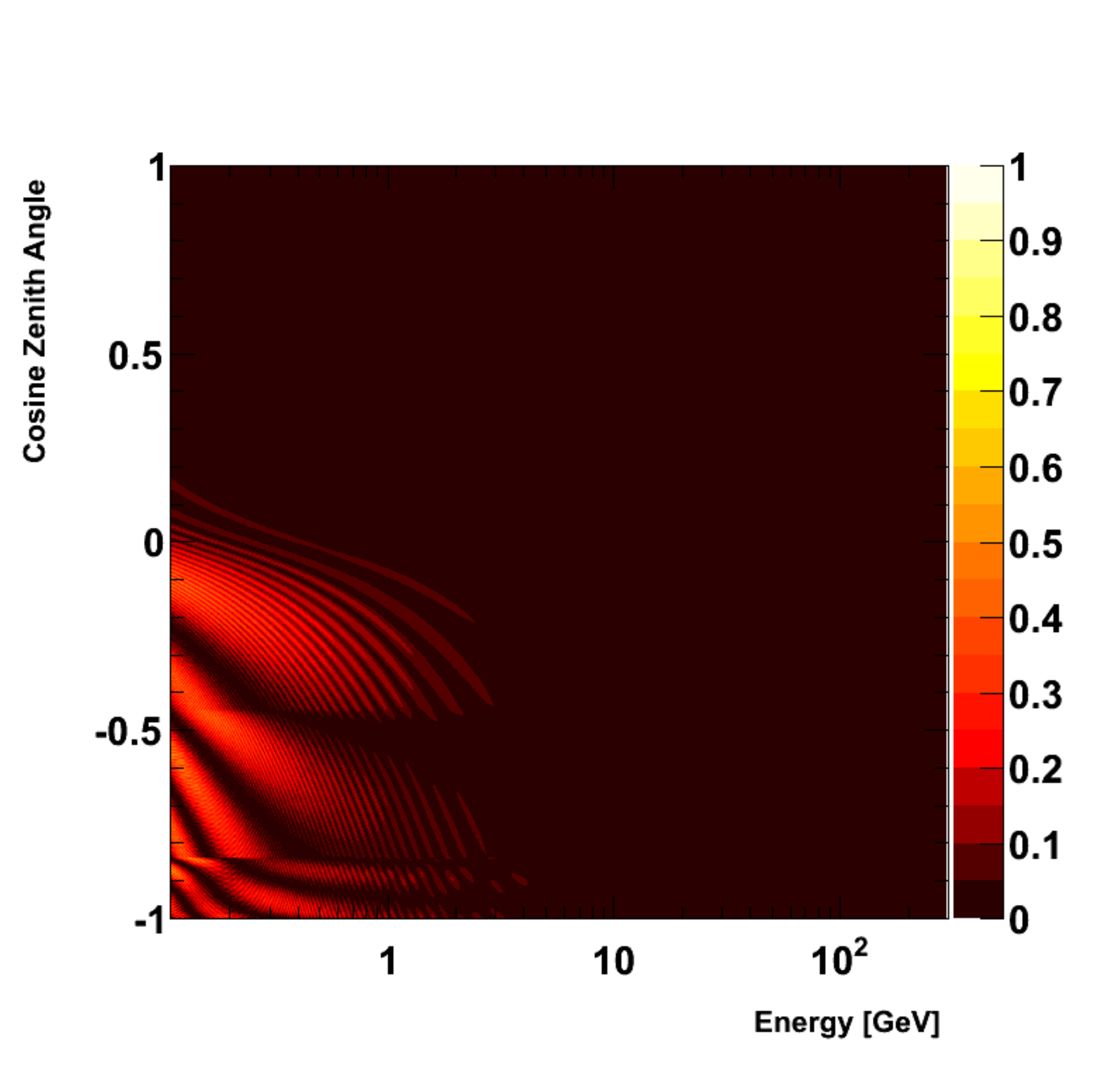} 
    \caption{Atmospheric neutrino oscillation probabilities as a function of energy 
             and cosine of the neutrino zenith angle (-1 corresponds to upward-going).
             The left (right) plot shows the $\nu_{\mu} \rightarrow \nu_{e}$ 
             ( $\bar{\nu}_{\mu} \rightarrow \bar{\nu}_{e}$ ) probability. 
             In this plot the normal mass hierarchy is assumed.
             For an inverted hierarchy the features in the neutrino plot 
             move to the antineutrino plot and vice versa.
            }
    \label{fig:atm_osc}
  \end{center}
\end{figure}

Though atmospheric neutrinos lack the precise timing and directional information 
afforded by the beam, they offer a high-statistics sample, roughly $150\times 10^3$events/Mton-year, 
with large matter effects. 
These matter effects provide mass hierarchy sensitivity in a manner analogous 
to the beam neutrinos, but more enhanced oscillations.
Neutrinos traversing the core of the Earth pass through a matter profile whose 
density varies from 1 to 13~g/cm$^{3}$, which induces a parametric 
oscillation resonance for energies between 2 and 10~GeV.
For neutrinos experiencing this effect the $\nu_{\mu} \rightarrow \nu_{e}$ (appearance channel)
oscillation probability can be as large as 50\% (Figure~\ref{fig:atm_osc}). 
This effect depends on both the sign of the mass hierarchy 
and whether the neutrino is a particle or antiparticle; for a normal (inverted) 
mass hierarchy only neutrinos (antineutrinos) undergo these oscillations.
Sensitivity to the mass hierarchy is obtained by studying the 
upward-going electron neutrino event rate in this energy region.
% with statistical separation of 
%the neutrino-like and antineutrino-like portions of the sample afforded by 
%a likelihood-based method utilizing the expected divia neutron tagging in the detector. 
It should be noted that this sample will provide a test of the mass hierarchy 
largely independent of the beam measurement. 

Resonant oscillations in the Earth also depend upon the value of $\sin^{2} \theta_{23}$.
Not only does it affect the appearance of electron neutrinos described above, 
but it also impacts the upward-going muon rate. 
Atmospheric neutrino sensitivity to the mixing parameters $\Delta m^{2}_{32}$ 
and $\theta_{23}$ is driven primarily by the oscillation of $\nu_{\mu} \rightarrow \nu_{\tau,e}$  
seen in this sample and these matter effects provide additional 
power to discriminate the octant of $\theta_{23}$.

At energies below a few GeV atmospheric neutrinos carry additional sensitivity 
to $\delta_{cp}$ again through an appearance channel.
However, with no precise knowledge of the incoming neutrino direction 
and poor correlation between the outgoing lepton from an interaction and its parent neutrino direction 
at these energies, the atmospheric neutrino sensitivity 
is weaker than that from the beam sample.

\subsection{Combination of beam and atmospheric neutrino data}

As described above atmospheric neutrinos provide complementary sensitivity to the same oscillation 
physics as the beam neutrino samples.
Though there are common systematic error effects between the two samples from the cross section and detector modeling, 
the disparate energy regimes and flux systematics allow for a nearly independent study of 
oscillations.
More importantly the atmospheric neutrino data are accumulated continuously and independently of the beam,
such that the combination of the two samples provides improved sensitivity on shorter time scales.
In this section we present a combined analysis of beam and atmospheric neutrino 
data assuming a Hyper-K detector in Japan (JD) and at the Mt. Bisul site (off-axis angle $1.3^{\circ}$ )
and compare with sensitivities assuming two detectors in Japan (JD$\times$ 2).
The treatment of the atmospheric neutrino samples and their systematic errors follows that of Super-Kamiokande, 
with no assumed improvements (c.f. the discussion in~\cite{Himmel:2013jva}).

\begin {figure}[p]
\begin{center}
    \includegraphics[width=0.70\textwidth]{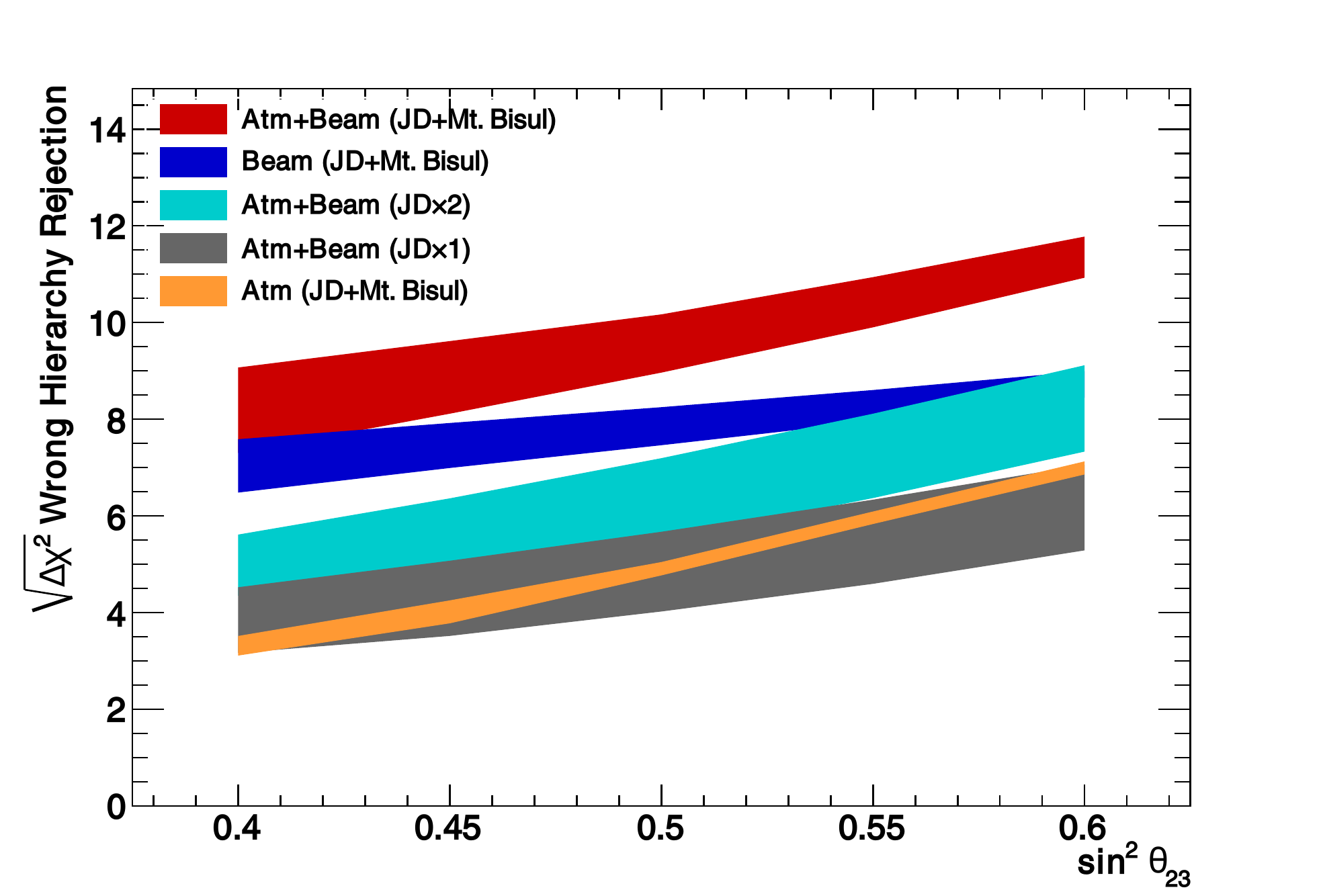}
    \caption{Sensitivity to the normal mass hierarchy 
             for components of a combined measurement of beam and atmospheric neutrinos for 
             a 10 year exposure.
             Here JD refers to a single Hyper-K detector in Kamioka, Japan, and JD$\times 2$ refers 
             two to such detectors operating simultaneously. 
             The horizontal axis shows the assumed value of $\sin^{2} \theta_{23}$ and the 
             width of the bands shows the variation in sensitivity with $\delta_{cp}$. }
    \label{fig:jdkd_hier_10}
  \end{center}
%\end {figure}
%
%
%\begin {figure}[htbp]
\begin{center}
    \includegraphics[width=0.70\textwidth]{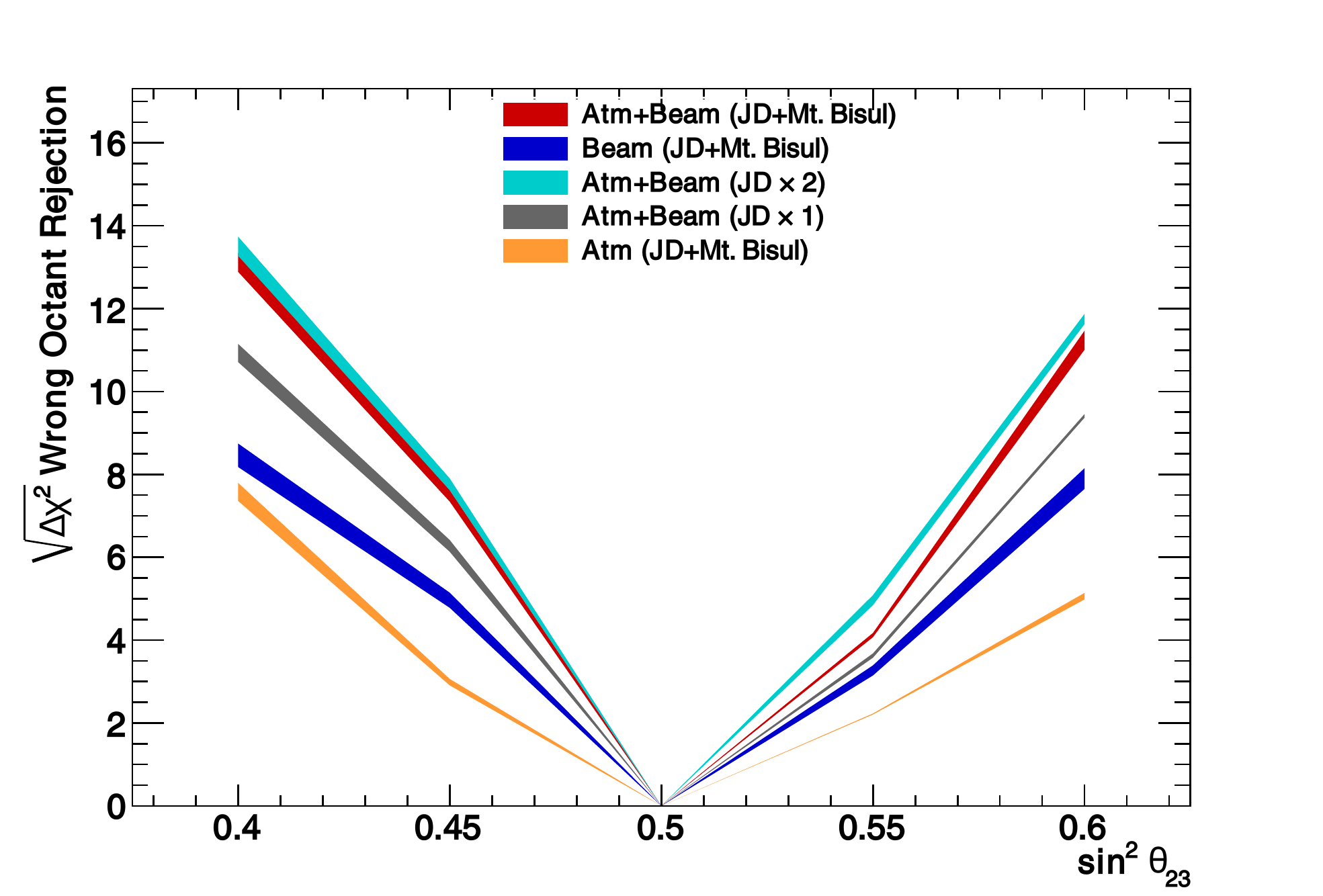}
    \caption{Sensitivity to the $\theta_{23}$ octant (right) assuming the normal mass hierarchy 
             for components of a combined measurement of beam and atmospheric neutrinos for 
             a 10 year exposure. The plot has been produced in the same manner as Figure~\ref{fig:jdkd_hier_10}.}
    \label{fig:jdkd_oct_10}
  \end{center}
\end{figure}

Figure~\ref{fig:jdkd_hier_10} shows the sensitivity to the mass hierarchy 
for the combined analysis using the same test statistic as Equation~\ref{eq:mh_test}.
After 10 years of running the expected ability to reject the wrong mass hierarchy 
is better than $\sqrt{\Delta \chi^{2}} = 7$.   
Atmospheric neutrinos by themselves provide sensitivity better than $\sqrt{\Delta \chi^{2}} > 3$ 
for all currently allowed values of $\sin^{2} \theta_{23}$ and have comparable sensitivity 
to the beam measurement at the Korean detector for the largest values of this parameter. 
Though the combined JD and Mt. Bisul beam measurement has better sensitivity than the atmospheric 
neutrino measurement alone, 
when all of the samples are combined the sensitivity improves further.
The power of this improvement manifests as an earlier realization of the hierarchy as shown in the left panel of Figure~\ref{fig:jdkd_jd_year}.
Within two years of operations the sensitivity will exceed $\sqrt{\Delta \chi^{2}} > 4$.

Sensitivity to $\sin^{2} \theta_{23}$ for the combined analysis and its components 
appears in Figure~\ref{fig:jdkd_oct_10}.
Here the test statistic reflects the ability to reject the incorrect octant as 
\begin{equation}
T_{octant} = \sqrt{\Delta\chi^2_{WO}-\Delta\chi^2_{CO}},
\label{eq:oct_test}
\end{equation}
\noindent where $\Delta\chi^2_{WO}$ and $\Delta\chi^2_{CO}$ represent 
the minimum likelihood value taken over the wrong and correct octants, respectively.
The minimum for the first (second) octant is taken over values of the likelihood 
in the range of parameters $\sin^{2} \theta_{23} < 0.5$  ($>0.5$).
If $\theta_{23}$ differs from maximal mixing by $2^{\circ}$ or more, 
the octant will be resolved at better that 3 units of the test statistic.
As shown in the figure, this marks a considerable improvement over 
the beam-only measurement.
 
\begin {figure}[p]
\begin{center}
    \includegraphics[width=0.70\textwidth]{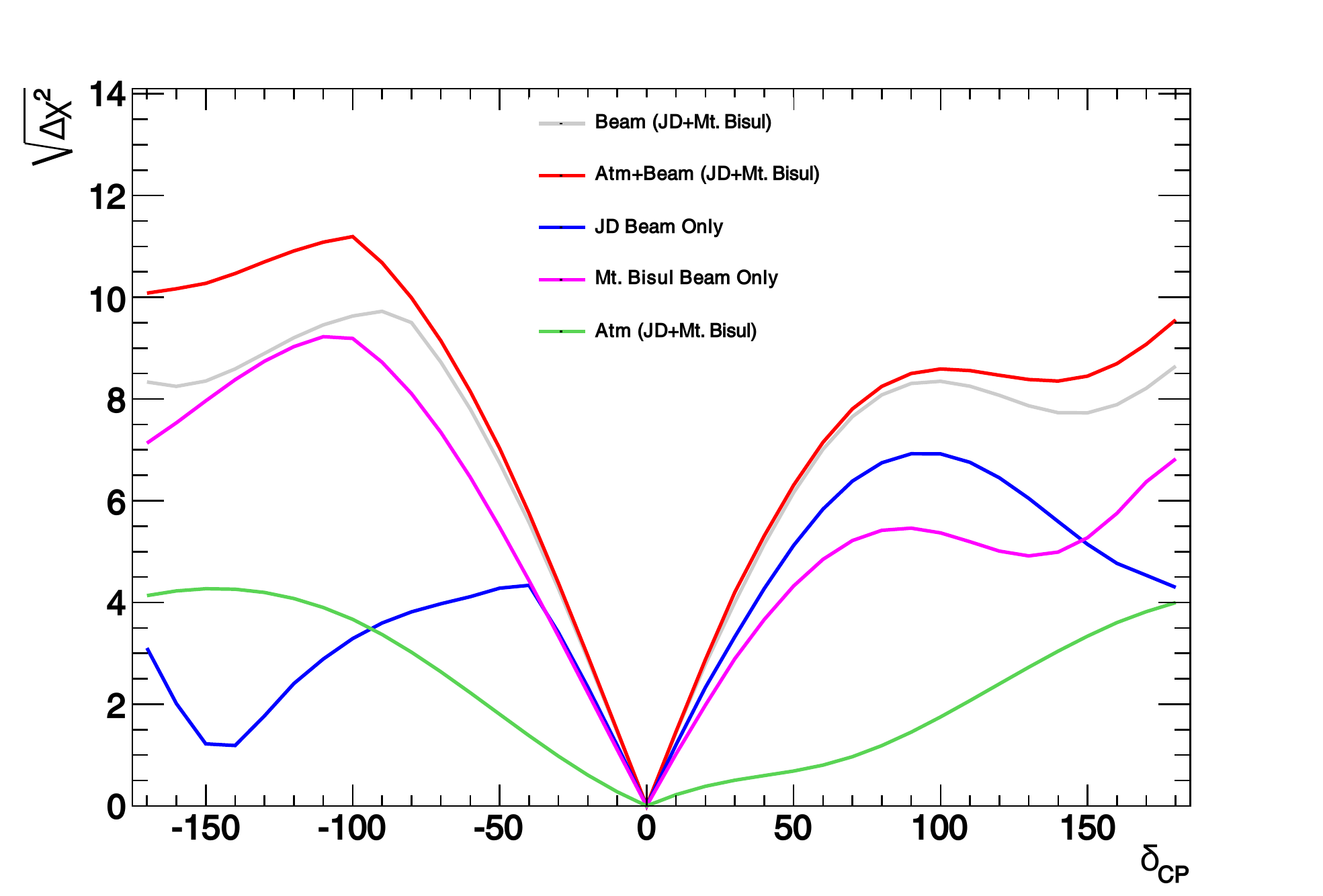}
    \caption{Sensitivity to $\delta_{cp}=0$ 
             for components of a combined measurement of beam and atmospheric neutrinos for 
             a 10 year exposure.  Here JD refers to a single Hyper-K detector in Kamioka, Japan.}
    \label{fig:jdkd_cp_10}
  \end{center}
%\end{figure}
%
%\begin {figure}[htbp]
\begin{center}
    \includegraphics[width=0.70\textwidth]{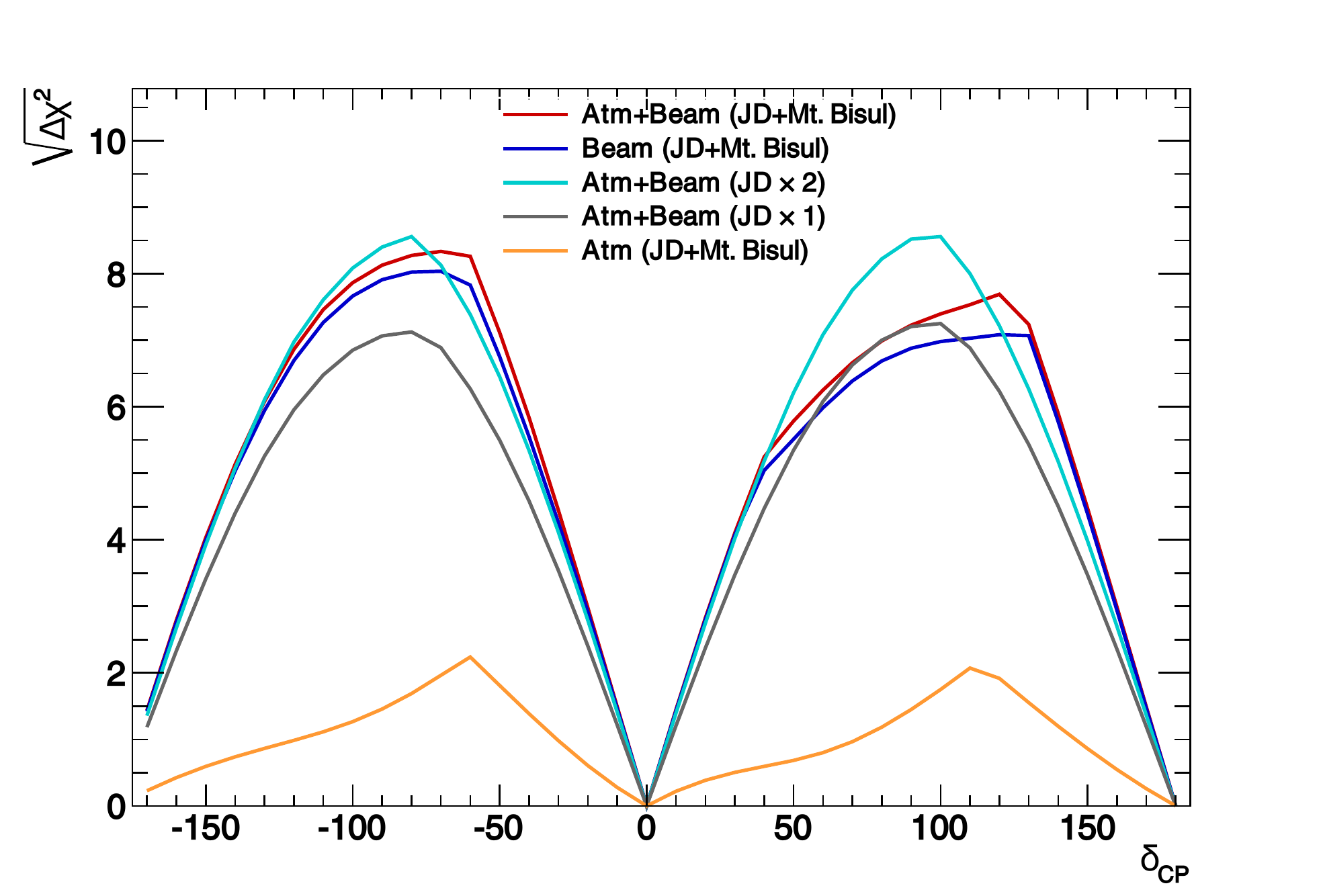}
    \caption{Sensitivity to CP violation ($\sin \delta_{cp} \neq 0$)
             for components of a combined measurement of beam and atmospheric neutrinos for 
             a 10 year exposure.
             Here JD refers to a single Hyper-K detector in Kamioka, Japan, and JD$\times 2$ refers 
             two to such detectors operating simultaneously. 
             The horizontal axis shows the assumed true value of $\delta_{cp}$.}
    \label{fig:jdkd_cpv_10}
  \end{center}
\end{figure}

\begin {figure}[htbp]
\begin{center}
    \includegraphics[width=0.49\textwidth]{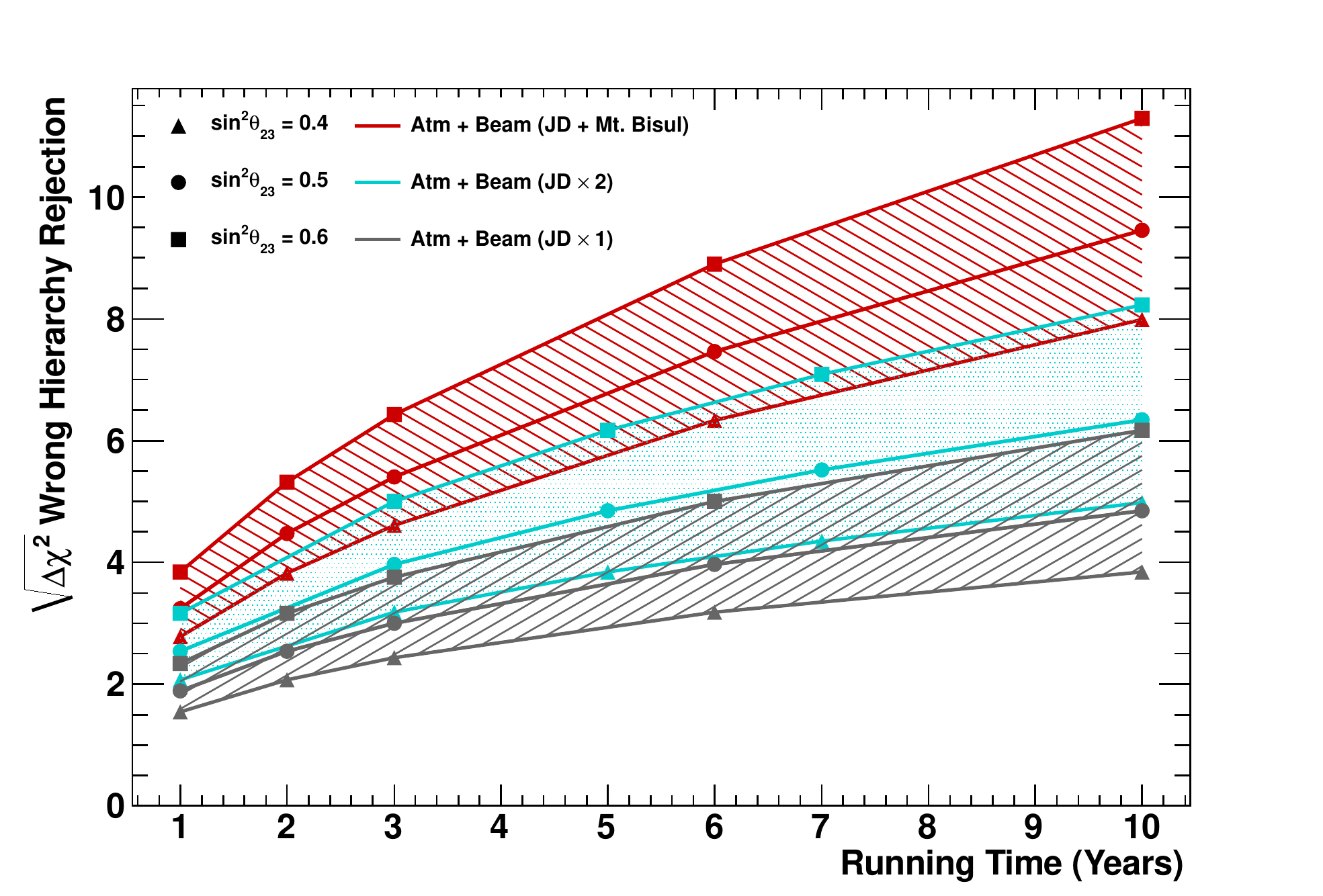} 
    \includegraphics[width=0.49\textwidth]{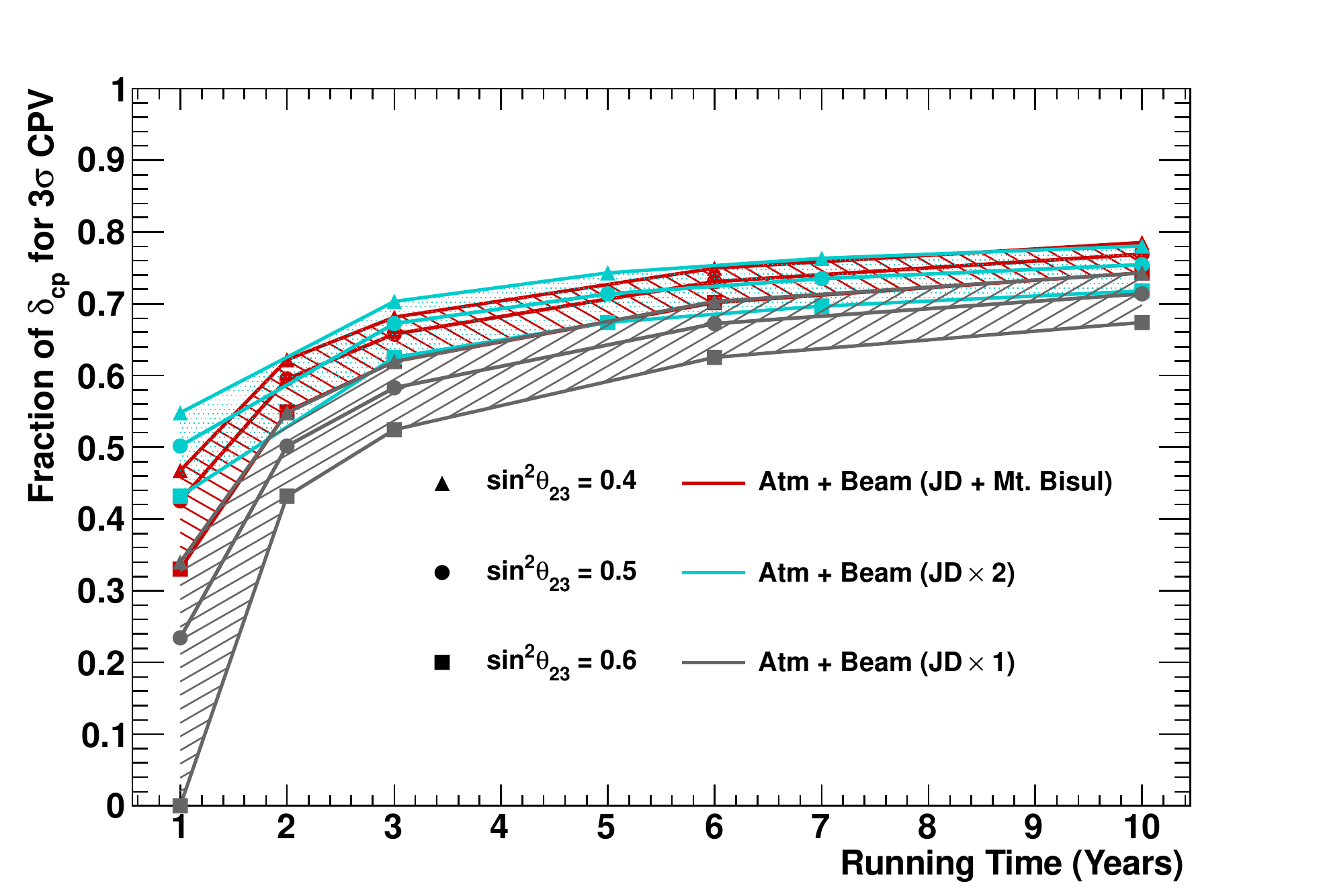} 
    \caption{Sensitivity as a function of year 
             to the mass hierarchy (left) and the fraction of $\delta_{cp}$ phase space 
             for which CP violation ($\sin \delta_{cp} \neq 0$) can be determined at 
             $3~\sigma$ or better.
             Red lines show a combined beam and atmospheric neutrino measurement 
             with one Hyper-K detector in Kamioka, Japan (JD) and one at the Mt. Bisul site 
             in Korea. Cyan lines show the same analysis assuming two detectors  
             in Kamioka (JD$\times$2) and grey lines assume only one detector in Kamioka (JD$\times$1). 
             Different symbols show the assumed value of $\sin^{2}\theta_{23}$.}
    \label{fig:jdkd_jd_year}
  \end{center}
\end {figure}

While atmospheric neutrinos themselves have less sensitivity to $\delta_{cp}$ 
than the beam measurement, they provide additional constraints on extreme 
values of the parameter as shown in the Figure~\ref{fig:jdkd_cp_10}.
Typically the atmospheric neutrino constraint covers about 50\% of 
the parameter space, such that one of the CP-conserving points 
$\sin \delta_{cp} = 0$ is weakly allowed regardless of the parameter's 
true value. 
For this reason the constraint on CP violation, shown Figure~\ref{fig:jdkd_cpv_10}, 
is weaker than that from the beam and 
provides only a slight improvement in sensitivity.
The test statistic used in this figure is the same as 
in Equation~\ref{eq:cpv_test}.
However, as with the other oscillation measurements 
the power of the combined beam and atmospheric measurement 
comes in the early realization of this sensitivity 
(c.f. the right panel Figure~\ref{fig:jdkd_jd_year}).

%======================================================================
%!TEX root =  ../t2hkk_white_paper.tex
%======================================================================
%\clearpage
\section{\label{sec:bene} Additional Benefits }
In addition to the long-baseline program with multiple baselines,
there are potential benefits to the Hyper-K physics program 
by placing a second detector in Korea.
There are two main benefits that arise from the second Hyper-K detector in Korea.
First, the candidate sites for a Korean detector are deeper than their 
Japanese counterparts, providing a greater over burden to reduce the flux of cosmic ray 
muons.
This translates into a reduced rate of spallation-induced isotopes that 
are backgrounds to lower energy physics, such as solar and supernova neutrino studies.
Second, the large geographical separation between the Japanese and Korean detectors provides 
two horizons for studying supernova neutrinos.
For a supernova burst, the likelihood of observing neutrinos 
below at least one of the horizons is increased, allowing broader study of the 
Earth-matter effect on these events.
This section explores the potential benefits to the Hyper-K physics program
beyond studies of PMNS mixing provided by a Korean detector.  
 
\subsection{Solar and supernova neutrino physics}

Observations of low energy neutrinos, such as those from the sun
($E < 10$~MeV) or the diffuse supernovae flux ($E < 30$~MeV), are complicated by 
backgrounds from natural sources.
Among these cosmic ray muon-induced spallation products, which 
decay to produce photons or electrons of similar energies, 
are a background that can be readily mitigated with the larger 
overburden afforded by a detector in Korea.
The rate of such spallation backgrounds at each of the candidate sites has 
been estimated with simulations based on their local tomographies.

\subsubsection{Estimate of muon spallation background}

\begin{table}[tb]
\caption{Position and altitude for simulated locations.}
\centering
\begin{tabular}{ccccc}
\hline \hline
 & Mt.~Bisul & Mt.~Bohyun \\ \hline
Latitude & 35$^\circ$43'00'' N & 36$^\circ$09'47'' N\\
Longitude & 128$^\circ$31'28'' E & 128$^\circ$58'26'' E\\
Altitude (820~m overburden) & 264~m & 304~m \\
Altitude (1,000~m overburden) & 84~m & 124~m \\
 \hline \hline
\end{tabular}
\label{tab:bgsim-location}
\end{table}%

\begin{table}[t]
\caption{\label{tab:muon_simulation} 
Calculated muon flux ($\Phi$) and average energy ($\overline{E}_{\mu}$) in Mt.~Bisul, Mt.~Bohyun, Hyper-K, and Super-K.}
\begin{tabular}{lcc}
\hline \hline
Detector site (overburden) & $\Phi$ ($10^{-7}\,{\rm cm}^{-2} {\rm s}^{-1}$) & $\overline{E}_{\mu}$ (GeV) \\
\hline
Mt.~Bisul (820~m) & 3.81 & 233 \\
Mt.~Bohyun (820~m) & 3.57 & 234 \\
Mt.~Bisul (1,000~m) & 1.59 & 256 \\
Mt.~Bohyun (1,000~m) & 1.50 & 257 \\
\hline
Hyper-K (Tochibora, 650~m) & 7.55 & 203 \\
\hline
Super-K & 1.54 & 258 \\
\hline \hline
\end{tabular}
\end{table}

The muon flux and average energy at each site are estimated using the muon simulation code \texttt{MUSIC}~\cite{Antonioli:1997qw, Kudryavtsev:2008qh}, 
a three-dimensional MC tool dedicated to muon transportation in matter.
Elevation data for the areas around Mt.~Bisul and Mt.~Bohyun have been extracted from the ``ALOS World 3D-30m'' database published by JAXA~\cite{ALOS} for input to the simulation.
The latitude, longitude and altitude of the simulated locations are summarized in Table~\ref{tab:bgsim-location}.
For both Mt.~Bisul and Mt.~Bohyun simulations assuming 820~m and 1,000~m overburdens have been performed 
using muons are generated at the surface following the parameterization in~\cite{Tang:2006uu}.
The rock type is assumed to be the same as the Super-K site (Inishi rock) with the density of 2.70~g/cm$^3$.
Table~\ref{tab:muon_simulation} summarizes the calculated muon flux ($\Phi$) and average energy ($\overline{E}_{\mu}$) at the Mt.~Bisul, Mt.~Bohyun, Hyper-K (Tochibora), and Super-K sites.
Based on the uncertainty of the exact rock composition, the uncertainty of muon flux is assumed to be $\pm$20\%.

\begin{figure}[tb]
\centering
\includegraphics[width=0.7\textwidth]{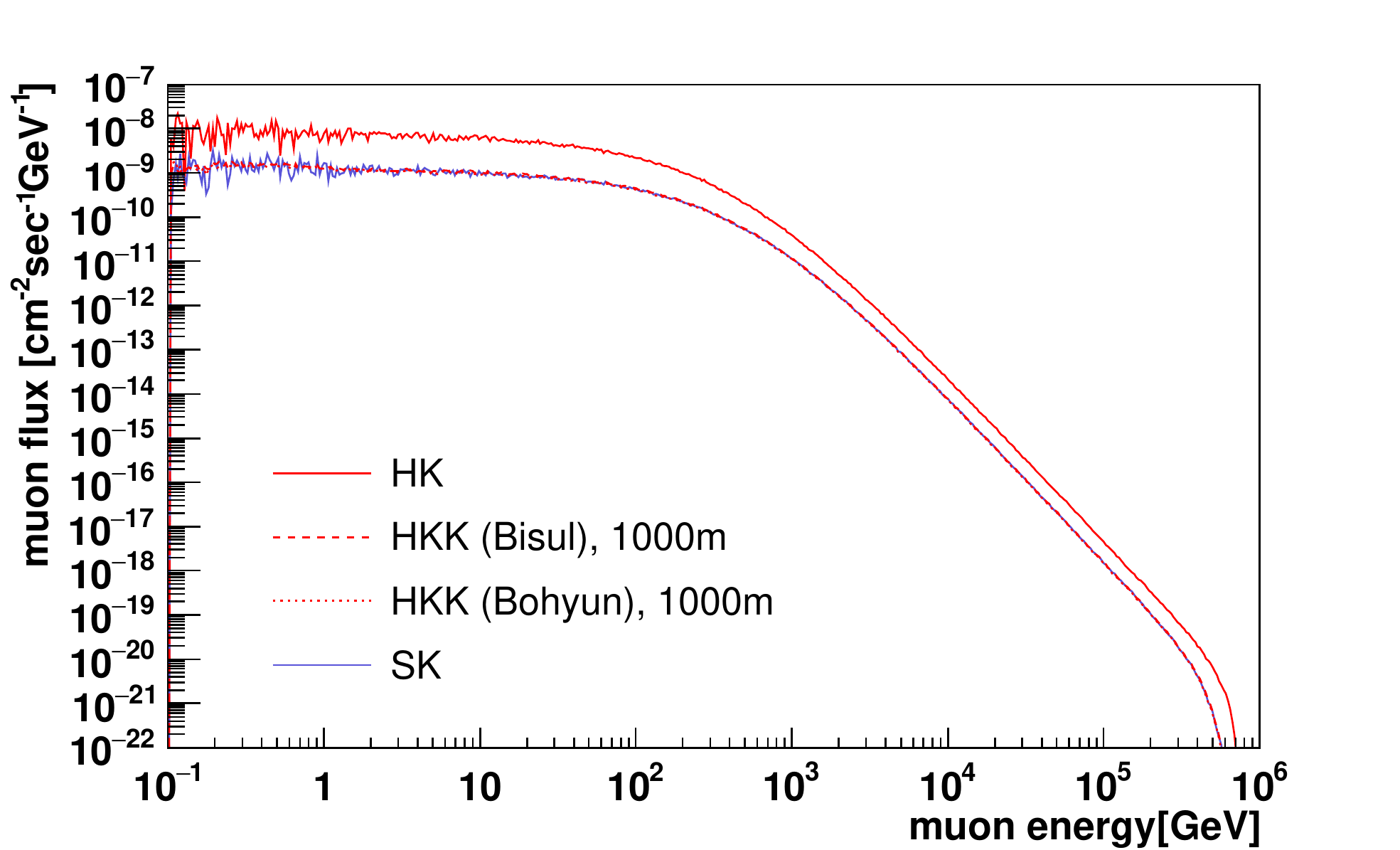}
\caption{Calculated muon energy spectra for HK (Tochibora), Mt.~Bisul (1,000~m overburden), Mt.~Bohyun (1,000~m overburden), and Super-K, based on the \texttt{MUSIC} simulation}
\label{fig:muon-spectrum}
\end{figure}

\begin{figure}[tb]
\centering
\includegraphics[width=0.7\textwidth]{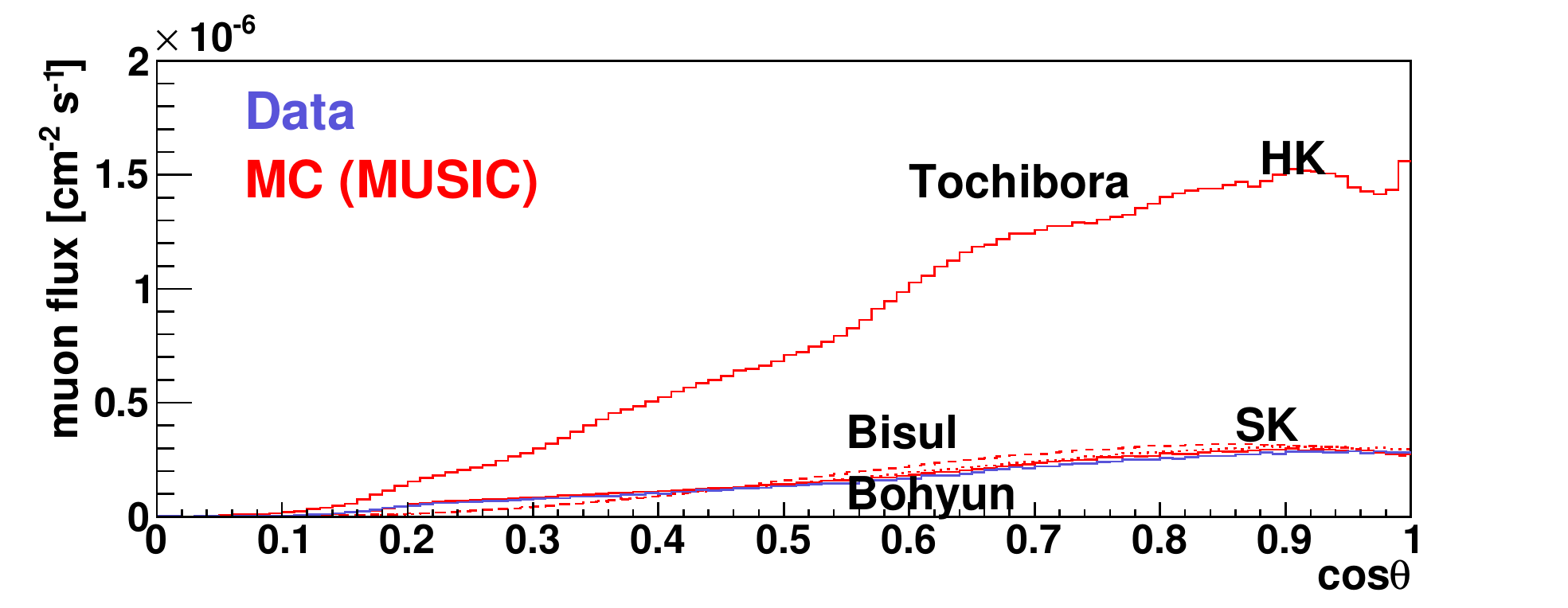}
\includegraphics[width=0.7\textwidth]{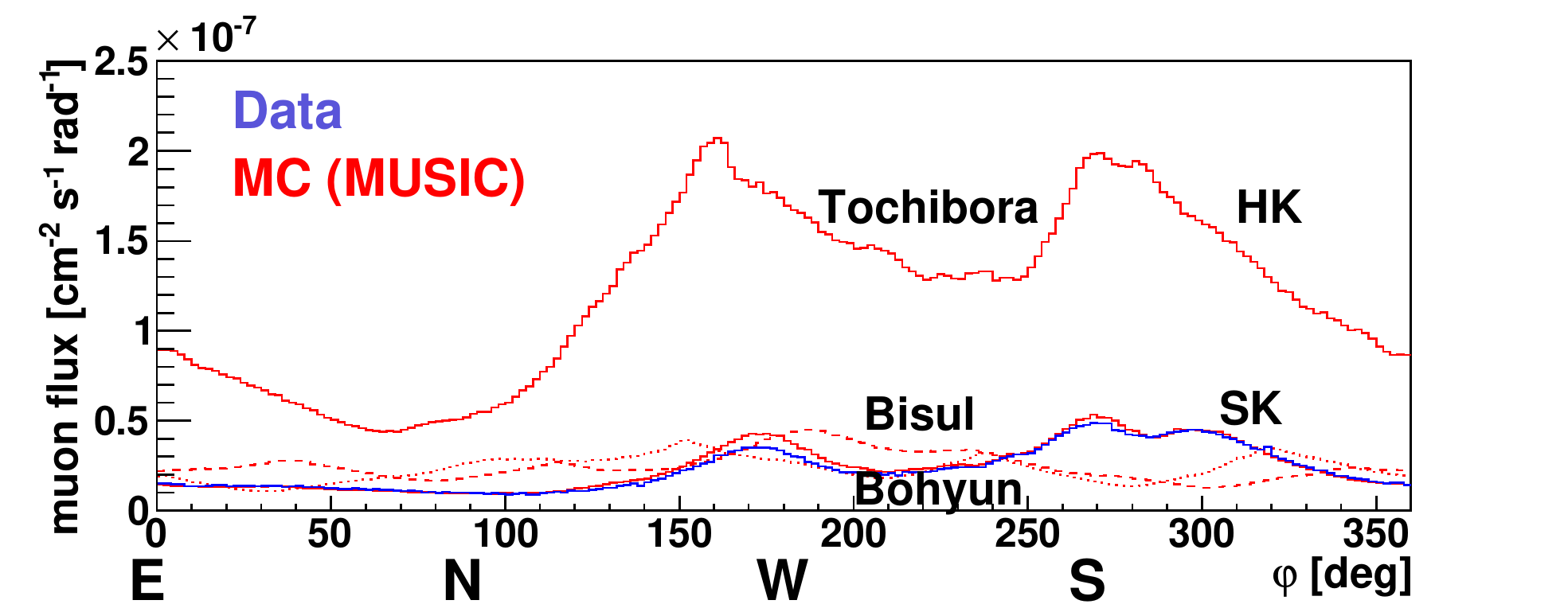}
\caption{Muon flux as a function of cosine of the zenith angle, $\cos\theta$, (upper) and azimuth angle $\phi$ (lower) for the Hyper-K (Tochibora), Mt.~Bisul (1,000~m overburden), Mt.~Bohyun (1,000~m overburden), and Super-K.
Here zero degrees represents the eastern direction.
Blue lines show the data from Super-K while the red lines show the MC prediction from the \texttt{MUSIC} simulation 
of the Tochibora (solid), Mt.~Bisul (dashed), and Mt.~Bohyun (dotted) sites. }
\label{fig:muon-direction}
\end{figure}

Figure~\ref{fig:muon-spectrum} shows the calculated muon energy spectra for Hyper-K (Tochibora, 650~m overburden), Mt.~Bisul (1,000~m overburden), Mt.~Bisul (1,000~m overburden), and SK.
The corresponding fluxes as a function of zenith and azimuthal angles are shown in Figure~\ref{fig:muon-direction}.
Note that the absolute flux and shape of the Super-K data are well reproduced by the simulation.
Further, with 1,000~m overburdens, the Korean sites are expected to have similar muon fluxes and energies as those observed at Super-K.
For the shallower 820~m overburden the flux is expected to increase by more than a factor of two according to Table~\ref{tab:muon_simulation}).
Using this information the muon flux ratios at Mt.~Bisul relative to Super-K, $\Phi (\textrm{HKK}_\textrm{Bisul})/\Phi (\textrm{Super-K})$,
is estimated to be $1.03\pm 0.21$ ($2.47 \pm 0.49$) assuming a 1,000~m (820~m) overburden.

Based on these calculated muon fluxes isotope production rates due to muon spallation have been calculated 
using \texttt{FLUKA}~\cite{Bohlen:2014buj, Ferrari:2005zk} version 2011.2b.
The isotope yield per muon track length, $Y$,  depends on the muon energy which increases with larger overburdens.
As a result, the isotope yield per muon becomes larger for deeper experimental sites.
For the 1,000~m overburden case the average muon energy is similar to that of Super-K 
and thus the ratio of their isotope yields per a muon are similar, $1.03 \pm 0.21$ .
The same ratio calculated for the Hyper-K (Tochibora) site is about 0.8~\cite{design-report}.
Interpolating from these two, the isotope yield per muon for the 820~m overburden is estimated to be 0.9
and similarly the isotope production rate for 820~m overburdens becomes about $2.22 \pm 0.44$ larger than that at Super-K.
In contrast Tochibora site is estimated to be about $4 \pm 1$ times larger.
For similar overburdens the isotope yield is not expected to differ largely between the two Korean sites.
Accordingly the yield is expected to be between two to four times smaller than at the Tochibora site in Japan.

\subsubsection{Potential benefits}
Lower spallation backgrounds at a Korean detector will result in improved sensitivity to solar neutrinos.
A day-night asymmetry in the rate of solar neutrinos due to MSW matter effects in the Earth~\cite{Wolfenstein:1977ue, Mikheev:1986gs, Mikheev:1986wj} 
is expected to be larger for the higher energy region of the $^8$B neutrino spectrum, where spallation is the dominant background source.
Neutrinos from the hep reaction chain fall in a similar energy region and as they are produced in a different region of the solar 
interior, can potentially provide new information on solar physics. 
With lower spallation backgrounds, the short time variability of the temperature in the solar core could be monitored more precisely with these neutrinos.
Further, lower backgrounds in the higher energy sample can improve resolution of the solar neutrino spectrum shape, whose lower 
energy region is a sensitive probe of matter effects, both standard and otherwise, in the sun.

Spallation backgrounds can be rejected based on their correlation with preceding muons.
For the solar neutrino analysis, the effect of spallation reduction is estimated keeping the signal efficiency to 80\%.
The remaining spallation fraction is estimated, based on a study using Super-K data, to be 1.2\%(2.3\%) for the 1,000~m(820~m) overburden.
This can be compared to 3.9\% estimated for the Hyper-K Tochibora site.
%In addition, one has to note that the production rate of spallation is a factor of four(two) smaller for 1,000(820)~m overburden site as discussed in the previous section.

The search for the diffuse flux of neutrinos produced by all supernova explosions since the beginning of the universe, the supernova relic neutrino (SRN) flux, 
similarly benefits from larger overburdens.
For the SRN analysis, because the signal flux is expected to be only a few tens/$\mbox{cm}^{2}$/the small signal flux, a more stringent event selection,
requiring a negligible spallation background contribution is used.
In this case, the signal efficiency is 79\%(56\\%) for neutrinos reconstructed 
with energies between 17.5 and 20~MeV and 90\% (75\%) between 20 and 26~MeV for the 1,000(820)~m overburden site.
These can be compared to 29\% and 54\%, respectively, for Hyper-K at Tochibora.
Above 26~MeV the spallation background decreases exponentially, but the signal is overwhelmed by backgrounds from atmospheric neutrinos.
In this way reduced spallation backgrounds will enhance the SRN detection capability particularly below 20~MeV.
After 10 years of operation, the number of events and the significance of non-zero observation of SRN are 100(90) events and 5.2(4.8)$\sigma$, respectively, for the 1,000(820)~m overburden site in Korea assuming the SRN flux of~\cite{Ando:2002ky}. 
Hyper-K at Tochibora is expected to observe 70 events with a corresponding significance of 4.2$\sigma$.

It is worth noting that the ability to observe neutrons via high-photon-yield photosensors or gadolinium doping will 
provide other physics opportunities for a Korean detector.
Among these the observation of neutrinos from nuclear reactors in Korea via their inverse beta decay reactions becomes 
possible. 
Similarly, such neutron tagging is expected to provide highly efficient suppression of atmospheric neutrino backgrounds 
to searches for proton decay.
Detailed studies of the expected sensitivity of such measurements is planned for a future document.

\subsection{Neutrino geophysics}

The inner Earth's chemical composition is one of the most important properties of our planet. 
While the matter density is well known through seismic measurements~\cite{Dziewonski:1981xy}, 
the chemical composition is much less understood~\cite{McDonough1995}. 
Neutrino oscillations depend on the electron density of the medium traversed by the neutrinos~\cite{Wolfenstein:1977ue,Mikheev:1986gs},
hence, the electron density distribution of the Earth can be reconstructed from the neutrino energy spectrum. 
Accordingly, the chemical composition of the Earth can be constrained for a given mass density distribution~\cite{Rott:2015kwa,Winter:2015zwx}.
Hyper-K is expected to be the first experiment that could experimentally confirm the Earth's core is composed of iron, 
ruling out lead or water scenarios at the $3~\sigma$ level~\cite{design-report}. 
The measurement relies on precisely measuring atmospheric muon neutrino disappearance 
and electron neutrino appearance in the energy range between $1$ and $8$~GeV as a function of the zenith angle.
This measurement is limited by the reduced neutrino flux at these energies, such 
that a detector in Korea will double the statistics available.

\begin{figure}[hbt]
\begin{center}
\includegraphics[width=0.45\textwidth]{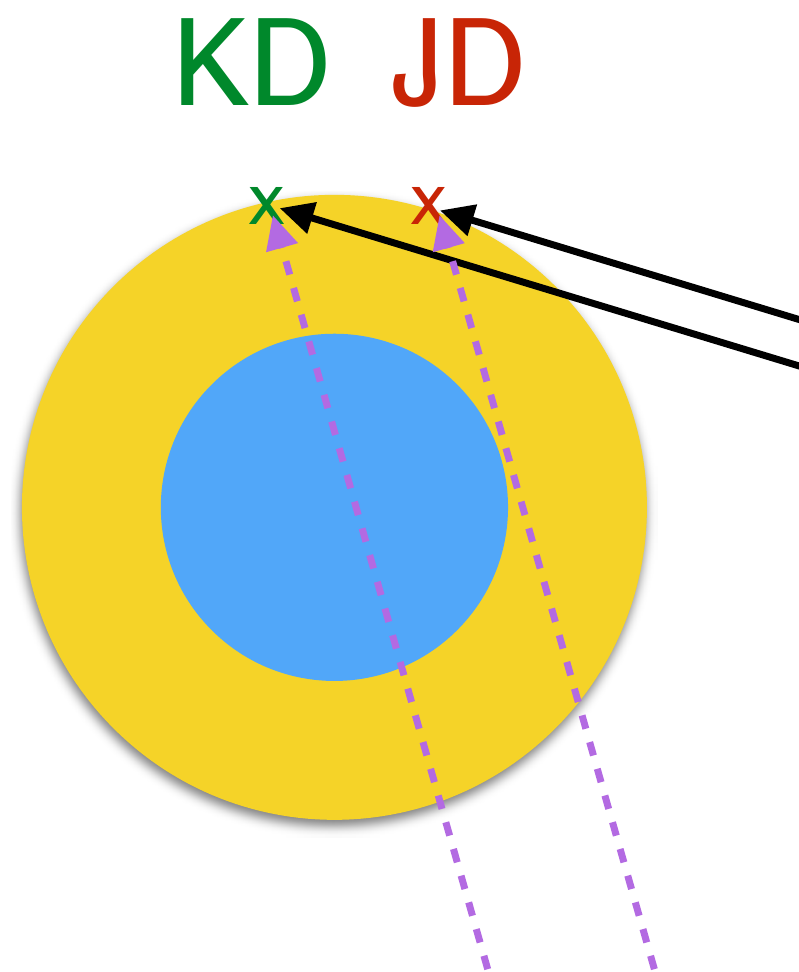}
\caption{Due to the distance between the Korean ("KD") and the Japanese ("JD") detector locations, both detectors could observe a supernova neutrino burst with different Earth shadowing. In scenario 1 (solid black arrows), one detector would observe an unshadowed flux, while neutrinos detected in the other detector would travel through Earth for up to 1800~km.
In scenario 2 (dashed purple arrows), one detector would be shadowed by the Earth`s mantle (yellow) only, while the other detector would be shadowed by the mantle and the outer core (blue). While the resulting difference in pathlength is small, up to 4400~km of the path of neutrinos reaching one detector would go through the outer core instead of the mantle. 
Since the matter density of the outer core is much larger than that of the mantle, the matter effect along these two paths would be markedly different.
Using an online tool for calculating Earth crossing probabilities for different detector locations, we find that the combined probability of these two scenarios is 6.4~\%~\cite{Mirizzi:2006xx}. (The difference between Mt.~Bohyun and Mt.~Bisul is $<$0.1~\%.)}
\label{fig:sn-shadowing}
\end{center}
\end{figure}

If a supernova occurs such that the neutrinos travel through the Earth before reaching the detector,
that is below the detector's local horizon, they will be subject 
energy-dependent matter effects.
This would manifest as a distortion of the energy spectrum of neutrinos or antineutrinos
depending on the mass hierarchy, which can be observed under favorable conditions~\cite{Dighe:1999bi,Borriello:2012zc}.
Having two geographically separated detector locations increases the
likelihood of the Earth shadowing the neutrino flux reaching one of the detectors.
In addition, comparing the energy spectra in two detectors that observe different shadowing scenarios (see
fig.~\ref{fig:sn-shadowing}) may make it easier to disentangle matter effects from supernova burst neutrino properties.

\subsection{Dark matter searches}

Hyper-K can search for physics beyond the standard model in the form
of self-annihilating dark matter captured in the Sun, Earth or from
the Galactic dark matter halo. Super-K has demonstrated this physics
potential through the world's best constraints on spin-dependent
scattering of dark matter with matter~\cite{Choi:2015ara}. Hyper-K can
improve upon Super-K's results and is expected to provide the best
indirect dark matter search sensitivities for masses below 100~GeV.
As the background to a neutrino signal from dark matter annihilation
in the Sun comes from atmospheric neutrinos a benefit from a second
site could come from reduced systematic uncertainties associated with
atmospheric neutrino fluxes.  
A neutrino signal originating from the decays of the dark matter annihilation products in the Sun is also
accompanied by a high multiplicity stopped meson decay low-energy
neutrino signal from hadronic showers of the annihilation products in
the center of the Sun~\cite{Rott:2012qb,Bernal:2012qh,Rott:2015nma}.
The expected signal consists of neutrinos of a few ten's of MeV from
muon decays at rest in the Sun as well as neutrino line signals at
29.8~MeV and 235.5~MeV from two-body charged pion and kaon decays at
rest.  The possible addition of gadolinium in
water~\cite{Beacom:2003nk} would reduce (invisible muon) backgrounds
significantly for this signal, which can very efficiently be detected
through the inverse beta decay reaction~\cite{Rott:2012qb}.

\subsection{Modifications to neutrino propagation}

T2HKK can also be a powerful probe of non-standard physics affecting
neutrino propagation, in particular effects observable as modifications to the 
standard PMNS survival or appearance probability.  
In Ref.~\cite{Ribeiro:2007jq}, various types
of non-standard physics scenarios for a beam experiment with 
detectors in both Korea and Japan has been considered,
including models of quantum decoherence,  violations of Lorentz
symmetry with and without CPT invariance, and non-standard neutrino
interactions with matter.  In most cases, configurations 
with a large detector in both countries 
have significantly improved sensitivity 
to such types of new physics, relative to having the equivalent 
detector mass in just one of them.
Not only do the two baselines provide a more complete measure of the 
neutrino spectrum and hence distortions arising from new physics,
but they also provide distance $L/E$ ranges to
constrain scenarios with non-oscillating (or with oscillating admixuters) $L/E$ effects. 
See Tables I, II and Fig. 6 of Ref.~\cite{Ribeiro:2007jq} for
more details. 
As one example the expected sensitivity to an enhanced matter effect caused by non-standard
interactions is presented in the next section. 

%======================================================================
%\subsection{Sensitivities with CC1pi Event Selection}
\subsection{Potential for improvement with a CC1$\pi$ event selection}
\label{subsec:cc1pi}

For Korean detector sites with a smaller off-axis angle, a harder
(anti)neutrino spectrum is present.  The (anti)neutrinos above 1 GeV
probe the first oscillation maximum and the region between the first
and second oscillation maxima.  These (anti)neutrinos are important
for determining the mass ordering, and measuring $\delta_{CP}$ if the
phase is near $\pi$/2 and 3$\pi$/2.  The quasi-elastic scattering cross
section is nearly constant above 1~GeV, while the rate of
other processes that include pion production increases.  
Higher statistics in the $>$ 1~GeV region may be achieved by
including  candidate events with evidence of pion production in
addition to the charged lepton. These additional candidate events may
include events where a Michel electron from a pion decay chain is
detected, or where a pion is directly detected by the reconstruction
of a second visible ring in the detector. The inclusion of these pion
production events will be the subject of future studies.

%======================================================================
\graphicspath{{nonSTD_nu/plots}}
%\linenumbers[1446]
%\setcounter{figure}{43}
%\setcounter{table}{9}
%\setcounter{equation}{8}
%======================================================================
\section{\label{sec:nonSTD} Sensitivity of T2HKK to non-standard interactions}

In this section we discuss the capability of the T2HKK experiment to
put constraints on Non-Standard Interactions (NSI) in
neutrino propagation. We present our results for
 only the $1.5^\circ$ off-axis and normal mass ordering. However at the end
we will comment about the results for other off-axis configurations of
T2HKK and also for inverted ordering.  The discussions in this
section are based on Refs.~\cite{Fukasawa:2016lew,Ghosh:2017ged}.

\subsection{Non-standard interactions}
In the presence of flavor changing neutral currents 
the standard neutrino-matter interaction potential is modified, allowing for 
neutrino flavor change via neutral current interactions with matter~\cite{Wolfenstein:1977ue,Guzzo:1991hi,Roulet:1991sm,Ohlsson:2012kf,Miranda:2015dra}.%\footnote{ 
The presence of such NSI effects can be studied in neutrino oscillation experiments,
especially in the long-baseline and atmospheric neutrino oscillation
experiments where the neutrinos experience the earth matter effect of long distances.
Theoretically this kind of interaction can arise from the following
four-fermion interaction:
\begin{eqnarray}
{\cal L}_{\mbox{\rm\scriptsize eff}}^{\mbox{\tiny{\rm NSI}}} 
=-2\sqrt{2}\, \varepsilon_{\alpha\beta}^{ff'P} G_F
\left(\overline{\nu}_{\alpha L} \gamma_\mu \nu_{\beta L}\right)\,
\left(\overline{f}_P \gamma^\mu f_P'\right),
\label{nsi:NSIop}
\end{eqnarray}
where $f_P$ and $f_P'$ are fermions with chirality $P$,
$\varepsilon_{\alpha\beta}^{ff'P}$ is a dimensionless constant and $G_F$
is the Fermi coupling constant. If these kinds of interactions exist
in nature, then the MSW matter potential looks like:
\begin{eqnarray}
{\cal A} \equiv
\sqrt{2} G_F N_e \left(
\begin{array}{ccc}
1+ \varepsilon_{ee} & \varepsilon_{e\mu} & \varepsilon_{e\tau}\\
\varepsilon_{\mu e} & \varepsilon_{\mu\mu} & \varepsilon_{\mu\tau}\\
\varepsilon_{\tau e} & \varepsilon_{\tau\mu} & \varepsilon_{\tau\tau}
\end{array}
\right),
\label{nsi:matter-np}
\end{eqnarray}
where $\varepsilon_{\alpha\beta}$ is defined by
\begin{equation}
\varepsilon_{\alpha\beta}\equiv\sum_{f=e,u,d}\frac{N_f}{N_e}\varepsilon_{\alpha\beta}^{f}\,.
\end{equation}
with $N_f~(f=e, u, d)$ representing the number density of the fermions $f$.
Here we define the NSI parameters as
$\varepsilon_{\alpha\beta}^f = \varepsilon_{\alpha\beta}^{ffL} + \varepsilon_{\alpha\beta}^{ffR}$.
The present 90\% C.L. bounds on the NSI parameters coming from non-oscillation 
experiments are compiled in Refs.\,\cite{Davidson:2003ha,Biggio:2009nt}.
Using the formula
\begin{eqnarray}
 \varepsilon_{\alpha\beta} \lesssim \left\{\sum_{P} 
      \left[\left(\varepsilon_{\alpha\beta}^{eP}\right)^2 +
      \left(3\varepsilon_{\alpha\beta}^{uP}\right)^2 +
      \left(3\varepsilon_{\alpha\beta}^{dP}\right)^2\right]
    \right\}^{1/2}
\label{nsi:epsf2eps}
\end{eqnarray}
for neutral Earth-like matter with an equal number of neutrons and
protons, Ref.\,\cite{Biggio:2009nt} gives the following bounds on
$\varepsilon_{\alpha\beta}$:
\begin{eqnarray}
\hspace{-25pt}
&{\ }&
\left(
\begin{array}{lll}
|\varepsilon_{ee}| < 4 \times 10^0 & |\varepsilon_{e\mu}| < 3\times 10^{-1}
& |\varepsilon_{e\tau}| < 3 \times 10^0\\
&  |\varepsilon_{\mu\mu}| < 7\times 10^{-2}
& |\varepsilon_{\mu\tau}| < 3\times 10^{-1}\\
& & |\varepsilon_{\tau\tau}| < 2\times 10^1
\end{array}
\right).
\label{nsi:epsilon-m}
\end{eqnarray}

The bounds on the NSI parameters $\varepsilon_{\alpha\beta}^{f}$ from
oscillation experiments are given in the Table 2 of
Ref.\,\cite{Miranda:2015dra} for $f=d$. 
%\footnote{ It is known that
%addition of the $3\times 3$ identity matrix to the Hamiltonian
%modifies only the phase of the probability amplitude and therefore
%does not change the oscillation probability.  The authors of
%Ref.\,\cite{Miranda:2015dra} discuss the constraints on the matter
%potential
%$\varepsilon_{\alpha\beta}-\varepsilon_{\mu\mu}\delta_{\alpha\beta}$ by
%subtracting the $3\times 3$ identity matrix with a coefficient
%$\varepsilon_{\mu\mu}$.}
%%%We note in passing that the constraint on
%%%$\varepsilon_{\alpha\beta}^f$ for $f=e$ from the solar neutrino data
%%%has not been given unlike in the cases for $f=u$ and $f=d$, so the
%%%formula (\ref{nsi:epsf2eps}) cannot be applied to get the constraint
%%%on $\varepsilon_{\alpha\beta}$ from oscillation experiments.  This is
%%%the reason why we use Eq. (\ref{nsi:epsilon-m}), which is obtained
%%%from non-oscillation experiments, as the constraints on
%%%$\varepsilon_{\alpha\beta}$.  
From Eq. (\ref{nsi:epsilon-m}) it is clear
that the bounds on $\varepsilon_{ee}$,
$\varepsilon_{e\tau}$ and $\varepsilon_{\tau\tau}$ are at least one
order of magnitude weaker compared to the other NSI parameters.  Thus,
in order to keep the number of parameter combinations to a manageable level,
this section uses the following anzatz:
\begin{eqnarray}
{\cal A} = \sqrt{2} G_F N_e\left(
\begin{array}{ccc}
1+\varepsilon_{ee}&0&\varepsilon_{e\tau}\cr
0&0&0\cr
\varepsilon_{e\tau}^\ast&0&\varepsilon_{\tau\tau}
\end{array}\right)\,.
\label{nsi:ansatz}
\end{eqnarray}
Therefore, the NSI parameters of interests are $\varepsilon_{ee}$,
$|\varepsilon_{e \tau}|$, $\varepsilon_{\tau\tau}$ and
arg($\varepsilon_{e\tau}) = \phi_{31}$.  
In the limit $\Delta m^2_{21}\to 0$ it is known \cite{Ota:2001pw,Kopp:2007mi,Yasuda:2007jp} that the oscillation
probability depends only on $\dcp+\phi_{31}$ 
and we therefore expect similar dependence on $\phi_{31}$ as on $\dcp$.

For the simulation of the T2HKK experiment, we have taken the
experimental configuration from the detector setup in
Section~\ref{sec:detector} and consider the highest energy (or least
off-axis) configuration at $1.5^\circ$.  
We also compare our results with the T2HK setup, that is the $2 \times$JD configuration.
We assume a 3:1 ratio of antineutrino and neutrino running. 
For this analysis we incorporate systematics by the method of pulls and
considered four pull variables including a signal normalization,
a background normalization, a signal tilt and a background
tilt.  Namely the numbers of signal ($S_j$)
and background ($B_j$) events are scaled as
$S_j\to S_j
[1+\sigma_{\mbox{\scriptsize\rm s}}\,\xi_{\mbox{\scriptsize\rm s}}
+\sigma^{\mbox{\scriptsize\rm tilt}}_{\mbox{\scriptsize\rm s}}\,
\xi^{\mbox{\scriptsize\rm tilt}}_{\mbox{\scriptsize\rm s}}
(E_j-E_{\mbox{\scriptsize\rm min}})/(E_{\mbox{\scriptsize\rm max}}-E_{\mbox{\scriptsize\rm min}})]$,
$B_j\to B_j
[1+\sigma_{\mbox{\scriptsize\rm b}}\,\xi_{\mbox{\scriptsize\rm b}}
+\sigma^{\mbox{\scriptsize\rm tilt}}_{\mbox{\scriptsize\rm b}}\,
\xi^{\mbox{\scriptsize\rm tilt}}_{\mbox{\scriptsize\rm b}}
(E_j-E_{\mbox{\scriptsize\rm min}})/(E_{\mbox{\scriptsize\rm max}}-E_{\mbox{\scriptsize\rm min}})]$,
where $\sigma_{\mbox{\scriptsize\rm s}}$,
$\sigma_{\mbox{\scriptsize\rm b}}$,
$\sigma^{\mbox{\scriptsize\rm tilt}}_{\mbox{\scriptsize\rm s}}$,
$\sigma^{\mbox{\scriptsize\rm tilt}}_{\mbox{\scriptsize\rm b}}$
($\xi_{\mbox{\scriptsize\rm s}}$,
$\xi_{\mbox{\scriptsize\rm b}}$,
$\xi^{\mbox{\scriptsize\rm tilt}}_{\mbox{\scriptsize\rm s}}$,
$\xi^{\mbox{\scriptsize\rm tilt}}_{\mbox{\scriptsize\rm b}}$)
are the systematic errors (the pull variables)
for a signal normalization,
a background normalization, a signal tilt and a background tilt,
respectively, $E_{\mbox{\scriptsize\rm max}}$ and
$E_{\mbox{\scriptsize\rm min}}$ are the maximum and minimum energy,
and $E_j$ is the energy of the $j$-th bin.
Hence we understand that the normalization errors affect the scaling of the
events, whereas the tilt errors affect their energy dependence.
Throughout the analysis we have fixed the tilt error to 10\%.
%The normalization errors assumed are listed in Table~\ref{nsi:tab1}
%and Table~\ref{tab:sys_error}. 
A total normalization error for each detector location and event
sample is used as listed in Table~\ref{nsi:tab1};
the magnitudes of these errors
are taken from Table~\ref{tab:sys_error}.
The same normalization error for signal and background are used, 
and all systematics are considered to be uncorrelated.
Unlike the PMNS-driven oscillation sensitivities described in the previous
sections, this analysis has been performed using 
the GLoBES~\cite{Huber:2004ka,Huber:2007ji} package 
with the MonteCUBES~\cite{Blennow:2009pk} NSI probability engine.

\begin{table}
\begin{center}
\begin{tabular}{|c|c|c|}
\hline
Set-up  & T2HKK at 1.5$^\circ$ & T2HK\\          
\hline
$\nu_e$ events &    3.84  & 4.71    \\
$\bar{\nu}_e$ events  & 4.11 & 4.47 \\
$\nu_\mu$ events & 3.83 & 4.13 \\
$\bar{\nu}_\mu$ events & 3.81 & 4.15 \\
\hline
\end{tabular}
\end{center}
\caption{Systematic uncertainties assumed for the sample normalization (in percentages) for a single detector at the T2HKK and T2HK experimental 
configurations.
The 'T2HK' column corresponds to the systematic errors of the Kamioka
detector (for T2HK the numbers are same for both the detectors) and
the 'T2HKK at 1.5$^\circ$' column corresponds to the systematic error of the
Korean detector (For T2HKK setup, the systematic error of the Kamioka
detector is given by the 'T2HK' column and the systematic error of the Korean
detector is given by the 'T2HKK at the 1.5$^\circ$' column).
}
\label{nsi:tab1} 
\end{table}

\subsection{Constraining NSI parameters}

\begin{figure*}[t!]
\begin{center}
\resizebox{1.1\textwidth}{!}{%
\hspace{-0.7 in}
\includegraphics{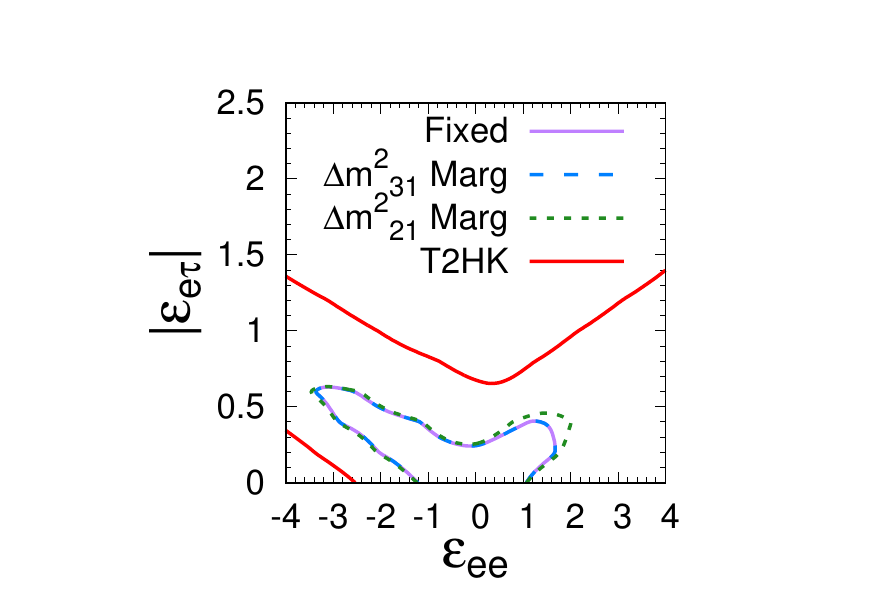}
\hspace{-1.4 in}
\includegraphics{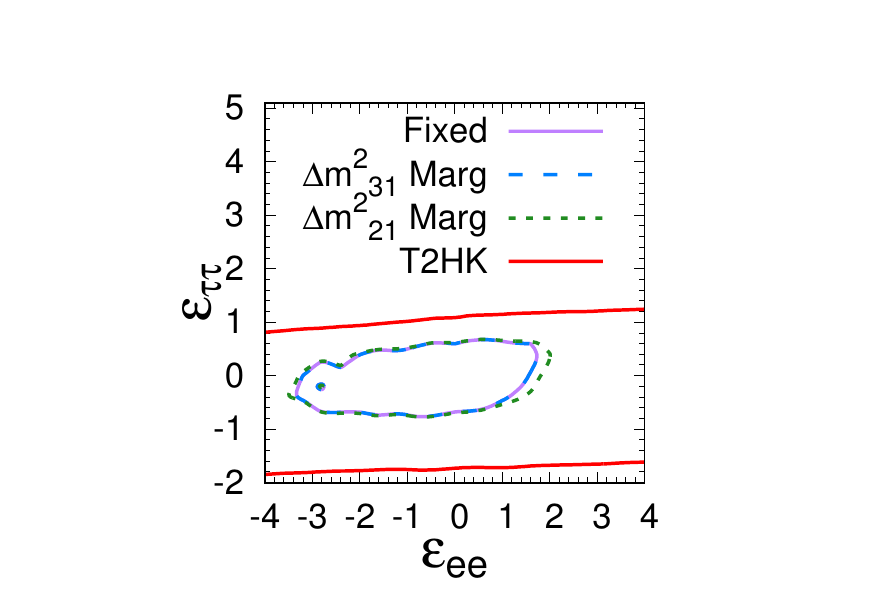}
\hspace{-1.4 in}
\includegraphics{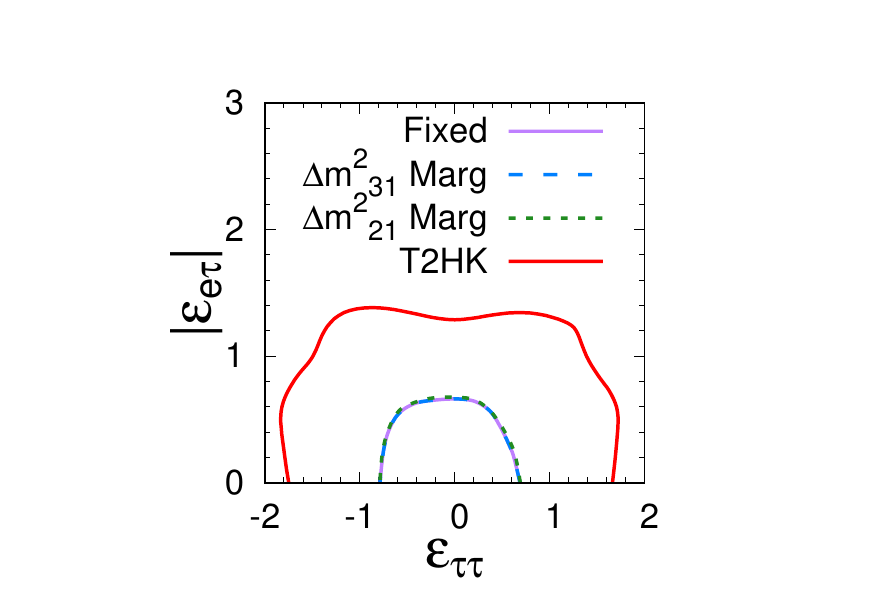}
\hspace{-10pt}
}
\caption{Ability of T2HKK to constrain the magnitude of the NSI parameters at 3$\sigma$ for $\theta_{23} = 45^\circ$ and $\delta_{CP}=270^\circ$ with Normal Hierarchy.  The red lines show the comparison for two detectors at the Kamioka site.  Other lines show the effect of the uncertainty on the value of the mass splittings. %The systematics are used as listed in Table \ref{nsi:tab1}.
} 
\label{nsi:fig1}
\end{center}
\end{figure*}

\begin{figure*}[t!]
\begin{center}
\resizebox{1.1\textwidth}{!}{%
\hspace{-0.7 in}
\includegraphics{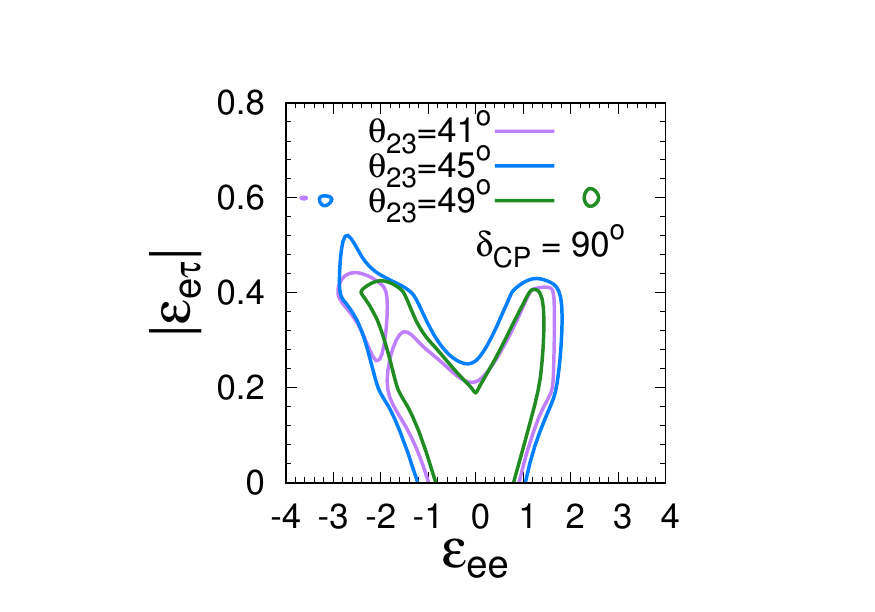}
\hspace{-1.4 in}
%\hspace{-10pt}
\includegraphics{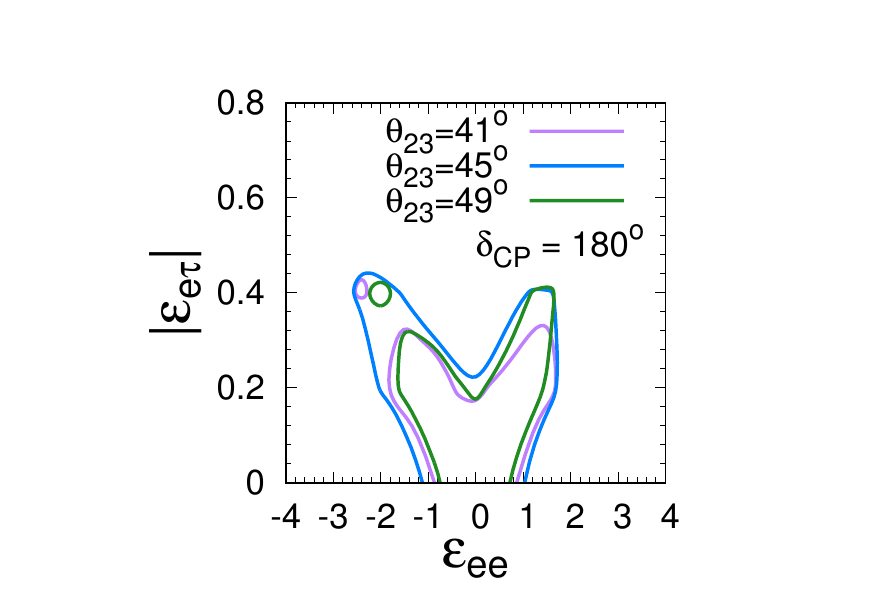}
\hspace{-1.4 in}
%\hspace{-10pt}
\includegraphics{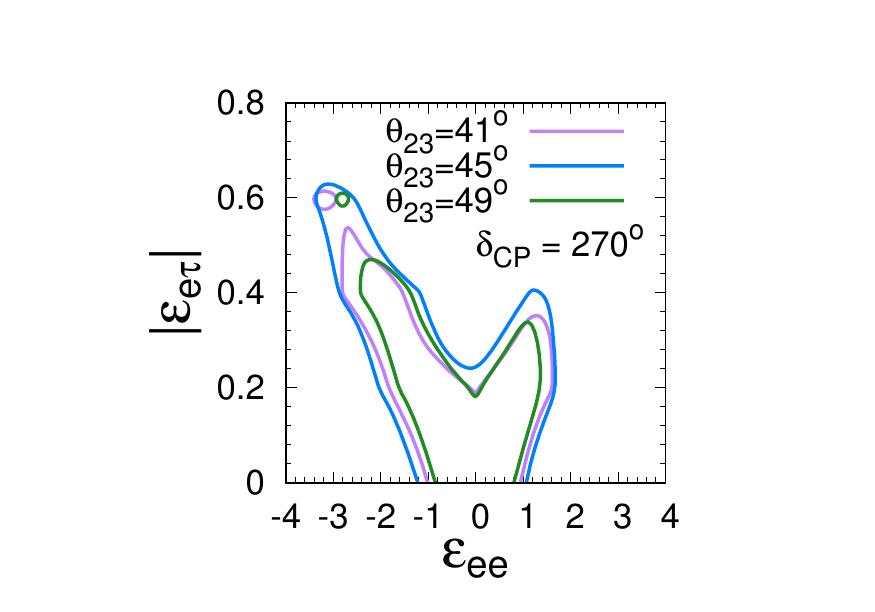}
\hspace{-10pt}
}
\caption{Ability of T2HKK to constrain the magnitude of the NSI parameters at 3$\sigma$ for different true values of $\theta_{23}$ and $\delta_{CP}$ with Normal Hierarchy.
} 
\label{nsi:fig2}
\end{center}
\end{figure*}

First we discuss the ability of the T2HKK experiment to constrain
the magnitudes of the NSI parameters $\varepsilon_{ee}$,
$|\varepsilon_{e\tau}|$ and $\varepsilon_{\tau\tau}$.  For this purpose we
assume the values of these parameters are zero and present our results
as limits in the $\varepsilon_{ee}$ -
$|\varepsilon_{e\tau}|$  plane. The assumed true value of
$\phi_{31}$ is zero and it has been marginalized over in the presentation of 
the results.
The PMNS oscillation parameters $\dcp$ and $\theta_{23}$ are
marginalized over as well,
whereas $\theta_{13}$, $\theta_{12}$, $\Delta
m^2_{21}$ and $\Delta m^2_{31}$ have been kept fixed close to their
globally preferred values~\cite{Forero:2014bxa,Esteban:2016qun,Capozzi:2013csa} and are
not varied unless otherwise mentioned.

In Fig. \ref{nsi:fig1}, we present the sensitivities with PMNS
parameters held at their currently favored values: $\theta_{23} = 45^\circ$ and
$\delta_{CP} = 270^\circ$. 
In the left, middle, and right panels the
$3 \sigma$ allowed region is given in the $\varepsilon_{ee}$ --
$|\varepsilon_{e\tau}|$, $\varepsilon_{ee}$ --
$\varepsilon_{\tau\tau}$, $\varepsilon_{\tau\tau}$ --
$|\varepsilon_{e\tau}|$ and planes, respectively. 
In each panel NSI parameters that are not plotted have been 
marginalized over. 
Here, and elsewhere unless otherwise noted, the marginalization is made over a range
of $-4$ to $+4$ for $\varepsilon_{ee}$, 0 to 2 for
$|\varepsilon_{e\tau}|$ and $-1$ to $+1$ for $\varepsilon_{\tau\tau}$.
We have additionally checked that the $\chi^2$ minima do indeed always
appear within these chosen ranges of $\varepsilon_{\alpha \beta}$.
The purple curve in each panel shows the allowed region when $\Delta
m^2_{21}$ and $\Delta m^2_{31}$ are kept fixed in the test spectrum.
The green dotted curves and blue dashed curves show the effect 
off marginalizing over these parameters.
It is clear that the uncertainty in $\Delta m^2_{31}$ has no effect on
the sensitivity, whereas the allowed region increases slightly when we
marginalize over $\Delta m^2_{21}$.  

To demonstrate how the sensitivity is improved by the use of a Korean detector, we also show
the equivalent result for the $2\times$JD configuration in Fig. \ref{nsi:fig1} (red curves).  
It can be seen that this configuration provides significantly weaker constraints. Indeed, in
this case it is necessary to extend the marginalization range of
$\varepsilon_{\tau \tau}$ out to $\pm4$ since there is still a significant posterior probability outside the original region.
From all the three panels, we note that the sensitivity of the KD+JD
(T2HKK) configuration is far better than the equivalent $2\times$JD
configuration, let alone the baseline single tank design.  
This is essentially the same effect as seen in the standard PMNS
oscillation model, with the longer baselines and the higher energies
both enhancing the matter effect.

Next we study the capability of T2HKK to constrain the NSI
parameters assuming different values of $\theta_{23}$ and
$\delta_{CP}$.  
Fig. \ref{nsi:fig2} shows how changes in
these parameters within their current allowed regions 
affect T2HKK's constraints in the $\varepsilon_{ee}$ --
$|\varepsilon_{e\tau}|$ plane.  
The left, middle, and right panels show 
$\delta_{CP} = 90^\circ$, $180^\circ$, and $270^\circ$, respectively. 
In each panel, the purple, blue, and green curves correspond to $\theta_{23} =
41^\circ$, $45^\circ$, $49^\circ$, respectively. 
From the figure we note that the sensitivity is best for $\delta_{CP}=180^\circ$ and worst for
$\delta_{CP} = 270^\circ$. 
The sensitivity for $\theta_{23}=45^\circ$ is the weakest in comparison to the 
other two tested values. 

\begin{figure*}[t!]
\centering
\resizebox{1.1\textwidth}{!}{%
\hspace{-0.7 in}
\includegraphics{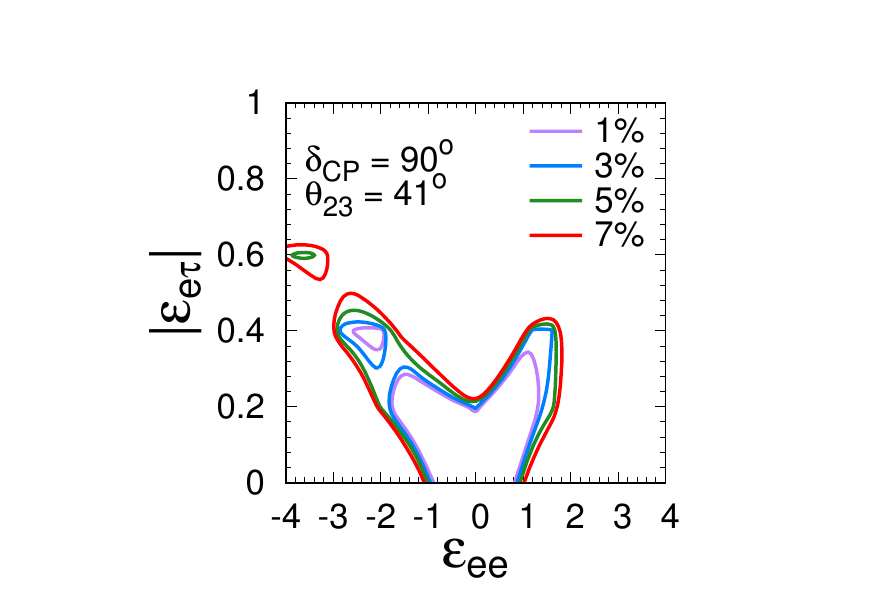}
\hspace{-1.4 in}
%\hspace{-10pt}
\includegraphics{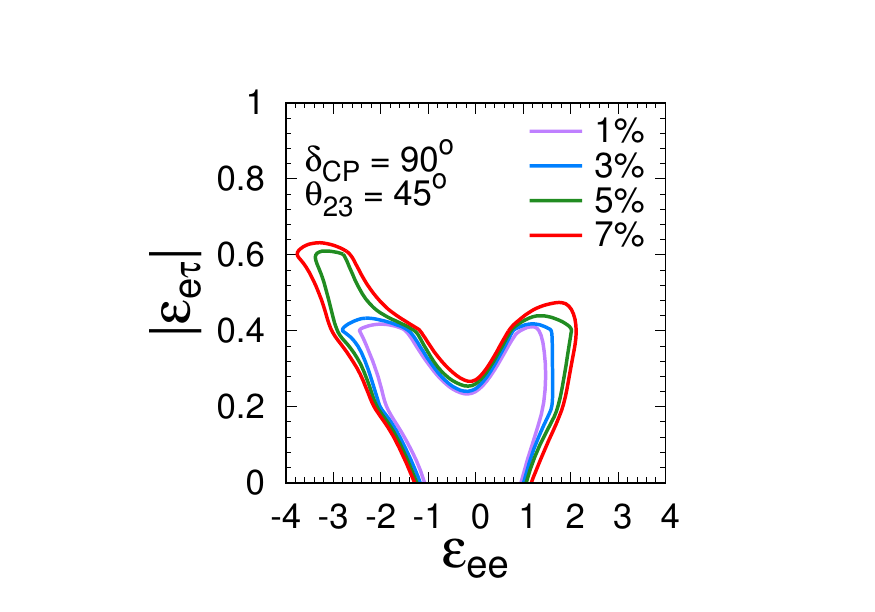}
\hspace{-1.4 in}
%\hspace{-10pt}
\includegraphics{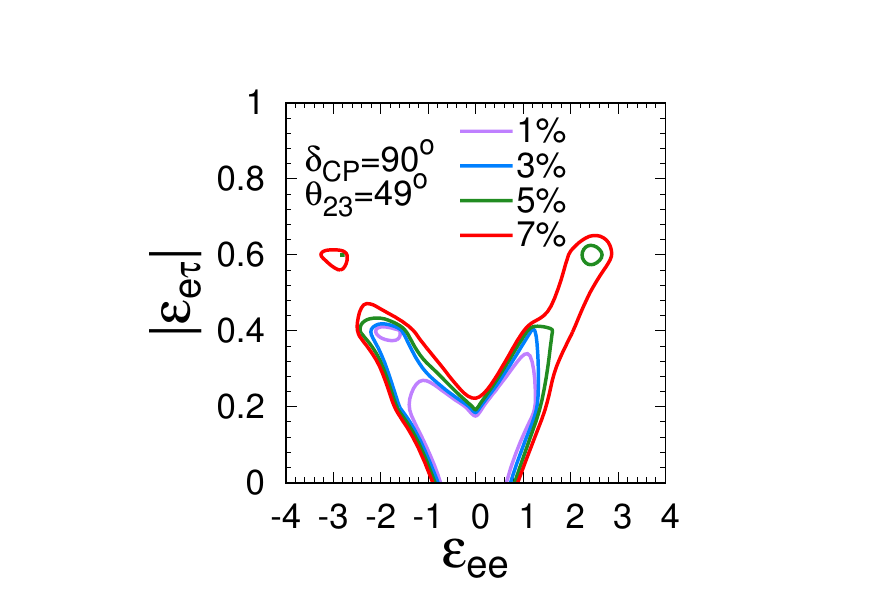}
\hspace{-10pt}
}
\resizebox{1.1\textwidth}{!}{%
\hspace{-0.7 in}
\includegraphics{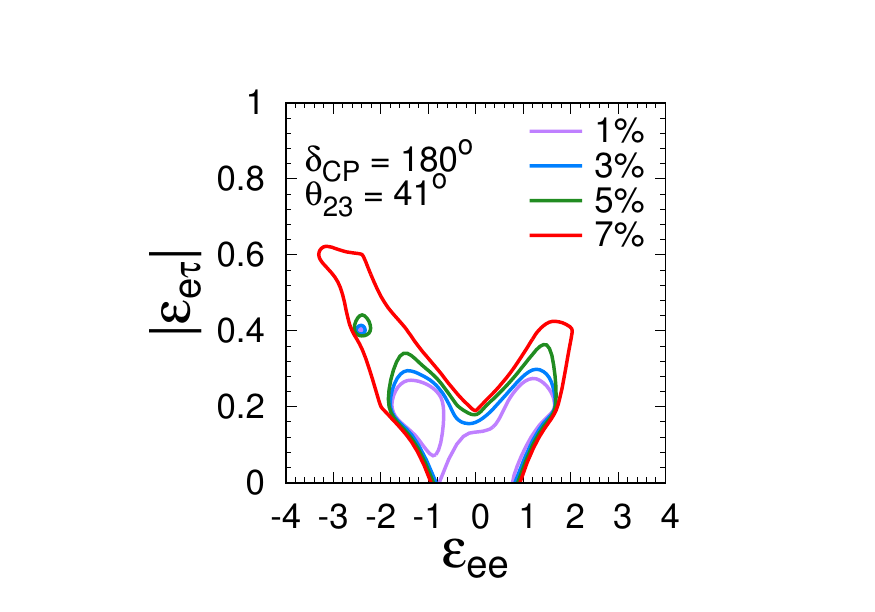}
\hspace{-1.4 in}
%\hspace{-10pt}
\includegraphics{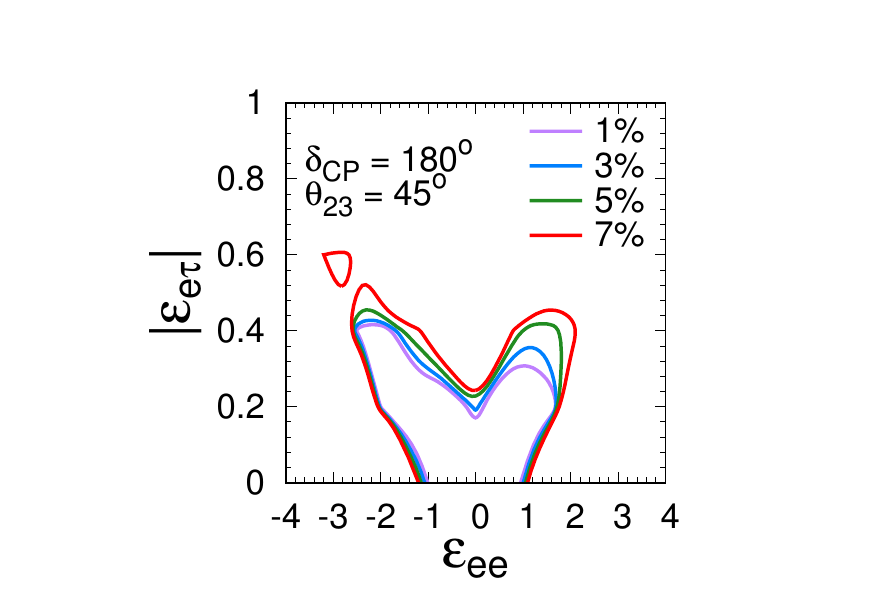}
\hspace{-1.4 in}
%\hspace{-10pt}
\includegraphics{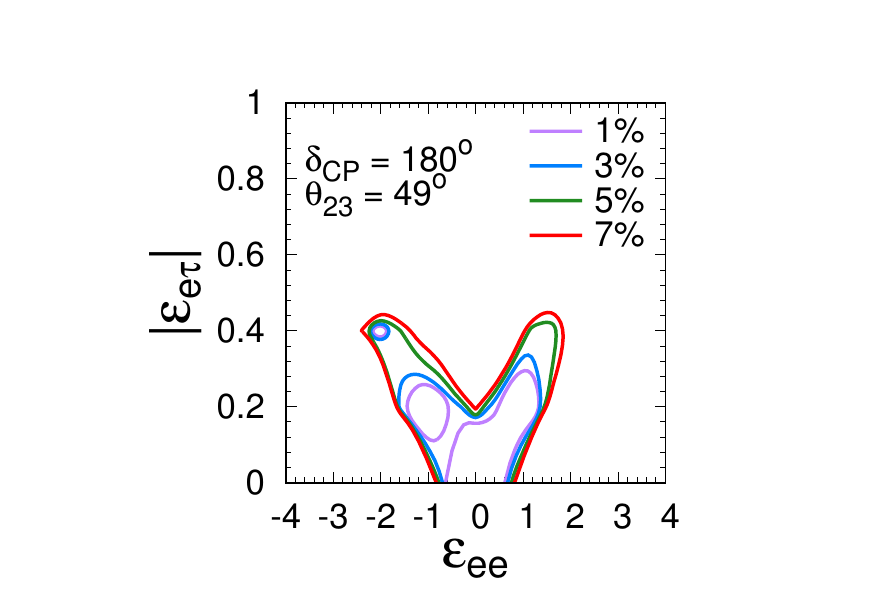}
\hspace{-10pt}
}
\resizebox{1.1\textwidth}{!}{%
\hspace{-0.7 in}
\includegraphics{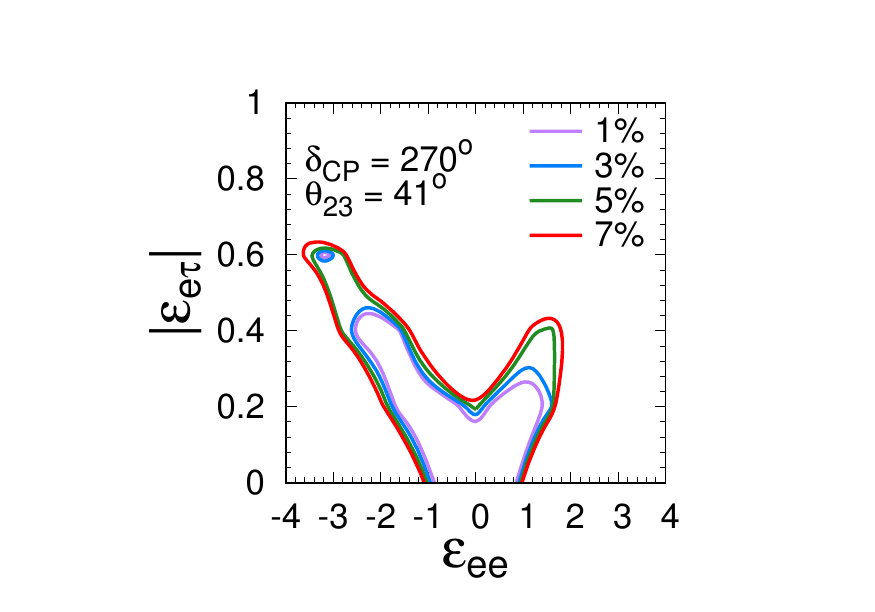}
\hspace{-1.4 in}
%\hspace{-10pt}
\includegraphics{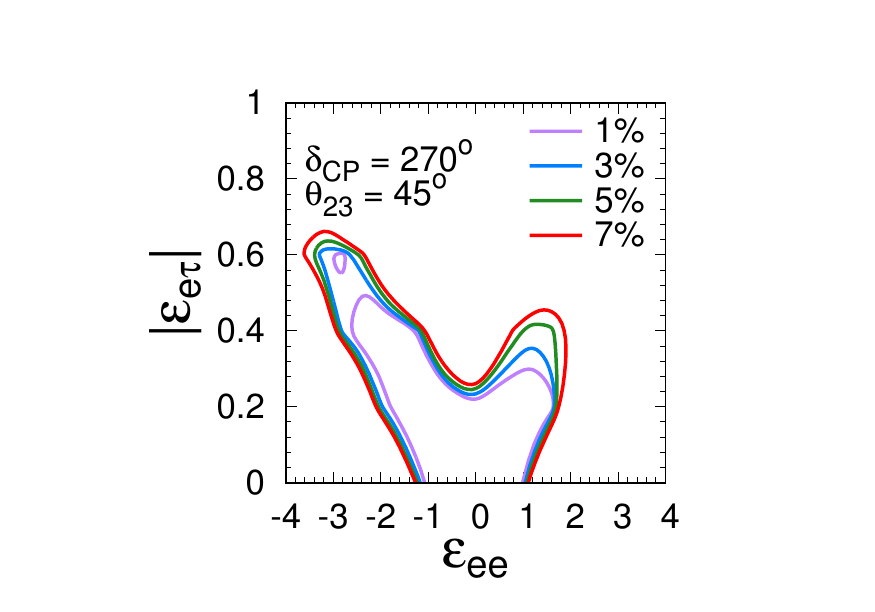}
\hspace{-1.4 in}
%\hspace{-10pt}
\includegraphics{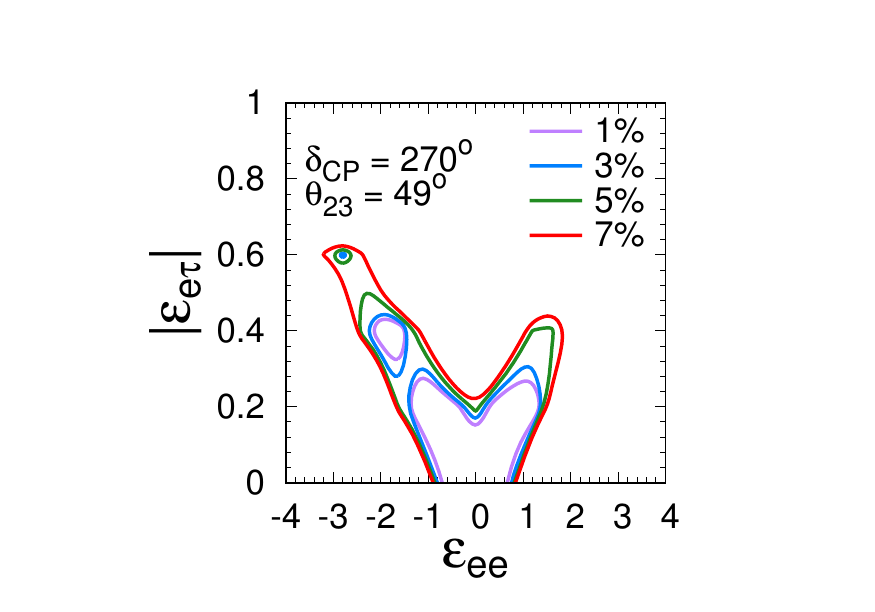}
\hspace{-10pt}
}
\caption{Ability of T2HKK to constrain the magnitude of the NSI parameters at 3$\sigma$ for different values of the PMNS parameters with Normal Hierarchy. The columns correspond to (from left to right) $\theta_{23}=41^\circ$, $45^\circ$ and $49^\circ$, and the rows are for (from top to bottom) $\delta_{CP} = 90^\circ$, $180^\circ$ and $270^\circ$.
}
\label{nsi:fig3}
\end{figure*}

Even at $O(1000)$\,km, there will be over a thousand events in the
data samples at a Korean detector, such that the statistical error on the event rate 
is a few percent.  
As a result systematic errors are expected to play an important role, so we next study 
their impact on T2HKK's ability to constrain the magnitude of the NSI parameters. 
To do this, we again examine the same $\varepsilon_{ee}$ --
$\left|\varepsilon_{e\tau} \right|$ space as in Fig.~\ref{nsi:fig2}
but for four different values of the systematic errors. 
In these plots a systematic error of $N\%$ implies a normalization error of $N\%$
applied to both signal and background events, both electron and muon events,
and both neutrinos and antineutrinos.  
The dependence can be seen in Fig.~\ref{nsi:fig3}. 
Here rows correspond to (from left to right)  $\delta_{CP} = 90^\circ$, $180^\circ$, and $270^\circ$.  
In each row the first, second and third panels corresponds to $\theta_{23} = 41^\circ$,
$45^\circ$ and $49^\circ$, respectively.
In all cases, we can see that the limits on the axes (i.e. with one of
the two parameters held at zero) are not substantially affected by
systematic uncertainties, but that the ability to rule out correlated
changes in $\left| \varepsilon_{ee} \right|$ and
$\left| \varepsilon_{e\tau} \right|$ are impacted by such a systematic
uncertainty.

\subsection{Constraining the CP phases}

\begin{figure*}[t]
\begin{center}
\resizebox{1.1\textwidth}{!}{%
\hspace{-0.7 in}
\includegraphics{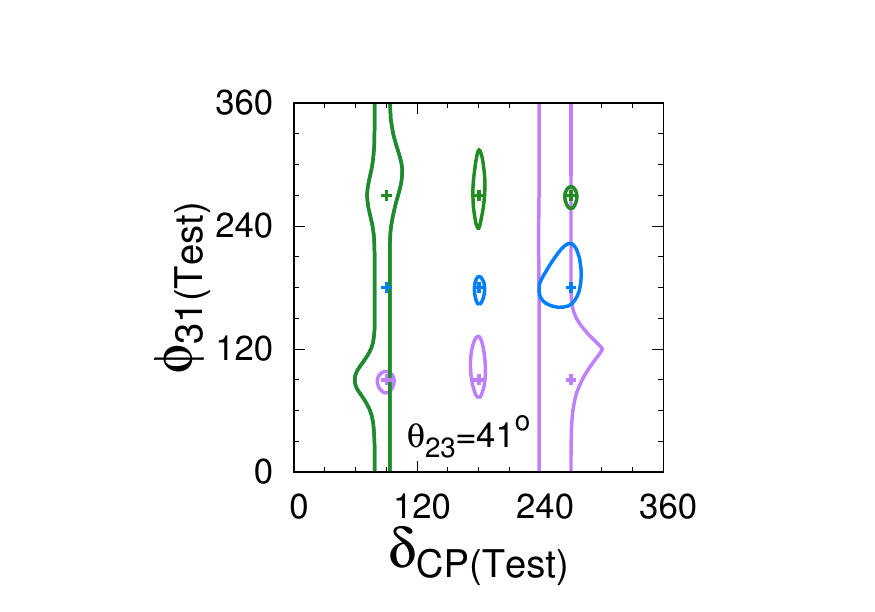}
\hspace{-1.4 in}
%\hspace{-15pt}
\includegraphics{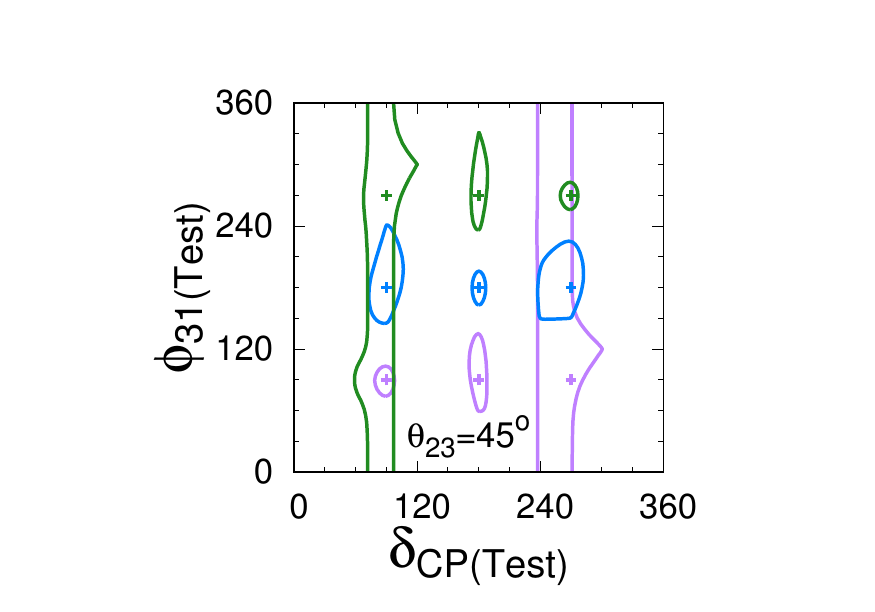}
\hspace{-1.4 in}
%\hspace{-15pt}
\includegraphics{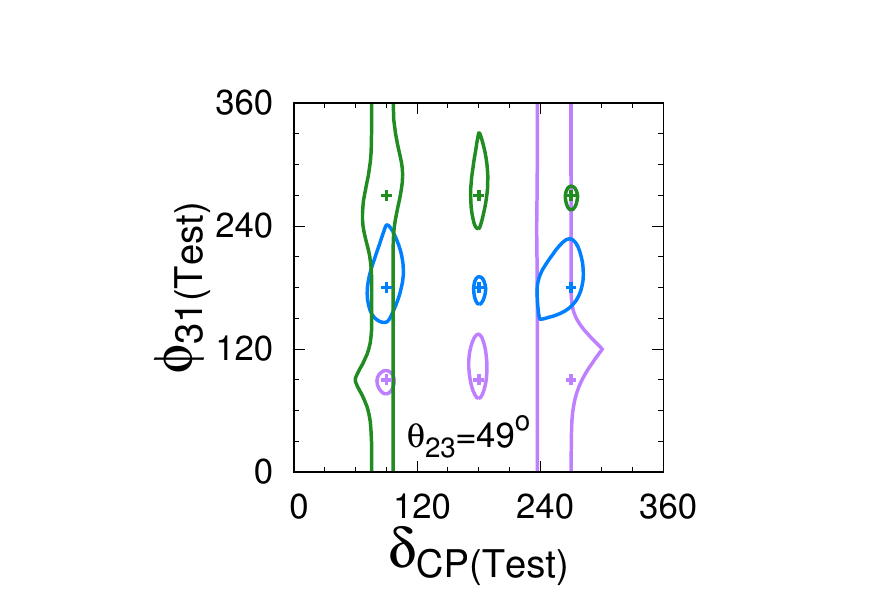}
\hspace{-20pt}
}
\caption{Ability of T2HKK to constrain the two CP phases at 90\% C.L. assuming Normal Hierarchy.
%The systematics are used as listed in Table \ref{nsi:tab1}.
The left, middle and right panels correspond to
$\theta_{23}=41^\circ$, $45^\circ$ and $49^\circ$ respectively, and
the purple, blue and green contours correspond to $\phi_{31}$ (true) $=
90^\circ$, $180^\circ$ and $270^\circ$ respectively.  The `\textbf{+}'
signs in all the three panels corresponds to the true values of ($\dcp$,
$\phi_{31}$). }
\label{nsi:fig4}
\end{center}
\end{figure*}

Assuming the NSI parameters are non-zero, there is an additional
source of CP violation from the argument of the off-diagonal
$\varepsilon_{e\tau}$ parameter.  
This would produce observable effects similar to those from the $\delta_{CP}$ parameter
of the PMNS formalism, so we next consider constraints in 
the $\delta_{CP}$ -- $\phi_{31}$ plane, with 
varying true values of $\theta_{23}$,
$\delta_{CP}$ and $\phi_{31}$. 
Here the  values of $\varepsilon_{ee}$,
$|\varepsilon_{e\tau}|$ and $\varepsilon_{\tau\tau}$ are assumed to
be 0.8, 0.2, and 0.0, respectively. 
These three parameters as well as $\theta_{23}$ are marginalized over in the analysis.  
As before, other PMNS parameters are held at their global best fit values.
Figure~\ref{nsi:fig4} shows the resulting allowed regions for T2HKK, 
with the left, middle, and right
panels showing results assuming  $\theta_{23} = 41^\circ$, $45^\circ$ and $49^\circ$,
respectively.
The purple, blue and green contours show the $\phi_{31}$ (true)
$= 90^\circ$, $180^\circ$ and $270^\circ$ cases.
From the panels it is clear that the best sensitivity comes when
$\phi_{31} \simeq \delta_{CP}$, and that the overall sensitivity is
not significantly affected by the true value of $\theta_{23}$.
On the other hand, for the ($\delta_{CP}$, $\phi_{31}$) combinations
($90^\circ$, $270^\circ$) and ($270^\circ$, $90^\circ$)
$\phi_{31}$ is entirely unconstrained.
%%Note also that the expected precision for $\delta_{CP}=180^\circ$ is better than $\delta_{CP} = 90^\circ$ and $270^\circ$ for $\phi_{31}
%%= 180^\circ$.

\begin{figure*}[t]
\begin{center}
\includegraphics[scale=1.2]{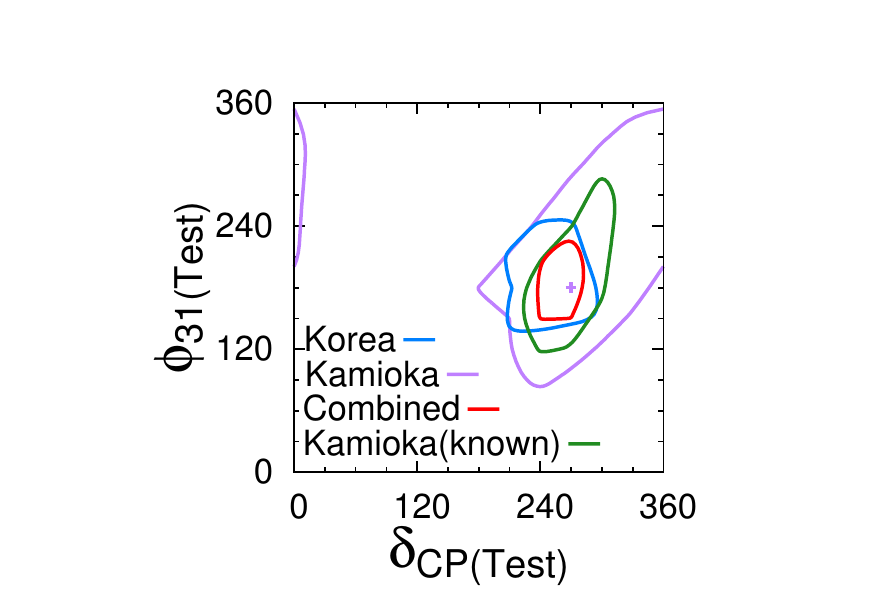}
\caption{Contribution of different detectors to constraints on the CP phases at 90\% C.L. assuming $\theta_{23}=45^\circ$, $\phi_{31} = 180^\circ$ and $\delta_{CP} = 270^\circ$ with Normal Hierarchy. See text for the full explanation.
} 
\label{nsi:fig5}
\end{center}
\end{figure*}

Sensitivity to constraining the CP phases benefits explicitly 
from having two detectors at different baselines in the T2HKK configuration.
To illustrate this Fig.~\ref{nsi:fig5} shows the contribution 
of each detector separately assuming 
$\theta_{23} = 45^\circ$, 
$\delta_{CP} = 270^\circ$, and 
$\phi_{31} = 180^\circ$.
The purple contour shows the contribution from the detector in Kamioka,
the blue contour shows that from the Korean detector,
 and the red is their combined sensitivity. 
For comparison the orange contour illustrates the expectation for two
detectors in Kamioka.
That there is not much difference between the contours with 
one and two detectors in Kamioka illustrates that the
increase in sensitivity seen in the T2HKK configuration 
comes mainly from its second baseline.

It is important to recognize that the combination of a detector in Kamioka
, which has higher statistics, with a detector in Korea, which
has a larger baseline, resolves parameter degeneracies and 
therefor allows for a simultaneous measurement of 
the NSI and CP parameters.
This is illustrated by the green contour, which shows the capability 
a single detector in Kamioka assuming that the NSI parameters
$\varepsilon_{ee}$, $|\varepsilon_{e\tau}|$, $\varepsilon_{\tau\tau}$ are non-zero, but known. 
This, firstly, demonstrates that a single measurement at
the first PMNS oscillation maximum is seriously limited by 
degeneracies in the extended model that cannot be untangled.  
Secondly, although the green contour shows a precise measurement, there is a degeneracy between the
two available CP parameters shown by the correlated nature of the allowed region. 
The degeneracy is lifted by measurements at the Korean detector. 
Indeed, the T2HKK configuration, shown in red, yields a more precise measurement 
even though the NSI parameters are considered unknown in the analysis.
In this sense an analysis that allows for the possibility of deviations from the
PMNS model can benefit much more from the extra information obtained
using multiple baselines than it would from simply improving the
available statistics for a single-baseline measurement.

\begin{figure*}[t!]
\centering
\resizebox{1.1\textwidth}{!}{%
\hspace{-0.7 in}
\includegraphics{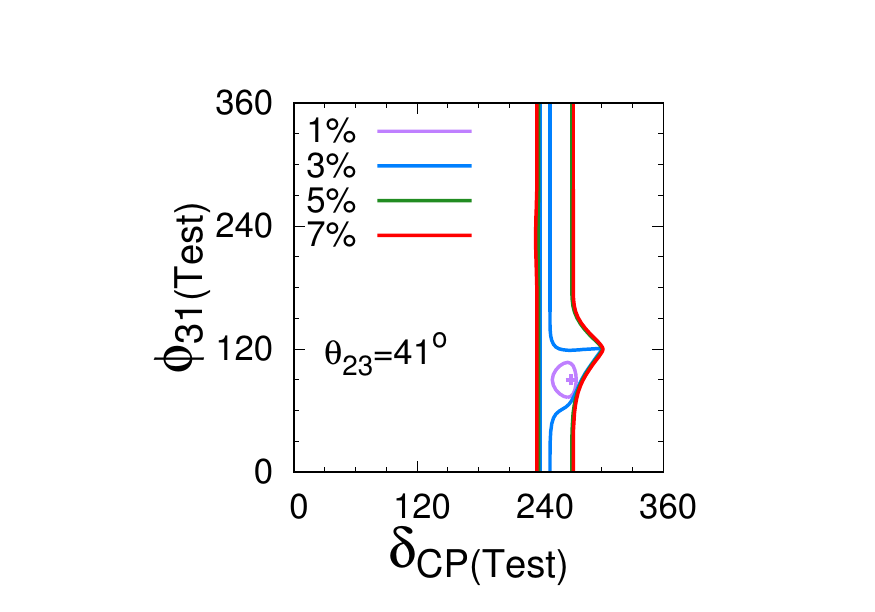}
\hspace{-1.4 in}
%\hspace{-15pt}
\includegraphics{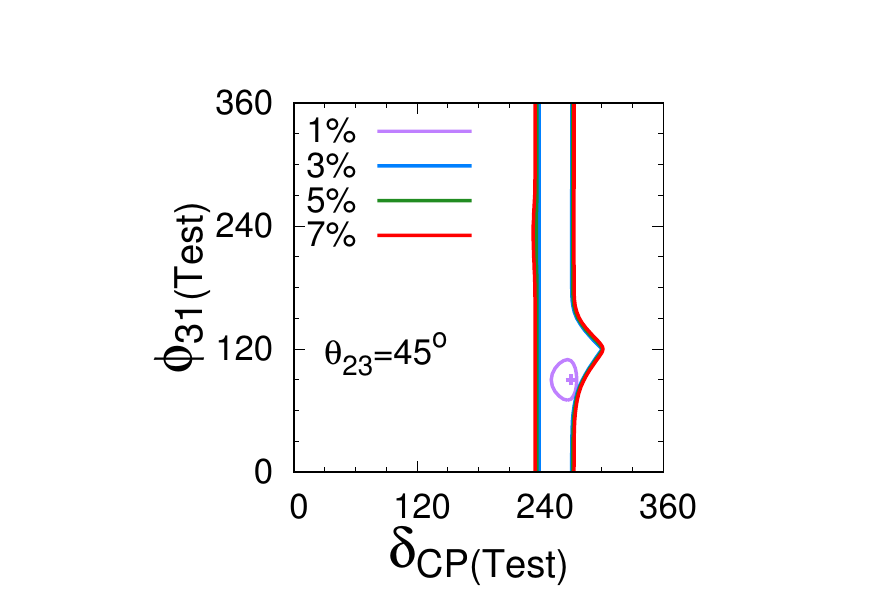}
\hspace{-1.4 in}
%\hspace{-15pt}
\includegraphics{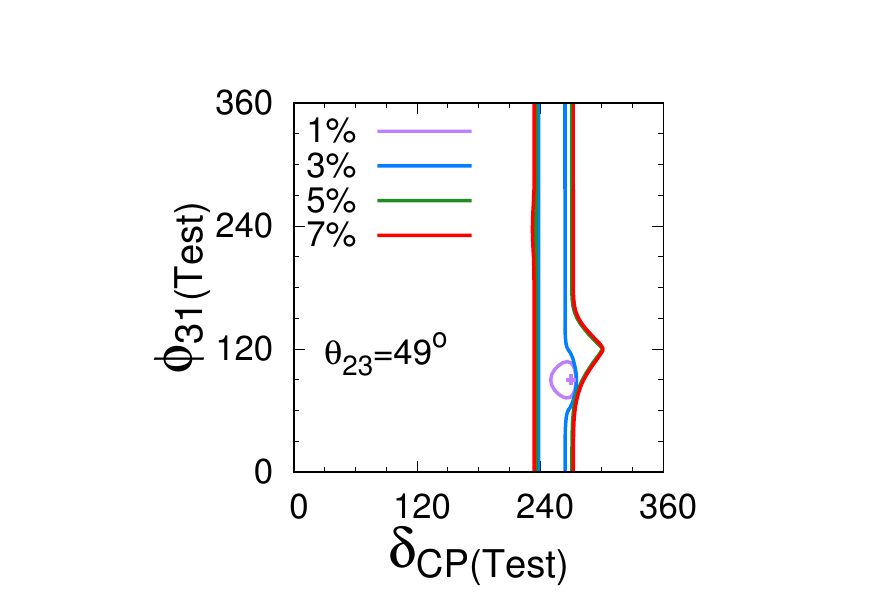}
\hspace{-20pt}
}
\resizebox{1.1\textwidth}{!}{%
\hspace{-0.7 in}
\includegraphics{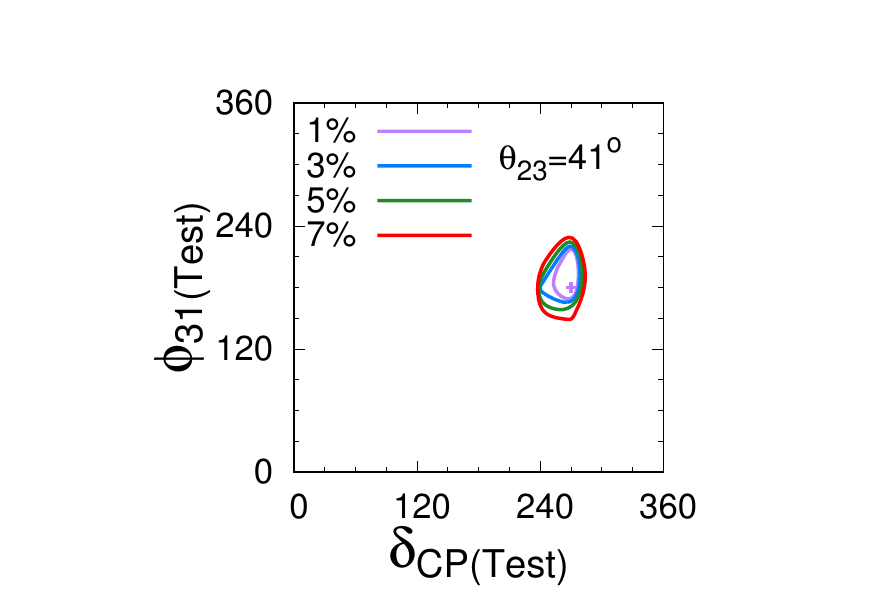}
\hspace{-1.4 in}
%\hspace{-15pt}
\includegraphics{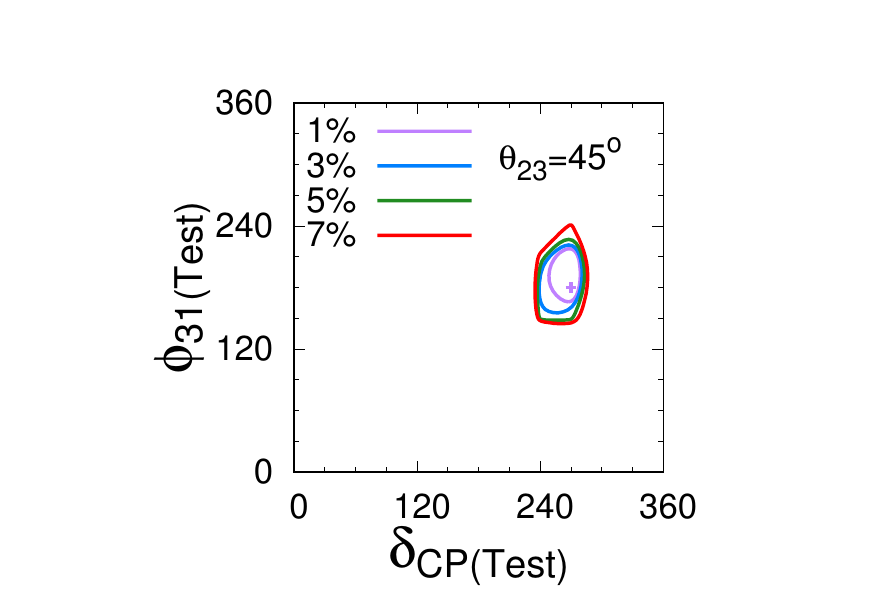}
\hspace{-1.4 in}
%\hspace{-15pt}
\includegraphics{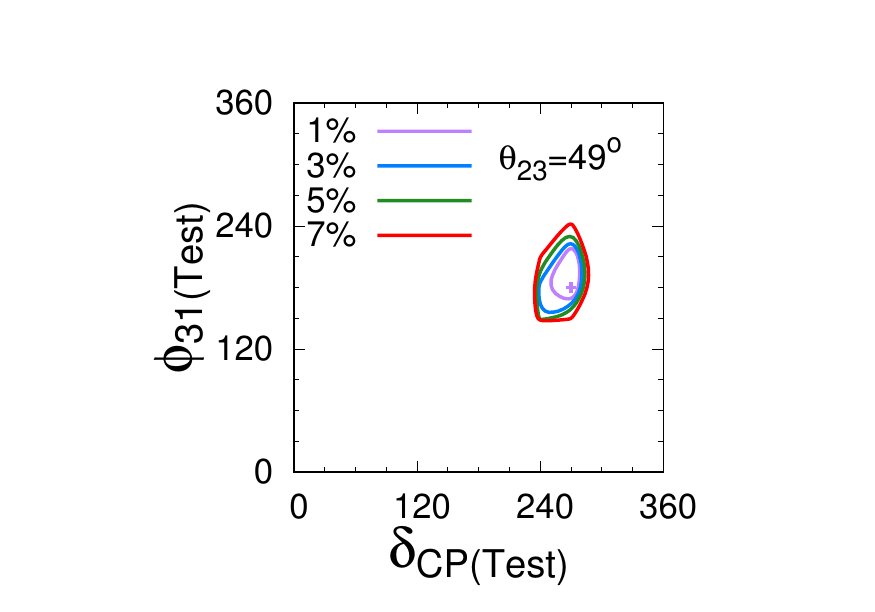}
\hspace{-20pt}
}
\resizebox{1.1\textwidth}{!}{%
\hspace{-0.7 in}
\includegraphics{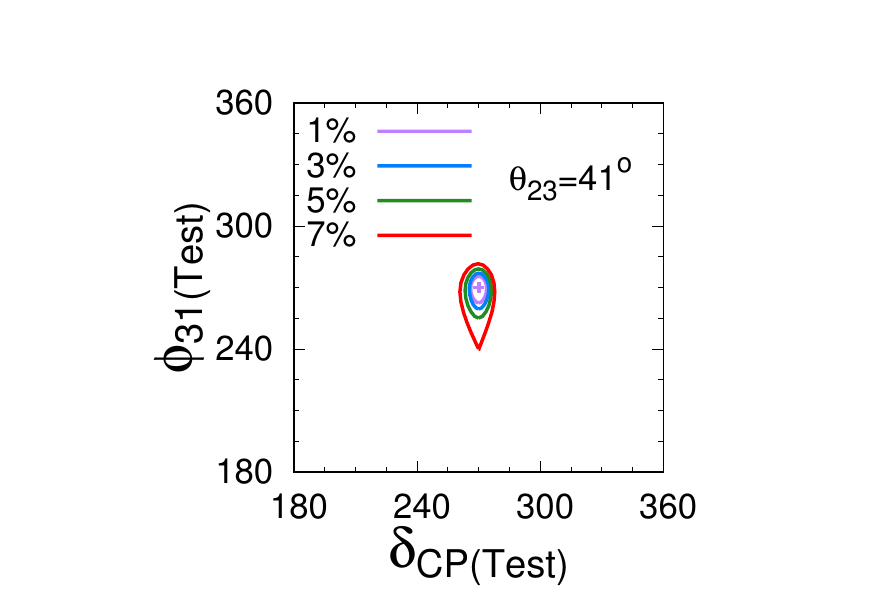}
\hspace{-1.4 in}
%\hspace{-15pt}
\includegraphics{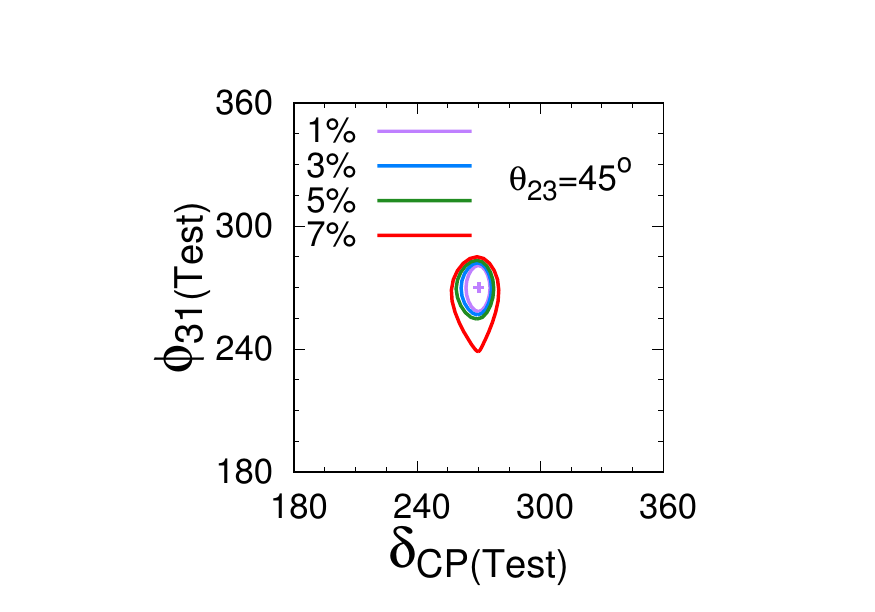}
\hspace{-1.4 in}
%\hspace{-15pt}
\includegraphics{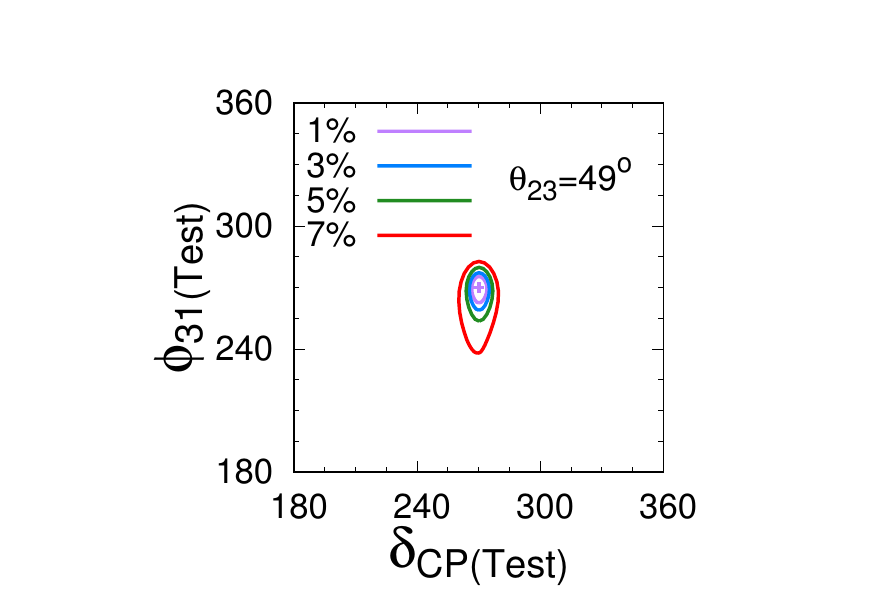}
\hspace{-20pt}
}
\caption{Ability of T2HKK to constrain the CP phases at 90\% C.L. with systematic uncertainties of 1\%, 3\%, 5\% and 7\% 
shown by the purple, blue, green, and red curves, respectively.
The columns correspond to $\theta_{23}=41^\circ$, $45^\circ$ and
$49^\circ$.  The rows are for $\phi_{31} = 90^\circ$, $180^\circ$ and
$270^\circ$.  The `\textbf{+}' signs correspond to the true values of
($\dcp$, $\phi_{31}$).  Normal Hierarchy is assumed.}
\label{nsi:fig6}
\end{figure*}

As before, we study how our systematic uncertainties affect our
ability to constrain the CP phases. 
Figure~\ref{nsi:fig6} shows the 
expected sensitivity under four systematic error assumptions, 1\%, 3\%, 5\% and 7\%,
using the same procedure as in Fig.~\ref{nsi:fig3}.
The first, second and third rows are for $\phi_{31} = 90^\circ$,
$180^\circ$ and $270^\circ$ respectively.  In each row, the left,
middle and right panels correspond to $\theta_{23}=41^\circ$,
$45^\circ$ and $49^\circ$ respectively.  
For $\delta_{CP} = 270^\circ$ and $\phi_{31} = 90^\circ$, 
it is possible to have a small closed contour in $\phi_{31}$ for small systematic error assumptions.
As the systematic error is increased from 1\%, it becomes 
impossible to constrain $\phi_{31}$ regardless of the assumed value of 
$\theta_{23}$.
For $\phi_{31}=180^\circ$ the sensitivity improves gradually when the
systematic uncertainty is reduced from 7\% to 1\% for all values of $\theta_{23}$. 
For $\phi_{31}=270^\circ$, the sensitivities evolve in a similar way to 
the $\phi_{31}=90^\circ$ case, but do not reach a
closed $\phi_{31}$ interval for $\delta_{CP} = 90^\circ$.

The above analyses have assumed a normal ordering and the $1.5^\circ$ degree 
off-axis detector configuration at T2HKK,  but here 
we discuss briefly the cases of an inverted ordering and other
off-axis angles.
For an inverted ordering we find that
the sensitivity to constrain the NSI parameters and the phases is
slightly weaker compared to normal ordering. 
For example, T2HKK can constrain $\varepsilon_{ee}$ in the region $-3.2
< \varepsilon_{ee} < 1.4$ assuming the normal ordering 
whereas the bound for inverted ordering is $-3.2 < \varepsilon_{ee} < 1.8$
assuming $\theta_{23} = 45^\circ$
$\delta_{CP} = 270^\circ$.

Among the three T2HKK off-axis detector configurations, 
$1.5^\circ$, $2.0^\circ$, and off-axis and $2.5^\circ$,
we find that the best sensitivity is obtained for $1.5^\circ$.
Indeed, the bounds on $\varepsilon_{ee}$ are
$-3.6 < \varepsilon_{ee} < 1.8$ and $-4 < \varepsilon_{ee} < 2.2$ for
$2.5^\circ$ and $2.0^\circ$ off-axis configurations of T2HKK,
respectively, assuming the normal ordering, $\theta_{23}=45^\circ$, 
and $\delta_{CP}=270^\circ$.
This is in line with naive expectation, since the NSI parameters
induce larger oscillation effects at higher energies. 
The $1.5^\circ$ configuration further benefits from 
having the largest number of events among the configurations
and a comparatively broad flux to provide more access to the 
neutrino energy spectrum~\cite{Fukasawa:2016lew}. 
For similar reasons, 
the sensitivity of the $2.0^\circ$ configuration is better than that of the 
$2.5^\circ$ one.

From the discussions above, we can conclude that the proposed
long-baseline T2HKK experiment would have good sensitivity to a
NSI in neutrino propagation and 
can be expected to place stronger bounds than that with
the two HK detectors in Kamioka (the T2HK setup) does.
In addition the sensitivity to constrain the NSI amplitudes does
not vary much with the assumed values of $\theta_{23}$ and $\delta_{CP}$.
The achievable precision on the phases does depend upon the true
values of $\delta_{CP}$ and $\phi_{31}$, and for particular
combinations it can be much harder to determine the value of
$\phi_{31}$.  
However,  the unique two-detector
configuration of the T2HKK setup is more powerful than a
single detector, and would be extremely helpful in measuring the neutrino CP
phases if NSI exist in nature. 
In studying the effect of systematics, it is found that T2HKK is not insensitive to the magnitude of
the systematic errors, but that while the overall measurement is improved by
reduction of the systematic uncertainties the systematics are most
important when considering sensitivity to specific degenerate
parameter combinations.

%======================================================================
%======================================================================
\section{Summary and Conclusion}
\label{sec:summ}

The design of the future Hyper-K experiment is to build two
identical water-Cherenkov detectors of 260~kt per detector in stage: 
one at the Tochibora mine in Japan at a 2.5$^\circ$ off-axis angle 
and a baseline of 295~km from the J-PARC neutrino target, and 
the other perhaps in Korea.  
The second detector improves physics sensitivities, 
from beam neutrino physics to astroparticle physics,
due to increased statistics.  
In particular searches for proton decay 
provide a strong motivation for having two detectors. 
According to our sensitivity studies, 
by locating the second detector in Korea
the physics sensitivities are further improved due to the longer baseline
($\sim$1100~km) and possibility of a larger overburden (1000~m) at the
candidate sites. 
These sites cover a range of possible off-axis
angles to the J-PARC beam, between $1^\circ$ and $3^\circ$,
depending on the site.  With the longer baseline in Korea both
the first and second oscillation maxima of the
PMNS neutrino appearance probability are reachable.  
The longer baseline of the 
Korean sites enhances the CP-violating component of the oscillation
probability, and resolves parameter combinations 
between the neutrino mass ordering and CP-violating phase that would be
nearly-degenerate when measuring only at the Japanese site with beam neutrinos. 
This is a unique opportunity afforded by the J-PARC neutrino beam.
By adding atmospheric neutrinos, the neutrino mass ordering determination gets more improved in both Japan and Korean sites. 
%Indeed, the NuMI experiments can only observe the first maximum, and other proposals (DUNE, ESSnuSB) reach beyond
%the the first maximum but without the ability to do a
%clean comparison to the $1^\text{st}$ maximum (due to very different neutrino
%energies for DUNE, and no 1$^\text{st}$ maximum measurement at all for
%ESSnuSB).

%There are several candidate sites in Korea with $\sim$1~km high mountains in
%Korea suitable for a second Hyper-K detector.

Assuming a relatively simplistic systematic uncertainty model based on
T2K systematic errors evaluations~\cite{Abe:2015awa}, sensitivity
studies of the long-baseline program of T2HKK have been performed. 
These have compared different configurations of Hyper-K detector(s) in Japan
and Korea for ten years of operation without staging and with
1.3~MW beam power for $\nu$:$\overline{\nu}$ = 1:3.  
In general, the configuration
with one detector in Japan and one in Korea at 
a smaller off-axis angle gives better sensitivity overall 
than two detectors in Japan.
Based on this systematic error model,
the benefits of a smaller off-axis angle seem to outweigh the extra
uncertainties of using a higher beam energy.
Overall the Mt. Bisul site,
with its 1088~km baseline, $1.3^\circ$ off-axis angle and 1084~m overburden, 
is the leading candidate location.
Although the smaller off-axis angle introduces more $\pi^{0}$-production 
high energy tail of the beam flux, the large value of $\theta_{13}$ makes this less important and
the sensitivity to the CP phase is improved over the nominal Hyper-K design.

According to our sensitivity studies, the precision at which
$\delta_{CP}$ can be measured improves from 22$^{\circ}$ (17$^{\circ}$) for 
one (two) detector in Japan, to at worst to 14$^{\circ}$ for T2HKK 
assuming CP is maximally violated.  
The coverage fraction for establishing CP violation at $5\sigma$ improves from 47\%
(55\%) to 60\% if the mass ordering is known from independent measurements
and the improvement is much larger otherwise.
The significance of a CP violation discovery is improved relative to 
having two detectors in Japan for $ 0 < \sin\delta_{CP} < 1 $, 
though the improvement is marginal for values near $1$ unless the mass ordering 
is still unknown at the time of the experiment. 
The significance at which the wrong mass ordering can be rejected for
any value of $\delta_{CP}$ improves from $0.7\sigma$ ($1\sigma$)
for the nominal Hyper-K design (two detectors in Japan) to
$5.5\sigma$ at T2HKK using beam neutrino data alone.  
In contrast, relative to two detectors in Japan the 
sensitivity of T2HKK to the atmospheric mixing parameters is 
weaker due to reduction in statistics over the longer baseline to Korea.
The addition of atmospheric neutrino improves the sensitivity overall, 
but is particularly useful for resolving the octant degeneracy.

T2HKK is also expected to have improved sensitivity to 
non-standard interactions in neutrino propagation.
Indeed, according to our study the sensitivity to the NSI parameters $\epsilon_{ee}$,
$\epsilon_{e\tau}$, and $\epsilon_{\tau\tau}$, is enhanced 
relative to Hyper-K configurations with only Japanese detectors, 
especially with the $1.5^\circ$ off-axis site in Korea, due to the
larger matter effects along its baseline.

With $\sim$1000~m overburdens sensitivities to solar neutrino and SRN
physics are further enhanced at the Korean candidate sites compared to
the Tochibora mine ($\sim$650~m overburden) due to a much lower muon
flux and spallation background rate.  Using a simple MC the expected
significance of a supernova relic neutrino search for 10 years 
of operation is 5.2 (4.2)$\sigma$
with the Korean (Tochibora) sites.

In this paper we have demonstrated the second detector in Korea 
provides enhanced sensitivity to Hyper-K's physics goals in broad physics programs. 

%Based on its ability to provide enhanced sensitivity to Hyper-K's physics 
%goals T2HKK provides with a viable and attractive means to upgrade the
%baseline project design.

%Further sensitivity studies with more realistic systematic
%uncertainties for the beam neutrino sample are expected to produce in better physics sensitivities.
%Sensitivity studies for low energy astroparticle physics will be also
%performed using a full MC simulation.  Based on the improved physics
%sensitivities from beam neutrino physics to astroparticle physics,

%

\hspace{0.5cm}

%--Acknowledgements
{\bf ACKNOWLEDGEMENTS}\\

This work was supported by MEXT Grant-in-Aid for Scientific Research
on Innovative Areas titled ``Unification and Development of the
Neutrino Science Frontier,'' under Grants No. 25105001, No. 25105004
and No. 25105009.  In addition, participation of individual
researchers has been further supported by funds from JSPS, Japan; the
European Union ERC-207282, H2020 RISE-GA644294-JENNIFER and H2020
RISE-GA641540-SKPLUS; SSTF-BA1402-06, NRF grants No. 2009-0083526,
NRF-2015R1A2A1A05001869, NRF-2016R1D1A1A02936965, NRF-2016R1D1A3B02010606 and
NRF-2017R1A2B4012757 funded by the Korean government (MSIP); RSF, RFBR
and MES, Russia; JSPS and RFBR under the Japan-Russia Research
Cooperative Program; Brazilian National Council for Scientific and
Technological Development (CNPq); STFC, UK.
%\end{acknowledge}
%======================================================================
\clearpage
\appendix
%\graphicspath{{plots}}
%======================================================================
\section{Construction Details for Bi-probability Plots }
\label{sec:biProb_appendix}

%The bi-probability plots in Section~\ref{sec:biProb} show three
%representative energies for each site.

Most common constructions of bi-probability plots show a single pair
of ellipses, for a given value of $L$ and $E$ and assumed oscillation
parameters.  This common practice is an over-simplification, although
long-baseline oscillation experiments have a negligible variation in
the baseline, the neutrino energy typically ranges over at least a
factor of two, and often more.  For the first generation of
$\nu_e$-appearance experiments (that is to say T2K and \nova) that use
a narrow-band beam peaking near the energy of the first oscillation
maximum this is tolerable, since the first period of the oscillation
runs from half the peak energy up to infinity.  When considering these
experiments there are two obvious `fixes':  Either use one
energy (typically the peak energy) as a stand in for the entire
spectrum of measured neutrinos or to integrate the
probability over the expected (without oscillations) spectrum of
events.

The latter method corresponds to an (idealized) rate-only measurement,
and provided backgrounds are accounted for could be compared to the
data in the form of number of neutrino and anti-neutrino events.  But
this is not often done, as integrating over the full spectrum (much of
which has lower appearance probabilities) significantly reduces the
sensitivity of the experiment.  On the other hand, the former method
does not have any problems with averaging, but the number of events
for which the ellipses is a good approximation is a small fraction of
the total.  This makes it difficult to summarize the overall
sensitivity of an experiment in a correct way.

For experiments where the event spectrum is broad compared to the
oscillation (wide band beams such as DUNE, and second-maxima
experiments such as T2HKK) approximating things as a single pair of
measurements is are even less suitable.  Part of the point of such
experiments is that they can make measurements at independent energies
and see the different $\delta_{CP}$ and mass ordering dependences.

Plotting a continuum of ellipses is not practical, so to give a sense
of how the energy affect the measurement of $\delta_{CP}$ some
representative energies have to be chosen.  The plots in
Section~\ref{sec:biProb} use three representative energies.  This is
still far from a complete summary of a real measurement, but it
provides a better illustration of what the configuration can measure.

\subsection{Choice of representative energies}
The energies used to summarize the interaction spectrum are chosen base
on a procedure that takes into account the interaction rates of
neutrinos but is independent of the oscillation probabilities.  Firstly
the interaction rate (i.e. flux $\times$ cross-section) spectrum is
calculated, in the absence of oscillations.  For water-Cherenkov
detectors such as Hyper-K and the proposed ESSnuSB the quasi-elastic
spectrum is used as their analyses use primarily quasi-elastic
events. For other experiments that can use any kind of neutrino
interaction, the inclusive CC cross section is more appropriate.  Note
that for ESSnuSB and any experiment that uses inclusive CC
cross-sections, the cross section grows roughly linearly with energy,
so the interaction rate spectrum is often substantially harder than
the corresponding flux.

From the interaction-rate spectrum, the blue ellipses represents the
peak energy, $E_\mathrm{P}$, the value which is typically taken as
representative in simpler bi-probability plots.  This divides the
interaction spectrum in two, with a fraction $f$ of events below (and
a fraction $1-f$ above) $E_\mathrm{P}$.  The green ellipses are drawn
for the median energy of the lower $f$ events, while the red are the
median energy of the upper $1-f$ events. In this way, 50\% of the
events lie between the energies represented by the green and red
ellipses.  This method of identifying a peak and central 50\% of the
spectrum is also used in Fig.~\ref{f:baseline_comparison}, where a
band covering the central 75\% of events is defined in a similar way.
 
The fact that measurements with a detector at Kamioka can be
reasonably approximated being `rate-only' is evident in
Fig.~\ref{f:bp_tochibora}.  Although the ellipses differ in size and
eccentricity, the separation between the two mass orderings, and the
dependence of the appearance probabilities on the value of $\delta$ is
similar for all three energies.  The most important difference is only
apparent on closer inspection: the $\delta=0$ CP conserving point
generates either higher or lower appearance probabilities than the
$\delta=\pi$ point.  Which point provides the larger appearance
probability depends on both the mass ordering and whether the
neutrino energy is above or below the energy of the oscillation
maximum.  For other configurations (Figs.~\ref{f:bp_bisul},
\ref{f:bp_bohyun} \ref{f:bp_minjuji}) the location, size and
orientation of the ellipses is dramatically different.

\subsection{Statistical sensitivity}
The grey ellipses give an indication of the statistical power of the
measurement made in each configuration. They use a simplified
background model to estimate a fractional error $\sqrt{(S+B)}/S$ from
the number of signal ($S$) and background ($B$) events expected in the
central 25\% of the events around the peak energy (i.e. 1/4 of the
total unoscillated flux).  The number of signal events is scaled
according to the appearance probabilities at the center of the
ellipse, while the background is assumed to be independent.  Five
ellipses are drawn, to show how the sensitivity will vary with the
actual oscillation probability.  This is a somewhat arbitrary measure
-- not least because the shape and location of the bi-probability
ellipse can vary over even this narrower energy range  -- it
enables some comparison between the statistical power of measurements
with different baselines and fluxes, and using different run lengths.

\end{document}